# Explainable Recommendation: Theory and Applications

Dissertation Submitted to

**Tsinghua University**

in partial fulfillment of the requirement

for the degree of

**Doctor of Philosophy**

in

**Computer Science and Technology**

by

**Zhang Yongfeng**

Dissertation Supervisor : Professor Ma Shaoping

**August, 2017**

# 个性化推荐的可解释性研究

（申请清华大学工学博士学位论文）

培 养 单 位：计 算 机 科 学 与 技 术 系
学　　　科：计 算 机 科 学 与 技 术
研　究　生：张　永　锋
指 导 教 师：马　少　平　教　授

二〇一七年八月

# Abstract

With the continuous growth of the Web, Personalized Recommender Systems (PRS) have been the important building blocks of many online web applications, which contribute to our daily lives in various manners. For example, the product recommendation engines in E-commerce websites recommend potentially interesting products to users, friend recommendation helps to find and connect users in social networks, video recommendation in video sharing websites help users to find favourite videos more quickly and efficiently, and news recommendation in news portals push the latest news to users according to their personalized information needs. In a way, personalized recommendation has become one of the most basic supportive techniques in the era of web intelligence.

Although personalized recommendation has been investigated for decades of years, the wide adoption of Latent Factor Models (LFM) has made the explainability of recommendations an important and critical issue to both the research community and practical application of recommender systems. For example, in many practical systems the algorithm just provide a personalized item recommendation list to the users, without persuasive personalized explanation about why such an item is recommended while another is not. Unexplainable recommendations introduce negative effects to the trustworthiness of recommender systems, and thus affect the effectiveness of recommendation engines. In this work, we investigate explainable recommendation in aspects of data explainability, model explainability, and result explainability, and the main contributions are as follows:

1. **Data Explainability**: Data input is the first step of typical recommender systems, and user-item rating matrix is the most basic data format for most personalized recommendation algorithms, especially for Matrix Factorization (MF)-based approaches. In this work, we propose Localized Matrix Factorization (LMF) framework based Bordered Block Diagonal Form (BBDF) matrices, and further applied this technique for parallelized matrix factorization. Traditional MF algorithms treat the original rating matrix as a whole for factorization, without specific understanding of the inherent structure embedded therein. In this work, however, we propose the (recursive) BBDF structure of sparse matrices, and formally prove its equivalence with community detection on bipartite graphs, with which to explain the inherent community structures and their relationships in sparse matrices. Based on this, we further propose the LMF framework, and prove its compatibility with most of the traditional MF algorithms, which makes it a unified parallelization framework for matrix factorization, that improves both the effect and efficiency at the same time.

2. **Model Explainability**: Based on user-item rating matrices, personalized recommendation algorithms attempt to model user preferences and make personalized recommendations. In this work, we propose Explicit Factor Models (EFM) based on phrase-level sentiment analysis, as well as dynamic user preference modeling based on time series analysis. For their prediction accuracy and scalability, Latent Factor Models (LFM) based on MF have achieved wide application in real-world systems. However, due to their inherently latent factors, it is usually difficult for LFM to provide intuitively understandable explanations to the recommendation algorithms and results, which reduces the persuasiveness of recommendations. In this work, we extract product features and user opinions towards different features from large-scale user textual reviews based on phrase-level sentiment analysis techniques, and introduce the EFM approach for explainable model learning and recommendation. Because user preference on features may change over time, we conduct dynamic user modeling based on time series analysis, so as to construct explainable dynamic recommendations.
3. **Economic Explainability**: Based on data analysis and user preference modeling, recommender systems actually manipulate the way that items are matched with users, and eventually affect the economic benefits of the online economic system. In this work, we propose the Total Surplus Maximization (TSM) framework for personalized recommendation, as well as the model specification in different types of online applications. More and more human activities are experiencing the continuous progressing from offline to online, and many commonly used online applications can be formalized into the 'producer–service–consumer' framework. For example, in E-commerce websites online retailers (producers) provide normal goods (services), and the users (consumers) thus make choices and purchases from the vast amount of online services. Based on basic economic concepts, we provide the definitions of utility, cost, and surplus in the application scenario of Web services, and propose the general framework of web total surplus calculation and maximization. Further more, we specific the total surplus maximization framework to different types of online applications, i.e., E-commerce, P2P lending, and online freelancing services. Experimental results on real-world datasets verify that our TSM framework is able to improve the recommendation performance and at the same time benefit the social good of the Web.

**Key words:** Personalized Recommendation; Collaborative Filtering; Sentiment Analysis; Explainability; Computational Economics; Artificial Intelligence

# 摘 要

随着互联网的迅速发展，个性化推荐系统已经逐渐成为各种网络应用中不可缺少的核心功能，并以各种各样的方式影响着人们日常生活的方方面面：电子商务网站中的购物推荐引擎为用户提供可能感兴趣的商品推荐；社交网络中的好友推荐为用户寻找潜在的好友关注；视频网站中的视频推荐为用户提供最可能点击的视频推荐；新闻门户网站中的内容推荐为用户提供最有信息量的新闻——个性化推荐技术已经是支撑互联网智能的基础技术之一。

个性化推荐系统已经经过了长达十几年的研究和发展，然而隐变量方法的大量使用使得个性化推荐算法及其推荐结果的可解释性仍然是困扰学术界重要问题之一，并且至今仍然没有在产业应用中得到很好的体现。举例而言，在很多实际推荐系统中，算法只为用户提供一份个性化的推荐列表作为结果，而难以向用户解释为什么要给出这样的推荐。缺乏可解释性的推荐降低了推荐结果的可信度，进而影响推荐系统的实际应用效果。考虑到推荐系统的应用范围之广和影响之大，可解释性推荐的研究具有其重要性和紧迫性。在本文中，我们从数据、模型和经济意义三个方面对推荐系统的可解释性进行研究，主要有贡献如下：

1. 数据的可解释性：数据输入是个性化推荐系统的第一步，而用户物品评分矩阵是个性化推荐算法，尤其是基于矩阵分解的个性化推荐算法最主要的数据输入形式。本文提出了基于双边块对角矩阵的局部化矩阵分解框架，并将其应用于矩阵分解的并行化。传统的矩阵分解算法将原始矩阵看做一个整体进行分解和预测，而缺乏对矩阵内在结构的理解。在本工作中，我们提出矩阵的双边块对角结构，并在理论上证明该结构与二部图上社区发现算法的数学等价性，从而解释矩阵内在的社区结构和社区关系。在社区结构的基础上，我们进一步提出了局部化的矩阵分解框架，并理论证明了它与传统矩阵分解算法的兼容性，从而为常用的矩阵分解算法提供了一个统一的并行化框架，在提高预测精度的同时大幅提高计算效率。
2. 模型的可解释性：在用户物品评分矩阵的数据基础上，个性化推荐模型对用户进行偏好建模并给出个性化推荐。本文提出了基于短语级情感分析的显式变量分解模型及其基于时间序列分析的动态化建模。基于矩阵分解的隐变量模型由于其较好的评分预测效果和可扩展性，逐渐成为了个性化推荐的基础算法并在实际系统中得到广泛的应用。然而由于变量在本质上的未知性，隐变量模型难以对推荐算法和推荐结果给出直观可理解的解释，进而降低了推

荐系统对用户的可信度。在本工作中，我们利用短语级情感分析技术从大规模的用户评论中抽取产品属性词及用户在不同属性上表达的情感，进而引入显式变量并提出基于显式变量分解模型的个性化推荐算法，一方面使得模型的优化过程具备了直观意义，另一方面给出在模型层面可解释的推荐结果和个性化推荐理由。由于用户在不同属性上的偏好具有时间周期性，我们利用时间序列分析对用户偏好进行动态建模和预测，从而实现动态时间意义上的可解释性推荐。

3. 推荐的经济学解释。推荐系统在用户行为数据和个性化偏好建模的基础上，以个性化推荐的方式隐式地调节商品在用户中的匹配和购买，从而在最终层面上影响所属系统的经济效益。本文提出基于互联网系统总福利最大化的个性化推荐框架并给出典型应用场景中的具体实现。随着人类传统线下活动的不断线上化，常见的互联网应用均可以形式化为“生产者—服务—消费者”模型，例如在电子商务网站中，网络商家（生产者）提供在线商品（服务），而网络用户（消费者）则在众多的商品中进行选择和购买。基于传统经济学的基本定义，本文首先给出了互联网环境下效用、成本和福利的基本概念与统一形式，并进一步给出了互联网应用中总社会福利的通用计算方法。在此基础上，我们以互联网服务分配为基本问题，提出基于网络福利最大化的个性化推荐框架。进一步，本文在典型的网络应用（电子商务、P2P 借贷、在线众包平台）中对该框架进行具体化，并进行个性化的网络服务推荐与评测。实验结果表明，该方法可以在为用户提供高质量服务推荐的同时提升社会总福利，即在提升用户体验的同时又增强了社会效益。

**关键词：**个性化推荐；协同过滤；情感分析；可解释性；计算经济学；人工智能

# 目　录

# 主要符号对照表

| | |
|---|---|
| MF | 矩阵分解（Matrix Factorization） |
| TF | 张量分解（Tensor Factorization） |
| FM | 分解机模型（Factorization Machine） |
| LFM | 隐式变量分解模型（Latent Factor Model） |
| EFM | 显式变量分解模型（Explicit Factor Model） |
| SVD | 奇异向量分解（Singular Value Decomposition） |
| NMF | 非负矩阵分解（Non-negative Matrix Factorization） |
| PMF | 概率矩阵分解（Probabilistic Matrix Factorization） |
| MMMF | 最大间隔矩阵分解（Maximum Margin Matrix Factorization） |
| LMF | 局部化矩阵分解（Localized Matrix Factorization） |
| PCA | 主成分分析（Principle Component Analysis） |
| ALS | 交替最小二乘迭代法（Alternative Least Square） |
| SGD | 随机梯度下降算法（Stochastic Gradient Descent） |
| CD | 坐标下降法（Coordinate Descent） |
| UGC | 用户生成内容（User Generated Content） |
| BDF | 块对角矩阵（Block Diagonal Form） |
| BBDF | 双边块对角矩阵（Bordered Block Diagonal Form） |
| ABBDF | 近似双边块对角矩阵（Approximate BBDF） |
| AR | 自回归（Auto Regressive） |
| MA | 移动平均（Moving Average） |
| ARIMA | 移动平均自回归（Auto-Regressive Integrated Moving Average） |
| RMSE | 根均方差（Root Mean Square Error） |
| MAE | 平均绝对误差（Mean Absolute Error） |
| MAPE | 平均百分比误差（Mean Average Percentage Error） |
| NDCG | 标准化折扣累计增益（Normalized Discounted Cumulative Gain） |
| AUC | ROC 特征曲线下面积（Area Under the ROC Curve） |
| CTR | 点击率（Click Through Rate） |
| AIC | 赤池信息量准则（Akaike Information Criterion） |
| AICc | 无偏赤池信息量准则（Akaike Information Criterion corrected） |
| TS | 社会总福利（Total Surplus） |
| CS | 消费者福利（Consumer Surplus） |
| PS | 生产者福利（Producer Surplus） |
| OSA | 网络服务分配（Online Service Allocation） |
| TSM | 总福利最大化（Total Surplus Maximization） |

# 第 1 章　引言

随着智能互联网时代的到来和发展，个性化推荐作为理解用户的核心技术之一成为智能网络的重要组成部分，并在各种实际系统中得到广泛应用。长期以来，个性化推荐技术的研究集中于如何为用户提供更为准确的被推荐物品，而在很大程度上忽视了推荐系统的可解释性，不利于推荐系统对用户的透明度和可信度。在本研究中，我们从数据、模型和结果三个方面对推荐系统的可解释性进行研究，力图做到不仅知其然更知其所以然，并在互联网真实用户场景下对理论模型进行实验验证。本章旨在阐述研究背景，简要回顾个性化推荐系统的主要技术与历史现状，给出本文的研究问题、面临的主要挑战及其实际意义和科学价值，并描述本工作的主要贡献和章节安排。

## 1.1　研究背景

互联网的快速发展开启了人类活动线上化的进程，越来越多传统上只能在线下完成的任务变得可以方便快捷地在互联网上完成。已经深入人们日常生活中的电子商务就是这一进程的典型代表，例如阿里巴巴[①]、京东商城[②]、亚马逊网络商城[③] 等电子商务网站的普及，使得人们不必走出家门即可购买自己所需要的商品，并且可以在更多的备选商品中进行挑选。不仅限于电子商务应用，社交网络平台如新浪微博[④] 和 Facebook[⑤] 的兴起使得人们可以在互联网上交友、沟通、获取实时资讯；在线叫车服务如滴滴[⑥] 和 Uber[⑦] 的发展使得用户不再需要线下街头打车；在线 P2P 借贷服务如宜信[⑧] 和 Prosper[⑨] 使得用户线上借贷和理财成为可能；在线房地产业务如 Zillow[⑩] 和 Airbnb[⑪] 的发展则使传统的房地产业务逐步线上化；在线自由职业平台如猪八戒网[⑫] 和亚马逊 MTurk[⑬] 的迅速发展甚至使得自由职业

① 阿里巴巴，中国最大的电子商务企业平台：http://www.alibaba.com，http://www.taobao.com
② 京东，中国最大的自主经营式电子商务平台：http://www.jd.com
③ 亚马逊，美国著名电子商务网站，在世界主要国家提供服务：http://www.amazon.com
④ 新浪微博，中国主要的微博客社交网络平台，http://www.weibo.com
⑤ Facebook（脸谱），全球最大的社交网络平台，http://www.facebook.com
⑥ 滴滴，中国最大的互联网在线打车服务：http://www.xiaojukeji.com
⑦ 优步，美国在线打车服务公司，在世界许多国家提供服务：http://www.uber.com
⑧ 宜信宜人贷，中国最早开始网络贷款服务的公司：http://www.creditease.cn
⑨ Prosper，美国在线 P2P 借贷服务平台：http://www.prosper.com
⑩ Zillow，美国主要的在线房地产买卖业务平台：http://www.zillow.com
⑪ Airbnb，全球主要的 P2P 租房服务平台：http://www.airbnb.com
⑫ 猪八戒，中国最大的在线自由职业平台：http://www.zbj.com
⑬ MTurk，亚马逊旗下自由职业平台：http://www.mturk.com

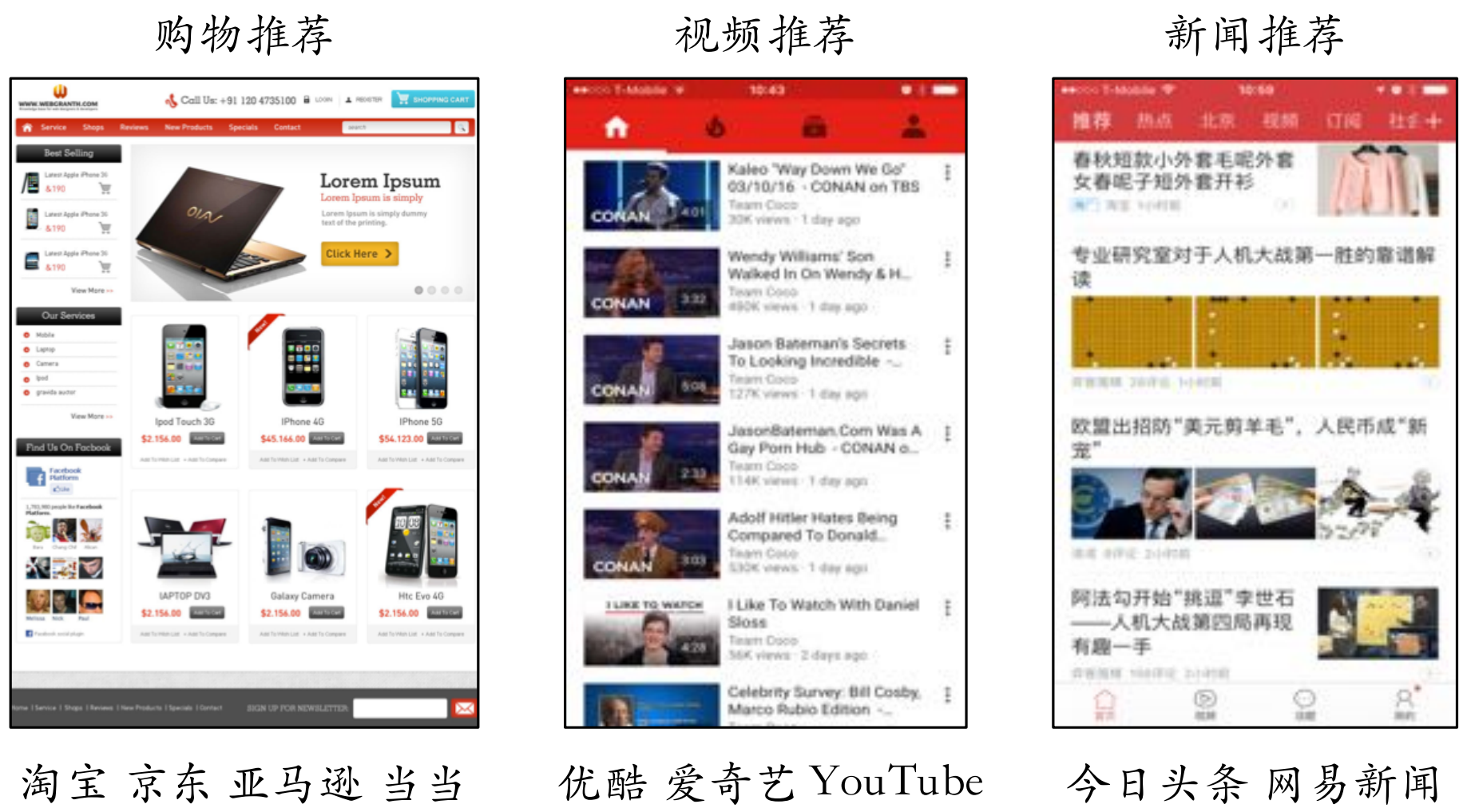

图 1.1　典型的个性化推荐系统及代表性实际产品应用示例

者在线工作和任务分配成为可能。

伴随着各种互联网应用的迅速发展，个性化推荐系统成为网络应用中不可缺少的重要组成部分，并在各种场景下以不同的方式影响着人们网络生活的方方面面，其研究也对国民生产生活的多方面具有重大意义：

第一，*个性化推荐技术的研究对互联网在线服务和信息系统具有重要的经济和市场价值*。随着人类线下活动的逐步线上化，互联网不再仅仅是一个信息流通和传播的平台，而是逐渐成为了一个完整的在线经济和社会系统，大量的社会生产生活和消费商贸活动以在线交易的方式在互联网上完成，而个性化推荐系统及其相关技术在这一过程中发挥着资源配置的核心作用。例如在电子商务网站中，个性化推荐将商家待销售的商品与具有相应需求的用户进行匹配，从而提高整个在线经济系统资源配置的效率，进而以提升消费的方式利于国民经济的发展。据京东商城推荐搜索部透露①，京东商城基于大数据的个性化推荐算法在 PC 端和移动端都已经为京东贡献了 10% 以上的订单；而据著名的科技咨询公司 VentureBeat 统计②，亚马逊的个性化推荐系统更是为其贡献了 35% 以上的销售额，推荐系统对在线经济的重要作用可见一斑。

不仅是在电子商务系统中，在线租房和房产业务中的房屋推荐通过综合考虑地理位置和价格等信息对用户需求进行精确定位，从而提高住房利用率和降低房产空置率；在线自由职业平台中的工作任务推荐系统更是综合考虑自由职业者的技术能力和预期报酬以及雇主的任务需求，通过精确的职业匹配令雇佣双方各得其所，使高效的在线劳务市场成为可能。除了显式的用户可见的个性化推荐系统之外，隐式的推荐系统也大量存在于网络之中，例如在线叫车服务系统对用户需

① 订单贡献率 10%——京东个性化推荐系统持续优化的奥秘，CSDN 资讯
② Aggregate Knowledge Raises $5M from Kleiner on a Roll, VentureBeat

求和付费意愿进行预测并对道路交通状况进行实施建模，从而为用户自动匹配最合适的司机，在满足双方出行需求的同时进行路线优化、缓解交通负担。

第二，个性化推荐系统的研究对国民生产和国家安全具有重大意义。在个性化推荐系统的信息匹配和资源配置过程中，恰当地融合经济效益及风险控制等因素的考量对互联网经济安全和网络环境的稳定可控具有重要作用。例如在网络借贷服务的理财产品推荐中，对理财产品的风险评估和用户风险承受能力的评估是产品匹配以及面向用户的推荐过程中所要考虑的重要因素，向不同用户推荐和展示合理的产品是网络金融服务实现风险可控的重要手段。又比如社交媒体的兴起和快速发展使得人们可以更加迅速快捷地发布、分享和传播信息，打破了长期以来新闻讯息由国家和新闻媒体机构垄断的局面。自媒体的兴起使信息传播更为高效、信息获取成本大大降低、社会生产生活更富于活力，但同时也为网络谣言、信息诈骗、极端思想、恐怖主义的传播带来了便利，而社交网络中的个性化信息排序和推荐技术则在满足用户个性化信息需求的同时，起到积极引导社会舆论的作用。

个性化推荐技术的研究不仅具有重要的实际应用意义，更具有重要的科学研究价值，如图1.2所示。

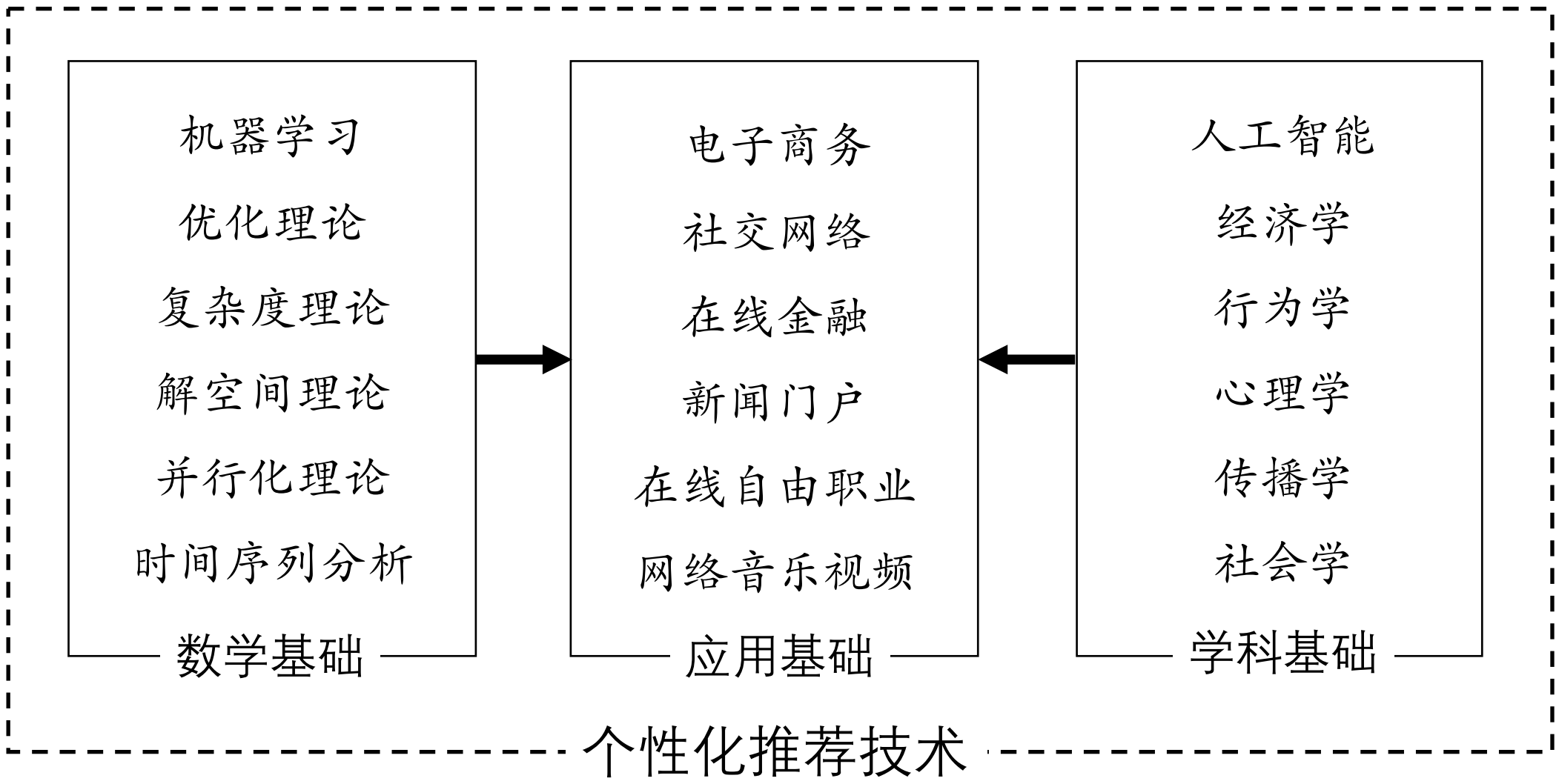

图 1.2　个性化推荐技术的研究在数学上可以拓展机器学习、尤其是优化理论等方面研究的深度和广度；同时在互联网经济系统建模和用户行为理解与分析等不同的场景下，涉及到经济学、行为学、心理学、传播学等诸多学科的交叉研究、相互补充与共同发展；不仅如此，个性化推荐技术的基础性、重要性和统一性奠定了其广泛的应用基础。

具体而言，其科学研究价值和意义主要包括如下三个方面。

第一，个性化推荐技术的研究涉及到多个重要的数学分支，有利于促进和拓展相关理论研究的深度和广度。充分理解用户行为模式和个性化的信息需求需要

对用户行为和偏好进行深入的数学建模，而互联网用户行为信息规模庞大且多种多样，例如电子商务网站中的用户浏览、购买、数值评分、文本评论等历史信息，在线视频音乐网站中的用户点击、观看、收听、时长等行为信息，以及社交网站中的用户好友关系、关注关系、地理位置、登录时间等社交信息等等。数据的多样性和异质性为用户行为分析和偏好建模带来了新的难度，而庞大的数据规模也为网络大数据的处理带来了极大的挑战——这些都为机器学习方法和相关数学模型理论的发展与应用提出了新的要求，对矩阵运算、并行化理论、解空间理论、时空信息处理、系统复杂度控制等相关理论的研究发展提供了重要的问题背景。在本研究中，我们将在矩阵分解、局部优化理论、解空间分析、并行化算法、时间序列分析等多个方面对理论前沿做出进一步的拓展。

第二，*个性化推荐技术的研究涉及到诸多学科的交叉综合，有利于促进跨学科学术研究的进一步发展*。个性化推荐技术的核心在于用户需求理解，只有对用户兴趣和需求进行精确的建模，才能给出具有针对性的个性化推荐，因此，个性化推荐技术的研究需要互联网用户行为学和心理学的支持；在社交网站、新闻门户等应用场景中，对好友、新闻、信息的推荐则依赖于对信息传播学和社会学的深入理解和应用；同时，电子商务、在线金融、在线职业网站、在线打车等网络业务的发展正不断将人类的线下经济学活动线上化，对互联网经济现象和用户在线经济行为的深入理解和正确建模对提供合时合地合情合理的推荐具有重要作用，而这依赖于对经济学相关理论的应用与发展。在本研究中，我们将借助经济学、心理学、行为学等学科的基本概念和主要结论，对互联网用户行为进行分析建模，并进一步给出个性化的推荐。

第三，*个性化推荐技术的研究涉及到众多互联网应用场景，有利于促进互联网整体的进一步个性化和智能化*。个性化推荐以其技术的基础性和方法的通用性已经成为诸多网络应用中不可或缺的组成部分，以显式或隐式的方式渗入到人们网络生活的方方面面。个性化推荐技术以其“理解用户”的核心思想，成为向用户提供智能服务的基础和关键的第一步，因此在未来以智能化为核心特征的下一代互联网的发展中具有重要的基础性意义，同时也是未来个性化生活和办公助理等平台化系统、以及智能家居等线下智能系统的核心技术之一。在本工作中，我们不仅限于理论研究和技术拓展，更进一步将相关理论应用到电子商务、在线娱乐、网络金融、在线自由职业等多个不同的各有特点的网络应用场景中，从而验证相关理论的实用性和有效性。

## 1.2　问题的提出

个性化推荐的研究及其实际应用包含三个依次递进的核心关切，分别是输入数据、模型算法，和线上应用的经济效益，如图1.3所示。一个典型的个性化推荐系统将用户在物品上的浏览、点击、购买、评论等行为信息作为输入数据，通过构建个性化推荐模型对用户进行偏好建模并给出个性化推荐列表，进而通过用户在推荐列表上的购买和消费行为产生经济效益。

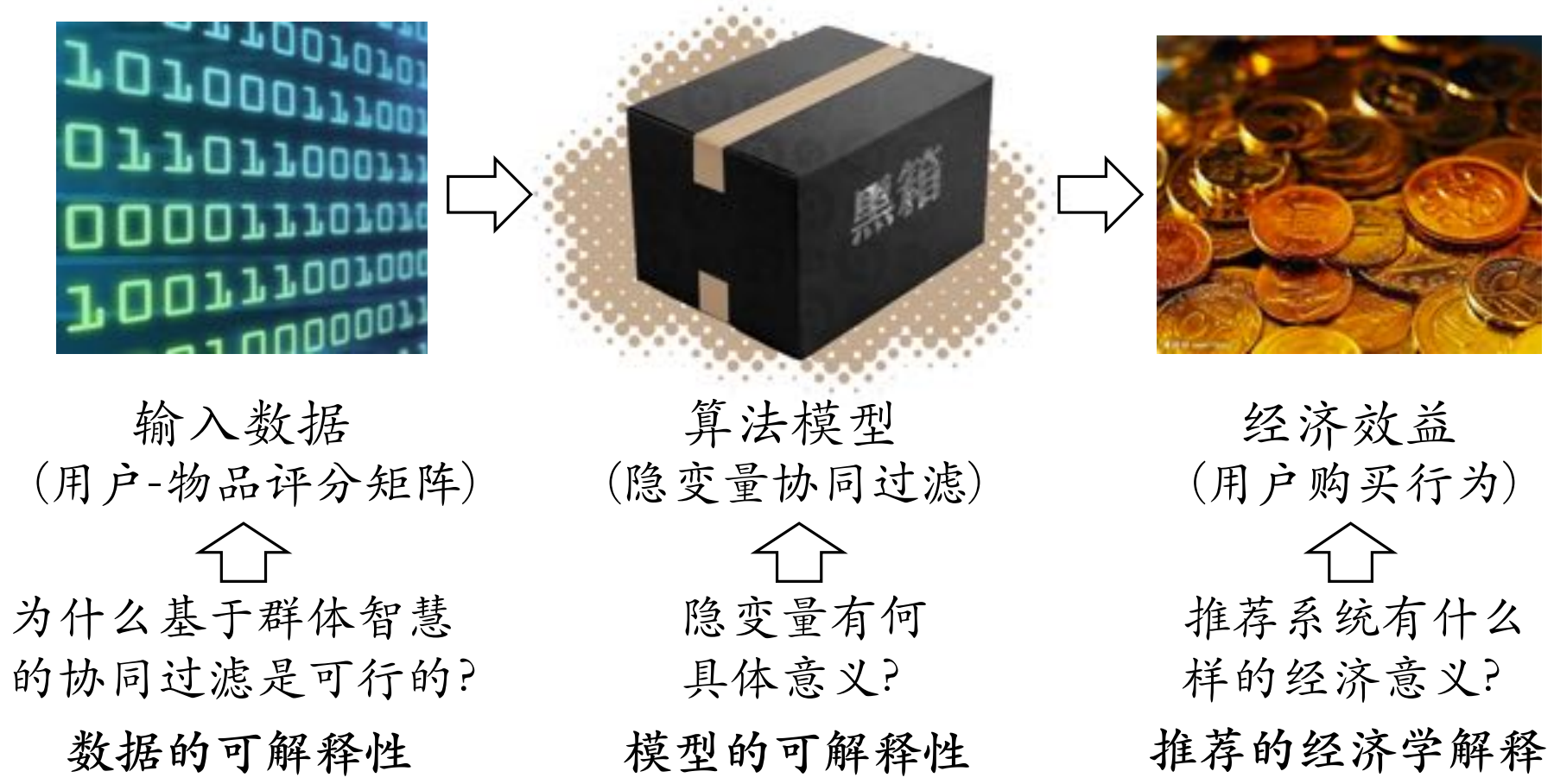

图 1.3　典型个性化推荐系统的主要构件及本研究需要解决的核心问题

输入数据是个性化推荐系统的基础，用户-物品评分稀疏矩阵是个性化推荐实际系统中最基本和最常用的输入数据形式。在用户物品评分矩阵中，每一行代表一个用户、每一列代表系统中的一个商品，而矩阵中的值表示相应用户在相应物品上的打分，例如电子商务网站中典型的 1~5 星数值打分，等等。由于系统中的商品往往有数百万甚至数亿个，而每一个用户只购买过少数的商品，因此矩阵中有大量的未知打分，因而打分预测成为个性化推荐的核心问题之一。近年来，矩阵分解等协同过滤技术在打分预测问题上取得了较好效果并得到广泛应用，然而“为何在稀疏矩阵上基于群体智慧的协同过滤得以可行”这一根本性的问题并未得到很好的回答，这涉及到数据层面上的可解释性问题。在本工作中，我们提出稀疏矩阵的迭代双边块对角型结构，指出和证明了稀疏矩阵内在的用户物品群组关系，并证明了矩阵分解协同过滤对群组的可拆分性这一重要的数学性质，从而为协同过滤的可行性找到了理论依据。同时，将杂乱无章的原始矩阵转化为内在的用户物品社区结构，一方面降低了数据稀疏性，另一方面将具有相似兴趣的用户及其历史评分聚在一起，从而提高协同过滤算法的预测效果。

在数据的基础上，个性化推荐模型试图对用户的个性化偏好进行建模，从而给出个性化推荐结果。基于隐变量的协同过滤（例如矩阵分解）技术是目前应用

最为广泛的推荐模型之一，然而隐变量模型将数据投影到一组未知的空间变量上，因此难以解释原有数据中用户的具体偏好，也难以据此为推荐结果给出直观的推荐理由，例如大多数电子商务推荐系统只是简单地给出“其他用户也购买了”等与真正模型无关的推荐理由，等等。在本工作中，我们提出与传统“隐变量分解模型”相对应的“显式变量分解模型”来解决模型的可解释性问题，通过引入属性词显式变量构建用户在属性词上的偏好矩阵，从而使得协同过滤算法具有明确的内在意义，并且可以为最终的推荐结果给出模型内生的推荐理由。

在数据和模型的基础上，个性化推荐算法通过用户在推荐列表上的点击、购买等行为为系统创造经济效益，因此个性化推荐实际上以推荐列表的方式隐形地控制着系统中物品与用户之间的匹配，从而影响推荐算法所在经济系统（如电子商务网站）中的资源分配。虽然电子商务、在线金融等网络应用已经是一个完整的线上经济系统，然而传统的个性化推荐研究主要从计算机科学家的视角出发关心点击率、购买率等性能指标，而对个性化推荐在整个系统中的经济学意义较少研究。在本工作中，我们将个性化推荐作为经济系统中资源分配的手段，提出了网络经济系统的生产者-消费者建模框架，并给出了系统最终所实现社会效益（系统总福利）的计算方法，从而对个性化推荐的经济学意义进行解释。在此基础上，我们进一步提出基于总福利最大化的推荐算法，从而在提高推荐系统用户体验的同时，也提高整个经济系统的社会效益。

## 1.3　本研究工作面临的主要挑战

个性化推荐技术的主要研究对象丰富多变，概括而言包括两大部分：其一是广泛存在于各种互联网应用中的被推荐物品，包括商品、视频、音乐、电影、新闻、金融产品、工作任务等方方面面；其二便是购买、消费和操作这些物品的网络用户。用户与物品之间交互方式的多样性、行为记录的丰富性、兴趣偏好的动态性，为个性化推荐技术的研究及其解释带来了诸多挑战，如图1.4所示，这主要包括如下几个方面：

**庞大的数据规模**：在典型的互联网应用中往往存在者数量极为庞大的用户和物品，例如据全球最大的用户评论网站 Yelp[①] 统计报告指出，截至 2016 年初其日活跃用户达 1.35 亿，并且拥有 9500 万用户评论历史记录；中国主要的电子商务网站淘宝网和京东商城的活跃用户同样数以亿计；而百度、谷歌、Facebook 等搜索和社交网站的活跃用户更是达到数以十亿计。庞大的数据规模对用户偏好建模和个性化推荐算法的可行性和实时性提出了较高的要求。在本工作中，我们提出了矩

① Yelp，全球最大的用户评论网站，http://www.yelp.com

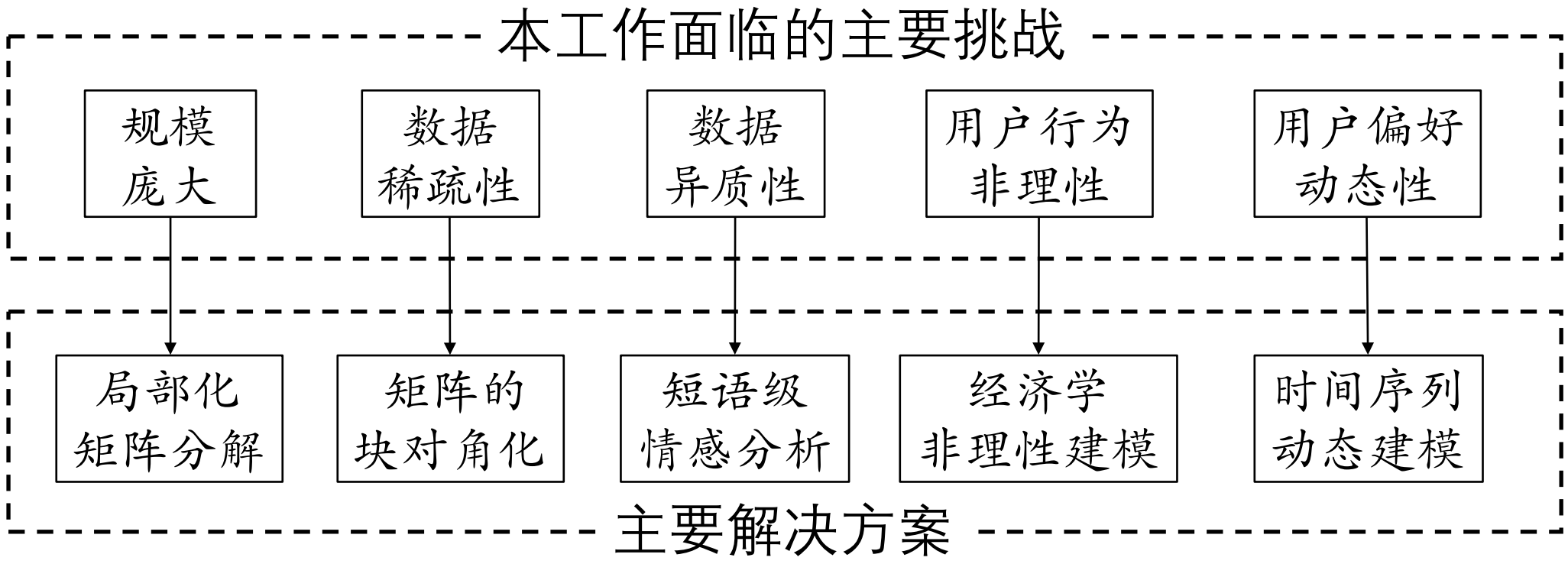

图 1.4　本工作面临的主要挑战及其对应的解决方案

阵的迭代双边块对角结构（Bordered Block Diagonal Form, BBDF），并证明了该结构与用户物品二部图社区发现的等价性，从而为用户物品行为数据的内在结构提供了解释框架。我们进一步证明了双边块对角矩阵在分解问题上的数学性质，并基于其性质提出了局部化矩阵分解框架（Localized Matrix Factorization, LMF），使得矩阵分解在数据层面上的并行化和局部化成为可能；不仅如此，我们还证明了该框架与许多常见矩阵分解算法的兼容性，从而为矩阵分解问题提供了一个统一的并行化解决方案。

**用户行为数据极其稀疏**：虽然网络应用中往往存在数以千万乃至亿计的用户和物品，然而相对而言单个用户却往往只与其中极少的一部分物品存在历史行为记录，这导致多数互联网用户行为日志存在严重的数据稀疏性。同样在 Yelp 中，有 49% 以上的用户仅仅只有一条历史评论，用户物品行为矩阵的平均密度仅有 0.043%，稀疏度达到了 99.957%。用户历史记录的稀疏性为用户行为的建模带来了挑战，使得我们不得不从有限的历史行为中估计用户的兴趣和偏好，并给出符合其个性化信息需求的推荐。在本工作中，我们利用矩阵的双边块对角结构来剔除其中不包含信息量的部分，在增强矩阵密度的同时提高用户偏好的预测精度。

**数据异质性**：由于用户行为方式和种类的多样性，网络应用所积累的用户行为历史数据往往具有明显的异质性，从而为异质信息的融合与协同处理带来了挑战。例如在电子商务等诸多网站的评论系统中，用户一方面可以对购买物品给出数值化的评分，另一方面可以同时以文本的方式给出评论，从而更为具体地表达自己的态度和偏好。长期以来，基于协同过滤的个性化推荐算法、尤其是基于矩阵分解的隐变量方法的相关研究只关注数值化评分的使用，而忽视了文本评论中所包含的丰富的用户个性化信息。这一方面是由于个性化的文本评论的异质性带来处理难度，另一方面也由于文本处理技术的瓶颈而难以发挥其应有的作用。近年来，随着短语级情感分析技术的发展和不断成熟，从用户评论文本中抽取结构化信息成为可能，从而为数值和文本异质信息的处理带来新的思路。在本工作中，

我们利用短语级情感分析技术从用户评论中进行产品属性和用户情感的抽取，并基于此提出显式变量分解模型（Explicit Factor Models, EFM），一方面同时处理用户数值评分和文本评论的异质信息，另一方面对推荐模型和结果给出属性级的个性化推荐理由与解释。

**用户行为的非理性：**用户历史行为数据虽然均为用户在实际环境下的真实行为记录，但并非全部都是用户在理性状态下的最优决策。经济学研究指出，用户在实际决策具有一定的非理性，这主要是由于人并非完全能够做到对目标对象的价值和效用进行精确估计并做出严格最优的决策。例如在购物网站中，用户常常由于时空和搜索量的限制不能找到最优的商品而选择了次优的替代品；用户往往也无法精确计算出最优的购买数量而选择了一个估计的数量，等等。非理性行为的存在为用户行为建模和模型估计带来了一定的偏差和挑战。在本工作中，我们引入经济学相关理论，对用户行为的非理性因素进行考虑和建模，从而消除由于用户非理性行为而带来的估计偏差。

**用户偏好的动态性：**用户的偏好并非一成不变，而是随着时间的推移发生兴趣的增强、减弱，或者转变。例如在化妆品购物领域，用户在夏季更加关注防晒相关的产品，而在秋冬季节则更为关注保湿相关的产品，不断变化的行为偏好为用户兴趣的建模提出了动态化和实时性的要求。在本工作中，我们采用经济学的时间序列分析理论对用户动态变化的兴趣进行建模，同时为了解决大数据环境下时间序列模型参数过多带来的计算可行性问题，我们提出了基于傅里叶级数辅助的时间序列分析模型，为大数据环境下用户偏好的动态跟踪和预测提供了解决方案。

## 1.4　本文的主要贡献

本文从数据、模型，以及经济学意义三个方面对个性化推荐的可解释性进行系统性的研究，研究工作的主要贡献分为三点，总结如下：

1. **数据的可解释性：**提出了基于双边块对角矩阵的局部化矩阵分解框架，并将其应用于矩阵分解的并行化。用户物品评分矩阵是个性化推荐算法、尤其是基于矩阵分解的个性化推荐算法的数据基础和主要输入。传统的矩阵分解算法将原始矩阵看做一个整体进行分解和预测，而缺乏对矩阵内在结构的理解。在本工作中，我们提出矩阵的双边块对角结构，并在理论上证明该结构与二部图上社区发现算法的数学等价性，从而解释矩阵内在的社区结构和社区关系。在社区结构的基础上，我们进一步提出了局部化的矩阵分解框架，并理论证明了它与传统矩阵分解算法的兼容性，从而为常用的矩阵分解算法提供了一个统一的并行化框架，在提高预测精度的同时大幅提高计算效率。我们

将在第3章介绍相关内容。

2. **模型的可解释性**：提出了基于短语级情感分析的显式变量分解模型及其基于时间序列分析的动态化建模。基于矩阵分解的隐变量模型由于其较好的评分预测效果和可扩展性，逐渐成为了个性化推荐的基础算法并在实际系统中得到广泛的应用。然而由于变量在本质上的未知性，隐变量模型难以对推荐算法和推荐结果给出直观可理解的解释，进而降低了推荐系统对用户的可信度。在本工作中，我们利用短语级情感分析技术从大规模的用户评论中抽取产品属性词及用户在不同属性上表达的情感，进而引入显式变量并提出基于显式变量分解模型的个性化推荐算法，一方面使得模型的优化过程具备了直观意义，另一方面给出在模型层面可解释的推荐结果和个性化的推荐理由。由于用户在不同属性上的偏好具有时间周期性，我们利用时间序列分析对用户偏好进行动态建模和预测，从而实现动态时间意义上的可解释性推荐。基于浏览器的大规模真实用户实验显示，我们的方法在点击率、购买率、一致性等多个线上指标上效果显著。该部分内容将在第4章进行介绍。
3. **推荐的经济学解释**：提出基于互联网系统总福利最大化的个性化推荐框架并给出典型应用场景中的具体实现。随着人类传统线下活动的不断线上化，常见的互联网应用均可以形式化为“生产者—服务—消费者”模型，例如在电子商务网站中，网络商家（生产者）提供在线商品（服务），而网络用户（消费者）则在众多的商品中进行选择和购买。基于传统经济学的基本定义，本文首先给出了互联网环境下效用、成本和福利的基本概念与统一形式，并进一步给出了互联网应用中总社会福利的通用计算方法。在此基础上，我们以互联网服务分配为基本问题，提出基于网络福利最大化的个性化推荐框架。进一步，本文在典型的网络应用（电子商务、P2P 借贷、在线众包平台）中对该框架进行具体化，并进行个性化的网络服务推荐与评测。实验结果表明，该方法可以在为用户提供高质量服务推荐的同时提升社会总福利，即在提升用户体验的同时又增强了社会效益。该部分内容将在第5章进行介绍。

# 第 2 章　研究现状与相关工作

个性化推荐系统作为一个重要的研究方向，其发展追根溯源已经经历了近二十年的时间。近年来，随着 Web2.0 时代的兴起，用户产生内容（User Generated Content，UGC）在互联网上得到了大量的积累，包括用户在搜索引擎中的搜索历史记录、在购物网站中的购买记录和评论、在社交网站中的图片文本等等，这为个性化的用户建模、意图理解和推荐带来了新的契机，也将个性化推荐系统的研究引向了新的高度。在本部分，我们从个性化推荐的发展历程、主要技术、以及研究现状进行归纳整理，并以矩阵分解算法为核心介绍个性化推荐系统的主要技术。另外，我们针对用户文本评论这一重要的用户生成内容形式之一介绍文本情感分析技术，该技术将在模型的可解释性部分得到使用。

## 2.1　个性化推荐

个性化推荐算法目前较为公认地可以分为如下三类[1,2]：基于内容的推荐（Content-based Recommendation）[3,4]、基于协同过滤的推荐（Collaborative Filtering-based Recommendation）[5–7]，以及混合型推荐系统（Hybrid Recommendation）[8–10]。其中，基于协同过滤的推荐因其对专家知识依赖度低以及可以利用群体智慧等特点，得到了最为深入也最为广泛的研究，它又可以被分为多个子类别，主要包括基于用户的协同过滤（User-based CF）[11]，基于物品的协同过滤（Item-based CF）[12]，以及基于模型的协同过滤（Model-based CF）[6]，等等。其中基于模型的推荐是一类方法的统称，它往指利用系统已有的数据和用户历史行为，学习和构建一个模型，进而利用该模型进行用户偏好建模、预测与个性化推荐，根据具体应用场景和可用数据的不同，这里的模型可以是常用的奇异值分解等矩阵分解模型[13]，也可以是主题模型、人工神经网络、概率图模型、组合优化甚至深度学习等机器学习模型[1]。在下面的部分，我们将在如上几个方面对个性化推荐系统的研究现状与相关工作进行具体的介绍。

**基于内容的推荐**：基于内容的推荐首先收集和标注特征信息并对用户和物品构建内容画像（Profile），例如电影的类型、导演、主演，用户的年龄、性别、内容偏好，等等。在此基础上，基于内容的推荐通过用户画像和物品画像的特征匹配算法进行个性化的推荐。

在理论与方法方面，Debnath 等研究了特征权重的选取方法及其对推荐效果的影响[14]；Martínez 等将语言学模型运用到基于内容的推荐当中，从而允许用户以

自然语言描述自身的兴趣爱好并获得个性化的推荐[15]；Blanco[16] 和 Gemmis[17] 等将语义网与基于内容的推荐相结合，利用语义网所蕴含的精确的特征关系为用户提供推荐；Noia 等进一步将最新的开放连接数据（Linked Open Data）项目语义网应用于个性化推荐[18]；Zenebe 等将模糊集理论应用于用户和物品特征集合的匹配过程从而为用户提供基于内容的推荐[19]；Cramer 等则在基于内容的推荐背景下研究了系统透明度对用户信任和接受度的影响[20]。

在实际应用方面，Mooney 等研究并推出了基于内容的图书推荐系统[21]；Cano 推出了基于内容的音乐推荐系统[22]；Basu 等研究了社交关系信息在推荐系统中的应用[23]，Cantador 等则进一步将基于内容的推荐应用于社会化标签系统（Social Tagging System），从而为用户推荐最可能感兴趣的对象进行标签标注[24]；Chen 等研究了基于内容的电子商务系统[25]；Phelan[26] 和 Kompan[27] 等则研究了基于内容的新闻推荐系统。

**基于协同过滤的推荐**：基于协同过滤的推荐是推荐系统中广泛使用的推荐技术，与基于内容的方法不同，协同过滤的核心思想在于借助其他用户的历史行为（群体智慧）来为当前用户给出推荐，而不仅仅是考虑当前用户自身的特征偏好。基于协同过滤进行推荐的思想一般认为最早出现在 GroupLens 的新闻推荐系统中[11]，该工作也就是后来人们所说的基于用户的协同过滤方法，除此之外，该工作也第一次提出了用户物品评分矩阵的补全预测问题，并且这一问题在 Herlocker[28] 中得到了进一步的形式化，并在 Breese[29] 中得到了实验验证，影响了推荐系统今后十几年的发展方向；Sugiyama 等将基于用户的协同过滤用于个性化搜索任务中并取得了不错的效果[30]；Ai 等将个性化和协同过滤的思想用户电子商务中的商品搜索问题中，提出了个性化商品搜索的研究问题及层次化表示学习方法，取得了不错的效果[31]。

Sarwar 等研究了协同过滤技术在电子商务网站中的应用[32]，并发现由于在基于用户的协同过滤中需要计算用户之间的两两相似度，使得在电子商务等用户数庞大的网站中计算量成为了一大瓶颈。为了解决该问题，Sarwar 等进一步提出了基于物品的协同过滤，利用物品的相似度来进行协同过滤式推荐[12]，该方法在亚马逊的个性化推荐系统中得到了重要的应用[33]，并且至今仍然是许多电子商务网站推荐系统的基础之一；由于基于用户和基于物品的协同过滤都涉及到用户和物品相似度的计算，两者一般可以统称为基于近邻的推荐方法（Neighbour-based Recommendation）[34]；Herlocker 等对通过选择不同的相似度计算函数，对基于用户的协同过滤方法的实际效果进行了分析和验证[35,36]；Karypis 则在 Top-N 推荐列表任务中对基于物品的协同过滤进行了实验验证和效果评价[37]；Huang 等对比

了不同的协同过滤算法在电子商务网站应用场景下的效果和效率[38]；Basu[23] 和 Kautz[39] 等最早讨论了社交网络与协同过滤的结合，从而使得社交推荐成为可能；Massa[40–43] 和 O'Donovan[44] 等研究了用户之间的信任关系在协同过滤相似用户选择过程中的应用，提出了信任敏感的（Trust-aware）协同过滤算法和研究方向，并开发了信任敏感的推荐系统实际应用模型 Moleskiing[45]。

为了进一步解决相似度计算量大的问题，Lemire 等提出了著名的 SlopeOne 系列算法将协同过滤的回归函数简化，在大大降低计算时间和存储需求的同时，取得与原始基于近邻算法相当甚至更好的效果[46]；O'Connor 等提出利用物品聚类来降低相似度计算的复杂度[47]；Gong 等尝试和比较了分别对用户和物品进行聚类的效果[48]；而 George 等则采用互聚类（Co-Clustering）的方法对用户和物品同时进行聚类，并在此基础上寻找近邻[49]；Ma 等[50] 基于相似度阈值过滤提出了一种寻找近邻并计算预测打分的加速算法；Zhou[51] 和 Zhao[52] 等则研究和实现了基于 Hadoop 的并行化相似度计算和协同过滤方法。

随着 2007 年 Netflix 矩阵预测大奖赛的兴起[53]，推荐系统的研究进入了一个新的高潮。由于在矩阵分解在预测效果上的明显优势，大量的矩阵分解算法得到深入的研究和扩展，这既包括对主成分分析（Principle Component Analysis）算法[54–57]、奇异值矩阵分解（Singular Value Decomposition）算法[58–60] 和非负矩阵分解（Non-negative Matrix Factorization）算法[61,62] 等已有矩阵分解算法的应用和扩展，也包括一些新算法的提出和研究，例如最大间隔矩阵分解（Maximum Margin Matrix Factorization）算法[63–67] 和概率矩阵分解（Probabilistic Matrix Factorization）算法[68–70]，等等。由于矩阵分解是很多个性化推荐算法的基础，我们将在下一小节对其进行详细的介绍。

冷启动问题（Cold-start）是协同过滤式推荐系统所面临的重要问题之一[71]。当新用户刚刚加入系统时，由于其只有很少甚至没有历史行为记录，使得协同过滤算法难以对其进行偏好建模，例如在基于用户的协同过滤当中，冷启动用户由于没有历史打分记录，造成无法为其计算相似近邻用户。同样的问题也存在于基于物品的协同过滤算法中，新加入的物品由于几乎没有用户打分，使得难以被算法推荐出来。Gantner 等通过学习属性特征映射来解决冷启动问题[72]；Zhang 等利用社会化标签来缓解冷启动问题[73]；Bobadilla 等则研究了神经网络学习算法在冷启动问题中的应用[74]；Leroy 等对冷启动的关联预测（Link Prediction）问题进行了研究[75]；Ahn 等提出了一种启发式的相似度计算方法来解决新用户冷启动的问题[76]；Zhou 等提出了功能矩阵分解模型（Functional Matrix Factorization），利用决策树和矩阵分解的结合在冷启动过程中为用户选择合适的物品进行打分，从而尽可能准

确地理解用户的偏好[77]。

与冷启动问题紧密相关的是协同过滤的数据稀疏性问题，相对于系统中规模庞大的物品总数，评价每个用户有过交互行为的物品只是很少的一部分，数据的稀疏性为用户偏好建模带来了挑战。Wilson 等通过实例研究了数据稀疏性问题在推荐系统中的影响[78]；Huang 等尝试利用关联规则挖掘来解决数据稀疏性问题[79]；Papagelis 等利用用户信任关系来缓解稀疏性[80]；Feng 等研究了神经网络在稀疏数据背景下推荐问题中的应用[81]；Zhang 等提出了矩阵的块对角结构，通过矩阵的块对角变换增加局部密度从而直接缓解稀疏性问题[82–84]；Zhang 等进一步分析了矩阵分解的解空间性质，并提出了增广矩阵分解算法用以解决数据稀疏性的问题[85]。

由于推荐系统是许多互联网应用中的重要部分，协同过滤也因此在各种应用场景下得到了丰富的应用。除了典型的电子商务推荐系统之外，Das 等利用协同过滤技术实现谷歌新闻推荐系统[86]；Ma 等利用协同过滤方法研究了社交网络推荐中的一系列重要问题，包括基于社交网络信任关系的推荐[87,88]、基于社会化正则项的推荐[89]、基于概率化矩阵分解的社交网络推荐[90]、基于上下文信息的社会化推荐[91]、以及显式和隐式信息在社会网络推荐中的应用[92,93]，等等；Lekakos[94]、Liu[95] 和 Jeong[96] 等研究了协同过滤技术在电影推荐中的应用；Celma[97]、Eck[98]、Wang[99] 等研究了音乐推荐技术及系统；Tewari[100]、Cui[101] 等研究了在线图书推荐；Zheng 等研究了在线服务推荐系统[102,103]；论文引用推荐是协同过滤推荐应用的另一个重要领域，He[104,105]、Caragea[106]、Zarrinkalam[107] 等对此进行了深入的研究。

**混合型推荐系统**：基于内容的推荐其优点是没有冷启动的问题，但是用户和物品画像的构建需要大量的时间和人力；而基于协同过滤的推荐通过利用群体的智慧对用户和物品进行画像和建模，但是也存在冷启动、数据稀疏性等不足之处。为了结合两者的优点而同时规避两者的缺点，研究界提出了混合型推荐系统[8,108]，对基于内容和基于协同过滤两种方法的结合成为混合型推荐系统的主流，在实际系统中得到了广泛的应用，现在大多数实际中的推荐系统都是综合多种推荐算法而构建的混合型推荐系统。根据算法融合方式不同，混合型推荐策略可以分为加权融合[109]、场景切换[110,111]、结果混合与重排序[112,113]、特征组合[114,115]、算法级联[111]、算法元层次融合[116] 等。

Burke 等将基于知识的专家系统与协同过滤结合，较早提出了混合型推荐系统的概念[117]；ClayPool 等进而将基于内容和协同过滤的推荐相结合用于新闻推荐的任务[118]；Wang 等基于相似度融合的方法对传统的用户协同过滤和物品协同过滤进行了结合[119]；Good 等提出结合个人助理（Personal Agents）的协同过滤框架[120]；

Pennock 等将基于近邻的协同过滤与基于模型的方法相结合[121]；Melville 等提出了基于内容增强（Content-boosted）的协同过滤方法[122]；Kim[123] 和 Cho[124] 等研究了基于决策树的混合推荐模型；Popescul[125] 和 Yoshii[126] 等研究了混合型推荐的概率化方法；近年来，Campos 等又将贝叶斯概率框架应用于混合型推荐系统中[116]；Burke 等研究了异构网络和数据环境下的混合型推荐算法[127]；Choi 等研究了用户隐式反馈与行为模式的结合[128]；Renckes 等考虑了用户隐私保护在混合型推荐中的体现[129]；Sun 等研究了基于排序学习的混合型推荐[130]；Huang 等基于用户物品关系图提出了一种融合内容和协同过滤的混合型推荐方法[131]。

在应用方面，斯坦福大学的研究人员首先推出了混合型推荐系统 Fab[10,132]，首次采用了内容和协同过滤结合的方法；Prasad[133] 和 Li[134] 等研究了电子商务网站背景下混合型推荐的应用；Yu 等利用混合型推荐实现了基于手机的上下文相关多媒体内容推荐系统[135]；Yoshii[126] 和 Donaldson[136] 等则对混合型推荐策略在音乐推荐中的应用；Lekakos[94] 和 Salter[137] 等基于内容和协同过滤研究了电影推荐；Vaz 等基于协同过滤和作者排序实现了一个在线图书推荐系统[138]；Lucas 等对在线旅游产品的推荐进行了研究[139]；Sobecki 等利用协同过滤和菜谱内容实现了在线菜谱教程推荐系统[140]；随着 MOOC 等在线学习平台的兴起，Chen[141]、Tang[142]、Khrib[143] 和 Bobadilla[144] 等研究了基于混合型推荐策略的在线课程推荐系统。

## 2.2 矩阵分解

矩阵分解类算法因其在 Netflix 大奖赛中打分预测任务上的优秀表现[145]，逐渐受到了学术界的广泛关注和深入研究。由于矩阵分解算法使用隐变量（Latent Factor）来描述用户偏好和物品属性，因此也常常被归为隐变量模型[146]。矩阵分解算法在为用户和物品计算偏好与属性向量时，本质上用到了其他用户和其它物品的历史记录，因此相关算法也属于基于协同过滤的推荐范畴。

严格数学意义上的矩阵分解是指将原始矩阵分为两个（或多个）子矩阵的乘积，使得子矩阵的乘积严格等于原始矩阵。上下三角矩阵分解（LU Decomposition）将原始矩阵分解为一个上三角矩阵和一个下三角矩阵的乘积[147]；正交矩阵分解（QR Decomposition）将原始矩阵分解为一个半正交矩阵和一个上三角矩阵的成绩[148]；奇异值矩阵分解（Singular Value Decomposition，SVD）则将原始矩阵分解为两个酉矩阵与一个奇异值对角阵的乘积[149]。它们构成了许多重要数学问题的计算基础，如离散优化、线性方程组求解、偏微分方程求解、信号处理、谱分析等等。

而在个性化推荐问题中，我们所处理的用户物品打分矩阵往往极其稀疏——矩阵中只有少数的观测值（用户对物品的打分），而大部分值是空缺的（相应的用

户没有对物品打分）。为了对用户潜在兴趣进行预测，我们并不关心原始矩阵的精确恢复，而更加关心利用矩阵分解实现对原始矩阵中空缺值的预测——近似矩阵分解应运而生。在个性化推荐背景下，矩阵分解一般指基于近似矩阵分解的矩阵补全（Matrix Completion）算法。

常见的矩阵分解算法是基于“低秩近似”假设的[13,150]，该类算法认为虽然一个矩阵的维度可以大至上千万乃至上亿级别，但是描述矩阵性质的内在因素其实只是少数有限的几个——虽然我们并不知道它们是什么（隐变量）。例如在一个包含上亿用户和上亿物品的购物网站用户评分矩阵中，决定用户打分偏好的可能只有价格、颜色、款式、流行度等几十或者数百个因素。因此，庞大的原始矩阵可以分解为两个低秩（几十或上百维）子矩阵的乘积，而两个矩阵分别描述了用户和物品在这些隐变量上的偏好。

以 SVD 为理论基础的主成分分析方法（Principle Component Analysis，PCA）首先得到了应用。Sarwar 等利用主成分分析方法研究了购物网站中用户打分行为特征[32]；Goldberg 等则进一步直接将其用于协同过滤的推荐问题当中[54]；与此同时，与 PCA 同源的截断式奇异值分解（Truncated SVD）算法被正式应用于协同过滤预测当中[32]，截断式矩阵分解通过舍弃较小的特征值来对原始矩阵进行近似，它具有较为直观的理论基础，即通过滤除原始矩阵中的噪声来达到降噪预测的目的；由于严格 SVD 算法具有较高的计算复杂度，为了适应线上推荐的需求，Sarwar 等提出了增量式 SVD 算法[151]、Brand 等提出了线上 SVD 算法[152]。

SVD 算法本身不对分解矩阵的值做额外的要求，但是在很多实际应用场景中，我们经常认为用户和物品在隐变量上的打分是正的，因此原本应用于图像处理的非负矩阵分解（Non-negative Matrix Factorization，NMF）[61,62] 开始在推荐系统中得到了关注。Zhang 等利用非负矩阵分解恢复评分矩阵中的缺失值[153]；Langville 等讨论了分解矩阵的初始化对非负矩阵分解效果的影响[154]；Chen 等提出了正交非负矩阵分解方法用以改进预测效果[155]；Arora 等[156] 对非负矩阵分解的性质进行了算法理论分析；Liu 等提出了分布式非负矩阵分解算法用以提高计算效率[157]。

Salakhutdinov 等研究了矩阵分解的概率意义，发现常见的奇异值分解和非负矩阵分解都可以等价地概率化并表达为最大似然估计的形式，由此提出了贝叶斯概率矩阵分解框架[68,69]。由于本质上的低秩近似假设，以上矩阵分解算法都面临同一个难题，即在实际应用中难以准确确定应该使用多少维的分解矩阵进行近似，在实际应用中，往往需要手动调试以确定效果最优的维度选择。为了避免该问题，Srebro 等避开了低秩近似假设，采用低范数假设提出了最大间隔矩阵分解（Maximum Margin Matrix Factorization，MMMF）算法[63]，通过最小化预测矩阵的核范数实现矩阵分

解，而在理论上允许无穷维的分解子矩阵；Rennie 等在此技术上分析了核范数的数学性质，提出了快速最大间隔矩阵分解（Fast MMMF）算法[64]；Xu 等研究了最大间隔矩阵分解算法的非参数化[158]；Weimer[65,67] 和 Decoste[66] 等研究了最大间隔矩阵分解在协同过滤和推荐系统中的应用。

Koren[159] 和 Bell[160] 等将矩阵分解作为主要算法之一应用于 Netflix 大奖赛中并取得了重要成功，并通过去掉 SVD 中的奇异值矩阵给出了 SVD 矩阵分解的双子矩阵形式，使其更适合实际应用中对效率和迭代更新的要求[59]；Koren 等进一步研究了矩阵分解与基于近邻方法的结合，提出了著名的 SVD++ 算法[58,161]，并研究了基于矩阵分解的系统过滤对用户动态临时兴趣的建模[162]，进而将其应用于雅虎音乐推荐系统[163,164]；Wu 等系统性研究了多矩阵分解算法的集成[165]；Takacs 等也研究了多种矩阵分解算法在大规模协同过滤系统中的效果[166]。

矩阵分解适用于数值化打分的显示反馈矩阵，而对于用户隐式行为反馈矩阵（如 0-1 矩阵）效果并不显著[167]。为了对隐式反馈进行更好地处理，Rendle 等将贝叶斯排序方法与矩阵分解相结合，提出了基于贝叶斯个性化排序的矩阵分解（Bayesian Personalized Ranking Matrix Factorization，BPRMF）算法[168]，取得了较好的效果。

近年来，随着互联网用户生成内容（User Generated Content，UGC）的不断丰富，为多种类异质信息的处理带来了新的要求，矩阵分解的高维形式——张量分解（Tensor Factorization）算法[169]——在推荐系统中的应用也逐渐受到了关注。Karatzoglou 等用张量描述不同的上下文信息，从而将张量分解用于上下文相关的推荐系统中，提出了一种上下文相关推荐的统一形式[170]；Rendle 等将张量的维度进行两两组合式的分解并用于标签预测和推荐任务中[171,172]；Xiong 等将用户临时和动态兴趣引入张量分解中，提出了动态张量分解模型[173]；Hidasi 等研究了张量分解对用户隐式反馈的处理[174]；Zheng 等将张量分解应用于 Flickr 图片分享网站中的小组推荐实际系统中，取得了理想的效果[175]。

同样为了处理大量的异质用户行为信息，Rendle 等进一步将两两维度组合的张量分解推广为任意维度组合的张量分解并提出了分解机（Factorization Machines，FM）模型[176,177]，分解机模型理论上具有更强的拟合能力，并且可以在一个统一的算法框架下进行模型求解，为实际系统中异质信息的处理提供了便利。目前，张量分解已经广泛用于谷歌的商业系统中，并提供 LibFM[178] 和 MyMedialite[179] 等多种开源实现。

在实际系统中，矩阵分解问题常用的求解算法有交替最小二乘法（Alternate Least Square，ALS）[59]、坐标下降法（Coordinate Descent，CD）[180,181] 和随机梯度

下降法（Stochastic Gradient Descent，SGD）[182]，等等。由于较好的收敛效果和计算效率上的优势，随机梯度下降算法在实际中得到了较为广泛的应用，为了适应产业级大规模数据的处理，Gemulla 等提出了分布式随机梯度下降算法[183]，Zhuang 等提出了并行化随机梯度下降算法[184]。

## 2.3　推荐的可解释性

除了直接展示推荐结果之外，推荐系统往往还要展示推荐恰当的推荐理由来告诉用户为什么系统认为这样的推荐是合理的。相关研究指出恰当的推荐理由可以提高用户对推荐结果的接受度[185,186]，同时也可以提高用户在系统透明度、可信度、可辨性、有效性和满意度等方面的体验[20,187,188]。但是，具体所采用的推荐算法可能影响推荐的可解释性和推荐理由的构建，并且隐变量模型的大量使用为推荐的可解释性带来了一定的难度[187]。

为了解决相关问题，学术界和产业界都进行了一定的探索，例如在亚马逊等电子商务推荐系统中往往简单地给出“购买了该产品的用户也购买了”等简单的模板式推荐理由[187]；在社交网站下相关的推荐系统中，则可以看到诸如“你的好友也查看了该内容”等基于社交关系的推荐理由[189,190]。然而，过度简化的一成不变的推荐理由难以为用户提供个性化的解释，降低了用户对推荐理由的信任度。

近年来，基于主题（Topic）对推荐算法进行一定的解释的方法得到了研究。例如，Mcauley 等将主题模型（Topic Models）与矩阵分解进行结合，从而将主题与矩阵分解中的隐变量进行映射和解释[191]；Ling 等同样基于主题与隐变量的对齐同时利用用户数值化评分和文本评论信息，并解决推荐系统的冷启动问题[192]；在此思想下，Bao 等进一步提出了主题矩阵分解（TopicMF）算法[193]。

然而在实际系统中，用户提及某一主题时并非一定是在表达正面情感，而在很多情况下恰恰相反是在表达负面情感，因此纯粹基于主题的方法往往在描述用户兴趣偏好时有所偏差[194–196]。在本研究中，我们基于短语级情感分析从评论中抽取产品属性词以及用户对其所表达的实际情感用于协同过滤，并将产品属性词作为具体的分解变量提出了显式变量分解模型，从而使推荐结果可以根据模型给出个性化的推荐理由。

## 2.4　文本情感分析

随着电子商务、社交网站、在线论坛等 Web2.0 平台的兴起和发展，互联网上积累了大量的评论、文章、帖子等用户文本信息，用以表达用户对产品和事件等对象丰富的观点和情感，文本情感分析也随之得到了重要的关注[197,198]。情感分析

在很多互联网应用扮演着重要的角色，例如情感检索[199]、口碑分析[200]，以及基于情感的文档摘要[201,202]，等等。

情感分析的核心任务之一是分析用户在文章、句子或者特定的产品属性上所表达的情感倾向性（Sentiment Orientation），它们分别对应于三个粒度上的情感分析，分别为篇章级（review/document-level）情感分析[203]，句子级（sentence-level）情感分析[204,205]，以及短语级（phrase-level）情感分析[206–211]。

篇章级情感分析的核心任务是对一篇文章或一条评论进行情感分类，一般包括正面、负面、中性等[197]。情感分类作为处理网络文本内容的重要技术之一，得到了学术界的广泛关注和研究，其中包括有监督学习方法[203,212–216]、无监督学习方法[202,217–221]，以及半监督学习方法[222–225]，等等。

短语级情感分析则试图在更为细粒度的产品属性或特征的水平上了解用户的情感倾向[202]。短语级情感分析的核心任务之一是情感词典（Sentiment Lexicon）的构建[207,210,211,226,227]，其中的每一条记录是一个“属性词-观点词-情感极性”三元组，例如在手机领域的评论中类似的三元组可以是“屏幕-清晰-正面”或者“噪声-大-负面”，等等。短语级情感分析利用特定领域的大规模用户评论语料抽取产品属性词和用户情感词并构建“属性-情感”词对，进一步采用机器学习等方法对词对进行情感极性标注，从而构建情感词典。高质量情感词典的构建是很多重要网络应用的基础，例如个性化推荐[194,228,229]和自动文档摘要[202,226]，等等。

虽然一些常见的情感词如“好”、“不错”、“不好”等在和不同的属性词匹配时经常表达确定的情感，但是有很多其它观点词的极性具有上下文相关性，即当它们和不同的属性词匹配时，会表达不同的情感极性，例如同样是情感词“高”，当和“质量”进行匹配时，表达的是一个正面的情感，而当和“噪声”匹配时，表达的却是负面的情感，这表明情感词典的构建是上下文相关的[206]，这为“属性-观点”词对情感极性的标注带来了挑战。在情感极性标注方面，研究人员设计和使用了各式各样的假设、启发式规则和优化算法。例如，一个常见的基本假设为，一条评论整体的情感极性是其内部各个“属性-观点”词对情感极性的综合[210,211]。一些工作采用了基于语言学的启发式规则[210,230,231]，一个常见的规则是“紧密相连的两个情感词往往具有相同的极性”，如在“声音清晰洪亮”中，“清晰”和“洪亮”被认为具备相同的极性，因为用户一个通顺的语言表达中几乎不可能将两个极性相反的观点词并列使用；类似的，另外一个规则是“被‘但是’连接的两个情感词具有相反的极性”，等等[232]。

## 2.5 本章小结

本章对于个性化推荐系统的发展历史和主要技术，以及本文将要用到的短语级情感分析主要技术进行了总结、归纳和概括，通过比较可知，本文的不同和创新之处主要包括如下几点：

1. *本文对个性化推荐的可解释性进行系统性理论分析和实验验证。*一直以来，推荐系统的研究着重于对预测精度和推荐效果等实际指标的提高，而忽略了对数据、模型和结果的可解释性研究。人机交互相关领域的一些研究成果分析了推荐理由及其展示方式对推荐系统效果提升的作用，但是相关研究仅限于非个性化的特定的推荐理由，并且未对推荐理由的实际产生机理进行深入的探讨。在本工作中，我们分别从数据和模型上对推荐的可解释性进行系统性的分析，通过对用户物品评分矩阵内在结构的分析解释数据内在的社区结构；通过对隐变量模型的可解释性分析提出显式变量分解模型，并进一步构建个性化可解释的推荐结果和推荐理由；最后，线上实验验证了相关技术算法的实际有效性。
2. *本文基于数据的可解释性提出局部化矩阵分解框架。*不同于以往对矩阵分解算法模型层面上的研究，本文从数据层面对矩阵分解算法的可拆分性进行理论分析，并据此提出局部化矩阵分解的框架。由于该框架是建立在对用户物品评分矩阵这一矩阵分解的基本输入数据进行预处理的基础上，因而与目前常见的矩阵分解算法兼容，实现了统一的矩阵分解并行化框架。在不同规模真实数据上的实验结果显示，局部化矩阵分解框架可以在提高矩阵分解预测精度的同时也提高并行化分解效率。
3. *本文对个性化推荐系统的经济学效益进行系统性分析。*随着互联网尤其是移动互联网的蓬勃发展，越来越多传统的线下人类活动被逐渐线上化，例如在线购物、在线社交、在线金融服务，等等。随着互联网的不断平台化和网络活动的逐渐丰富，互联网也越来越成为一个网络社会系统，而不再仅仅是简单的信息流通平台，推荐系统在这一过程中所起到的资源分配作用因而具有其潜在的网络经济效益。本文将常见的互联网应用形式化为“生产者—服务—消费者”模型，基于传统经济学的基本定义给出了互联网环境下效用、成本和福利的基本概念与统一形式，并进一步给出了网络推荐系统中总社会福利的通用计算方法，进而对个性化推荐模型的经济学意义进行系统的解释。

通过分析个性化推荐在数据、模型和经济学意义上的内在规律并由此建立模型，本文实现了对推荐系统基本输入数据内在结构的解释、对隐变量模型相关算法推荐机制的解释、对推荐结果的个性化推荐理由解释，以及对推荐系统经济学

效益的解释，相比于传统方法，具有理论基础坚实、直观意义显著、运行速度高效、学科前景广阔等优势，突出了本研究工作的独特性。

# 第 3 章　数据的可解释性

用户历史行为数据是个性化推荐系统的第一个环节，本章从数据层面对个性化推荐系统的可解释性进行分析和建模。首先介绍稀疏矩阵的双边块对角化与二部图社区发现之间的关系，从而指出稀疏矩阵内在的群组结构；在此基础上，进一步分析双边块对角矩阵的数学分解性质，并提出局部化矩阵分解算法以提高预测精度和并行化效率。第四小节介绍性能评测结果，并在最后总结本章内容。

## 3.1　矩阵的群组结构

本节介绍个性化推荐系统的主要输入用户物品评分稀疏矩阵内在的群组结构，分析和证明稀疏矩阵的迭代双边块对角结构与基于点割集的图分割社区发现算法内在的等价性，并提出基于块对角矩阵的协同过滤模型。内容包括双边块对角矩阵及其性质、矩阵的双边块对角化算法，以及基于块对角矩阵的协同过滤。

### 3.1.1　本节引言

协同过滤作为个性化推荐的重要技术已经得到学术界广泛的关注和产业界的实际应用，但是目前仍然少有对用户物品评分矩阵的内在结构的理解和研究，并且在此基础上协同过滤也面临着几大重要问题：一是矩阵的稀疏性，这往往会对评分预测算法的精度带来负面影响；二是由规模庞大的输入矩阵所带来的算法可扩展性问题；三是算法缺乏对矩阵中来自不同兴趣社群的用户的区分，这一方面不利于对特殊用户群兴趣的发掘，另一方面降低了协同过滤算法发现和利用具有相似兴趣的用户物品群组提高预测精度的能力。

为了解决这些问题，传统方法主要集中于矩阵聚类（Matrix Clustering）[47,49,233–236] 和社区发现（Community Detection）[237–243] 相关方法的研究和应用。

基于矩阵聚类的方法将用户和/或物品进行聚类并以类别为单位进行协同过滤，然而在实际系统中聚类结果往往难以给出直观的解释；另外，相关方法一般假设每个用户或物品属于且只属于一个类别，而在实际应用中用户往往拥有多方面的兴趣爱好，因此该假设未必成立。

基于社区发现的方法则利用用户对物品的打分、评论等历史行为关系构建用户物品二部图，在该图上通过社区发现来提高物品推荐的精度，并利用所挖掘的兴趣社区及其关系来提高推荐的多样性。通过社区发现来判断和利用群体用户的

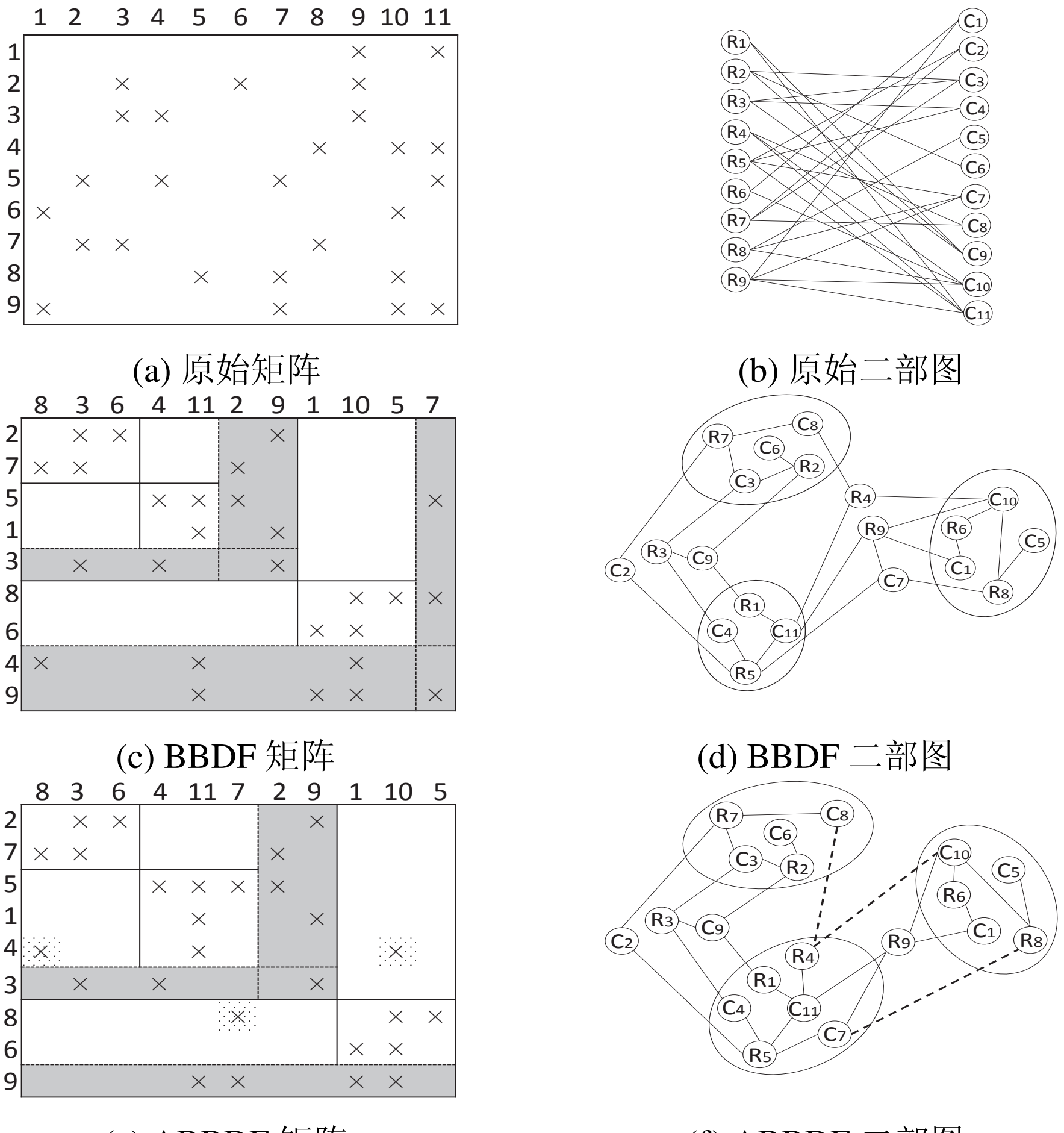

(a) 原始矩阵　(b) 原始二部图

(c) BBDF 矩阵　(d) BBDF 二部图

(e) ABBDF 矩阵　(f) ABBDF 二部图

图 3.1　矩阵的双边块对角化及其所对应的二部图社区发现结构示例

一般偏好和小众用户的特殊偏好确实有助于推荐效果的提升，然而基于图的方法难以充分利用已被证明非常成功的协同过滤算法的优势。

实际上，用户物品评分矩阵和二部图具有数学上的等价关系。在具体的问题形式化之前，图3.1利用一个直观的示例来说明本工作中将要用到的矩阵与二部图结构及其对应关系。图3.1(a)表示一个用户物品评分矩阵，其中矩阵的每一行对应一个用户、每一列对应一个物品、矩阵中的每一个非零值（记为‘×’符号）表示相应的用户对相应物品的打分；该矩阵可以等价地表示为一个用户物品二部图，如图3.1(b)所示，其中原矩阵的每一行（用户）表示为一个 R 节点，每一列（物品）表示对一个 C 节点，原矩阵中的非零值对应于二部图中的一条边。图3.1(c)表示由原始矩阵得到的一个双边块对角矩阵（Bordered Block Diagonal Form，BBDF），其中原始矩阵的第 4 行、第 9 行，以及地 7 列被排列到了矩阵的下边和右边（双边），而剩下的部分可以被排列为两个块对角子矩阵；进一步其中第一个子矩阵中的第 3 行、第 2 列和第 9 列又可以被迭代地排列到矩阵的双边，该过程可以如此迭代执行。该过程对应于原始二部图上的一个点分割过程，如图3.1(d)所示，当我们去掉

二部图中的节点 $R_4$、$R_9$ 和 $C_7$ 时，二部图可以被分割为两个子图，而当我们在其中第一个子图中进一步去掉节点 $R_3$、$C_2$ 和 $C_9$ 时，该子图又可以被分割为两个子图，如此过程可以迭代执行。类似地，图3.1(e)描述了一个由原始矩阵得到的近似双边块对角矩阵（Approximate Bordered Block Diagonal Form，ABBDF），它表示当我们去掉矩阵中 $X_{8,7}$ 和 $X_{4,10}$ 两个非零值时，剩下的部分可以通过将第 9 行排列到下边而得到一个双边块对角型，当进一步将第一个对角块中的非零值 $X_{4,8}$ 去掉时，该对角块又可以排列为一个双边块对角型。同理，该过程也可以由矩阵的二部图等价表达，如图3.1(f)所示，其中虚线表示被删掉的边，它们和矩阵中被删掉的非零值一一对应。

基于矩阵的（近似）双边块对角型，我们一方面可以充分利用在矩阵预测领域取得显著效果的各种协同过滤算法，另一方面可以充分利用二部图社区发现算法所发掘的用户兴趣社区关系，从而一举两得。需要指出的是，由于本质上仍然是在用户物品评分矩阵上进行协同过滤，因此（近似）双边块对角矩阵上的矩阵运算不仅限于某种特定的协同过滤算法，而是适用于各种常见的协同过滤算法，例如基于用户/物品的协同过滤以及各种矩阵分解算法，等等。

在本章，我们提出矩阵的（近似）双边块对角结构，并设计相关算法将原始矩阵排列为对应结构；进一步，我们在该结构上进行协同过滤式矩阵预测，并对其效果进行实验验证。

### 3.1.2　相关工作

个性化推荐的输入数据形式多种多样，包括用户-物品评分矩阵[244]，搜索引擎日志[245]，用户在浏览器中的浏览日志[246,247]，等等。其中，用户-物品评分矩阵因其较好的结构和较为清晰的问题定义，在推荐系统的研究中占据了主要的位置。

目前，基于用户物品评分矩阵的协同过滤算法[244,248] 已经在矩阵预测任务上得到了广泛的应用，不同于基于内容的推荐方法[4]，协同过滤式推荐算法利用用户的群体智慧对用户偏好进行分析建模，而无需对用户/物品大量属性的人工构建，因而在实际系统中得到了广泛的应用。

基于近邻的协同过滤（Nearest Neighbor Based）是协同过滤的基本方法之一，而基于用户的协同过滤[11] 和基于物品的协同过滤[12] 是其中最为常用的两种协同过滤算法。在相关算法中，一个用户或物品的近邻可以通过多种相似度指标来获得，例如 Pearson 相关性系数或余弦相似度，等等。基于用户的方法首先为目标用户（矩阵中的行向量）寻找最相似的用户群，并基于这些用户在物品上的打分历史记录来预测目标用户的评分；类似地，基于物品的方法则计算物品（列向量）之

间的两两相似度，并在预测目标用户在目标物品上的打分时，利用该用户在历史物品上的打分与目标物品的相似度加权平均来获得预测打分。基于近邻的协同过滤原理简单、易于操作，然而往往难以发现用户或物品中隐含的协同性，并且相似度的大量计算使其在实际应用中效率较低。

基于矩阵分解（Matrix Factorization，MF）[59] 的方法则将原始矩阵分解为子矩阵的乘积并进行矩阵补全和预测。常用的方法有 SVD 奇异值分解 [13,249] 和 NMF 非负矩阵分解 [62,153,250]，等等。然而相关方法在训练环节对持续密集计算的需求降低了算法的可扩展性，并且由于实际系统中用户行为数据的不断变化，对模型的不断重复训练提出了较高的要求。为了解决相关问题，研究人员提出了增量式或分布式版本的 SVD 或 NMF 算法 [151,152,157,183,249]，然而考虑到矩阵局部的变化（如用户增加了一个打分）对分解结果的影响并不是局部性的，导致算法仍然需要对全局模型进行不断的重新分解。

为了解决相关问题，科研人员研究了各种各样的矩阵聚类算法来提高算法的效率、可扩展性，以及应对数据稀疏性的问题。基于用户或物品聚类的方法 [47] 首先将用户行向量或物品列向量进行聚类，并进一步将相似用户或物品的计算限制在特定的类别中，从而大大降低计算量；同时，其它一些聚类方法试图对用户和物品同时进行聚类，例如互聚类（Co-Clustering）[49,251,252]、乒乓聚类 [233]、基于聚类的低秩近似算法 [235]，等等。借助用户物品聚类的协同过滤往往可以提高算法的可扩展性，然而算法往往难以为聚类结果给出直观的解释。另外，相关算法强制要求每一个用户或物品属于且仅属于同一个聚类，然而这在实际系统和数据中往往并非一个正确的假设。

近年来，随着社交网站的不断发展，基于图的社区发现算法得到了广泛的关注和研究 [253,254]。实际上，用户物品评分矩阵和用户物品评分二部图之间具有数学上的等价关系，并且基于二部图的社区发现算法和结果可以等价地表示为矩阵上（近似）双边块对角结构，在该结构上，任何已有的协同过滤算法仍然可以不加修改地得到执行，但是如果我们将矩阵中所体现出来的用户物品社区结构考虑在内，就可以进一步提高相关协同过滤算法的精度和可扩展性。

### 3.1.3 双边块对角矩阵及其性质

在本节，我们对矩阵的双边块对角型及其数学性质进行介绍，这将成为下一节基于矩阵双边块对角型协同过滤算法的基础 [82–85]。

**定义 3.1：** 双边块对角型矩阵（Bordered Block Diagonal Form，BBDF）

我们称矩阵 $X$ 是一个双边块对角型矩阵当且仅当它可以仅通过行交换和列交

换两种操作被重新排列成如下的形式：

$$X = \begin{bmatrix} A_{11} & A_{12} & \cdots & A_{1k} & A_{1B} \\ A_{21} & A_{22} & \cdots & A_{2k} & A_{2B} \\ \vdots & \vdots & \ddots & \vdots & \vdots \\ A_{k1} & A_{k2} & \cdots & A_{kk} & A_{kB} \\ A_{B1} & A_{B2} & \cdots & A_{Bk} & A_{BB} \end{bmatrix} = \begin{bmatrix} D_1 & & & & C_1 \\ & D_2 & & & C_2 \\ & & \ddots & & \vdots \\ & & & D_k & C_k \\ R_1 & R_2 & \cdots & R_k & B \end{bmatrix} \tag{3-1}$$

即 $X_{ij} = \mathbf{0}\ (i \neq j, 1 \leq i, j \leq k)$。其中每一个子矩阵 $D_i\ (1 \leq i \leq k)$ 被称为一个“对角块”（Diagonal Block）；$R = [R_1 \cdots R_k B]$ 和 $C = [C_1^T \cdots C_k^T B^T]^T$ 为矩阵的“双边”（Borders）。每个对角块 $D_i$ 可以被迭代式地再次排列为一个双边块对角型。□

可见，矩阵的双边块对角型是块对角矩阵（Block Diagonal Form，BDF）在概念上扩展，当 $R = C = \mathbf{0}$ 时，$X = \mathrm{diag}(D_1 D_2 \cdots D_k B)$ 退化为一个块对角矩阵。

**定义 3.2：　近似双边块对角型矩阵**（Approximate BBDF，ABBDF）

我们称矩阵 $X$ 为近似双边块对角矩阵，当且仅当 $X$ 可以在去掉一个或多个非零值的条件下，通过行交换和列交换转化为一个双边块对角矩阵：

$$X = \begin{bmatrix} A_{11} & A_{12} & \cdots & A_{1k} & A_{1B} \\ A_{21} & A_{22} & \cdots & A_{2k} & A_{2B} \\ \vdots & \vdots & \ddots & \vdots & \vdots \\ A_{k1} & A_{k2} & \cdots & A_{kk} & A_{kB} \\ A_{B1} & A_{B2} & \cdots & A_{Bk} & A_{BB} \end{bmatrix} = \begin{bmatrix} D_1 & A_{12} & \cdots & A_{1k} & C_1 \\ A_{21} & D_2 & \cdots & A_{2k} & C_2 \\ \vdots & \vdots & \ddots & \vdots & \vdots \\ A_{k1} & A_{k2} & \cdots & D_k & C_k \\ R_1 & R_2 & \cdots & R_k & B \end{bmatrix} \tag{3-2}$$

即与双边块对角矩阵相比，在近似双边块对角矩阵中 $X_{ij}\ (i \neq j, 1 \leq i, j \leq k)$ 可以被允许包含非零值散点。类似的，对角块 $D_1 D_2 \cdots D_k$ 可以迭代地被调节为近似双边块对角型。□

近似双边块对角矩阵是双边块对角矩阵的概念扩展，即允许双边块对角矩阵在非对角块部分存在少量的非零值散点。与近似双边块对角矩阵相对应，我们把双边块对角矩阵也称为精确的双边块对角矩阵。

**定义 3.3：　社区发现**（Community Detection，CD）

给定一个图 $\mathcal{G} = (\mathcal{V}, \mathcal{E})$，图 $\mathcal{G}$ 中的一个社区 $C_i$ 被定义为一个顶点集合 $C_i \subseteq \mathcal{V}$；图 $\mathcal{G}$ 上一个包含了 $k(k \geq 1)$ 个社区的社区发现结果 $C = \{C_1 C_2 \cdots C_k\}$ 是满足如下性质的社区的集合：

i. $C_i \neq \emptyset\ (1 \leq i \leq k)$

ii. $C_i \cap \left( \bigcup_{j=1, j\neq i}^{k} C_j \right) \neq C_i\ (1 \leq i \leq k)$

其中第二个性质保证每一个社区至少有一个自己独有的节点，而不是整个社区被完全包含在其它社区之中。□

需要指出的，由于社区发现的定义往往与具体的应用和算法有关，因此迄今为止还没有一个被学术界广泛接受的通用的社区发现的数学定义[255]。因此，本文给出的定义只是用于分析和介绍（近似）双边块对角矩阵内在的群组结构与二部图社区发现之间深刻的数学关系。然而虽然如此，定义3.3所给出的定义可以囊括大部分社区发现算法所给出的社区发现结果。根据 $C_i \cap (\bigcup_{j=1, j\neq i}^{k} C_j) = \emptyset \ (1 \leq i \leq k)$ 的成立与否，这既包括社区之间相互独立的情形（即社区之间不共享节点，每一个用户或物品只属于一个社区），也包括同一个用户或物品属于多个不同社区的情形（即社区之间存在共享节点）。

**定理 3.1：** 在二部图 $\mathcal{G} = (\mathcal{V}, \mathcal{E})$ 上，任何一个包含 $k$ 个社区的社区发现结果 $\mathcal{C} = \{C_1 C_2 \cdots C_k\}$ 都可以在该二部图所对应的矩阵 $X$ 上被表示为一个包含 $k$ 个对角块的近似双边块对角形式。

**证明** 不失一般性，我们假设 $\bigcup_{i=1}^{k} C_i = \mathcal{V}$，即每个节点至少属于一个社区。如果该条件不成立，我们可以首先简单地把 $\mathcal{V} - \bigcup_{i=1}^{k} C_i$ 中的节点所对应的行或列排列到矩阵的双边，则剩下的部分一定保证该条件成立。我们使用数学归纳法进行证明。

当 $k = 1$ 时，结论自然成立。此时，我们把整个矩阵看做一个只包含单个对角块的近似双边块对角矩阵。

当 $k = 2$ 时，我们把 $\mathcal{S} = C_1 \cap C_2$ 中的节点排列到矩阵的双边，并进一步把 $C_1' = C_1 - \mathcal{S}$ 和 $C_2' = C_2 - \mathcal{S}$ 中的节点分别排列为矩阵的两个对角块，如图3.2所示。

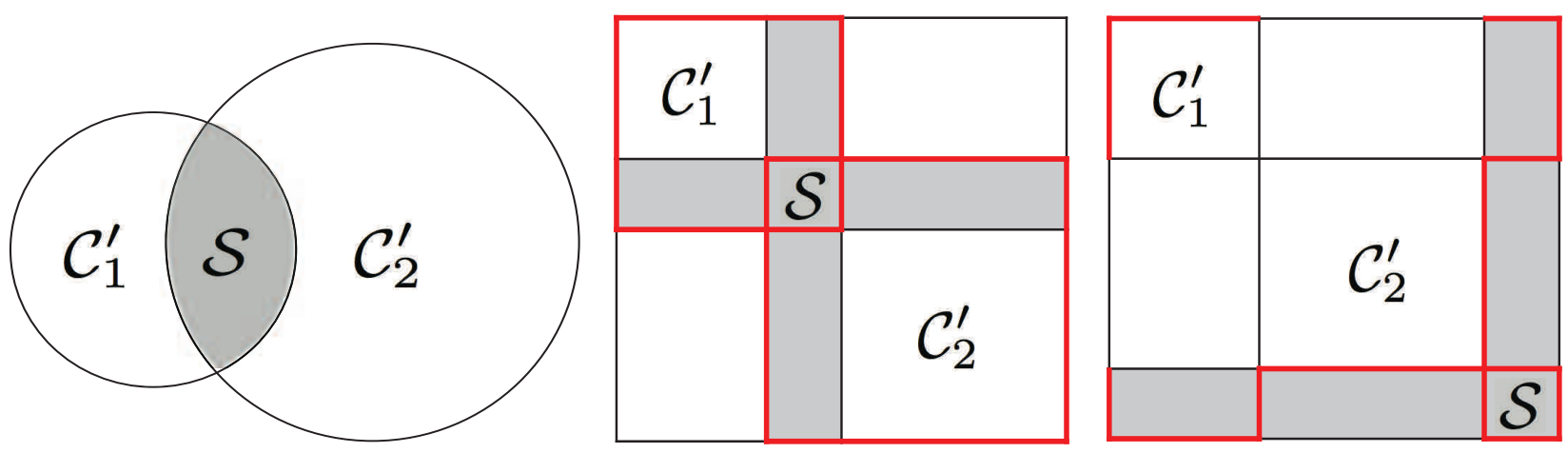

(a) 社区发现　　(b) 中间结果　　(c) ABBDF 结构

图 3.2　$k = 2$ 条件下社区发现结果与其对应的双边块对角型示例

假设该结论对 $k = n - 1$ 成立，当 $k = n$ 时，我们令 $\mathcal{S}_1 = C_1 \cap (\bigcup_{i=2}^{n} C_i)$ 以及 $C_i' = C_i - \mathcal{S}_1 \ (1 \leq i \leq n)$，如图3.3所示。

需要注意的是，根据定义3.3我们有 $C_i' \neq \emptyset \ (1 \leq i \leq n)$。此时，将 $\mathcal{S}_1$ 中的节点排列到双边，将 $C_1'$ 中的节点排列为左上对角块，将 $\bigcup_{i=2}^{n} C_i'$ 中的节点排列为右下对角块。根据归纳假设，$\bigcup_{i=2}^{n} C_i'$ 部分可以用同样的方式被排列为一个包含 $n - 1$ 个

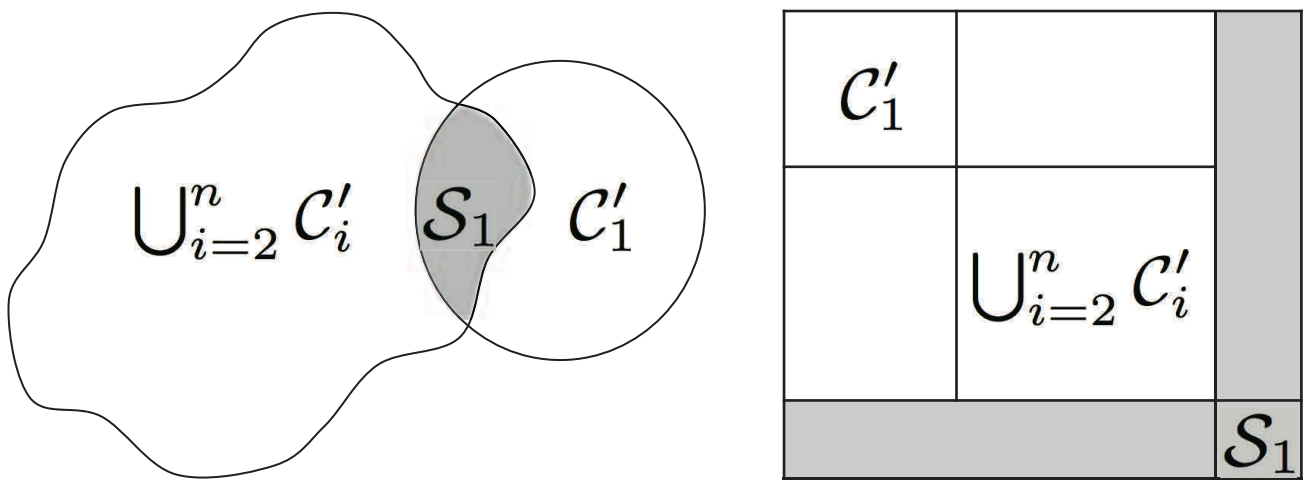

图 3.3　$k = n$ 条件下社区发现结果与其对应的双边块对角型示例

对角块的近似双边块对角型。因此，该矩阵最终可以被排列为一个包含 $n$ 个对角块的近似双边块对角型矩阵，并且其中的双边所对应的节点为 $\mathcal{S} = \bigcup_{i=1}^{n-1} \mathcal{S}_i$。 □

为了进一步帮助理解，我们以 $k = 3$ 的情形为例作进一步的考察，如图3.4所示。首先，令 $\mathcal{S}_1 = C_1 \cap (C_2 \cup C_3)$ 以及 $C_i' = C_i - \mathcal{S}_1$ $(i = 1, 2, 3)$。当我们将 $\mathcal{S}_1$ 中的节点排列到双边时，矩阵中存在两个对角块，分别对应于 $C_1'$ 和 $C_2' \cup C_3'$ 中的节点。进一步，我们令 $\mathcal{S}_2 = C_2' \cap C_3'$ 以及 $C_i'' = C_i' - \mathcal{S}_2$ $(i = 2, 3)$，通过迭代式地将 $\mathcal{S}_2$ 中的节点排列到双边，其中右下对角块可以进一步被排列为近似双边块对角型，并生成两个新的对角块，它们分别对应于 $C_2''$ 和 $C_3''$ 中的节点。最终，矩阵包含三个对角块，且最终的双边所对应的节点为 $\mathcal{S} = \mathcal{S}_1 \cup \mathcal{S}_2$。

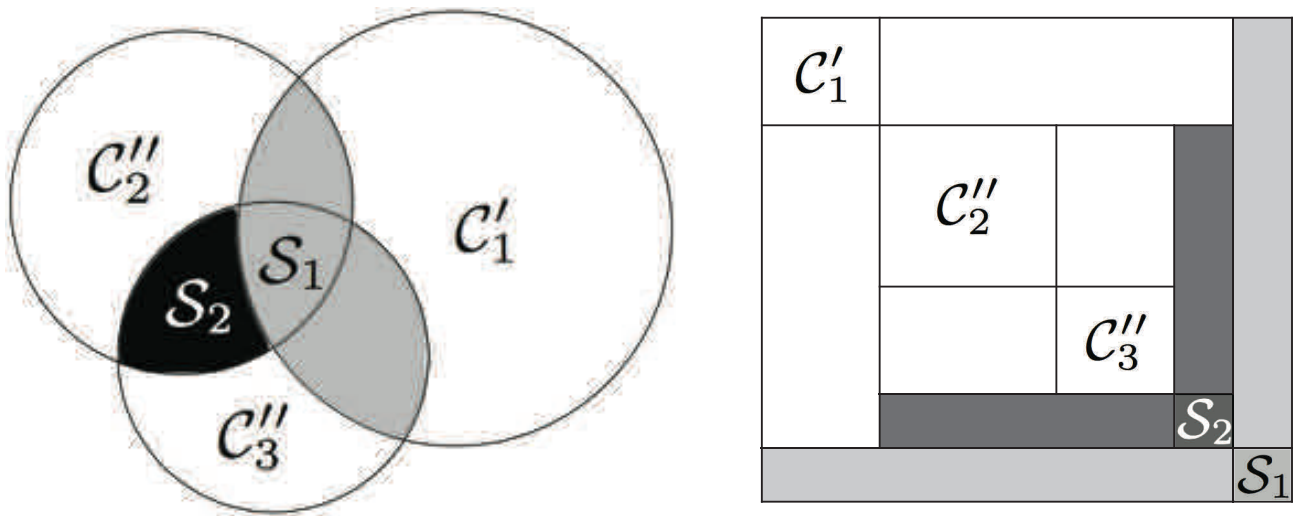

图 3.4　$k = 3$ 条件下社区发现结果与其对应的双边块对角型示例

**推论 3.1：** 给定二部图上的一个社区发现结果 $C = \{C_1 C_2 \cdots C_k\}$；对于每一个 $1 \le i \le k$，令 $\mathcal{S}_i = C_i \cap (\bigcup_{j=1, j\neq i}^{k} C_j)$ 以及 $C_i' = C_i - \mathcal{S}_i$。那么 $C$ 对应于一个精确的双边块对角型矩阵当且仅当对于所有的 $i \neq j$，二部图中不存在从 $C_i'$ 到 $C_j'$ 的边。

**证明** 该推论是如下事实所导出的自然结论：在近似双边块对角型中，非对角块上的非零值对应于连接两个节点 $u$ 和 $v$ 的一条边，其中 $u$ 和 $v$ 分别来自于这样两个社区 $C(u)$ 和 $C(v)$，其中 $u$ 和 $v$ 分别是 $C(u)$ 和 $C(v)$ 所独有的节点。 □

直观来看，（近似）双边块对角矩阵中的每一个对角块是一个*用户物品群组*，其所对应的行和列该社区的*主流用户*和*物品*；矩阵中的行边和列边则是跨社区的

桥接用户和桥接物品，其中桥接用户是那些兴趣比较广泛而在多个群组中均有历史行为记录的用户，而桥接物品则是那些受到广泛欢迎因而被来自多个群组的用户采纳或购买过的物品。在近似双边块对角矩阵中，用户兴趣确实主要集中在各自的群组内，但是也会不时对来自其它群组的物品或桥接物品发生兴趣并产生购买行为。接下来的定义和定理主要阐述如何将一个原始稀疏矩阵排列为（近似）双边块对角型矩阵。

**定义 3.4：基于点割集的图分割**（Graph Partitioning by Vertex Separator, GPVS）

考虑一个无向图 $\mathcal{G} = (\mathcal{V}, \mathcal{E})$，$Adj(v)$ 表示节点 $v$ 的邻接节点，对于一个节点子集 $\mathcal{V}' \subset \mathcal{V}$，我们有 $Adj(\mathcal{V}') = \{v_j \in \mathcal{V} - \mathcal{V}' : \exists v_i \in \mathcal{V}' s.t.\ v_j \in Adj(v_i)\}$。

我们称 $\mathcal{V}_S \subset \mathcal{V}$ 是一个点割集当且仅当由节点 $\mathcal{V} - \mathcal{V}_S$ 所张成的子图包含 $k \geq 2$ 个连通分支。形式上，基于点割集的图分割被定义为 $\Gamma_v = \{\mathcal{V}_1\mathcal{V}_2 \cdots \mathcal{V}_k; \mathcal{V}_S\}$，其中 $\mathcal{V}_i \neq \emptyset$，$\mathcal{V}_i \cap \mathcal{V}_S = \emptyset$，对于每一个 $1 \leq i \leq k$ 我们有 $Adj(\mathcal{V}_i) \subset \mathcal{V}_S$，对于 $1 \leq i < j \leq k$ 我们有 $\mathcal{V}_i \cap \mathcal{V}_j = \emptyset$，以及 $(\bigcup_{i=1}^{k} \mathcal{V}_i) \cup \mathcal{V}_S = \mathcal{V}$。需要指出的是，在这里我们允许 $\mathcal{V}_S = \emptyset$ 的情况。

基于点割集的图分割其直观意义在于去掉一个点集合（点割集）及与其相连的边，剩下的部分可以被分割为 $k$ 个互不相连的连通分支。基于点割集图分割的社区发现对应于一个精确的矩阵双边块对角结构，这可以由推论3.1和定义3.4的结合导出，因为此时我们有 $\mathcal{S}_i = \mathcal{C}_i \cap (\bigcup_{j=1, j\neq i}^{k} \mathcal{C}_j) = \mathcal{V}_S$，以及当 $i \neq j$ 时，$\mathcal{C}'_i = \mathcal{C}_i - \mathcal{S}_i = \mathcal{V}_i$ 是互不相连的。

图3.1(c)和3.1(d)给出了基于点割集的图分割及其对应的双边块对角矩阵示例。在该示例中，通过移除点割集 $\{R_4, R_9, C_7\}$，剩下的节点被分为两个连通分支，分别对应于矩阵中的两个对角块；进一步，通过移除点割集 $\{R_3, C_2, C_9\}$，其中一个对角块又可以被排列为包含两个子对角块的双边块对角矩阵。

**定义 3.5：基于边割集的图分割**（Graph Partitioning by Edge Separator, GPES）

考虑无向图 $\mathcal{G} = (\mathcal{V}, \mathcal{E})$，我们称 $\mathcal{E}_S \subset \mathcal{E}$ 是一个边割集当且仅当删除 $\mathcal{E}_S$ 中的边时，我们有 $\Gamma_e = \{\mathcal{V}_1\mathcal{V}_2 \cdots \mathcal{V}_k\}\ (k \geq 2)$，其中，对于 $1 \leq i \leq k$ 有 $\mathcal{V}_i \neq \emptyset$；对于 $1 \leq i < j \leq k$ 有 $\mathcal{V}_i \cap \mathcal{V}_j = \emptyset$；$\bigcup_{i=1}^{k} \mathcal{V}_i = \mathcal{V}$；并且对于任意的 $i \neq j$，由点集 $\mathcal{V}_i$ 和 $\mathcal{V}_j$ 所导出的子图是非连通的。

我们称 $\mathcal{C} = \{\mathcal{C}_1\mathcal{C}_2 \cdots \mathcal{C}_k\}$ 为基于点割集图分割的社区发现结果，其中 $\mathcal{C}_i = \mathcal{V}_i\ (1 \leq i \leq k)$。

考虑到 $\mathcal{S}_i = \mathcal{C}_i \cap (\bigcup_{j=1, j\neq i}^{k} \mathcal{C}_j) = \emptyset$，基于点割集图分割的社区发现结果对应于一个无双边的近似块对角矩阵，在必要的情况下，该矩阵可以被用来进一步构建

一个近似双边块对角矩阵，在接下来的双边块对角矩阵构建算法的介绍中，我们将利用该性质进行研究。

**定义 3.6：　密度**（Density）

令 $X$ 是一个 $m \times n$ $(m, n \geq 1)$ 的矩阵，$\mathrm{n}(X)$ 表示 $X$ 中非零点的个数，$\mathrm{area}(X) = m \times n$ 表示 $X$ 的面积，则矩阵 $X$ 的密度定义为 $\rho(X) = \frac{\mathrm{n}(X)}{\mathrm{area}(X)}$，$k$ 个矩阵 $X_1 \cdots X_k$ 的平均密度为 $\bar{\rho}(X_1 \cdots X_k) = \frac{\sum_{i=1}^{k} \mathrm{n}(X_i)}{\sum_{i=1}^{k} \mathrm{area}(X_i)}$。令 $\mathcal{G}$ 表示矩阵 $X$ 所对应的二部图，则 $\rho(\mathcal{G}) = \rho(X)$，$\bar{\rho}(\mathcal{G}_1 \cdots \mathcal{G}_k) = \bar{\rho}(X_1 \cdots X_k)$。

我们将矩阵中的一个行或列统称为一个*向量*，向量 $x$ 限制在子矩阵 $B$ 中的密度定义为 $\rho(x(B))$，其中 $x(B)$ 表示向量 $x$ 在子矩阵 $B$ 中的部分，如图3.5中虚线部分所示。

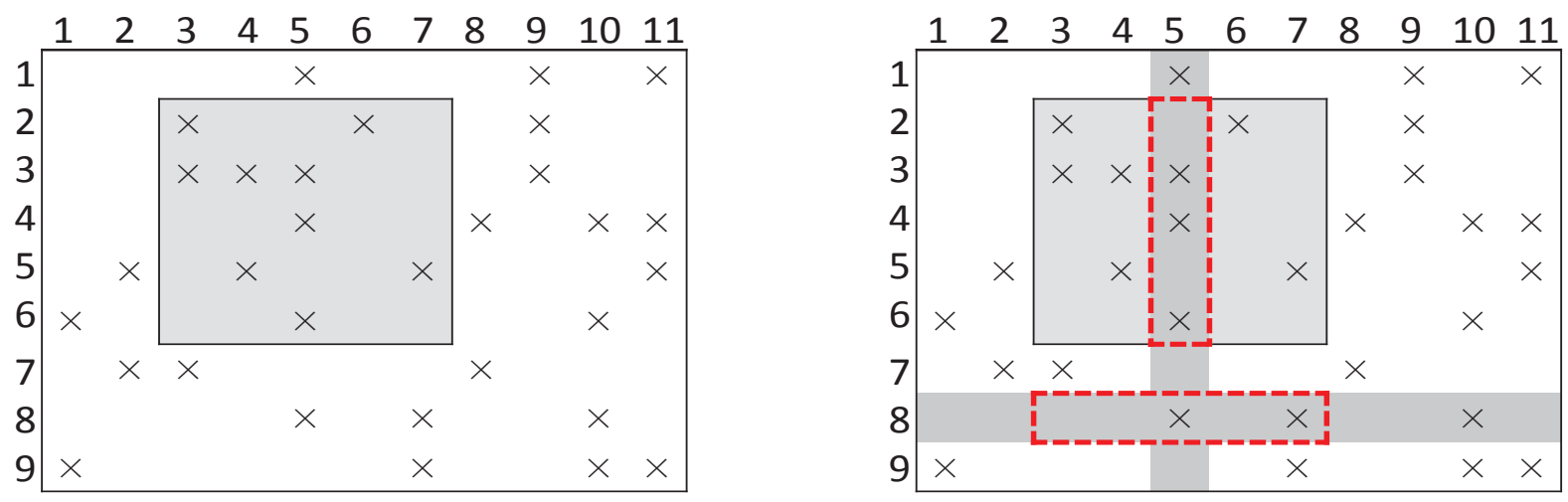

(a) 子矩阵　　(b) 向量在子矩阵中的部分

图 3.5　子矩阵和向量，以及向量在子矩阵中的部分示例

矩阵或图的密度是在各种社区发现算法中常用的概念[255]。如图3.5(a)的示例中，阴影部分的密度为 $\frac{9}{25}$；在图3.5(b)中，第 5 列的密度为 $\frac{5}{9}$，而第 8 行限制在阴影子矩阵中的密度为 $\frac{2}{5}$。

### 3.1.4　矩阵的双边块对角化算法

本部分介绍矩阵的（近似）双边块对角化算法，从而将一个稀疏矩阵转化为一个（近似）双边块对角型矩阵。我们直接使用密度作为启发式条件：首先，在社区发现相关算法中高密度聚集性的社区往往意味着偏好相似、趣味相投的群组，在具有相似兴趣的用户物品群组中进行协同过滤可以提高预测效果；其次，通过直接提高对角块子矩阵的非零值密度，协同过滤算法也可以降低数据稀疏性和散点过拟合的影响，从而进一步提高预测效果。

#### 3.1.4.1　矩阵的精确双边块对角化算法

在矩阵的精确双边块对角化算法过程中，一个被迭代执行的子过程是将某些向量（行或列）排列到双边，并将剩下的部分排列为一系列的对角块。这样的迭代执

行框架一般被称为 George 矩阵嵌套解析（George's Nested Dissection）框架[254,256]，并在基于重排列的矩阵算法中得到了广泛的应用。我们首先介绍这个子过程所对应的算法实现。

该子过程等价于用户物品二部图上基于点割集的图分割（GPVS）过程[253]。一般而言，一个图往往存在不止一个点割集，因此一个恰当的图分割算法试图寻找最小点割集，即通过删掉最少个数的点而将图分为几个连通分支。寻找图的最小点割集已被证明是 NP 难问题[257]，然而，该问题由于其在实际应用中的基础性和重要性而得到了广泛的研究，学术界提出了多种高效的启发式方法进行图分割[258]，例如基于层次分析（multilevel）的图分割、基于谱分析（spectral-based）的图分割，以及基于核方法（kernal-based）的图分割，等等。理论和实验研究结果指出，基于层次分析的图分割在计算效率和分割质量方面都具有明显的优势[253,254,258,259]，在本工作中，我们选择著名的基于层次分析的图分割开源工具 Metis[260] 来进行基本的点割集图分割。

如前所述，我们采用密度来直接控制迭代式矩阵块对角化过程以获得高密度的群组和子矩阵，因为高密度的社区在实际应用中往往代表了一定的用户兴趣社群，并在社区发现相关研究中被广泛应用[255]。算法1描述了迭代的子过程。

**Algorithm 1:** BASIC-BBDF-PERMUTATION($X, \mathcal{G}$) //BBDF 迭代子过程

**Input:** 用户物品评分矩阵 $X$，$X$ 所对应的二部图 $\mathcal{G} = (\mathcal{V}, \mathcal{E}) = (\mathcal{R} \cup \mathcal{C}, \mathcal{E})$;

//$\mathcal{R}$ 和 $\mathcal{C}$ 分别为 $\mathcal{V}$ 中的行节点集合和列节点集合;

**Output:** 单层双边块对角矩阵及其对角块平均密度 $\bar{\rho}$

$\Gamma_v \leftarrow \{\mathcal{V}_1 \mathcal{V}_2 \cdots \mathcal{V}_k; \mathcal{V}_S\} \leftarrow \text{GPVS}(\mathcal{G})$;

将矩阵 $X$ 的行按照 $\mathcal{R}_1 \mathcal{R}_2 \cdots \mathcal{R}_k \mathcal{R}_S$ 的顺序进行排列;

将矩阵 $X$ 的列按照 $\mathcal{C}_1 \mathcal{C}_2 \cdots \mathcal{C}_k \mathcal{C}_S$ 的顺序进行排列;

**return** $\bar{\rho}(D_1 D_2 \cdots D_k)$ //$D_i$ 表示第 $i$ 个对角块，对应于顶点集 $\mathcal{V}_i = \mathcal{R}_i \cup \mathcal{C}_i$;

在该过程中，我们希望矩阵被重新排列之后对角块的平均密度高于原始矩阵的密度，接下来，我们讨论该性质的满足情况。基于定义3.6的符号体系有：

$$\rho(X) = \frac{\mathrm{n}(X)}{\mathrm{area}(X)}, \quad \bar{\rho}(D_1 \cdots D_k) = \frac{\sum_{i=1}^{k} \mathrm{n}(D_i)}{\sum_{i=1}^{k} \mathrm{area}(D_i)} \tag{3-3}$$

令 $n = \mathrm{n}(X)$，$n_1 = \sum_{i=1}^{k} \mathrm{n}(D_i)$，$n_2 = n - n_1$；$s = \mathrm{area}(X)$，$s_1 = \sum_{i=1}^{k} \mathrm{area}(D_i)$，

$s_2 = s - s_1$，我们有：

$$\rho_X = \rho(X) = \frac{n}{s}, \quad \rho_1 = \bar{\rho}(D_1 \cdots D_k) = \frac{n_1}{s_1}, \quad \rho_2 = \frac{n_2}{s_2} \tag{3-4}$$

其中 $\rho_2$ 表示矩阵中非对角部分与双边部分的平均密度。令 $\bar{\rho}(D_1 \cdots D_k) > \rho(X)$，即 $\rho_1 > \rho_X$，那么我们有：

$$\frac{n_1}{s_1} > \frac{n}{s} = \frac{n_1 + n_2}{s_1 + s_2} \Leftrightarrow \frac{n_1}{s_1} > \frac{n_2}{s_2} \Leftrightarrow \rho_1 > \rho_2 \tag{3-5}$$

该式表明，要使排列结果中对角块的平均密度相对于原矩阵的密度将提升，当且仅当排列之后对角块的平均密度高于矩阵其它部分的平均密度即可。根据点割集图分割算法总是试图寻找最小点割集的特性，该条件在实际中很容易得到满足；另一方面，矩阵中对角块的总面积往往小于矩阵其它部分的面积之和[253]。因此，我们选择算法1中所返回的对角块平均密度作为如下算法2中迭代式双边块对角化算法的重要控制变量。

**Algorithm 2:** BBDF-PERMUTATION($X, \mathcal{G}, \rho$) //迭代双边块对角化算法

**Input:** 用户物品评分矩阵 $X$ 及其对应的二部图 $\mathcal{G} = (\mathcal{V}, \mathcal{E})$，目标密度 $\rho$

**Output:** 被排列为双边块对角型的矩阵 $X$

```
ρ_X ← ρ(X);
if ρ_X < ρ then
    ρ̄ ←BASIC-BBDF-PERMUTATION(X, G);
    if ρ̄ > ρ_X then
        for X 的每一个对角块 D_i do
            BBDF-PERMUTATION(D_i, G_{V_i}, ρ); //V_i 为 D_i 的顶点集合，G_{V_i} 是
              由 V_i 所导出的子图
        end
    end
end
```

在如上基于密度的矩阵双边块对角化算法中，参数 $\rho$ 是预定的对角块目标平均密度；该算法在原始矩阵上执行块对角化子过程 BASIC-BBDF-PERMUTATION，并进一步迭代式地在生成的每一个对角块上继续执行该子过程，直到某一个子矩阵

上的对角块平均密度达到了 $\rho$、或者再也无法通过 Basic-BBDF-Permutation 子过程提高密度，则在该对角块上停止迭代块对角化。

需要指出的是，在算法的第 4 行，即使对角块平均密度还没有达到目标密度 $\rho$，我们也不对对角块平均密度无法进一步提升的情况做更多的处理；这样的对角块被看做一个稀疏的群组，这种情况在目标密度 $\rho$ 取值过高时比较容易出现。利用平均密度的控制来防止这样的对角块无限制迭代下去的目的之一是防止出现大量小而离散的社区。因此，合理的目标密度设置对构建合适的块对角矩阵有重要意义，在实验部分，我们将对目标密度的设置对结果的影响进行具体的分析。

#### 3.1.4.2　矩阵的近似双边块对角化算法

近似双边块对角可以由基于边割集的图分割算法 GPES 得到，同样，我们仍然使用 Metis 中基于层次分析的边割集图分割算法。近似双边块对角化过程同样需要制定目标密度 $\rho$，在该算法中，我们首先计算原始二部图的一个 GPES 图分割，并在忽略边割集中的边所对应的矩阵非零值的基础上对原始矩阵进行排列。

与精确双边块对角化中对无法继续提升密度的对角块不做进一步处理的做法不同，在近似双边块对角化中，我们通过从对角块中抽取向量排列到边上，从而进一步提升其密度，直到对角块的平均密度达到我们的密度要求，如算法3所示。

**Algorithm 3:** ABBDF-Permutation($X, \mathcal{G}, \rho$)　//近似迭代双边块对角化算法

**Input:** 用户物品评分矩阵 $X$ 及其对应的二部图 $\mathcal{G} = (\mathcal{V}, \mathcal{E})$，目标密度 $\rho$

**Output:** 被排列为近似双边块对角型的矩阵 $X$

**if** $\rho_X \geq \rho$ **then**
    **return**;
**end**
**else**
    $\Gamma_e \leftarrow \{\mathcal{V}_1\mathcal{V}_2\cdots\mathcal{V}_k\} \leftarrow$ GPES($\mathcal{G}$);
    将矩阵 $X$ 中的行按照 $\mathcal{R}_1\mathcal{R}_2\cdots\mathcal{R}_k$ 的顺序进行排列;
    将矩阵 $X$ 中的列按照 $C_1C_2\cdots C_k$ 的顺序进行排列;
    $\{\mathcal{V}_1'\mathcal{V}_2'\cdots\mathcal{V}_k';\mathcal{V}_S'\} \leftarrow$Improve-Density($A, \mathcal{G}, \Gamma_e$);
    **for** $X$ 中的每一个对角块 $D_i$ **do**
        ABBDF-Permutation($D_i, \mathcal{G}_{\mathcal{V}_i'}, \rho$);
    **end**
**end**

其中的 Improve-Density 子程序每次从对角块中选择移除之后能最大限度提升平均密度的向量，并将其移动到边上，如算法4所示。

需要指出的是，在实际系统实现中我们不需要逐一检查对角块中的每一个向量来寻找能够最大化提升密度的那一个向量。实际上，我们只需要依次尝试检测那些在对角块中具有最小受限密度的向量，即在对角块中含有最少个数非零值的那些向量。

**Algorithm 4:** Improve-Density$(X, \mathcal{G}, \Gamma_e)$ //对角块平均密度提升算法

**Input:** 用户物品评分矩阵 $X$ 及其对应的二部图 $\mathcal{G} = (\mathcal{V}, \mathcal{E}) = (\mathcal{R} \cup \mathcal{C}, \mathcal{E})$，
图 $\mathcal{G}$ 上的 GPES 结果 $\Gamma_e = \{\mathcal{V}_1 \mathcal{V}_2 \cdots \mathcal{V}_k\}$

**Output:** 提升密度之后的图分割结果 $\Gamma'$

$\{\mathcal{V}_1' \mathcal{V}_2' \cdots \mathcal{V}_k'; \mathcal{V}_S'\} \leftarrow \{\mathcal{V}_1 \mathcal{V}_2 \cdots \mathcal{V}_k; \emptyset\}$;
**while** $\bar{\rho}(D_1 D_2 \cdots D_k) < \rho(X)$ **do**
　$x', i' \leftarrow 0, \bar{\rho}' \leftarrow 0$;
　**for** 对于每一个对角块 $D_i$ **do**
　　**for** 对于 $D_i$ 中的每一个向量 $x$(包括行和列) **do**
　　　$\bar{\rho} \leftarrow \frac{\sum_{j=1}^{k} \mathrm{n}(D_j) - \mathrm{n}(x(D_i))}{\sum_{j=1}^{k} \mathrm{area}(D_j) - \mathrm{area}(x(D_i))}$;
　　　**if** $\bar{\rho} > \bar{\rho}'$ **then**
　　　　$x' \leftarrow x, i' \leftarrow i, \bar{\rho}' \leftarrow \bar{\rho}$;
　　　**end**
　　**end**
　**end**
　将向量 $x'$ 排列到双边;
　$\mathcal{V}_{i'}' \leftarrow \mathcal{V}_{i'}' - \{\mathrm{node}(x')\}$;
　$\mathcal{V}_S' \leftarrow \mathcal{V}_S' \cup \{\mathrm{node}(x')\}$;//node$(x')$ 表示 $\mathcal{V}_{i'}'$ 中对应于向量 $l'$ 的节点
**end**
**return** $\Gamma' \leftarrow \{\mathcal{V}_1' \mathcal{V}_2' \cdots \mathcal{V}_k'; \mathcal{V}_S'\}$;

### 3.1.5　基于块对角阵的协同过滤

基于（近似）双边块对角矩阵的协同过滤其一大优势是，任何一个已有的协同过滤算法都可以在由对角块和边块组成的子矩阵上得到执行，从而使得双边块对角矩阵的适用范围非常广泛。在本节，我们利用矩阵的双边块对角型构建协同

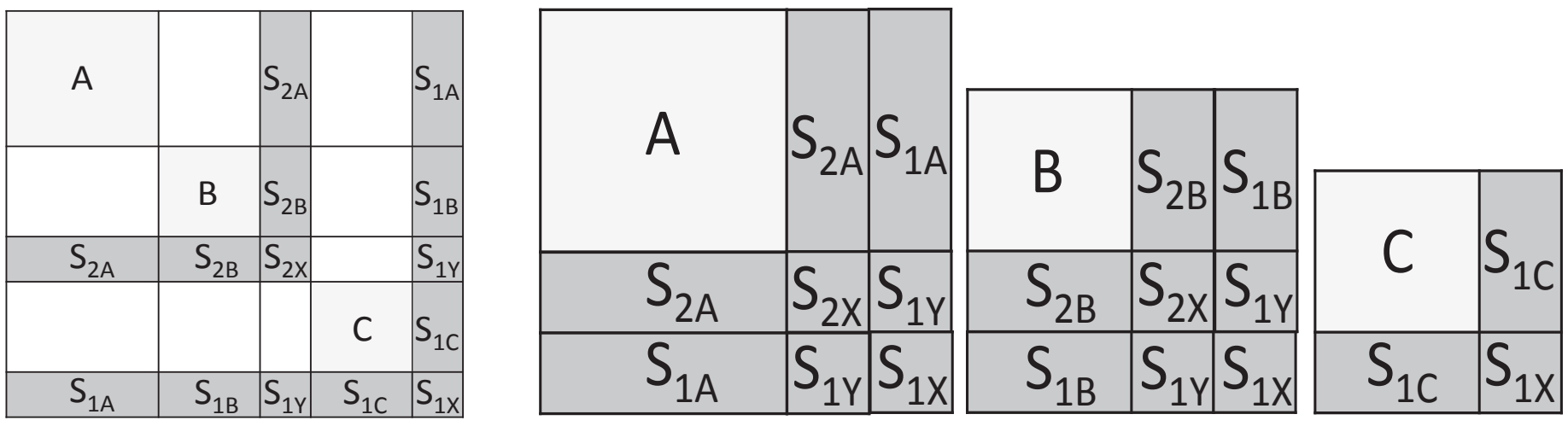

(a) (A)BBDF 结构矩阵　　　　　　(b) 抽取和构建的子矩阵

图 3.6　利用（近似）双边块对角矩阵抽取和构建协同过滤子矩阵示例

过滤的框架，从而使得现在有协同过滤算法都可以在该框架下得到使用。由于矩阵中通过内在的群组结构将具有相似行为和兴趣的用户聚在一起，使得我们可以借助群组内的协同过滤提高已有算法的预测精度；另一方面通过将高维庞大矩阵上的协同过滤转化为小而密的子矩阵上的协同过滤，我们的框架大大提高协同过滤算法的可扩展性。

图3.6给出了基于块对角矩阵的协同过滤的直观示例。对于每一个对角块，我们通过将其与来自各个层次的双边拼接来重建它所对应的用户物品群组，直观而言，就是我们利用群组的主流用户和物品以及系统中的桥接用户和桥接物品共同进行协同过滤式的预测和推荐，通过这一方式，我们既利用了群组内用户的共有兴趣、也利用了整个数据集合中的热门用户和热门物品，从而在为小众用户提供推荐时兼顾大众兴趣。在图3.6中，与图3.6(a)中的 $A, B, C$ 三个对角块相对应，我们在图3.6(b)中构建了三个子矩阵。需要指出的是，在近似双边块对角矩阵的子矩阵构建过程中，我们忽略出于非对角块上的非零值散点，并将其看做用户在社区外的临时特殊爱好，虽然如此，这些散点有助于我们发现用户的特殊兴趣从而对提高推荐系统的新颖性和惊喜度有重要意义。

最后，我们在每一个拼接子矩阵上执行任何一个协同过滤算法，由于原矩阵中行边和列边的交叉块（如图3.6中的 $S_{1X}$ 和 $S_{1Y}$ 等）出现在多个拼接子矩阵中，它们所对应的预测值也有多个。在协同过滤的基础上，我们把来自多个子矩阵的预测值进行平均作为该交叉块的最终预测结果，以表示桥接用户和桥接物品在与不同的社区匹配后进行预测的平均情况。

以上基于双边块对角矩阵的矩阵分解和协同过滤具有严格数学基础，在下面一节，我们将介绍双边块对角矩阵的分解性质，并以此为基础进一步对基于块对角矩阵的协同过滤进行形式化，给出局部化矩阵分解协同过滤框架。

## 3.2 局部化矩阵分解算法

本节分析双边块对角矩阵在矩阵分解任务上的数学特性，并基于矩阵的双边块对角结构提出局部化矩阵分解框架。内容包括双边块对角矩阵的分解性质、矩阵分解算法在块对角矩阵上的可拆分性、局部化矩阵分解框架、矩阵的块对角块化算法及其收敛性证明。

### 3.2.1 本节引言

由于拥有较高的预测精度，基于低秩近似的矩阵分解算法在协同过滤个性化推荐中获得了广泛的应用。矩阵分解算法利用较少个数（例如数十个）的隐变量将高维的原始矩阵表示为分解因子矩阵的乘积，从而达到降维和预测的目的。

然而在实际系统中，矩阵分解仍然面临着几个方面的重要问题，主要包括数据稀疏性、由矩阵变动而带来的频繁的模型重复训练、以及模型的可扩展性，等等。例如在电子商务系统中，每个用户所购买的物品数量（少则几个，多则几十个上百个）相对于系统中的物品总数量（数千万甚至上亿个）而言少之又少，由此所带来了数据稀疏性问题使得矩阵分解算法难以较好建模用户偏好并给出更为准确的推荐；由于实际系统处于不断的动态变化之中，每时每刻都有大量的打分在矩阵中被新增、删除或者修改，使得矩阵分解算法必须定期或不定期地重复执行并更新分解结果，而频繁地重新分解庞大的原始矩阵为系统带来了较大的计算负荷，并同时降低了算法的可扩展性。

在本节，我们提出基于双边块对角矩阵的局部化矩阵分解（Localized Matrix Factorization，LMF）框架，该框架可以与矩阵的奇异值分解（SVD）、非负矩阵分解（NMF）、概率矩阵分解（PMF）、最大间隔矩阵分解（MMMF）等常见的矩阵分解算法兼容，从而一方面提高这些算法的预测精度，另一方面为这些算法提供了一个通用的并行化框架。在具体的介绍之前，我们先用图3.7中的直观的例子来回顾一下矩阵的双边块对角型结构，其中图3.7(a)表示一个原始稀疏矩阵，经过行和列的调节之后，图3.7(b)表示一个单层双边块对角矩阵，进一步将其中一个对角块进行迭代的调节，我们得到一个多层双边块对角矩阵，如图3.7(c)所示。

局部化矩阵分解框架首先将原始稀疏矩阵转化为一个多层双边块对角矩阵，进一步从中抽取和拼接子矩阵构建一个对角块矩阵，并最终利用对角块矩阵上的预测结果来对原始矩阵给出打分预测。该框架拥有如下的优势：首先，局部化矩阵分解从原始矩阵中抽取密度较高的子矩阵（局部化），通过利用矩阵内的相似用户群组结构并降低数据稀疏性来提高预测精度；其次，该框架的局部性使得我们在原始矩阵得到更新（新增、修改或删除数据点）时，不需要重新对整个庞大的原

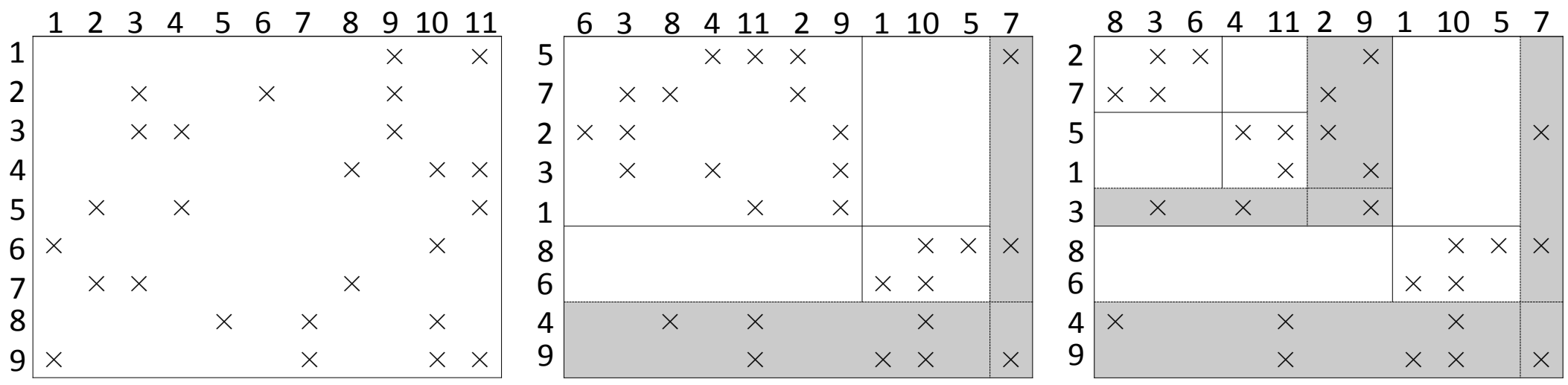

(a) 原始稀疏矩阵　(b) 单层双边块对角矩阵　(c) 多层双边块对角矩阵

图 3.7　矩阵的双边块对角型结构示例

始矩阵进行重新分解，而只需要在变化的子矩阵上重新分解即可，从而提高算法在实际应用中的可扩展性；最后，由于子矩阵各自的分解过程互不影响，使得算法可以很容易的并行化，从而提高矩阵分解的效率。

在接下来的部分，我们首先介绍本部分的相关工作，接下来介绍矩阵的双边块对角型在矩阵分解任务上的数学性质，并进一步提出和分析我们的局部化矩阵分解框架及其与已有矩阵分解算法的关系。

### 3.2.2　相关工作

矩阵分解作为协同过滤的重要手段在实际系统中得到了广泛的应用，为了统一 SVD、NMF 等各种常用的矩阵分解算法，Singh 等从机器学习的角度出发，将矩阵分解算法描述为损失函数、正则化项、约束条件等几个主要组成部分，给出了一个矩阵分解的统一形式化表示形式[261]，不同的矩阵分解算法本质上是，该形式在后续工作中为矩阵分解的相关理论分析带来了很大的方便。

尽管矩阵分解在实验中获得较好的预测精度，然而在实际应用中，矩阵分解仍然面临着几个重要的问题，包括数据稀疏性、模型可扩展性、模型重复训练的问题，等等。为了解决数据稀疏性的问题，传统方法依赖于向矩阵中填充虚拟值（imputation）来降低原始矩阵的稀疏性[13]。然而虚拟值填充往往需要向矩阵中加入大量的非零值才能达到较为理想的效果，而这大大增加了矩阵分解的计算量和复杂度；同时虚拟值填充需要较好的先验知识和启发式规则来尽可能生成较为合理的虚拟值，然而这样的先验知识往往并不容易总结归纳；如果向矩阵中加入了错误的虚拟值，则反而会降低矩阵分解的效果[59]。模型重复训练的问题则源于在实际系统中用户物品评分矩阵的动态变化，例如新用户的注册、新打分的生成、以及用户对历史评分的修改或删除行为，等等。

在提高矩阵分解的可扩展性方面，传统方法主要集中于各种矩阵聚类方法的使用[234,235,262–266] 以及对现有矩阵分解算法设计相应的并行化或分布式版本[151,157,183]。这些方法相对于在原始矩阵上直接进行矩阵分解而言，只能获得近似

的预测结果，而不能等价地恢复原始算法的给出的结果；另外，它们往往也只限于某种特定的矩阵分解算法，并需要针对不同的方法设计和实现不同的并行化算法。与这些方法不同，基于块对角矩阵良好的数学性质，我们在理论上证明了局部化矩阵分解框架并不是原始算法的近似，而是可以精确地等价恢复非并行化算法给出的预测结果；另外，该框架与常见的矩阵分解协同过滤算法兼容，从而为矩阵分解提供了一个统一的并行化框架。

### 3.2.3　双边块对角矩阵的分解性质

为了符号的一致性和叙述的方便，本节首先给出如下符号化的示例和定义。我们仍然考虑通过矩阵的行交换和列交换对稀疏矩阵的结构进行重新排列，并定义如下的块对角矩阵 $X$：

$$X = \begin{bmatrix} D_1 & & & \\ & D_2 & & \\ & & \ddots & \\ & & & D_k \end{bmatrix} \triangleq \mathrm{diag}(D_i) \tag{3-6}$$

然而并非所有的稀疏矩阵都可以被调节为块对角矩阵，但一般都可以被调节为一个（单层）双边块对角矩阵，如下面的例子所示，其中包含 $k$ 个对角块：

$$X = \left[\begin{array}{c:c} D & C \\ \hdashline R & B \end{array}\right] = \left[\begin{array}{ccc:c} D_1 & & & C_1 \\ & \ddots & & \vdots \\ & & D_k & C_k \\ \hdashline R_1 & \cdots & R_k & B \end{array}\right] \tag{3-7}$$

矩阵中每个 $D_i (1 \leq i \leq k)$ 为一个对角块，$R_b \triangleq [R_1 \cdots R_k B]$ 和 $C_b \triangleq [C_1^T \cdots C_k^T B^T]^T$ 分别为矩阵的行边和列边。矩阵中的每一个对角块都有可能进一步被迭代地进行双边块对角化，从而得到多层的双边块对角结构。为了便于理解，我们给出如下的简单示例，其中 $\mathcal{I}_*$ 和 $\mathcal{J}_*$ 分别为行标号集合和列标号集合。

$$X = \begin{array}{c} \begin{array}{ccc} \mathcal{J}_1 & \mathcal{J}_2 & \mathcal{J}_B \end{array} \\ \left[\begin{array}{cc:c} D_1 & & C_1 \\ & D_2 & C_2 \\ \hdashline R_1 & R_2 & B \end{array}\right] \begin{array}{c} \mathcal{I}_1 \\ \mathcal{I}_2 \\ \mathcal{I}_B \end{array} \end{array} = \begin{array}{c} \begin{array}{cccccc} \mathcal{J}_{11} & \mathcal{J}_{12} & \mathcal{J}_{B_1} & \mathcal{J}_2 & \mathcal{J}_B \end{array} \\ \left[\begin{array}{ccccc} \boxed{\begin{array}{cc:c} D_{11} & & C_{11} \\ & D_{12} & C_{12} \\ \hdashline R_{11} & R_{12} & B_1 \end{array}} & & \begin{array}{:c} C_1^1 \\ C_1^2 \\ C_1^3 \end{array} \\ & D_2 & C_2 \\ \hdashline R_1^1 \; R_1^2 \; R_1^3 & R_2 & B \end{array}\right] \begin{array}{c} \mathcal{I}_{11} \\ \mathcal{I}_{12} \\ \mathcal{I}_{B_1} \\ \mathcal{I}_2 \\ \mathcal{I}_B \end{array} \end{array} \tag{3-8}$$

在公式 (3-8) 中，第一个对角块 $D_1$ 被迭代地排列为一个双边块对角型结果。对 $D_1$ 的重排列影响边矩阵 $R_1$ 和 $C_1$，但是该过程值只影响其中行和列的顺序，而不影响其数值。进一步，对角块子矩阵 $D_{11}$、$D_{12}$ 或 $D_2$ 可能进一步被排列为双边块对角形式。

块对角矩阵或双边块对角矩阵具有较好的分解性质，对于块对角矩阵，我们首先有如下的直观性质。

**命题 3.1：** 对于如公式 (3-6) 所示的块对角矩阵 $X = \mathrm{diag}(D_i)$，如果对于每一个对角块 $D_i$ 我们有 $D_i = U_iV_i^T$，那么对于原始矩阵 $X$ 我们有 $X = \mathrm{diag}(U_i) \cdot \mathrm{diag}(V_i^T)$。

该命题表示了块对角矩阵中各个对角块在矩阵分解上的独立性。如前所述，我们并不能保证每一个稀疏矩阵都可以被排列和表示为块对角矩阵的形式，但是对于更为一般的双边块对角矩阵，我们有如下的性质。

**命题 3.2：** 对于如公式 (3-7) 所示的双边块对角矩阵，令拼接块 $\tilde{X}_i$ 的分解为：

$$\tilde{X}_i \triangleq \begin{bmatrix} D_i & C_i \\ R_i & B \end{bmatrix} = U_iV_i^T = \begin{bmatrix} U_{i1} \\ U_{i2} \end{bmatrix} \begin{bmatrix} V_{i1}^T & V_{i2}^T \end{bmatrix} \tag{3-9}$$

则我们有：

$$D_i = U_{i1}V_{i1}^T \quad R_i = U_{i2}V_{i1}^T \quad C_i = U_{i1}V_{i2}^T \quad B = U_{i2}V_{i2}^T \tag{3-10}$$

进一步，令：

$$U = \begin{bmatrix} U_{11} & & & \\ & U_{21} & & \\ & & \ddots & \\ & & & U_{k1} \\ U_{12} & U_{22} & \cdots & U_{k2} \end{bmatrix} \quad V = \begin{bmatrix} V_{11} & & & \\ & V_{21} & & \\ & & \ddots & \\ & & & V_{k1} \\ V_{12} & V_{22} & \cdots & V_{k2} \end{bmatrix} \tag{3-11}$$

则我们有：

$$UV^T = \begin{bmatrix} U_{11}V_{11}^T & & & U_{11}V_{12}^T \\ & U_{21}V_{21}^T & & U_{21}V_{22}^T \\ & & \ddots & \vdots \\ & & U_{k1}V_{k1}^T & U_{k1}V_{k2}^T \\ U_{12}V_{11}^T & U_{22}V_{21}^T & \cdots U_{k2}V_{k1}^T & \sum_{i=1}^{k} U_{i2}V_{i2}^T \end{bmatrix} = \begin{bmatrix} D_1 & & C_1 \\ & \ddots & \vdots \\ & D_k & C_k \\ R_1 \cdots & R_k & kB \end{bmatrix} \tag{3-12}$$

$UV^T$ 与原始双边块对角矩阵 $X$ 的唯一区别在于交叉块 $B$ 被乘以对角块个数 $k$。□

在命题3.2中，我们实际上是在分解一个块对角矩阵 $\tilde{X} = \mathrm{diag}(\tilde{X}_i) = \mathrm{diag}\left(\left[\begin{smallmatrix} D_i & C_i \\ R_i & B \end{smallmatrix}\right]\right) (1 \le i \le k)$；根据命题3.1，如果有 $\tilde{X}_i = U_i V_i^T$，则我们有 $\tilde{X} = \mathrm{diag}(U_i) \cdot \mathrm{diag}(V_i^T)$，通过对重复预测的子矩阵 $B$ 求平均，我们就可以利用每个对角块矩阵各自的分解结果 $\tilde{X}_i = U_i V_i^T$ 来恢复原始矩阵 $X$。

该结论可以进一步推广到多层双边块对角矩阵上，为了符号表示上的清晰，我们仍然使用公式 (3-8) 中的例子来进行说明。为了将原始矩阵 $X$ 转化为块对角矩阵，我们首先将第一个对角块 $D_1$ 转化为块对角形式，得到如下的中间结果：

$$
\tilde{X}_{int.} = \begin{array}{c} \begin{array}{cccccc} \mathcal{J}_{11} & \mathcal{J}_{B_1} & \mathcal{J}_{12} & \mathcal{J}_{B_1} & \mathcal{J}_{2} & \mathcal{J}_{B} \end{array} \\ \left[\begin{array}{ccccc:c} D_{11} & C_{11} & & & & C_1^1 \\ R_{11} & B_1 & & & & C_1^3 \\ & & D_{12} & C_{12} & & C_1^2 \\ & & R_{12} & B_1 & & C_1^3 \\ & & & & D_2 & C_2 \\ \hdashline R_1^1 & R_1^3 & R_1^2 & R_1^3 & R_2 & B \end{array}\right] \end{array} \begin{array}{l} \mathcal{I}_{11} \\ \mathcal{I}_{B_1} \\ \mathcal{I}_{12} \\ \mathcal{I}_{B_1} \\ \mathcal{I}_{2} \\ \mathcal{I}_{B} \end{array} \tag{3-13}
$$

该矩阵包含 3 个对角块，通过在该矩阵上进一步执行如上的拆分过程，我们可以得到如下的块对角矩阵 $\tilde{X}$，其中对于非对角 $i \neq j$，有 $\tilde{X}_{ij} = \mathbf{0}$。

$$
\tilde{X} = \begin{array}{c} \begin{array}{cccccccc} \mathcal{J}_{11} & \mathcal{J}_{B_1} & \mathcal{J}_{B} & \mathcal{J}_{12} & \mathcal{J}_{B_1} & \mathcal{J}_{B} & \mathcal{J}_{2} & \mathcal{J}_{B} \end{array} \\ \left[\begin{array}{ccc} \boxed{\begin{array}{ccc} D_{11} & C_{11} & C_1^1 \\ R_{11} & B_1 & C_1^3 \\ R_1^1 & R_1^3 & B \end{array}} & \tilde{X}_{12} & \tilde{X}_{13} \\ \tilde{X}_{21} & \boxed{\begin{array}{ccc} D_{12} & C_{12} & C_1^2 \\ R_{12} & B_1 & C_1^3 \\ R_1^2 & R_1^3 & B \end{array}} & \tilde{X}_{23} \\ \tilde{X}_{31} & \tilde{X}_{32} & \boxed{\begin{array}{cc} D_2 & C_2 \\ R_2 & B \end{array}} \end{array}\right] \end{array} \begin{array}{l} \mathcal{I}_{11} \\ \mathcal{I}_{B_1} \\ \mathcal{I}_{B} \\ \mathcal{I}_{12} \\ \mathcal{I}_{B_1} \\ \mathcal{I}_{B} \\ \mathcal{I}_{2} \\ \mathcal{I}_{B} \end{array} \triangleq \mathrm{diag}(\tilde{X}_1, \tilde{X}_2, \tilde{X}_3) \tag{3-14}
$$

同样，我们可以分别独立地分解每一个对角块 $\tilde{X}_1$，$\tilde{X}_2$ 和 $\tilde{X}_3$，并将重复预测的子矩阵进行平均从而恢复原始矩阵。

实际上，我们也可以绕开对中间结果的构建，而从原始的多层双边块对角矩阵直接构造块对角矩阵。由于 $\tilde{X}$ 中的每一个对角块 $\tilde{X}_i$ 分别对应于原始矩阵中的每一个对角块 $D_i$，我们可以将每一个 $D_i$ 与其所对应的所有行边部分和裂变部分、以及这些行边和列边所对应的交叉块进行拼接，从而直接得到 $\tilde{X}_i$。进一步，将公式 (3-8) 中的任何一个对角块进一步迭代地执行双边块对角化不会影响已有的对角块所拼接成的子矩阵。

### 3.2.4　近似矩阵分解算法及其可拆分性质

在实际系统中，我们所使用的矩阵分解算法往往是基于机器学习迭代优化的近似矩阵分解算法，而非如上所述的精确恢复原始矩阵。Singh 等对已有的矩阵分解算法进行了总结，将矩阵分解算法归纳为损失函数、正则化项、约束条件等主要组成部分，并给出了如下的矩阵分解的通用表示形式[261]。

**定义 3.7：　近似矩阵分解算法**

令 $X \in \mathbb{R}^{m\times n}$ 为一个稀疏矩阵，并令 $U \in \mathbb{R}^{m\times r}, V \in \mathbb{R}^{n\times r}$ 为该矩阵的分解因子矩阵，则一个矩阵分解算法 $\mathcal{P} = (f, \mathcal{D}_W, \mathcal{C}, \mathcal{R})$ 可以表示为如下主要部分的组合：

1. 预测函数 $f : \mathbb{R}^{m\times n} \to \mathbb{R}^{m\times n}$
2. 数据权重矩阵 $W \in \mathbb{R}_+^{m\times n}$，用来衡量损失函数中对不同观测值采用的权重
3. 损失函数 $\mathcal{D}_W(X, f(UV^T)) \geq 0$，用来描述将 $X$ 近似为 $f(UV^T)$ 时的误差
4. 对分解因子矩阵的硬性约束 $(U, V) \in \mathcal{C}$
5. 正则化项 $\mathcal{R}(U, V) \geq 0$

对于一个矩阵分解模型 $X \approx f(UV^T) \triangleq X^*$，我们求解如下的优化目标：

$$\underset{(U,V)\in\mathcal{C}}{\operatorname{argmin}} \left[\mathcal{D}_W(X, f(UV^T)) + \mathcal{R}(U, V)\right]. \tag{3-15}$$

其中损失函数 $\mathcal{D}(\cdot,\cdot)$ 往往在第二个参数项上满足凸性，并且该函数往往可以表示为矩阵 $X$ 中各个元素上的损失的（加权）和。例如加权奇异值分解算法（Weighted SVD）的损失函数为[60]：

$$\mathcal{D}_W(X, f(UV^T)) = \|W \odot (X - UV^T)\|_{Fro}^2 \tag{3-16}$$

其中 $\odot$ 表示矩阵的元素对应相乘。在本工作中，我们称 $X \approx f(UV^T) \triangleq X^*$ 为 $X$ 的近似矩阵分解。

$$X = \begin{bmatrix} X_1 & & & \\ & X_2 & & \\ & & \ddots & \\ & & & X_k \end{bmatrix} \approx f(UV^T) = f\left(\begin{bmatrix} U_1 \\ U_2 \\ \vdots \\ U_k \end{bmatrix} \begin{bmatrix} V_1^T & V_2^T & \cdots & V_k^T \end{bmatrix}\right) = f\left(\begin{bmatrix} U_1V_1^T & U_1V_2^T & \cdots & U_1V_k^T \\ U_2V_1^T & U_2V_2^T & \cdots & U_2V_k^T \\ \vdots & \vdots & \ddots & \vdots \\ U_kV_1^T & U_kV_2^T & \cdots & U_kV_k^T \end{bmatrix}\right) \tag{3-17}$$

我们在近似矩阵分解的意义下考察块对角矩阵 $X = \mathrm{diag}(X_i)(1 \leq i \leq k)$，如公式 (3-17) 所示，并给出矩阵分解可拆分性的定义和性质，作为局部化矩阵分解框

架的基础。在下面的定义和定理中，我们用 $X_{ij} = \begin{cases} X_i & (i=j) \\ \mathbf{0} & (i \neq j) \end{cases}$ 来表示公式 (3-17) 中 $X$ 的各个子矩阵，并用 $W_{ij}$ 表示 $X_{ij}$ 所对应的权重矩阵。$f(UV^T)_{ij}$ 表示 $f(UV^T)$ 中用来近似 $X_{ij}$ 的子矩阵，即 $X_{ij} \approx f(UV^T)_{ij}$。为了简化，对于 $i = j$ 的情形，我们用 $f(UV^T)_i$ 来表示 $f(UV^T)_{ii}$，并用 $W_i$ 来表示 $W_{ii}$。

**定义** 3.8：　**可拆分的预测函数**

当预测函数 $f : \mathbb{R}^{m\times n} \to \mathbb{R}^{m\times n}$ 满足如下性质时，我们称它为可拆分的：

$$f(UV^T)_{ij} = f(U_iV_j^T) \ \ (1 \leq i, j \leq k) \tag{3-18}$$

大部分矩阵分解算法采用分别独立作用在矩阵中各个元素上的预测函数，即预测矩阵 $Y = f(X)$ 可以通过单独地分别计算 $y_{ij} = f(x_{ij})$ 来求得。例如在奇异值分解 SVD 中，预测函数 $f(x) = x$；而在非负矩阵分解 NMF 中，预测函数为 $f(x) = \log(x)$。元素级的预测函数自然满足如上的可拆分性质。

**定义** 3.9：　**可拆分的损失函数**

对于损失函数 $\mathcal{D}_W(X, f(UV^T))$，当满足如下性质时，我们称它为可拆分的：

$$\mathcal{D}_W(X, f(UV^T)) = \sum_{i=1}^{k} \mathcal{D}_{W_i}(X_i, f(UV^T)_i) \tag{3-19}$$

在实际应用中，我们可以从两个角度来解读该性质。

首先，大部分的矩阵分解算法将 $\mathcal{D}$ 定义为矩阵中每一个预测点上的损失之和。例如奇异值分解（SVD）、非负矩阵分解（NMF）、概率矩阵分解（PMF）、以及最大间隔矩阵分解（MMMF），等等。它们的损失函数都可以归结为某种 Bregman 测度[267] 的实例。元素级的损失函数具有如下的性质：

$$\mathcal{D}_W(X, f(UV^T)) = \sum_{i,j} \mathcal{D}_{W_{ij}}(X_{ij}, f(UV^T)_{ij}) \tag{3-20}$$

其次，在实际系统中我们所处理的矩阵往往非常稀疏，其中包含大量的零值，而矩阵中的一个零值仅仅表示该打分点未观测，而非用户在对应的物品上打了零分。因此，实用的矩阵分解算法只在观测值上计算损失函数，具体到块对角矩阵中，我们有 $W_{ij} = \mathbf{0}\ (i \neq j)$，以及：

$$\mathcal{D}_{W_{ij}}(X_{ij}, f(UV^T)_{ij}) = 0 \ \ (i \neq j) \tag{3-21}$$

综合公式 (3-20) 和公式 (3-21) 可知，在实际应用中矩阵分解算法满足定义3.9中损失函数的可拆分性。

**定义 3.10：　可拆分的硬性约束**

当一个分解因子矩阵硬性约束 $C$ 满足如下的性质时，我们称其为可拆分的：

$$(U,V)\in C \text{ 当且仅当 } (U_i,V_i)\in C \ \ (1\le i\le k) \tag{3-22}$$

多数矩阵分解算法在实际应用中并不对分解因子矩阵的性质进行硬性约束，但也存在一些常用的硬性约束，主要包括非负性约束（分解因子矩阵 $U$ 和 $V$ 中的值均为非负数）、正交性约束（$U$ 和 $V$ 各自的列向量之间互相正交）、概率化约束（$U$ 和 $V$ 中的每一个元素非负且行向量一范数为 1）、稀疏性约束（$U$ 和 $V$ 的行向量满足稀疏性阈值约束）、基数约束（$U$ 和 $V$ 每一个行向量中的非零值个数满足给定的约束条件）。在这个意义下，非负性、概率化、稀疏性以及基数约束均为可拆分的硬性约束。例如，$(U,V)$ 中的每一个行向量一范数为 1 当且仅当同样的性质在每一个分解子矩阵 $(U_i,V_i)(1\le i\le k)$ 上都成立。然而，正交性约束并非一个可拆分的约束：$(U,V)$ 的正交性并不能保证每一个 $(U_i,V_i)$ 满足同样的性质。在本工作中，我们只关注可拆分的硬性约束。

**定义 3.11：　可拆分的正则化项**

对于一个正则化项 $\mathcal{R}(U,V)$，当它满足如下的性质时，我们称其为可拆分的：

$$\mathcal{R}(U,V)=\sum_{i=1}^{k}\mathcal{R}(U_i,V_i) \tag{3-23}$$

最常见的正则化项为 $\ell_p$-norm 正则化，它对于任意的 $p\ge 1$ 都是可拆分的：

$$\mathcal{R}(U,V)=\lambda_U\|U\|_p^p+\lambda_V\|V\|_p^p=\sum_{i=1}^{k}\left(\lambda_U\|U_i\|_p^p+\lambda_V\|V_i\|_p^p\right)=\sum_{i=1}^{k}\mathcal{R}(U_i,V_i) \tag{3-24}$$

常用的 Frobenius 正则化项为 $\ell_p$-norm 在 $p=2$ 时的情形；原始的最大间隔矩阵分解 MMMF[63] 采用矩阵的核范数 $\|X\|_\Sigma$（矩阵 $X$ 所有奇异值之和）作为正则化项，而核范数不是可拆分的正则化项；然而，基于核范数与 Frobenius 范数之间重要的关系式 $\|X\|_\Sigma=\min\limits_{X=UV^T}\frac{1}{2}(\|U\|_F^2+\|V\|_F^2)$，Rennie 等[64] 提出了快速矩阵最大间隔矩阵分解算法，由于最终采用了 Frobenius 范数，使得快速最大间隔矩阵分解中的正则化项也是可拆分的。

**定义 3.12：　可拆分的矩阵分解算法**

我们称一个矩阵分解算法$\mathcal{P} = (f, \mathcal{D}_W, C, \mathcal{R})$是可拆分的，当且仅当$f, \mathcal{D}_W, C, \mathcal{R}$均满足可拆分性质，即公式 (3-18)~3-23均得到满足。我们用 $(U, V) = \mathcal{P}(X, r)$ 来表示对矩阵 $X$ 在算法 $\mathcal{P}$ 下使用 $r$ 个分解因子进行分解时所得到分解结果。

需要指出的是，虽然需要满足四个可拆分条件使得矩阵分解算法的可拆分性看上去很难得到满足，但是很多在实际应用中经常使用的矩阵分解算法实际上都满足可拆分性质，例如奇异值分解、非负矩阵分解、概率矩阵分解、最大间隔矩阵分解，在本工作中，我们也主要考虑这些满足可拆分性的矩阵分解算法。

**定理 3.2：**　令 $X$ 为一个块对角矩阵，如公式 (3-17) 所示，$\mathcal{P} = (f, \mathcal{D}_W, C, \mathcal{R})$ 是一个满足可拆分性质的矩阵分解算法；令 $(U, V) = \mathcal{P}(X, r)$ 以及 $(U_i, V_i) = \mathcal{P}(X_i, r)(1 \leq i \leq k)$，那么我们有：

i. $U = [U_1^T U_2^T \cdots U_k^T]^T, V = [V_1^T V_2^T \cdots V_k^T]^T$

ii. $X_{ij} \approx f(U_i V_j^T)$ $(1 \leq i, j \leq k)$

**证明** i. 考虑如公式 (3-15) 所定义的矩阵分解优化问题，并且其中的预测函数 $f$、损失函数 $\mathcal{D}_W$，硬性约束 $C$，以及正则化项 $\mathcal{R}$ 都满足可拆分性，那么我们有：

$$
\begin{aligned}
&(U, V) = \mathcal{P}(X, r) \\
&= \underset{(U,V)\in C}{\operatorname{argmin}} \left[\mathcal{D}_W(X, f(UV^T)) + \mathcal{R}(U, V)\right] \\
&= \underset{(U,V)\in C}{\operatorname{argmin}} \sum_{i=1}^{k} \left[\mathcal{D}_{W_i}(X_i, f(UV^T)_i) + \mathcal{R}(U_i, V_i)\right] \\
&= \underset{(U,V)\in C}{\operatorname{argmin}} \sum_{i=1}^{k} \left[\mathcal{D}_{W_i}(X_i, f(U_i V_i^T)) + \mathcal{R}(U_i, V_i)\right] \\
&= \bigwedge_{i=1}^{k} \left\{\underset{(U_i,V_i)\in C}{\operatorname{argmin}} \left[\mathcal{D}_{W_i}(X_i, f(U_i V_i^T)) + \mathcal{R}(U_i, V_i)\right]\right\} \\
&= \bigwedge_{i=1}^{k} \left\{\mathcal{P}(X_i, r)\right\} = \bigwedge_{i=1}^{k} \left\{(U_i, V_i)\right\}
\end{aligned}
\tag{3-25}
$$

因此我们有 $U = [U_1^T U_2^T \cdots U_k^T]^T$ 以及 $V = [V_1^T V_2^T \cdots V_k^T]^T$。

ii. 该结论可以从预测函数 $f$ 的可拆分性直接导出：

$$
X_{ij} \approx f(UV^T)_{ij} = f(U_i V_j^T) \tag{3-26}
$$

并且对于任意的 $1 \leq i, j \leq k$ 都成立。 □

由定理3.2可知，我们可以独立地对每一个对角块进行分解，并通过每个对角块的结果获得对原始矩阵最终的预测结果。

### 3.2.5 局部化矩阵分解框架

基于如上定义定理，我们提出面向可拆分矩阵分解算法的局部化矩阵分解（Localized Matrix Factorization，LMF）框架，在该框架中我们首先将一个稀疏矩阵调节为如公式 (3-8) 所示的（多层）双边块对角矩阵的形式，并进一步将其排列为一个块对角矩阵，如公式 (3-14) 所示；我们在块对角矩阵上执行局部化矩阵分解算法，并利用分解结果对原始矩阵进行预测。

假设一个多层双边块对角矩阵 $X$ 被转化为块对角矩阵 $\tilde{X} = \mathrm{diag}(\tilde{X}_i)(1 \le i \le k)$，并且 $X_{\mathcal{I}_* \sim \mathcal{J}_*}$ 以及 $\tilde{X}_{\mathcal{I}_* \sim \mathcal{J}_*}$ 分别表示 $X$ 和 $\tilde{X}$ 中的子矩阵，例如在公式 (3-8) 中，$R_{12} = X_{\mathcal{I}_{B_1} \sim \mathcal{J}_{12}}$，并且它在公式 (3-14) 中以 $\tilde{X}_{\mathcal{I}_{B_1} \sim \mathcal{J}_{12}}$ 的形式被重复了两次。局部化矩阵分解框架利用如下三个步骤将矩阵 $\tilde{X}$ 的预测结果还原为原始矩阵 $X$ 的预测：

i. 利用可拆分的矩阵分解算法 $\mathcal{P} = (f, \mathcal{D}_W, \mathcal{C}, \mathcal{R})$ 获得对每一个对角块 $\tilde{X}_i$ 的分解因子矩阵 $(U_i, V_i) = \mathcal{P}(\tilde{X}_i, r)$，则有：

$$\tilde{X}_i \approx f(U_i V_i^T) \triangleq \tilde{X}_i^* \triangleq \tilde{X}_{ii}^* \tag{3-27}$$

其中 $\tilde{X}_i^*$ 表示对 $\tilde{X}_i$ 的近似。

ii. 利用 $\tilde{X}_i$ 和 $\tilde{X}_j$ 的分解结果预测 $\tilde{X}$ 中的零块 $\tilde{X}_{ij}(i \neq j)$：

$$\tilde{X}_{ij} \approx f(U_i V_j^T) \triangleq \tilde{X}_{ij}^* \tag{3-28}$$

现在我们有 $\tilde{X}$ 的近似结果 $\tilde{X}^* \triangleq \{\tilde{X}_{ij}^* | 1 \le i, j \le k\}$。

iii. 将被重复预测的子矩阵 $\tilde{X}^*$ 求平均得到对原始矩阵 $X$ 中对应子矩阵的预测。假设 $X_{\mathcal{I}_* \sim \mathcal{J}_*}$ 在 $\tilde{X}$ 中被重复预测了 $k$ 次，且第 $t$ 个预测对应于子矩阵 $\tilde{X}_{i_t j_t}$，而该子矩阵的近似矩阵为 $\tilde{X}^{*(i_t j_t)}_{\mathcal{I}_* \sim \mathcal{J}_*}$。那么，对 $X_{\mathcal{I}_* \sim \mathcal{J}_*}$ 的最终预测结果为：

$$X^*_{\mathcal{I}_* \sim \mathcal{J}_*} = \frac{1}{k} \sum_{t=1}^{k} \tilde{X}^{*(i_t j_t)}_{\mathcal{I}_* \sim \mathcal{J}_*} \tag{3-29}$$

为了更好理解，我们仍以公式 (3-8) 中的子矩阵 $R_{12} = X_{\mathcal{I}_{B_1} \sim \mathcal{J}_{12}}$ 为例。$X_{\mathcal{I}_{B_1} \sim \mathcal{J}_{12}}$

在公式 (3-14) 中被重复了两次，其中一个在 $\tilde{X}_{12}$ 中，而另一个在 $\tilde{X}_{22}$ 中。因此：

$$X^*_{\mathcal{I}_{B_1}\sim\mathcal{J}_{12}} = \frac{1}{2}(\tilde{X}^{*(12)}_{\mathcal{I}_{B_1}\sim\mathcal{J}_{12}} + \tilde{X}^{*(22)}_{\mathcal{I}_{B_1}\sim\mathcal{J}_{12}}) \tag{3-30}$$

按照类似的方式，我们对 $X$ 中的每一个子矩阵 $X_{\mathcal{I}_*\sim\mathcal{J}_*}$ 构建近似矩阵 $X^*_{\mathcal{I}_*\sim\mathcal{J}_*}$。通过将这些预测结果进行拼合，我们最终得到原矩阵 $X$ 的预测结果 $X^* = \{X^*_{\mathcal{I}_*\sim\mathcal{J}_*}\}$。

### 3.2.6　局部化矩阵分解框架

局部化矩阵分解算法的一个重要优势在于，从原始矩阵中抽取出来的对角块子矩阵可以分别独立地进行矩阵分解，因而可以方便地并行化。而子矩阵的并行化同时分解要求不同的子矩阵（对应于一个线程或进程）具有相似的规模，从而不同的线程可以在相似的时间内完成，进而方便对重复预测的子矩阵进行平均并预测原始矩阵，否则，较早完成的线程不得不等待最后完成的线程。为了达到这一要求，我们需要进一步改进与设计合适的矩阵双边块对角化算法，使得算法所构造的对角块子矩阵具有相似的规模。

为此，我们在上一节矩阵精确块对角化算法的基础上，提出了基于矩阵规模启发式规则的平衡矩阵块对角化算法。对于一个包含 $k$ 个对角块 $D_1D_2\cdots D_k$ 的矩阵 $X$（例如公式 (3-8) 中的双边块对角矩阵包含三个对角块，分别为 $D_{11}D_{12}$ 和 $D_2$），我们用 $\tilde{X} = \mathrm{diag}(\tilde{X}_1\tilde{X}_2\cdots\tilde{X}_k)$ 表示 $X$ 所对应的块对角矩阵（如公式 (3-14) 所示）。算法5描述了平衡矩阵块对角化算法的基本流程，该算法通过调用 Balanced-Permutation$(X, \hat{\rho}, 1)$ 来启动。

算法5同样需要目标对角块平均密度 $\hat{\rho}$ 作为输入，与算法2和算法3不同的是，为了保证最终用于矩阵分解的各个块对角矩阵的规模相似，我们采用最终块对角矩阵 $\tilde{X}$ 中各个对角块 $\tilde{X}_1\cdots\tilde{X}_k$ 的平均密度，而不再是双边块对角矩阵中每个对角块 $D_1\cdots D_k$ 的平均密度，因为 $D_1\cdots D_k$ 并不用于最终的矩阵分解，而用于矩阵分解的是 $\tilde{X}_1\cdots\tilde{X}_k$。在每一轮递归中，算法按照矩阵面积由大到小依次检查矩阵 $X$ 的每一个对角块 $D_i$，如果发现可以通过分割一个对角块 $D_i$ 来提升最终 $\tilde{X}$ 中拼接而成的各个对角块的平均密度，那么算法就采纳这一分割并按照分割结果对被分割的对角块进行重排序并递归，直到算法达到了目标平均对角块密度 $\hat{\rho}$ 而停止，或者因无论对哪一个对角块进行进一步的分割都无法提升平均密度而停止。

需要指出的是，在该算法中我们采用了矩阵面积作为启发式规则来获得较好的分割平衡性，接下来我们分析该启发规则的合理性及其作用。

图3.8(a)展示了对角块 $D_i$ 及其对应的边块，这些子矩阵拼接在一起实际上就是块对角矩阵 $\tilde{X}$ 中对应于 $D_i$ 的对角块子矩阵 $\tilde{X}_i$。当 $D_i$ 被分割为两个对角块 $D_i^1$

**Algorithm 5:** BALANCED-PERMUTATION($X, \hat{\rho}, k$) //平衡双边块对角化算法

**Input:** 用户物品评分矩阵 $X$，最终拼接而成用于分解的子矩阵目标平均密度 $\hat{\rho}$，$X$ 中当前对角块个数 $k$

**Output:** 被排列为（多层）双边块对角型的矩阵 $X$ 及其块对角矩阵 $\tilde{X}$

$\rho \leftarrow \bar{\rho}(\tilde{X}_1\tilde{X}_2\cdots\tilde{X}_k)$;<br>
**if** $\rho \geq \hat{\rho}$ **then**<br>
　**return** $\tilde{X}$; //已经达到密度要求<br>
**end**<br>
**else**<br>
　$[D_{s_1}D_{s_2}\cdots D_{s_k}] \leftarrow \mathrm{Sort}([D_1D_2\cdots D_k])$; //将对角块按大小降序排列<br>
　**for** $i \leftarrow 1$ *to* $k$ **do**<br>
　　$[D_{s_i}^1 D_{s_i}^2] \leftarrow \mathrm{MetisNodeBisection}(D_{s_i})$;<br>
　　//利用 Metis 的核心图分割进程将 $D_{s_i}$ 分割为两个对角块<br>
　　**if** $\bar{\rho}(\tilde{X}_{s_1}\cdots\tilde{X}_{s_{i-1}}\tilde{X}_{s_i}^1\tilde{X}_{s_i}^2\tilde{X}_{s_{i+1}}\cdots\tilde{X}_{s_k}) > \rho$ **then**<br>
　　　$X' \leftarrow$ 对 $X$ 将 $D_{s_i}$ 排列为 $[D_{s_i}^1 D_{s_i}^2]$;<br>
　　　BALANCED-PERMUTATION($X', \hat{\rho}, k+1$); //递归<br>
　　　**break**; //不再需要检查下一个对角块<br>
　　**end**<br>
　**end**<br>
　**return** $\tilde{X}$; //不再有任何对角块可以提升平均密度<br>
**end**

和 $D_i^2$ 时，我们得到两个新的对角块矩阵 $\tilde{X}_i^1$ 和 $\tilde{X}_i^2$，如图3.8(b)中虚线所包围的部分所示，而图中实线所包围的部分构成原始矩阵 $\tilde{X}_i$。因此，将矩阵 $\tilde{X}_i$ 转化为 $\tilde{X}_i^1$ 和 $\tilde{X}_i^2$ 的过程实际上从原始矩阵中删掉了两个空白子矩阵，并将其替换为一些重复的非零矩阵，如图3.8(b)所示。

令 $s = \sum_{t=1}^{k} s(\tilde{X}_t)$、$n = \sum_{t=1}^{k} n(\tilde{X}_t)$；并令 $\Delta s_1$ 为被删掉的空白矩阵的总面积，$\Delta s_2$ 为它们所替换成的非零矩阵的总面积，以及 $\Delta n$ 为 $\Delta s_2$ 中所包含的非零值的个数，则在对 $D_i$ 进行分割之后，平均密度的增量为：

$$\Delta\rho = \rho' - \rho = \frac{n + \Delta n}{s - \Delta s_1 + \Delta s_2} - \frac{n}{s} = \frac{s\Delta n + n\Delta s}{s(s - \Delta s)} \tag{3-31}$$

其中 $\rho$ 和 $\rho'$ 为对 $D_i$ 进行分割前和分割后 $\tilde{X}$ 中对角块的平均密度，$\Delta s \triangleq \Delta s_1 - \Delta s_2$。

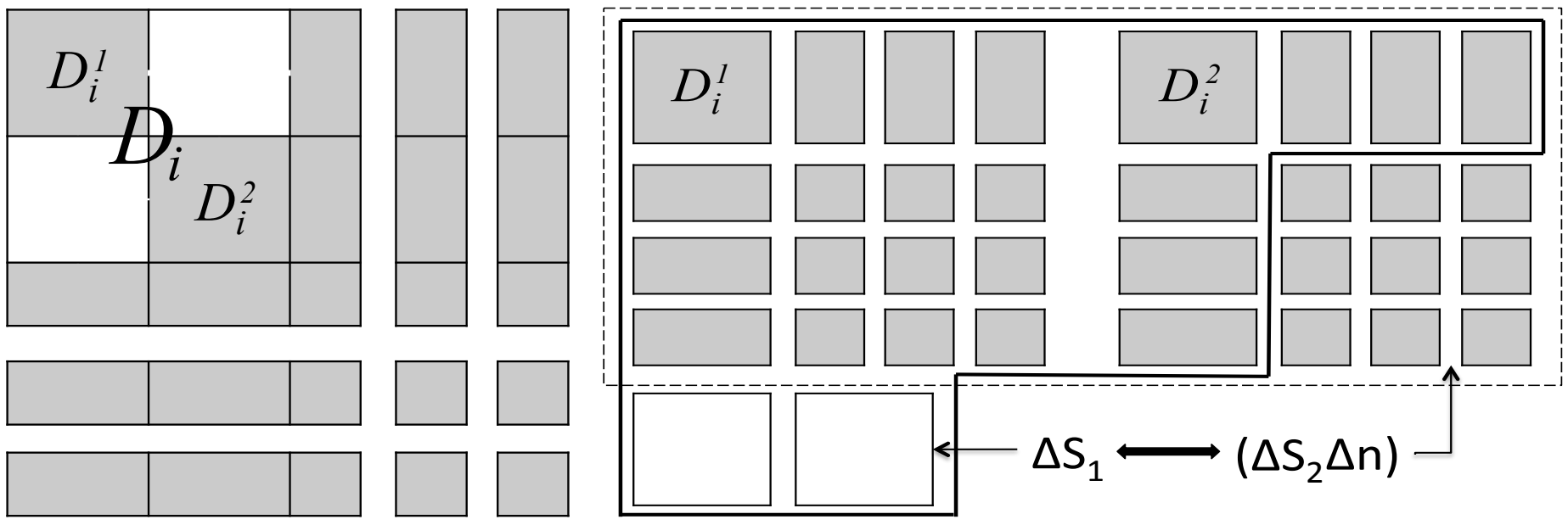

(a) $D_i$ 对应的子矩阵 $\tilde{X}_i$　　(b) $D_i^1$ 和 $D_i^2$ 所对应的子矩阵 $\tilde{X}_i^1$ 和 $\tilde{X}_i^2$

图 3.8　对角块分割、重排序，以及新的子矩阵拼接过程示例，其中阴影为非零值部分

由于 $s - \Delta s > 0$，我们有如下的性质：

$$\Delta\rho > 0 \leftrightarrow s\Delta n + n\Delta s = s\Delta n + n(\Delta s_1 - \Delta s_2) > 0 \tag{3-32}$$

如果 $\Delta s > 0$, 那么公式 (3-32) 自然得到成立，否则我们需要保证如下关系成立：

$$\frac{n}{s} < \frac{\Delta n}{\Delta s_2 - \Delta s_1} \tag{3-33}$$

虽然在数学上并不能得到绝对的保证，但是由于如下的性质在实际应用中往往是成立的，因此公式 (3-33) 在实际应用中也往往是成立的：

$$\frac{n}{s} < \frac{\Delta n}{\Delta s_2} < \frac{\Delta n}{\Delta s_2 - \Delta s_1} \tag{3-34}$$

直观而言，(3-34) 意味着在分割之后多出来的非零块（$\Delta s_2$）的密度要高于分割之前的 $k$ 个对角块子矩阵 $\tilde{X}_1\tilde{X}_2\cdots\tilde{X}_k$ 的平均密度，因为后者包含很多不含任何打分的零矩阵。进一步根据公式 (3-31) 可知，$\Delta s$ 越大则密度增量 $\Delta\rho$ 也越大，且优先分割面积较大的对角块使得算法最终尽可能地生成面积较为平均的子矩阵，这也是我们采用矩阵面积作为启发式规则的原因之一。在实验部分，我们将进一步指出该启发式规则几乎在所有的情况下都能保证第一个尝试分割的对角块就能够提高平均密度。

Metis 中基于点割集图分割将二部图分割为两个社区的算法其复杂度为 $O(n)$，其中 $n$ 为矩阵中非零值的个数[256]。假设矩阵 $X$ 最终被排列为包含 $k$ 对角块的多层双边块对角矩阵，其中 $k \ll n$，且算法在每轮递归时均选择面积最大的对角块进行分割，那么算法递归树的深度为 $O(\lg k)$，且算法递归树每层的计算复杂度均为 $O(n)$，因此算法5的复杂度为 $O(n \lg k)$。

## 3.3　性能评测

在本节，我们对基于双边块对角矩阵的数据协同过滤进行性能评测，主要包括双边块对角矩阵与群组结构的定性研究，以及局部化矩阵分解算法性能和预测精度的定量研究。

### 3.3.1　双边块对角矩阵与群组结构的定性研究

在本节，我们对双边块对角矩阵的群组结构进行定性研究，从而对实际用户物品评分矩阵数据中所隐含的内在群组结构给出直观的展示。我们采用了公开的 MovieLens-100K 和 MovieLens-1M 数据集[①]，以及雅虎音乐（Yahoo! Music）数据集[②]，其中雅虎音乐是目前个性化推荐研究领域规模最大的数据集，因此我们使用该数据集验证算法在较大数据规模下的性能表现。除此之外，我们还从国内著名的点评网站“大众点评”收集了一年的用户评分数据。相关统计数据如表3.1所示，其中雅虎音乐的用户打分区间为 1 ~ 10，除此之外其它数据集的用户打分区间为 1 ~ 5。

表 3.1　数据集基本统计信息

| | MovieLens-100K | MovieLens-1M | Dianping | Yahoo! Music |
|---|---|---|---|---|
| 用户数 | 943 | 6,040 | 11,857 | 1,000,990 |
| 物品数 | 1,682 | 3,952 | 22,365 | 624,961 |
| 打分数 | 100,000 | 1,000,209 | 510,551 | 262,810,175 |
| 打分数/用户 | 106.045 | 165.598 | 43.059 | 262.550 |
| 打分数/物品 | 59.453 | 253.089 | 22.828 | 420.523 |
| 矩阵密度 | 0.0630 | 0.0419 | 0.00193 | 0.000421 |

与其它数据集不同的是，大众点评是一个基于位置的服务系统，因此每一个用户物品评分都伴随着被评分物品的位置（经纬度）信息，利用该信息，我们在接下来的实验中对矩阵的块对角型给出直观的解释。

在矩阵的精确双边块对角化算法2和近似双边块对角化算法3中，一个最重要的参数为目标平均块密度 $\rho$，设置较低的目标密度将会得到较少但较大的对角块，而设置较高的目标密度将会得到较多但规模较小的对角块。图3.9给出了大众点评数据上的直观结果示例，其中我们分别在算法中设置目标密度 $\rho = 0.005$ 和 $\rho = 0.01$ 来获得最终结果。

① http://grouplens.org/datasets/movielens

② https://webscope.sandbox.yahoo.com

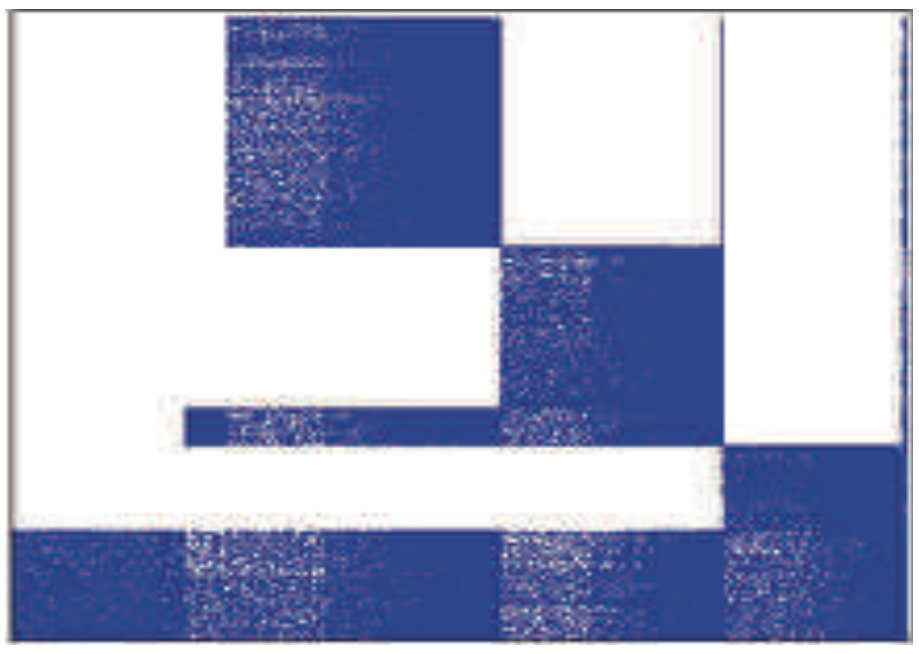

(a) 精确块对角矩阵 $\rho = 0.005$

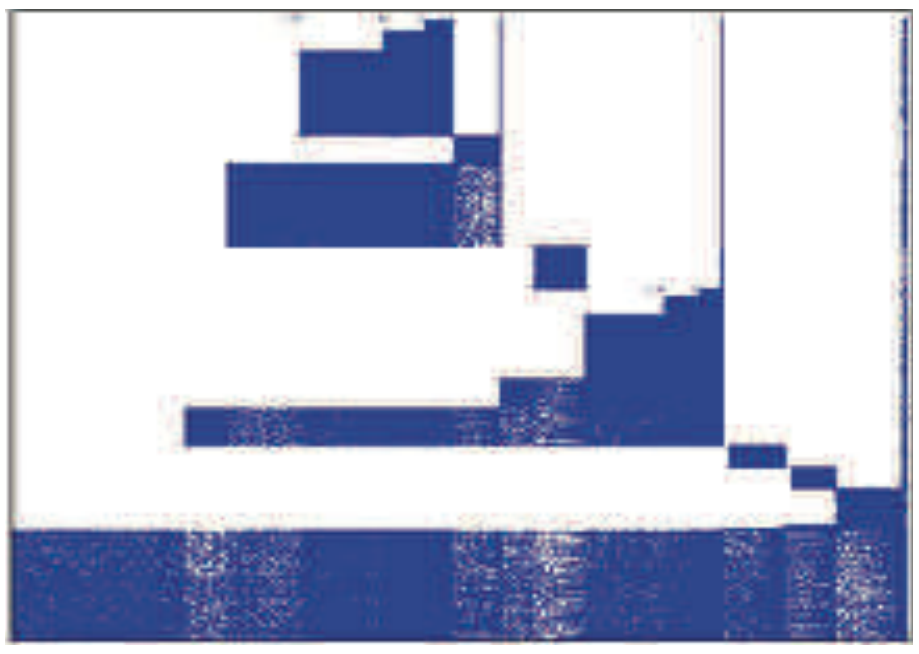

(b) 精确块对角矩阵 $\rho = 0.01$

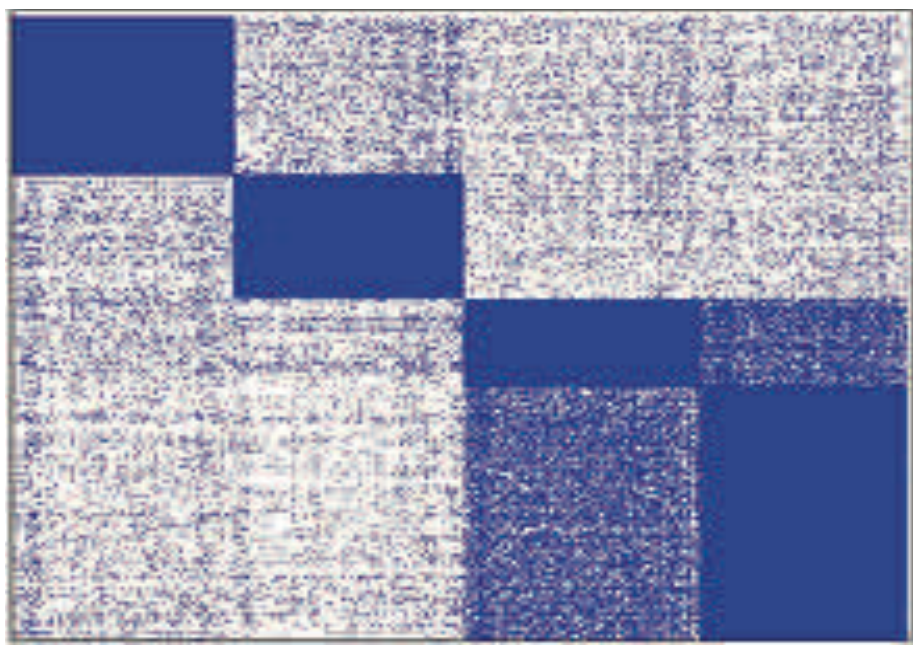

(c) 近似块对角矩阵 $\rho = 0.005$

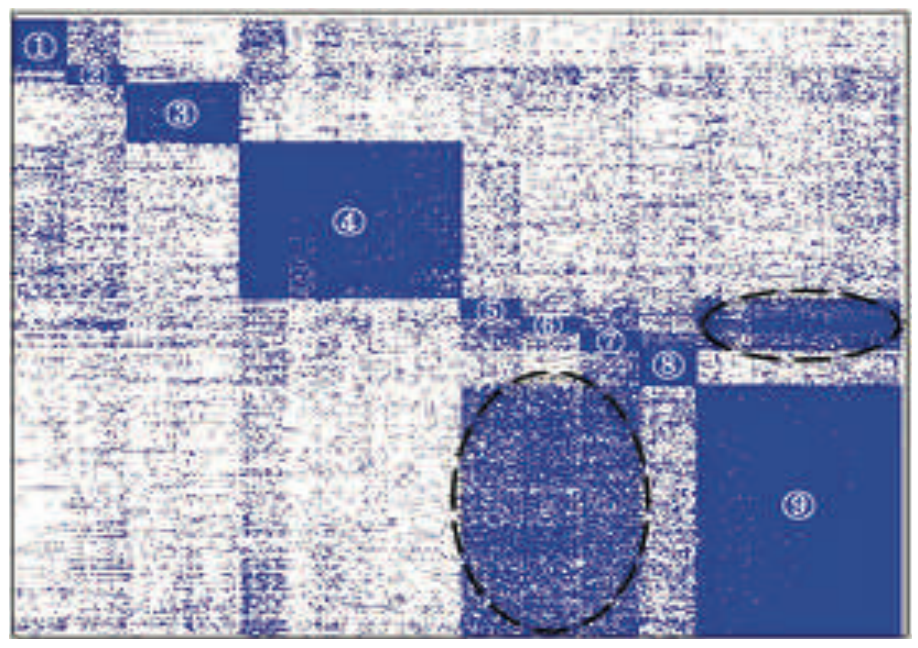

(d) 近似块对角矩阵 $\rho = 0.01$

图 3.9　大众点评数据在不同的目标密度要求下算法所给出的精确和近似双边块对角结构示例，其中图 (d) 共包含 9 个对角块，从左上角到右下角依次编号为 ① ~ ⑨，虚线所示的跨领域密集行为对应于第⑤⑥⑦和⑨个对角块

选择恰当的目标密度对于生成合适的双边块对角矩阵及在该结构上的协同过滤预测效果具有重要的作用。如果目标密度过低，则隐藏在矩阵中的用户物品群组结构就无法被完全挖掘出来；而如果目标密度过高，则会形成很多小而杂的零散群组。不管是哪种情况，都可能造成无法正确抽取和构造具有相似行为和兴趣偏好的用户群，从而无法提高甚至是降低协同过滤预测效果。

恰当的目标密度选择可以使算法给出有合理实际意义的用户物品群组。如图3.9(d)的近似双边块对角矩阵中，我们的算法从原本混为一体的用户物品评分中抽取和构建了 9 个对角块，从左上角到右下角分别用编号①到⑨表示。有趣的是，我们发现这 9 个对角块分别对应于中国的 9 个主要城市，分别为北京、上海、广州、深圳等等，如表3.2所示，其中每个对角块的精度均接近或超过 90%。

可以发现群组⑤⑥⑦和⑨之间相比于其它群组之间具有更为密集的跨群组的行为历史记录：矩阵中横向的跨群组用户行为表示来自某一城市的用户经常前往其它城市进行消费，而矩阵中的纵向跨群组用户行为则表示某餐厅经常被来自其它城市的用户所访问。有趣的是，我们通过实际分析发现该现象所涉及的四个城市分别为上海、苏州、杭州和南京，如图3.10所示。由于同为长三角经济区的主要城市，这些城市之间的交通非常便利、经济关系也非常紧密，因而带来了更多的跨

表 3.2　大众点评数据 $\rho = 0.005$ 条件下近似双边块对角矩阵各社区统计分析数据，其中精度为每个对角块中确实属于相应城市的物品（餐厅）的比例

| 社区编号 | 1 | 2 | 3 | 4 | 5 | 6 | 7 | 8 | 9 |
|---|---|---|---|---|---|---|---|---|---|
| 对应城市 | 成都 | 深圳 | 天津 | 北京 | 南京 | 苏州 | 杭州 | 广州 | 上海 |
| 用户数 | 323 | 288 | 922 | 2903 | 684 | 262 | 295 | 845 | 4531 |
| 物品数 | 1189 | 1359 | 2548 | 5011 | 1327 | 1443 | 1309 | 1274 | 4586 |
| 精度 | 89.6 | 90.4 | 92.6 | 94.4 | 91.6 | 88.7 | 89.2 | 90.7 | 91.1 |

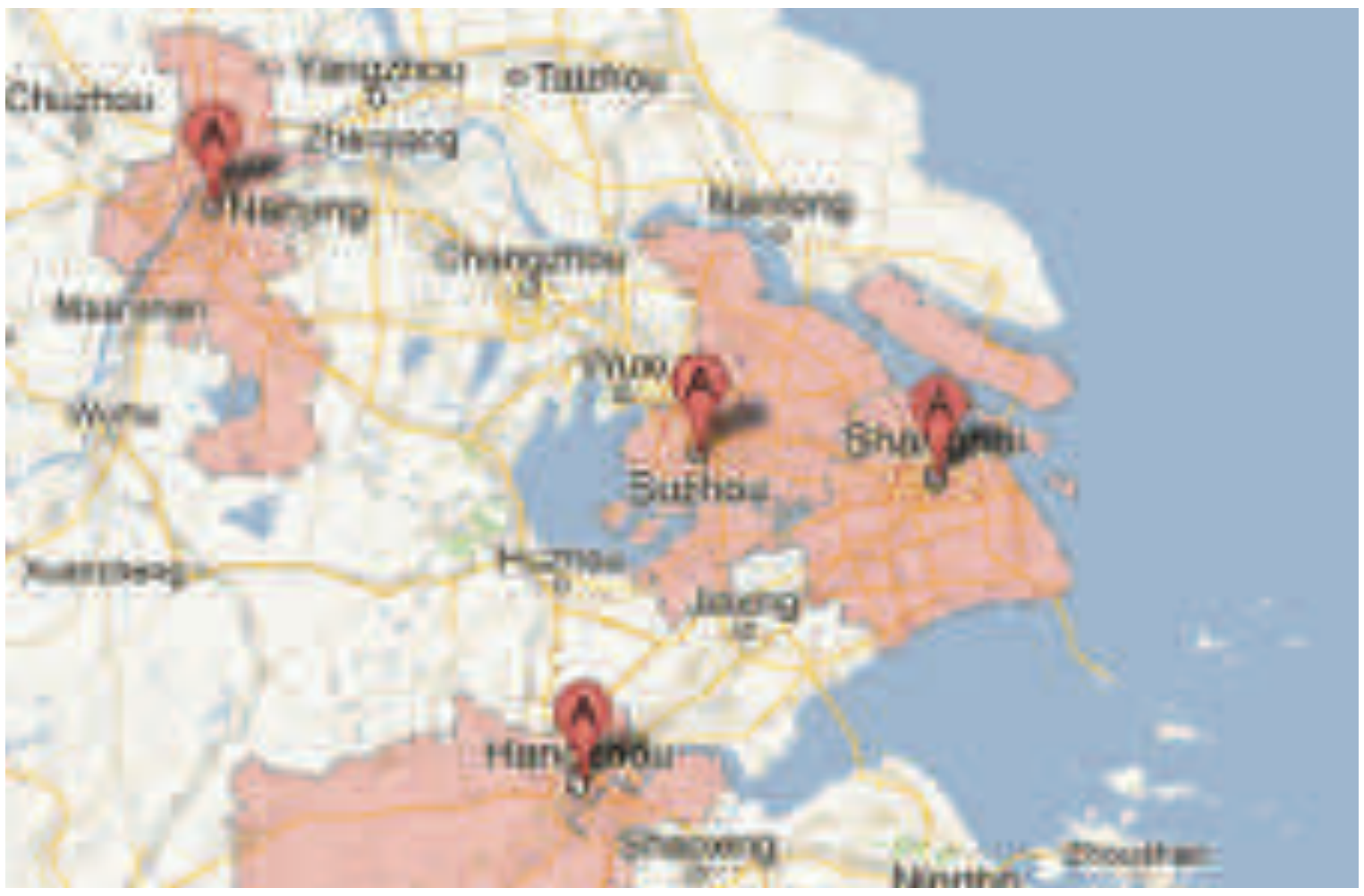

图 3.10　长三角地区城市群示例，其中标红部分分别为上海、苏州、杭州、南京四城市

群组用户行为。

图3.11描述了四个数据集上执行精确双边块对角化（算法2）和近似双边块对角化（算法3）时，所设定的目标密度 $\rho$ 与最终所获得的对角块个数的关系。由图可见，随着目标密度的提升，精确双边块对角化算法所生成的对角块个数首先会增加，然而当目标密度达到某一水平时，对角块个数会趋于稳定，这是由于当矩阵的对角块密度无法进一步被提升时，无论目标密度为多少算法都不再将对角块进行分割。然而对于近似双边块对角算法而言，对角块个数随着目标密度的提升而持续增加，这是由于即使对角块的密度无法通过分割而得到提升，算法也可以通过将低密度的行或列向量移除到矩阵的双边来进一步提升密度，从而使算法持续执行，并最终达到目标平均密度。

### 3.3.2　局部化矩阵分解算法性能及预测精度

本节对基于双边块对角矩阵的协同过滤进行定量的评测。具体地，我们对局部化矩阵分解框架在不同矩阵分解算法下的效果进行评测。

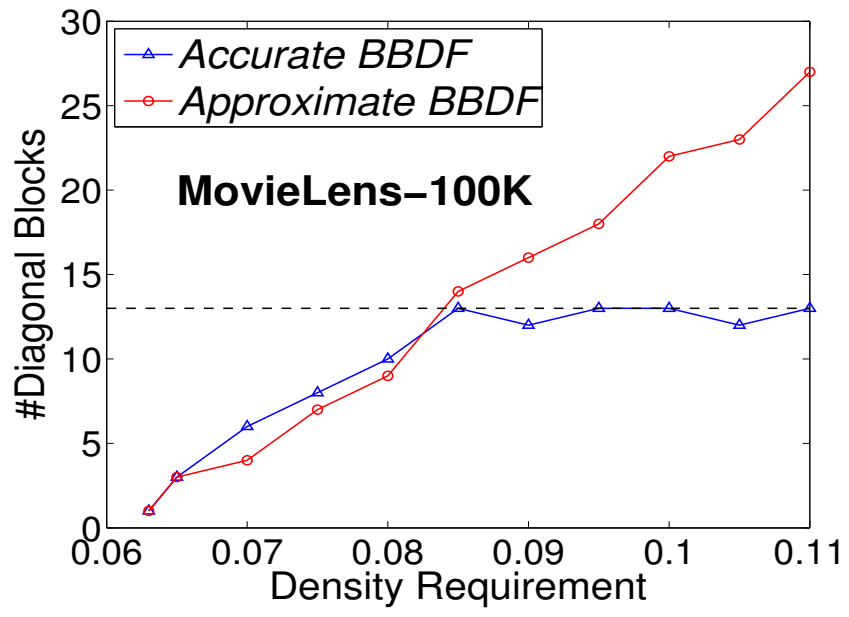

(a) MovieLens-100K 数据集

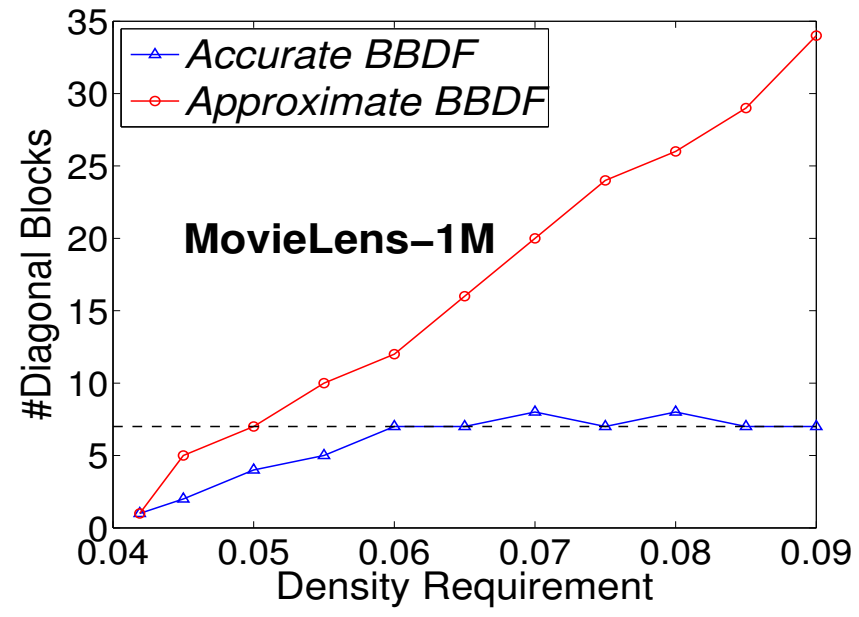

(b) MovieLens-1M 数据集

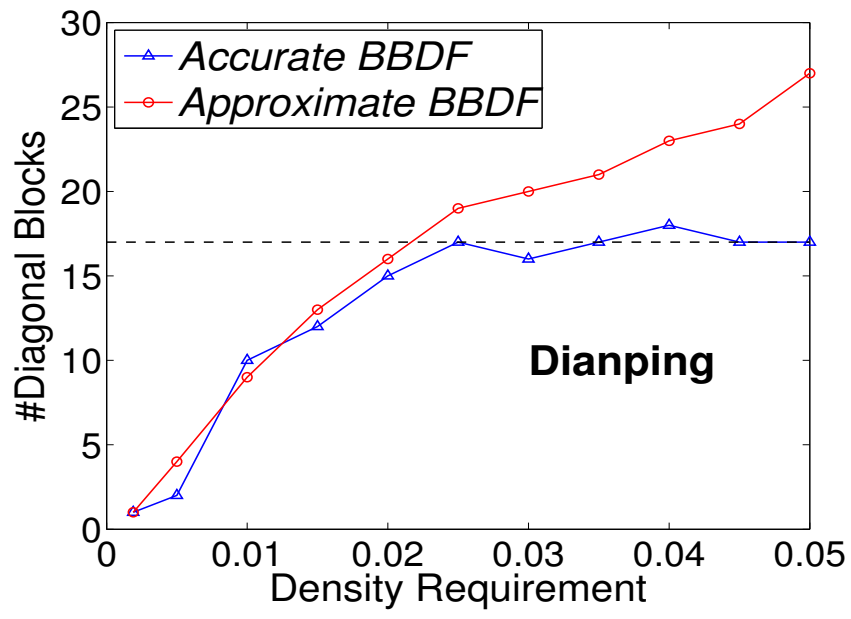

(c) 大众点评数据集

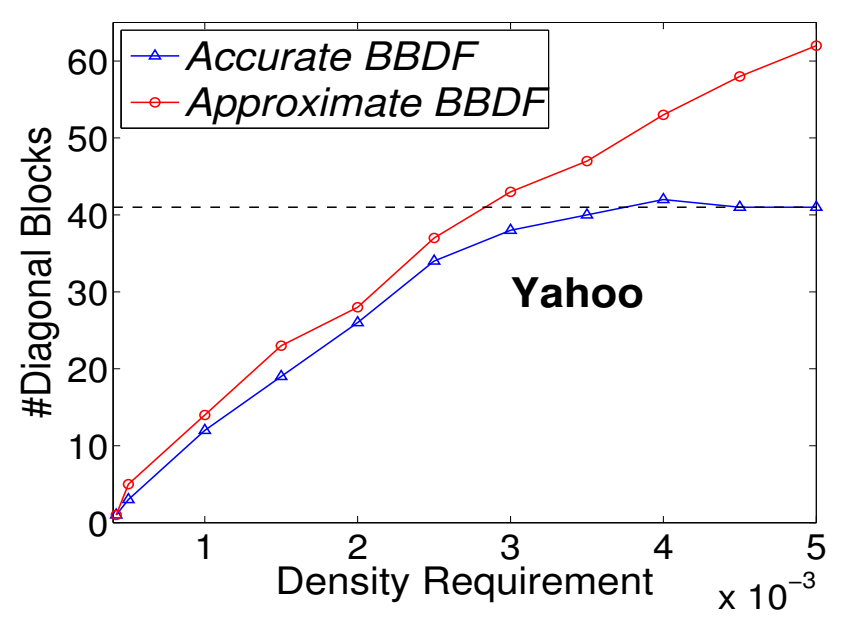

(d) 雅虎音乐数据集

图 3.11　不同数据集下精确和近似双边块对角化算法中目标密度与最终对角块个数关系图，其中横坐标表示算法所设定的目标密度 $\rho$，纵坐标表示最终形成的对角块个数

### 平衡双边块对角化算法分析

首先，我们分析平衡双边块对角化过程（算法5）中基于矩阵面积的启发式规则的合理性。我们采用首次命中率 FCHR（First Choice Hit Rate）作为评测指标：

$$\text{FCHR} = \frac{\text{所有递归中}D_{s_1}\text{被最终用来分割的次数}}{\text{算法总的递归次数}} \tag{3-35}$$

它表示在每轮递归按照面积排序的对角块列表 $[D_{s_1}D_{s_2}\cdots D_{s_k}]$ 中，我们直接选择第一个对角块进行分割就能够提升平均块密度的情况的百分比，在这些情况下，算法不需要对剩下的对角块进行分割，FCHR 越高，表明我们的启发式规则越有效；FCHR 的最大值为 1，表示在所有的情况下算法只需要对第一个对角块直接分割就可以提升平均密度，此时算法没有任何额外的无效计算。

FCHR 与目标密度之间的关系如图3.12中虚线所示，图中实线表示相应生成的对角块个数。需要指出的是，与图3.11不同，由于平衡双边块对角算法试图保证最终拼接得到的用于矩阵分解的对角块子矩阵规模相似，因此算法所使用的目标密度是指从生成的双边块对角矩阵中拼接而得到的对角块子矩阵 $\tilde{X}_i$ 的平均密度，而不是双边块对角矩阵中对角块 $D_i$ 的平均密度，因此该图中对角块个数随着目标密度的提高而持续增加。

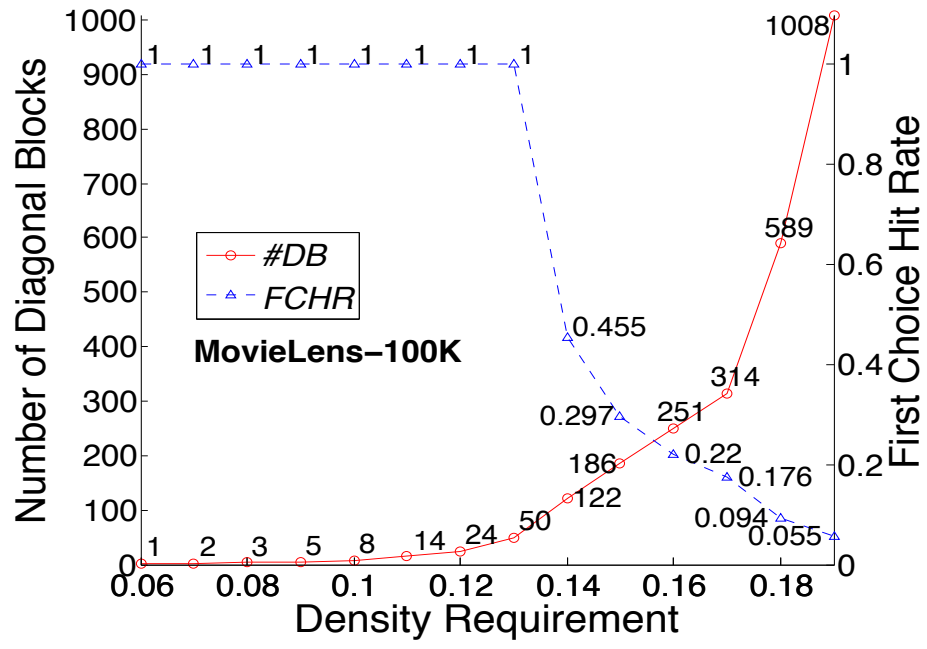

(a) MovieLens-100K 数据集

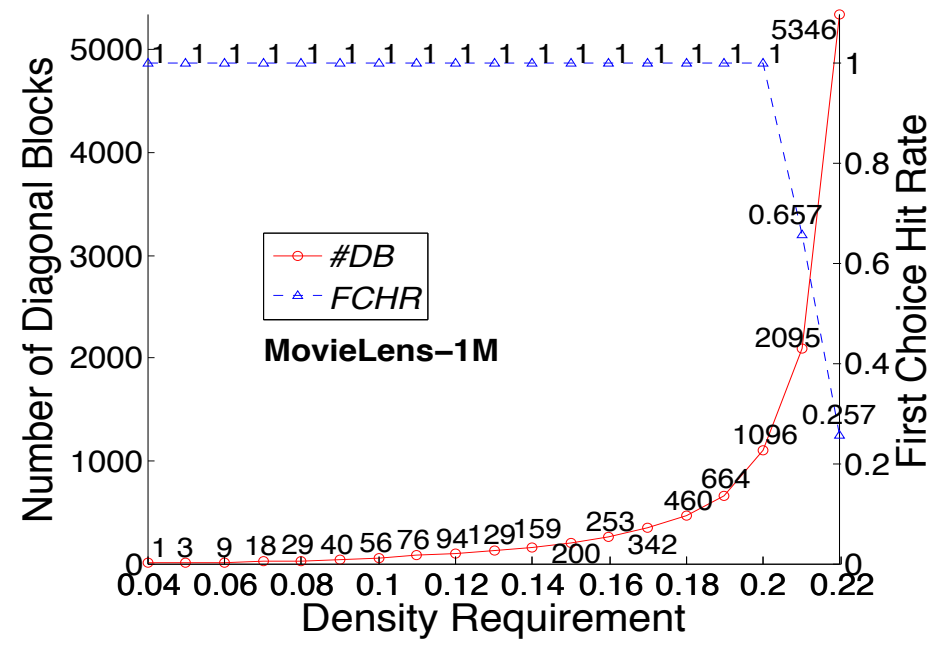

(b) MovieLens-1M 数据集

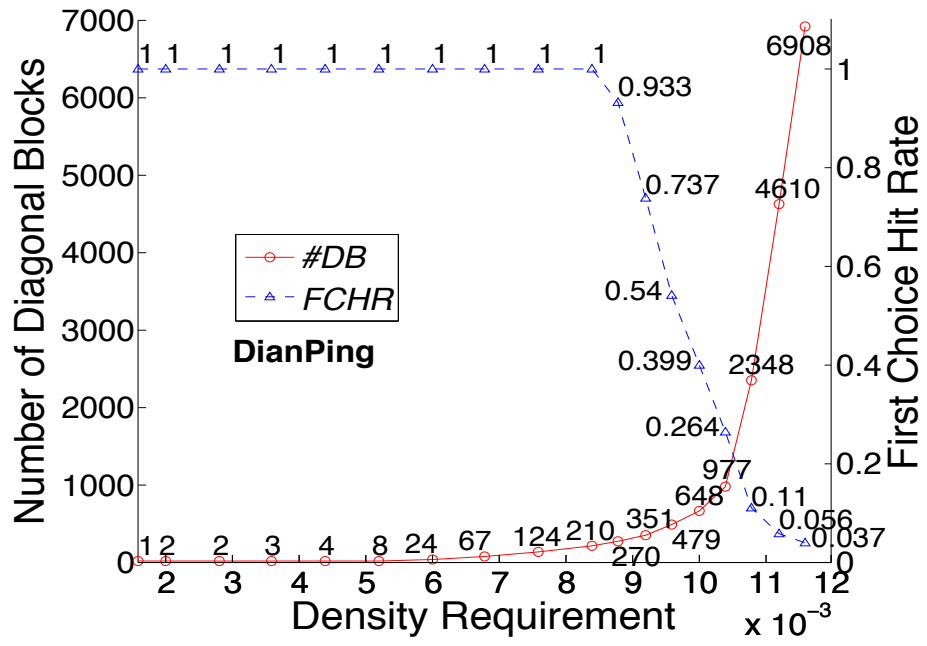

(c) 大众点评数据集

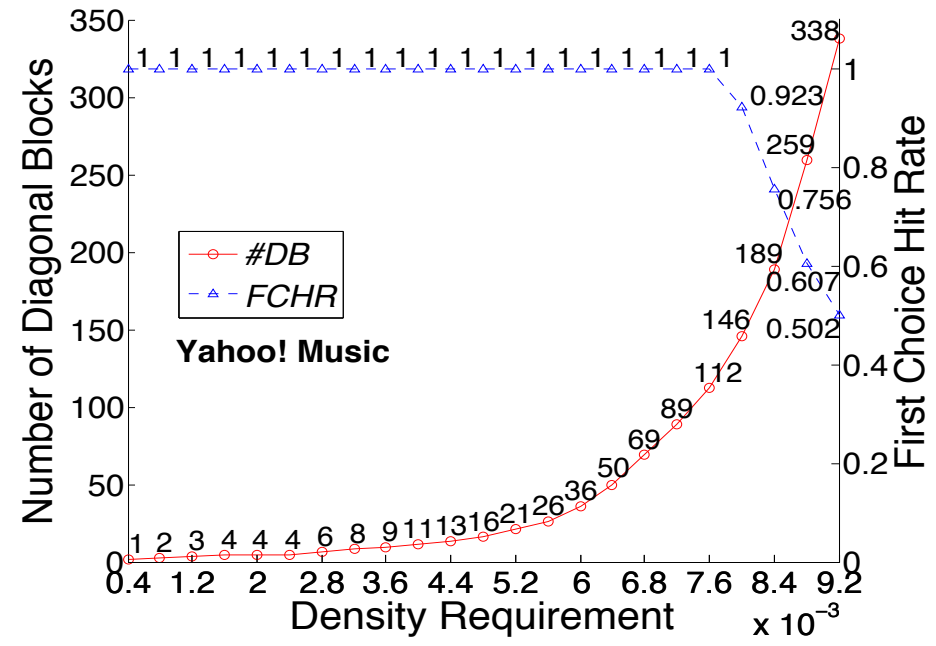

(d) 雅虎音乐数据集

图 3.12　各个数据集中目标密度与对角块个数和首次命中率之间的关系，其中横坐标为目标密度、左侧纵坐标为最终对角块密度、右侧纵坐标首次命中率；与图3.11不同，这里的目标密度是指从生成的双边块对角矩阵中拼接而得到的对角块子矩阵 $\tilde{X}_i$ 的平均密度，而非双边块对角矩阵中对角块 $D_i$ 的平均密度，因此该图中对角块个数随着目标密度的提高而持续增加。

由图3.12可见，在所有四个数据集上，当目标密度较小时我们有 FCHR = 1，而当目标达到一定值之后随着目标目标密度的进一步增加 FCHR 开始下降，这与3.2.6小节中对平衡矩阵块对角化算法的理论分析相一致。然而在实际应用中，我们一般只需要将原始矩阵分割为若干个或十几个对角块进行并行化矩阵分解，而不需要上百个甚至上千个对角块，有图3.12可见，在我们关心范围内，FCHR = 1 总是可以得到保证，因此算法所使用的启发式规则在实际应用中是有效的。

表 3.3　平衡双边块对角化算法计算时间与最终块对角个数 $k$ 之间的关系

| 对角块个数 $k$ | 5 | 10 | 15 | 20 | 50 | 100 | 150 | 200 |
|---|---|---|---|---|---|---|---|---|
| ML-100K / ms | 160 | 180 | 196 | 208 | 224 | 340 | 422 | 493 |
| ML-1M / s | 4.45 | 5.61 | 6.25 | 6.76 | 8.31 | 9.51 | 10.25 | 10.74 |
| DianPing / s | 6.01 | 9.69 | 11.61 | 12.84 | 14.64 | 15.06 | 16.18 | 16.95 |
| Yahoo! Music / min | 8.03 | 9.54 | 10.95 | 12.08 | 17.67 | 21.83 | 23.35 | 24.73 |

我们进一步分析平衡双边块对角化算法的计算性能，实验在 64G 内存 8 核 3.1GHz 服务器上进行，我们调节目标密度以获得期望的对角块个数，并记录模型

训练的时间，结果如表3.3所示。在该实验中我们可以发现，随着对角块个数的增多，算法运行时间逐渐增加，但同时对角块的规模也不断减小，因此算法分割每一个对角块的时间也逐渐减小、算法运行越来越快。这为算法处理大规模矩阵数据时的可扩展性提供了保证。

### 局部化矩阵分解预测精度分析

为了对局部化矩阵分解框架的协同过滤预测精度进行评价，我们采用如下基线算法进行研究：

- SVD：奇异值矩阵分解，我们采用 Frobenius 范数，并采用交替最小化方差（Alternating Least Squares）算法进行模型训练和优化[59]。
- NMF：非负矩阵分解[62]，同样采用 Frobenius 范数正则化。
- PMF：概率化矩阵分解，我们采用基于马尔科夫蒙特卡洛（MCMC）的贝叶斯概率分解模型[68]。
- MMMF：最大间隔矩阵分解，我们采用了快速最大间隔矩阵分解算法[64]。

我们采用最常用的根均方差（Root Mean Square Error，RMSE）作为评测指标对矩阵打分的预测精度进行评测，对于测试集中的 $N$ 个真实值 $r_i$，设它们各自的预测值为 $\hat{r}_i$，则 RMSE 为：

$$\mathrm{RMSE} = \sqrt{\frac{\sum_{i=1}^{N}(r_i - \hat{r}_i)^2}{N}} \tag{3-36}$$

对于 MovieLens 和点评数据集，我们采用五折交叉验证的方法进行实验，每次选择 80% 的数据作为训练集，并在剩下的 20% 数据上进行测试并计算 RMSE，最终对五次实验结果计算平均值。由于雅虎音乐数据集已经被分割为了训练集和测试集，我们直接用相应的训练集进行模型训练，并在测试集上进行测试。

分解因子的个数 $r$ 在矩阵分解算法中起着重要的作用，当 $r$ 过小时，模型复杂度不足以较好地近似原始矩阵，而当 $r$ 过大时，模型训练的耗时显著增加，且容易带来过拟合的问题。由于抽取出来的子矩阵 $\tilde{X}$ 和原始矩阵 $X$ 的规模显著不同，因此我们需要对分解因子个数 $r$ 对预测精度的影响进行深入的分析。

在本实验中，我们利用 MovieLens-1M 数据集（其它数据集上具有相似的观测结果），并设置目标密度 $\hat{\rho} = 0.055$，此时原始矩阵被排列为包含 4 个对角块的多层双边块对角矩阵，并从中抽取和拼接成包含 4 个对角块的块对角矩阵 $\tilde{X} = \mathrm{diag}(\tilde{X}_1, \tilde{X}_2, \tilde{X}_3, \tilde{X}_4)$。表3.4描述了四个对角块的基本统计信息。

我们以步长 5 为单位从 5 到 100 调节 $r$ 的取值，对于每一个矩阵分解算法，我

表 3.4　MovieLens-1M 矩阵所抽取的四个对角块矩阵相关统计数据

| | $\tilde{X}_1$ | $\tilde{X}_2$ | $\tilde{X}_3$ | $\tilde{X}_4$ |
|---|---|---|---|---|
| 用户数 | 1,507 | 1,683 | 1,743 | 1,150 |
| 物品数 | 2,491 | 3,108 | 3,616 | 3,304 |
| 打分数 | 118,479 | 259,665 | 462,586 | 192,267 |
| 密度 | 0.0316 | 0.0496 | 0.0734 | 0.0506 |

们执行如下的两组实验。首先，我们在原始矩阵 $X$ 上直接利用 $r$ 个分解因子来进行矩阵分解，并计算相应的 RMSE 值；其次，我们在局部化矩阵分解框架下在如上所述的 4 个对角块上分别利用 $r$ 个维度进行矩阵分解，并对重复的矩阵块求平均得到最终的预测结果，同样，我们计算预测值的 RMSE。我们在矩阵 $X$ 上采用交叉验证确定每个矩阵分解算法的最优正则化系数参数，对于 SVD[59] 和 NMF[62] 算法我们最终设定正则化系数 $\lambda = 0.065$；对于 PMF[68] 有正则化系数 $\lambda_U = \lambda_V = 0.002$；对于 MMMF[64]，正则化常数 $C = 1.5$（各符号与相应算法论文原文相一致）。RMSE 与分解因子个数 $r$ 的关系如图3.13所示。

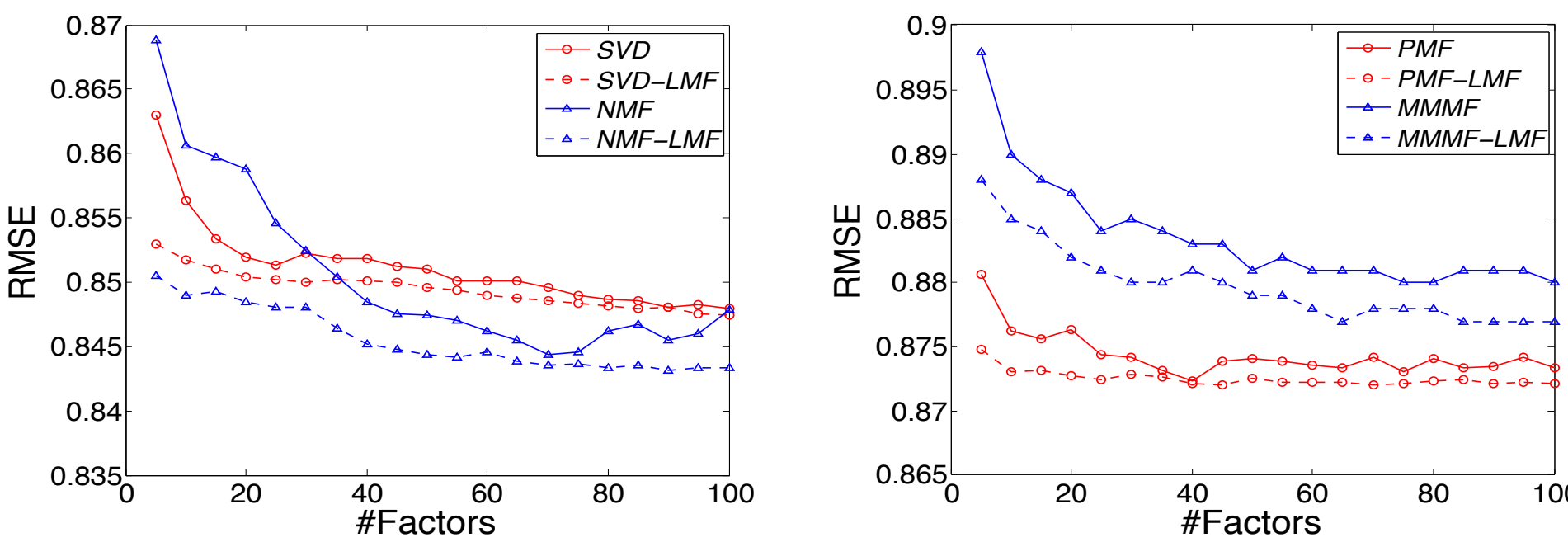

(a) SVD 与 NMF 算法　　(b) PMF 与 MMMF 算法

图 3.13　原始矩阵与局部化矩阵分解框架（LMF）下 RMSE 与分解因子个数 $r$ 的关系示意，其中对于每个算法，实线表示在原始矩阵上执行相应算法对应的预测精度，虚线表示在局部化矩阵分解框架下的预测精度

实验结果显示，局部化矩阵分解框架能够帮助矩阵分解算法获得更高的预测精度，进一步，当我们所使用的分解因子个数越少时，预测效果的提升越明显，这是由于当使用较少的分解因子时，算法不能较好地拟合原始的大规模矩阵，然而对于局部化矩阵分解框架下抽取出来的小规模矩阵却足以给出较好的拟合。由于在使用较少分解因子的情况下就能够获得同样甚至更好的预测效果，使得算法的模型复杂度和训练时间大大降低，这也是局部化矩阵分解框架的优势所在。

我们进一步分析不同的目标密度设置对预测精度的影响。在本实验中，我们设置不同的目标密度 $\hat{\rho}$，从而得到包含不同个数对角块的块对角矩阵 $\tilde{X}$，同样在局部化矩阵分解框架下进行预测并计算 RMSE。根据上一实验的结果，我们发现使

用 60 个分解因子就足以获得较为理想的预测精度，因此在本实验中我们将分解因子的个数固定为 60。正则化系数的设定与上一实验相同。

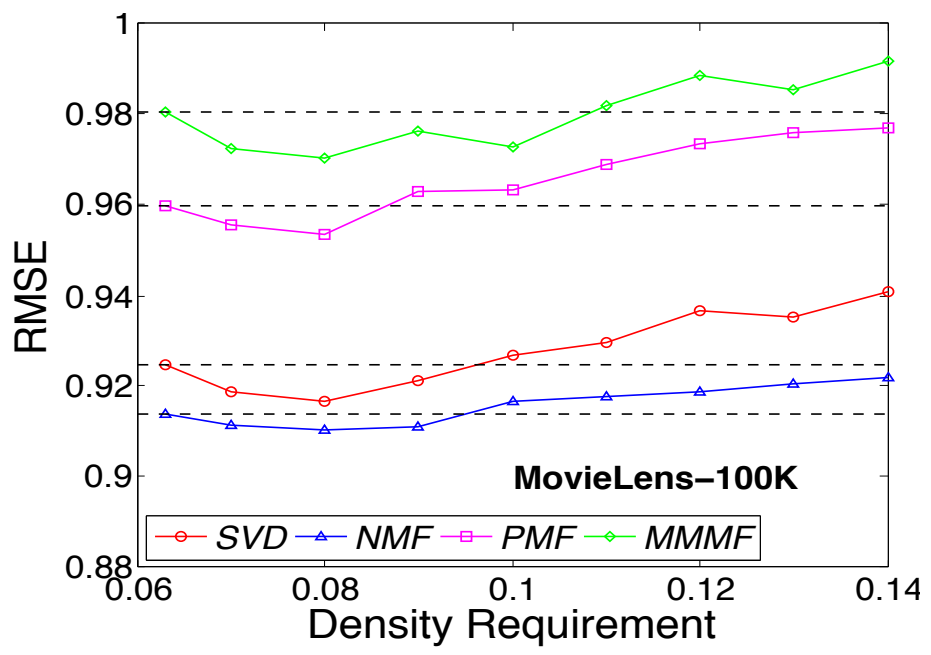

(a) MovieLens-100K 数据集

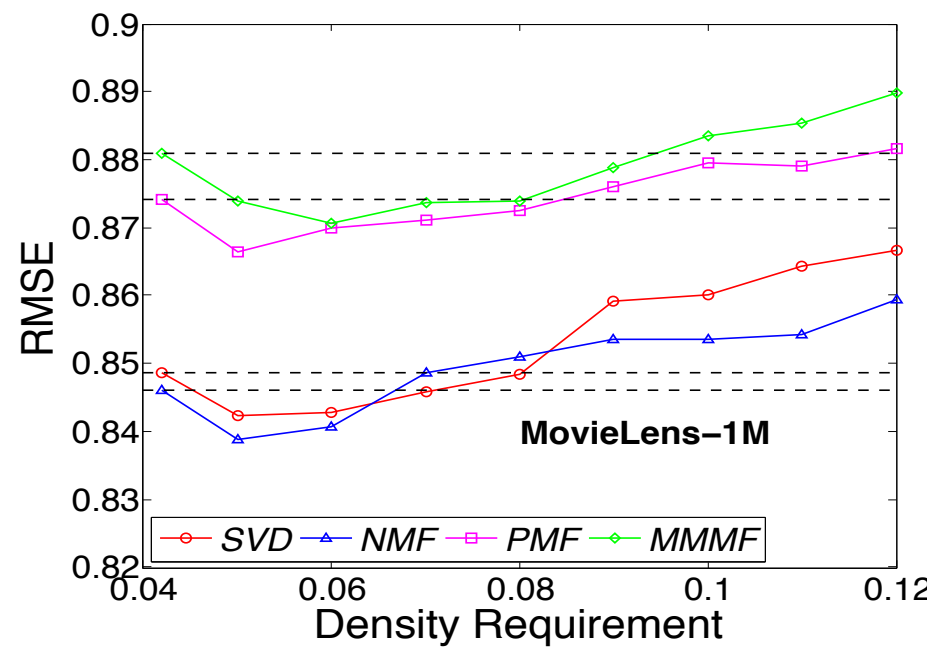

(b) MovieLens-1M 数据集

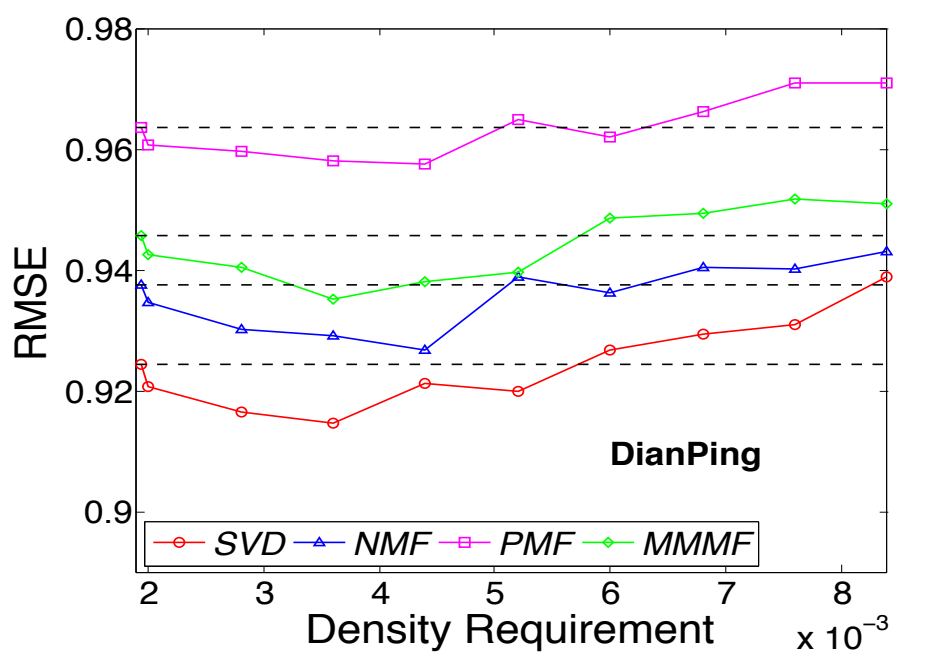

(c) 大众点评数据集

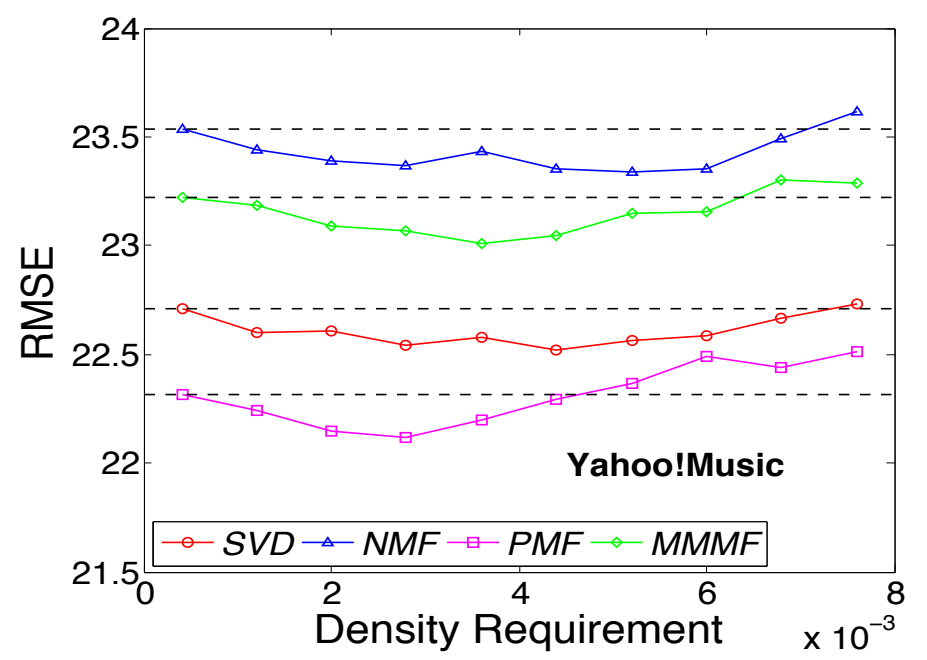

(d) 雅虎音乐数据集

图 3.14　各个数据集上 RMSE 随平衡双边块对角化算法目标密度设置的变化示意图，其中的虚线表示直接在原始矩阵上进行矩阵分解所对应的 RMSE，此时等价于将目标密度设为原始矩阵本身的密度，因此原始矩阵不需要重排列就可以满足密度要求。

不同目标密度设定 $\hat{\rho}$ 条件下各个数据集上 RMSE 的最终值如图3.14所示。在每一个子图中，四条曲线分别代表四种矩阵分解算法，包括 SVD、NMF、PMF、MMMF，每条曲线的第一个点所对应的目标密度 $\hat{\rho}$ 为该矩阵本身的密度，在该点上，平衡双边块对角化算法实际上没有对矩阵做任何调节，因此该点所对应的 RMSE（每条曲线所伴随的虚直线）是不使用我们的局部化矩阵分解框架而在原始矩阵上直接进行矩阵分解的基线效果。在图中，每条曲线在虚线以下的部分表示 RMSE 的减小，亦即预测精度的提升，反之依然。

实验结果显示，在选择合适的目标密度时，我们的局部化矩阵分解框架可以帮助矩阵分解算法提升预测精度，而当目标密度过高时，预测精度相对于基线算法反而会降低。具体而言，当目标密度较小时，预测精度随着目标密度的增加而提高，而当目标密度的设置过高时精度开始下降，这主要是由于此时平衡双边块对角化算法会形成过多小而零散的对角块矩阵，从而影响矩阵分解算法对用户偏好和物品属性的正确建模。表3.5显示了 MovieLens-1M 数据集在不同目标密度 $\hat{\rho}$ 下

所形成的对角块矩阵 $\tilde{X}$ 的平均用户和物品数，可见当 $\hat{\rho} \geq 0.1$ 时对角块的平均用户数仅有几百个。

表 3.5　MovieLens-1M 数据集在不同目标密度 $\hat{\rho}$ 下对角块 $\tilde{\mathbf{X}}$ 平均用户数和物品数统计

| 目标密度 $\hat{\rho}$ | 0.045 | 0.052 | 0.060 | 0.069 | 0.081 | 0.102 | 0.129 | 0.160 |
|---|---|---|---|---|---|---|---|---|
| 对角块个数 $k$ | 2 | 4 | 8 | 16 | 32 | 64 | 128 | 256 |
| 平均用户数 | 3020 | 1520 | 779 | 409 | 220 | 128 | 82 | 61 |
| 平均物品数 | 3170 | 3129 | 3055 | 3064 | 3015 | 3007 | 3015 | 3030 |

然而，在目标密度较大的取值范围内，我们的局部化矩阵分解框架可以帮助原始矩阵分解算法提高预测精度。将该实验结果与图3.12的实验结果相结合可见，在实际应用中我们既不需要也没有必要设定过高的目标密度，较低的合适的目标密度一方面可以保证基于矩阵面积的启发式规则的有效性，另一方面可以生成合适个数的（几个或十几个）对角块矩阵以方便并行化处理，同时进一步还可以提高矩阵分解算法的预测精度。

表 3.6　局部化矩阵分解框架下算法所能取得的最好预测效果及其所对应的目标密度 $\hat{\rho}$ 和对角块个数 $k$

| 分解算法 | MovieLens-100K | | | | MovieLens-1M | | | |
|---|---|---|---|---|---|---|---|---|
| | baseline | $\hat{\rho}$ | $k$ | RMSE | baseline | $\hat{\rho}$ | $k$ | RMSE |
| SVD | 0.9249 | 0.08 | 3 | 0.9165 | 0.8487 | 0.05 | 3 | 0.8423 |
| NMF | 0.9138 | 0.08 | 3 | 0.9102 | 0.8461 | 0.05 | 3 | 0.8388 |
| PMF | 0.9598 | 0.08 | 3 | 0.9534 | 0.8741 | 0.05 | 3 | 0.8664 |
| MMMF | 0.9807 | 0.08 | 3 | 0.9703 | 0.8810 | 0.06 | 9 | 0.8740 |
| 分解算法 | DianPing | | | | Yahoo!Music | | | |
| | baseline | $\hat{\rho}$ | $k$ | RMSE | baseline | $\hat{\rho}$ | $k$ | RMSE |
| SVD | 0.9244 | 0.0036 | 3 | 0.9145 | 22.713 | 0.0044 | 13 | 22.519 |
| NMF | 0.9376 | 0.0044 | 4 | 0.9267 | 23.538 | 0.0052 | 21 | 23.335 |
| PMF | 0.9636 | 0.0044 | 4 | 0.9575 | 22.312 | 0.0028 | 6 | 22.121 |
| MMMF | 0.9457 | 0.0036 | 3 | 0.9352 | 23.218 | 0.0036 | 9 | 23.007 |

表3.6描述了在每一个数据集上我们的局部化矩阵分解框架所达到的最好预测精度及其对应的目标密度和对角块个数。在实验中，我们仍然在 MovieLens 和 DianPing 数据集上采用五折交叉验证的方法，在雅虎音乐数据集上，由于已经分好了训练集和测试集，我们在同样的数据集上执行五次算法，并求结果的平均值。最终，MovieLens 和 DianPing 数据集（打分范围为 1~5）上的标准差 $\leq 0.002$，在雅虎音乐数据集（打分范围为 1~10）上标准差 $\leq 0.01$。由结果可见，在目标密度

合适的最好情况下我们的局部化矩阵分解算法都可以提升预测精度，且原始矩阵越稀疏，提升效果越好。这主要是由于两方面的原因：首先，从原始矩阵抽取出的子矩阵具有更高的密度，因而在一定程度上缓解矩阵分解算法的数据稀疏性问题，提高预测效果；其次，由于矩阵内在的结构，使得抽取出的子矩阵中用户具有更为相似的兴趣偏好，因而提升协同过滤的效果。

### 并行化效率分析

局部化矩阵分解框架的一个重要优势在于，一旦块对角矩阵 $\tilde{X} = \mathrm{diag}(\tilde{X}_i)$ 被构建好，那么各个对角块 $\tilde{X}_i$ 的分解和预测就可以同时并行地执行。根据定理3.2的可拆分性，在各个对角块上学习各自的分解因子矩阵 $(U_i, V_i)$ 是互相独立的，因此我们不需要设计复杂的并行化和进程通信算法，而只需要简单地使用独立的线程去分解每一个对角块即可，这大大降低了系统复杂度、提高算法的可扩展性。

本部分实验包含三个步骤：首先，我们将矩阵 $X$ 调节为一个多层双边块对角结构，并从中构建一个块对角矩阵 $\tilde{X} = \mathrm{diag}(\tilde{X}_i)(1 \le i \le k)$，由于我们的实验机包含 8 个内核，我们将原始矩阵分割为 8 个对角块；进而，我们利用多线程将每一个对角块在一个独立的内核上进行分解，使得它们彼此之间互不干扰；最后，我们利用每一个对角块的分解结果 $(U_i, V_i)$ 来近似和预测原始矩阵。

我们记录每一个步骤所消耗的时间（在第二个步骤中，我们记录耗时最长的一个对角块所使用的时间），并将三个步骤的总耗时作为我们的局部化矩阵分解框架的耗时。各个矩阵分解算法所使用的分解因子个数同样为 60，算法正则化系数的取值与上一节相同，实验结果如表3.7所示，其中“Base”表示直接对原始矩阵

表 3.7　八对角块并行下算法的耗时及加速比

| 分解算法 | MovieLens-100K | | | MovieLens-1M | | |
|---|---|---|---|---|---|---|
| | Base | LMF | 加速比 | Base | LMF | 加速比 |
| SVD | 23.9s | 7.7s | 3.10 | 184.9s | 43.4s | 4.26 |
| NMF | 8.7s | 3.9s | 2.23 | 86.6s | 22.1s | 3.92 |
| PMF | 43.8s | 11.6s | 3.78 | 265.1s | 60.1s | 4.41 |
| MMMF | 19.6min | 4.71min | 4.16 | 1.73h | 21.5min | 4.83 |
| 分解算法 | DianPing | | | Yahoo!Music | | |
| | Base | LMF | 加速比 | Base | LMF | 加速比 |
| SVD | 143.7s | 35.7 | 4.03 | 6.22h | 1.21h | 5.14 |
| NMF | 64.4s | 16.6s | 3.88 | 4.87h | 1.05h | 4.64 |
| PMF | 190.5s | 44.1s | 4.32 | 7.91h | 1.48h | 5.34 |
| MMMF | 48.5min | 10.2min | 4.75 | 38.8h | 6.22h | 6.24 |

$X$ 进行分解的耗时，“LMF”表示我们的局部化矩阵分解框架的耗时，加速比为两者的比值。实验结果显示，基于简单多线程并行化的局部化矩阵分解框架可以显著节省模型的计算时间，并且在大规模稀疏矩阵上加速效果更为显著，这在实际应用中具有重要的意义。

## 3.4　本章小结

本章首先对稀疏矩阵内在的用户物品群组结构进行分析，提出了矩阵的（近似）双边块对角型结构，并证明了该结构与二部图社区发现的等价性，在此基础上，我们进一步给出了基于图分割的矩阵精确和近似双边块对角化算法。基于 MovieLens、大众点评和雅虎音乐数据集的定性实验分析给出了矩阵（近似）双边块对角型在真实数据上的直观性质。

在定量分析方面，本节对双边块对角型矩阵的分解性质进行了理论分析，研究了矩阵分解算法的可拆分性，并在此基础上给出了基于双边块对角型矩阵的局部化矩阵分解算法，从而一方面充分利用社区内用户相似的兴趣偏好和行为模式而提高预测精度，另一方面通过对角块的并行分解提高计算效率。为了便于并行处理时各个对角块计算时间的平衡，我们进一步提出了平衡双边块对角化算法，从而生成规模相似的对角块子矩阵。

在四个真实数据集上的实验结果验证了目标密度参数和分解因子个数参数对预测精度的影响，实验结果显示，在合适的目标参数设定下，局部化矩阵分解框架可以在提高预测精度的同时也提高计算效率，在大规模数据集下的预测性能表现验证了基于双边块对角矩阵的协同过滤算法具有较好的数据可扩展性。

本章相关内容发表于 CCF-A 类国际会议 WWW 2013 和 SIGIR 2013，以及 CCF-B 类会议 CIKM 2014 上，被来自中国、美国、加拿大、欧洲以及新加坡的高校和科研机构引用，并作为基线算法进行比较和改进。相关成果应用于搜狗推荐、豆瓣 FM 音乐推荐等实际产品中。

# 第 4 章　模型的可解释性

在用户物品评分矩阵这一基本输入的基础上，基于矩阵分解的隐变量模型在实际系统中得到了广泛的应用。然而隐变量实际意义的缺乏使得模型学习得到的用户偏好向量与物品属性向量难以有直观的解释，而这进一步影响了推荐结果的可解释性。在本章，我们对模型的可解释性进行研究，并以此为基础给出个性化的推荐理由。本章内容主要包括显式变量分解模型及其可解释性、个性化推荐理由的构建、用户兴趣的动态化建模，以及相关的性能评测。

## 4.1　显式变量分解模型

本节介绍基于短语级情感分析的显式变量分解模型及其可解释性，内容包括基于短语级情感分析的物品属性词和用户情感词挖掘、情感词典的构建、基于属性词显式变量的多矩阵分解模型与可解释性，及属性级个性化推荐理由的构建。

### 4.1.1　本节引言

推荐的可解释性及推荐理由的构建对推荐系统具有重要作用，向用户恰当地解释为何一个物品被推荐出来，首先有助于提高推荐系统的对用户的透明度和可信性，其次可以提高推荐结果的可辨性和说服力，最后可以帮助用户更快地做出正确的决定，从而提高用户与系统交互的效率。目前，学术界和产业界已经对推荐理由的构建进行了研究和实际使用，而目前常见的推荐理由多是基于统计的非个性化推荐理由，例如在亚马逊等购物网站中常见的“购买了该产品的用户也购买了”，等等。

另一方面，基于矩阵分解的隐变量模型（如图4.1所示）由于其较高的打分预测精度和模型可扩展性而受到学术界和产业界的广泛关注和使用，目前已经成为多数实际推荐系统的基础算法。然而基于矩阵分解隐变量模型的推荐算法具有几个方面的重要问题：首先，由于变量的隐性，我们很难理解用户是如何将自己对产品各方面的态度和看法融合成一个单一的数值化打分的；其次，这进一步使得我们难以为用户给出契合特定偏好的和需求的推荐结果；最后，变量的隐性使得我们难以向用户解释为何一个物品被推荐了出来，甚至更难解释为什么其它物品没有被推荐出来，这使用户在实际系统中倾向于认为被推荐物品仅仅是为了满足商业利益的广告。可解释性的缺乏降低了系统的可信度和说服力，也降低了系统满足用户需求的潜在能力。

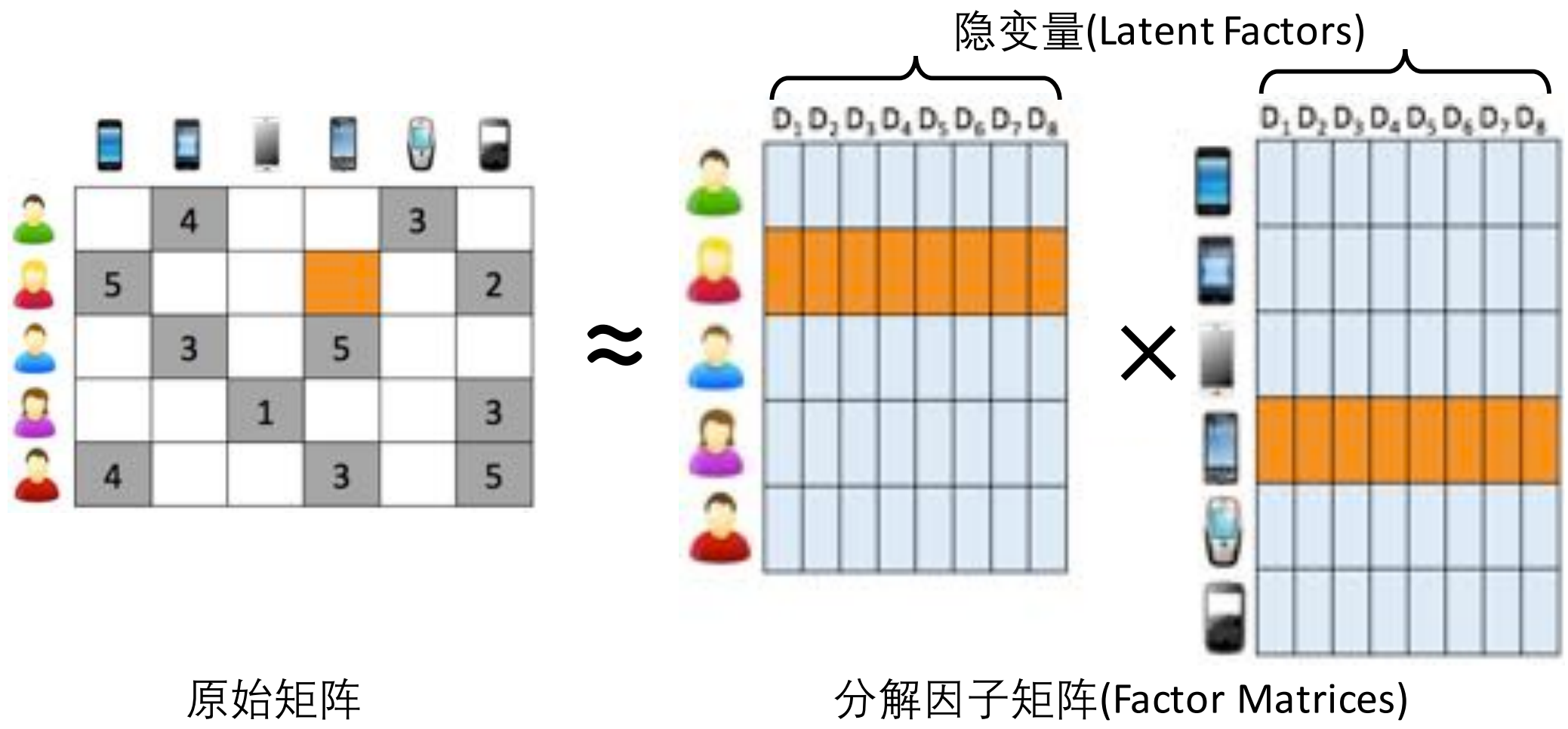

图 4.1　基于矩阵分解的隐变量模型示意，其中原始矩阵被分解为两个因子矩阵的乘积，在两个矩阵中，用户和物品分别被投影到一组共同但未知的维度（隐变量）上，从而得到用户和物品的表示向量，并利用向量乘积获得打分预测

由于如上的原因，系统人员在实践中往往面临着如下的两难困境：直观可解释的简单模型无法给出较好的预测精度和推荐结果，而复杂的隐变量模型虽然可以给出较好的预测精确、却难以对推荐结果进行直观的解释。因此，如何构造既具有高预测精度、又具有直观可解释性的模型成为重要的研究问题。

幸运的是，互联网上越来越丰富的用户文本评论数据以及情感分析技术在这些数据上的发展为相关问题的解决带来了一些新的契机。目前，多数在线购物网站（如淘宝网、亚马逊等）和在线评论系统（如大众点评、Yelp 等）允许用户在给出数值化打分的同时也给出文本评论以描述自己的观点和看法。文本评论包含了用户在对应商品上丰富的情感、观点和偏好信息[210]，而这为可解释的推荐给出了新的数据源。例如在如下的评论示例中：

**例** 4.1：　手机的样式很漂亮，但是续航时间有点短。

我们可以利用短语级情感分析技术抽取出诸如（样式，漂亮，+1）和（续航时间，短，-1）等具有（属性词，情感词，情感极性）结构的三元组[202,210,230]，其中“属性词”描述产品的特定方面，“情感词”用来描述用户在这些方面的态度观点，“情感极性”则为特定的属性词和情感词组合在一起时所对应的情感倾向性，它是一个落在 $[-1,+1]$ 区间内的数值，其中“+1”表示明确的积极正面情感、“-1”表示明确的消极负面情感，“0”表示中性情感。

即使对于同一产品，不同的用户也可能关注产品的不同属性。在实际的用户评论中，我们发现不同的用户倾向于关注和评论不同的产品属性词，即使他们为同一款产品给出了相同的打分，也有可能是在考虑不同的产品属性的基础上给出的，例如有的用户更加关注手机的屏幕性能，而有的用户则更加关注手机的续航

时间，等等。从产品评论中挖掘恰当的产品属性词和用户情感词，一方面能够帮助我们更好地了解用户不同的偏好，另一方面帮助我们设计更为智能的推荐算法和个性化的推荐理由，从而使得我们不仅可以为用户提供恰当的产品推荐结果，还可以为用户提供恰当的推荐理由来解释为什么推荐或不推荐某一个产品。

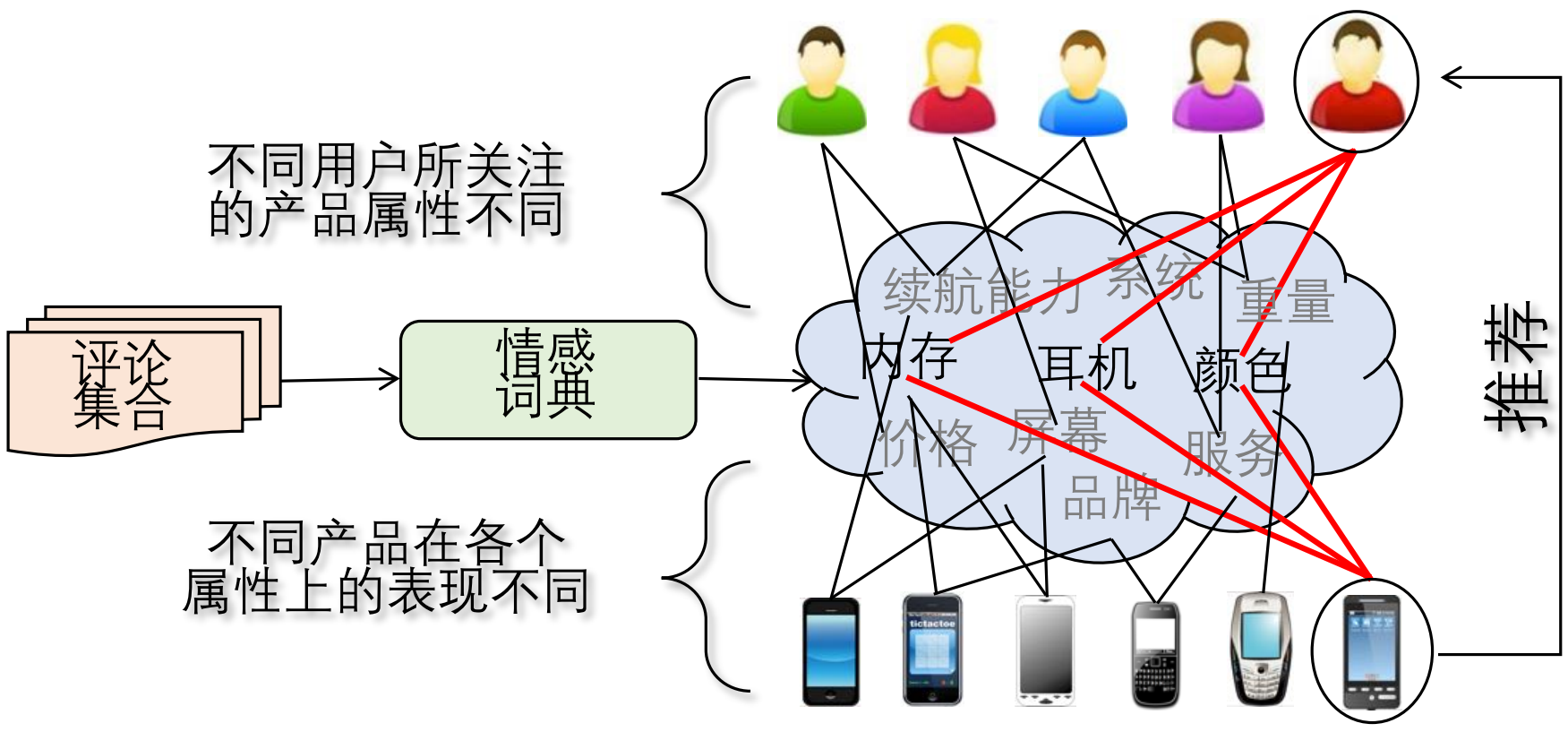

图 4.2　我们首先基于短语级情感分析构建情感词典，并从中获得属性词集合作为模型的特征空间，进而分析和预测用户对属性的关注情况和物品在属性上的表现情况，进而为用户给出在他关心的属性上表现好的商品作为推荐。

在这一理念下，本节提出了显式变量分解模型以在获得高预测精度的同时给出可解释的推荐结果[194,228]。图4.2给出了基于显式变量分解模型的可解释性推荐基本流程。首先，我们利用短语级情感分析技术从特定领域（如手机）的大规模用户评论中构建情感词典，即（*属性词*，*情感词*，*情感极性*）的三元组集合，并将其中的属性词作为显式特征空间。进而，我们利用每个用户的历史评论构建用户属性关注度矩阵、以及利用每个产品的历史评论构建产品属性好评度矩阵，并结合原始的用户物品评分矩阵，构建基于多矩阵分解的显式变量分解模型来获得预测及个性化推荐。在图4.2的例子中，系统推断出用户对*内存*、*耳机*和*颜色*等属性词比较感兴趣，则会向用户提供在这些属性上表现优秀的产品作为推荐。

在接下来的部分，我们首先对相关工作进行简要的介绍，进而详细描述基于多矩阵分解的显式变量分解模型及个性化推荐理由的构建。在性能评测部分，我们将介绍算法在线下预测性能和推荐列表排序性能上的表现，以及在线上实验中点击率等指标上的表现。

### 4.1.2　相关工作

在包括文本评论在内的互联网用户生成内容中，往往包含着丰富的产品属性和用户偏好信息，而这些信息可以借助情感分析进行挖掘和结构化[197,198]。如第2章所介绍的，情感分析可以在三个层面上进行，分别为篇章级情感分析、句子级情感

分析和短语级情感分析。篇章级[203,213] 和句子级[204,205,268] 情感分析对文章或句子进行情感分类，并将其映射为预定义的情感极性，包括常见的正面、负面、中性等情感极性，以及喜怒哀乐等更为具体的情绪表达；而短语级情感分析[206,209–211] 试图从文本中抽取具体显式的产品属性，并进一步通过寻找和匹配用户在该属性上所使用的情感词来分析用户所表达的情感极性[202]。短语级情感分析的核心任务为情感词典的构建[210,211,226,227]，其中每一个词条为一个（属性词，情感词，情感极性）三元组，除推荐任务之外，情感词典还在多种实际系统中起重要作用，例如文本摘要、情感检索，等等。需要指出的是，在短语级情感分析中，情感词的情感极性具有上下文相关性，即同一个观点词与不同的属性词匹配时，可能表达不同甚至相反的情感。例如情感词好往往具有固定的情感极性，而情感词高用来描述属性词质量时具有正面的情感极性，而当和噪声匹配时却表达负面的情感。

随着文本评论的逐渐积累，如何利用文本数据提高推荐性能在近年来越来越受到研究人员的关注[191,269–278]。最初，研究人员将自由文本结构化并映射到人工定义的本体（ontology）上[269]，然而概念本体往往是领域相关的，需要对不同的领域分别构建相应的本体结构，费时费力且成本较高，另外也无法与被广泛认可的协同过滤算法很好地融合。近年来，研究人员尝试对用户评论进行篇章级情感分析，从而将每条评论映射为一个情感极性得分，进而利用该信息提高打分预测的精度[273,275–278]，然而相关方法无法挖掘用户在各个产品属性上的具体情感。为了解决相关问题，一些工作采用话题模型在文本评论中抽取话题，并将评论映射为话题分布，从而帮助提高打分预测的精度[191,192,272]。由于缺乏对用户情感的抽取，相关方法无法了解和使用用户的情感极性：用户提到产品的某一话题时并非意味着对该话题的正面偏好，而实际上恰恰有可能是在该话题上表达负面情感，单纯依赖话题的出现与否并不能准确把握用户的偏好，相反却有可能引入误差。与此不同，我们采用短语级情感分析具体挖掘用户在特定属性上表达的情感，从而对用户偏好进行更为细致的建模，并给出准确的推荐和直观的解释。

研究发现在推荐系统中为用户提供恰当的推荐理由有助于提高用户对推荐的接受度[185,186]，同时有助于提升系统的透明度、可信度、有效性、推荐结果的可辨性，以及系统使用效率[20,187,188]。由于不同推荐算法的推荐机制不同，因而系统所使用的具体推荐算法不同在实质上影响推荐理由的生成。总体而言，大量使用的隐变量模型在设计初衷上恰恰不考虑算法流程的可解释性，而是通过参数的自由化提升打分预测的精度，这为个性化推荐理由的构建带来了困难[187]。

为了解决相关问题，目前的方法多采用模型无关的推荐理由，即在算法生成推荐之后再指定合适的推荐理由，而该推荐理由往往与物品被算法推荐出来的机

制没有直接关系，例如在购物网站中常见的“80% 的用户也查看了”，以及在社交网站中“您的好友也查看了”等推荐理由[189,190]，然而类似的推荐理由往往过于简化了物品得到推荐的真正原因。与此不同的是，在本工作中我们提出由模型直接生成的短语级个性化推荐理由，在此基础上我们不但可以给出个性化推荐，还可以给出个性化“不”推荐，即告诉用户当前查看的物品可能并不适合购买，从而提高系统的可用性和可信度。

### 4.1.3　基于用户评论的情感词典构建

我们首先利用大规模评论语料构建情感词典 $\mathcal{L}$，为此我们设计和开发了短语级情感分析工具包，其中包括基于统计自然语言处理的属性观点词对挖掘系统[208]，以及基于迭代优化的情感极性自动标注算法[207,209]，同时该工具包支持共享和下载①。短语级情感词典的构建主要包括如下三个步骤：首先，我们基于语法和词法分析从大规模评论语料中抽取属性词集合 $\mathcal{F}$；其次，我们抽取观点词集合 $\mathcal{O}$，并基于语法规则对评论中的每一个属性词匹配相对应的情感词，从而构建属性情感词对；最后，我们基于标签扩散的方法对属性观点词对进行极性标注，得到每一个词对的情感得分 $S$，并最终得到情感词典 $\mathcal{L}$ 中的各个情感词条 $(F, O, S)$。在性能评测部分，我们将对本节所构建的情感词典的统计属性和和主要技术指标进行评测和报告。

给定情感词典 $\mathcal{L}$ 和一条文本评论，我们将该评论结构化为一个属性情感对 $(F, S')$ 的集合，其中 $S'$ 是在这条评论中用户在属性词 $F$ 上所表达的情感极性。

为了更直观地描述该过程，我们给出图4.3所示的例子。在图中所示的用户物品行为矩阵中，阴影部分表示相应用户对相应物品的评论，包括一个数值化的打分和一条文本。自此基础上，我们首先判断该文本所命中的情感词条，在本例中，

表 4.1　本节符号表

| | |
|---|---|
| $\mathcal{L}$ | 上下文相关的情感词典 |
| (F,O,S) | $\mathcal{L}$ 中的一个词条，其中 $S \in [-1, 1]$ |
| $\mathcal{F}$ | $\mathcal{L}$ 的属性词集合，其中 $\lvert\mathcal{F}\rvert = p$ |
| $\mathcal{O}$ | $\mathcal{L}$ 的情感词集合，其中 $\lvert\mathcal{O}\rvert = q$ |
| $A \in \mathbb{R}_+^{m\times n}$ | 用户物品数值评分矩阵 |
| $m, n$ | 用户数和物品数 |
| $X \in \mathbb{R}_+^{m\times p}$ | 用户属性关注度矩阵 |
| $Y \in \mathbb{R}_+^{n\times p}$ | 物品属性好评度矩阵 |
| $N$ | 矩阵 $A, X, Y$ 中的最高打分范围，常见系统中为 5 星（N=5） |

① http://yongfeng.me/software

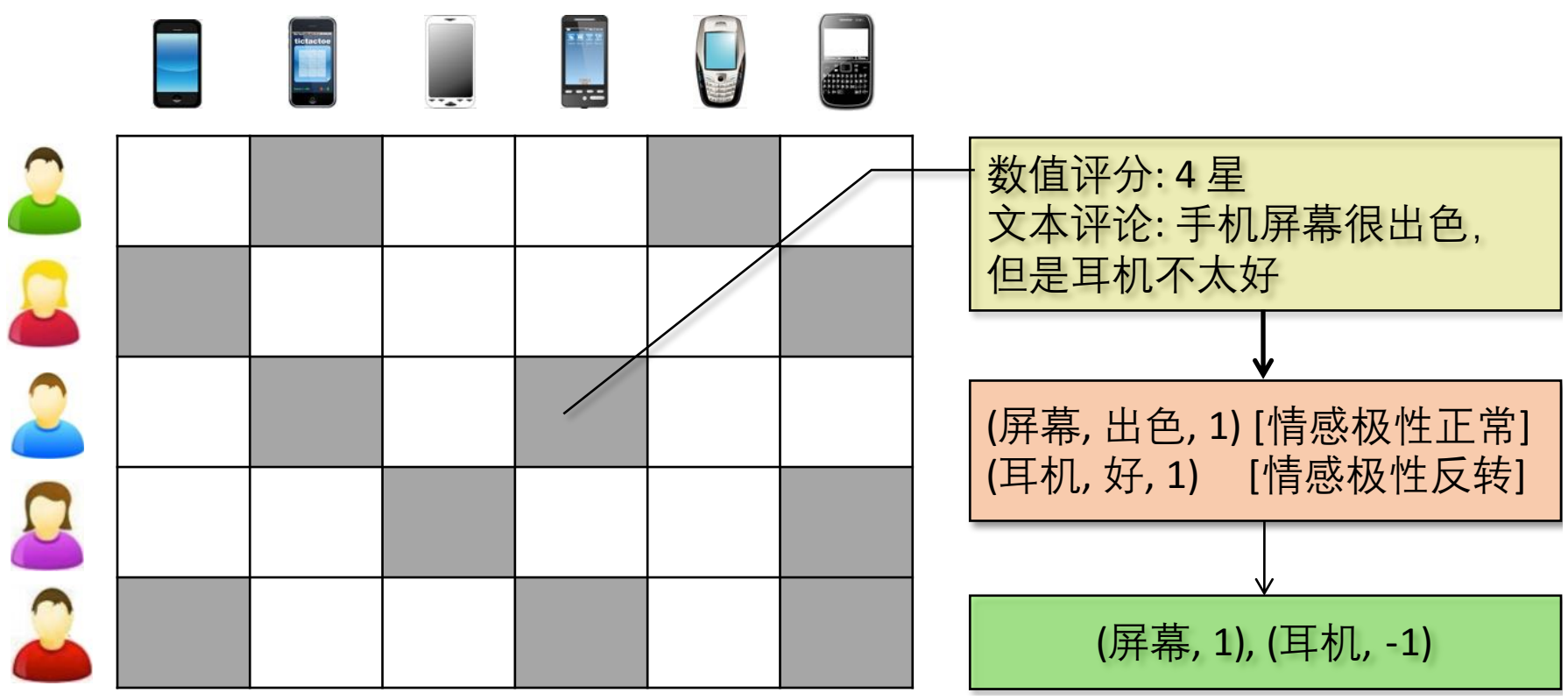

图 4.3　用户物品行为矩阵及评论文本的结构化表示示例，其中矩阵中的阴影值表示相应的用户对相应的物品有过评论行为，每条评论包括一个数值化的打分和一条文本；算法首先从评论文本中抽取命中的（属性词，情感词，情感极性）词条，进一步考察每一个词条是否被否定词反转，并进一步计算出用户在每一个属性词上所表达的最终情感极性。

包括（屏幕，出色，+1）和（耳机，好，+1）两个词条；进一步，我们判断每一个词条是否被否定词极性反转，发现第二个词条的进行被否定词“不”反转，对于被反转的情感词，我们对其原有的情感极性取相反数作为用户表达的真实情感：

$$S' = \begin{cases} S，\text{如果情感词}O\text{没有被否定词修饰和反转} \\ -S，\text{如果情感词}O\text{被否定词修饰和反转} \end{cases} \tag{4-1}$$

在此基础上，我们最终得到用户在每一个属性词上表达的最终情感，并得到该文本评论的结构化表示，在图4.3的例子中，这包含（屏幕，+1）和（耳机，-1）两个部分。

### 4.1.4　显式变量分解模型及其可解释性

在结构化文本评论的基础上，我们首先构造用户属性关注度矩阵来描述用户对不同属性的关注程度，以及物品属性好评度矩阵来描述物品在不同属性上的性能表现。

#### 用户属性关注度矩阵

由于不同的用户对产品不同属性的关注程度不同，他们在历史评论中对不同属性提及的频率也各不相同，一般而言，用户对自己所关心的属性往往提及频率更高。因此，我们构造用户属性关注度矩阵 $X$，矩阵中的每一个元素表示相应的用户对属性的关心程度，如图4.4所示。

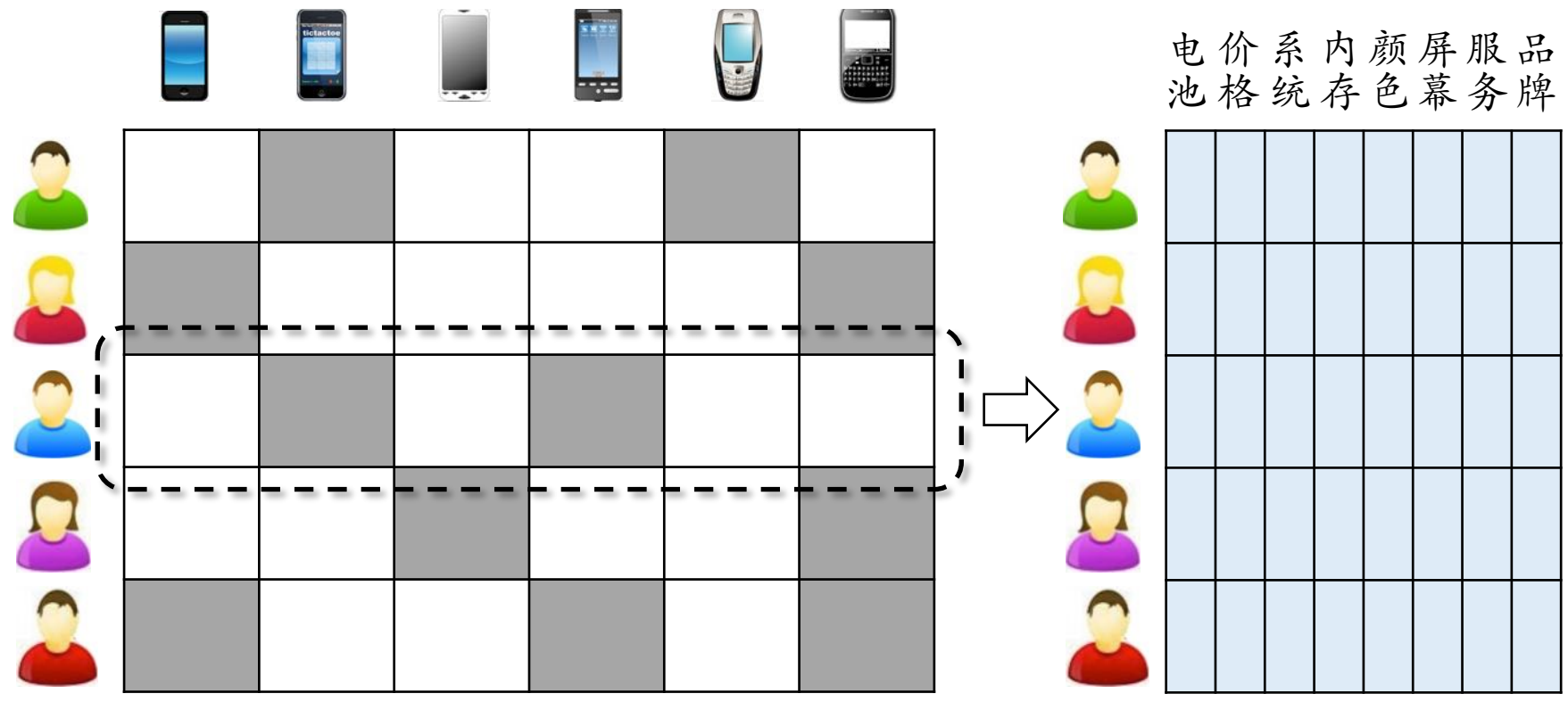

图 4.4　通过分析每一个用户的历史评论文本构建用户-属性关注度矩阵

令 $\mathcal{F}=\{F_1,F_2,\cdots,F_p\}$ 为属性词集合，$\mathcal{U}=\{u_1,u_2,\cdots,u_m\}$ 为用户集合。我们考察给定用户 $u_i$ 的所有历史评论，并从中抽取所有命中的属性词及考虑否定词后用户在指定属性词上表达的最终情感 $(F,S')$。设属性词 $F_j$ 被用户 $u_i$ 提及 $t_{ij}$ 次，则我们定义用户属性关注度矩阵的每一个元素值如下：

$$X_{ij}=\begin{cases}0\text{，如果用户}u_i\text{没有提及过属性词}F_j\\1+(N-1)\left(\frac{2}{1+e^{-t_{ij}}}-1\right)\text{，其它情况}\end{cases}\tag{4-2}$$

公式 (4-2) 的主要作用在于将用户的提及频率 $t_{ij}$ 经由 Sigmoid 函数转化到与评分矩阵 $A$ 相同的范围 $[1,N]$ 上，在大多数实际系统中 $N$ 的取值为 5，也就是常见的五星评分体系，例如亚马逊、淘宝、京东商城，等等。

### 物品属性好评度矩阵

类似地，我们构建物品属性好评度矩阵 $Y$，其中矩阵的每一个元素描述相应的物品在属性上的表现，如图4.5所示。令 $\mathcal{P}=\{p_1,p_2,\cdots,p_n\}$ 表示 $n$ 个物品，对于每一个物品 $p_i$，我们考察它历史上所获得的所有评论，并同样从中抽取所有的 $(F,S')$ 结构。设在产品 $p_i$ 的评论中属性词 $F_j$ 被提 $k$ 次，且这些提及的平均情感极性为 $s_{ij}$，则我们定义 $Y_{ij}$ 为：

$$Y_{ij}=\begin{cases}0\text{，如果物品}p_i\text{的评论中没有提及过属性}F_j\\1+\frac{N-1}{1+e^{-k\cdot s_{ij}}}\text{，其它情况}\end{cases}\tag{4-3}$$

该定义一方面考虑用户在该属性上表达的整体情感极性（通过 $s_{ij}$），另一方面也考虑该物品在每个属性上的流行度（通过 $k$）。同样，$Y$ 矩阵中的元素值通过

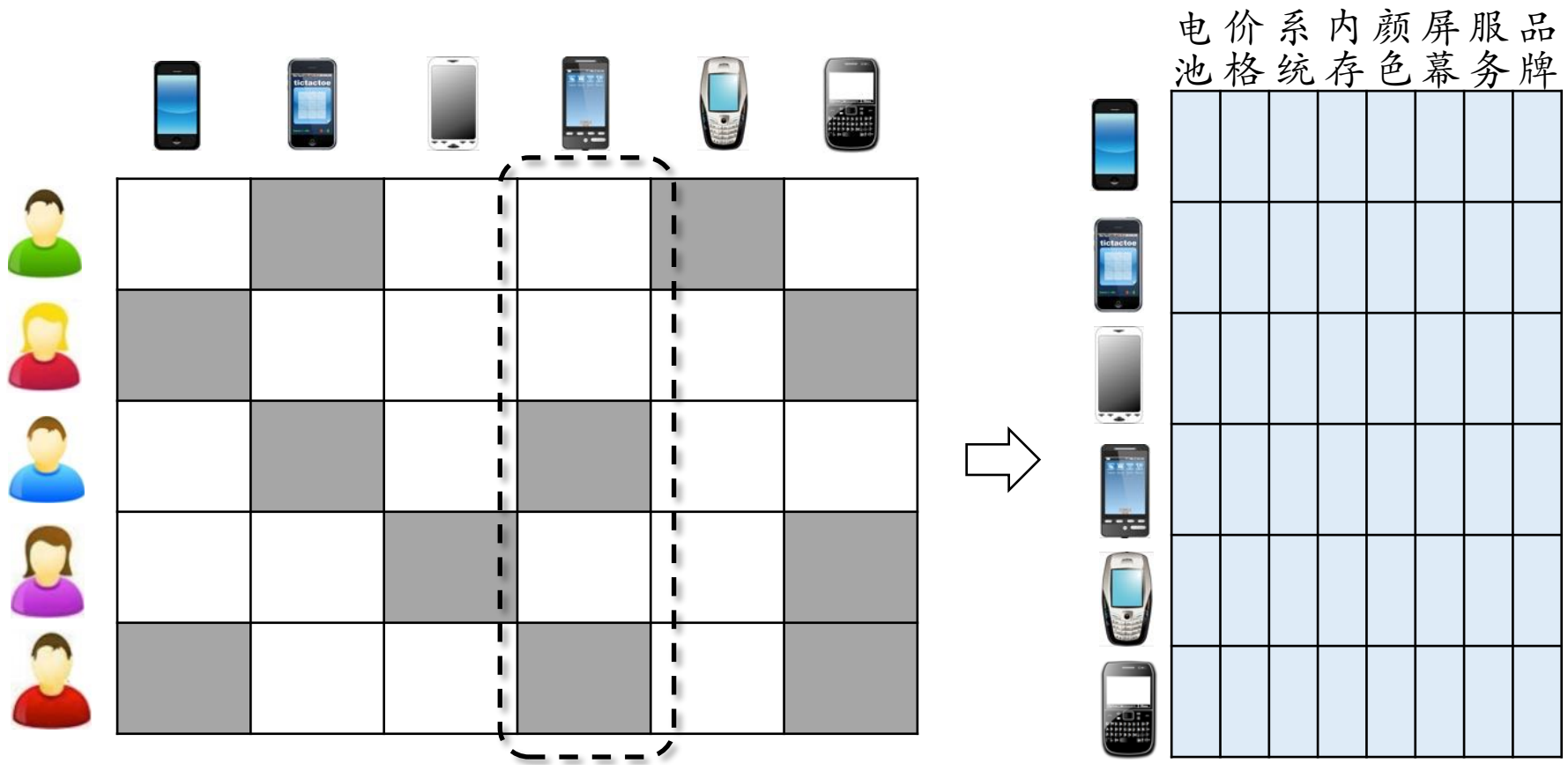

图 4.5 通过分析每一个物品所获得的历史评论文本构建物品-属性好评度矩阵

Sigmoid 函数转化到 $[1, N]$ 区间内。

### 显式变量分解模型

矩阵 $X$ 和 $Y$ 中的非零值表示已经在历史评论文本中观测到的用户、物品、属性词关系，接下来我们描述如何将这些信息整合为多矩阵分解模型，从而在获得高预测精度和推荐效果的同时提高模型的可解释性。

与用户物品评分矩阵 $A$ 上直接的分解模型相似，我们对用户属性关注度矩阵 $X$ 和物品属性好评度矩阵 $Y$ 进行分解，即根据矩阵中已经观测到的非零值构建用户、物品和属性的表示，从而对矩阵中未观测到的值（以零值表示）进行预测和估计，如公式 (4-4) 下所示：

$$\begin{aligned}&\underset{U_1,U_2,V}{\text{minimize}}\left\{\lambda_x\|U_1V^T-X\|_F^2+\lambda_y\|U_2V^T-Y\|_F^2\right\}\\&\text{s.t. } U_1\in\mathbb{R}_+^{m\times r}, U_2\in\mathbb{R}_+^{n\times r}, V\in\mathbb{R}_+^{p\times r}\end{aligned} \tag{4-4}$$

其中 $\lambda_x$ 和 $\lambda_y$ 为正则化系数，$r$ 表示显式变量分解因子的个数。

我们假设用户对物品的总体评价（矩阵 $A$ 中相应的打分）是由用户在产品各个属性上的评价综合得出的，因此我们利用公式 (4-4) 中所使用的参数 $U_1$ 和 $U_2$ 来近似评分矩阵 $A$，其中包含 $r$ 个维度的 $U_1$ 以及 $U_2$ 分别为用户和物品在显式属性上的表示向量。然而我们所挖掘出来的显式属性词并非一定能囊括用户所考虑的全部可能属性，很有可能有一些潜在属性是我们没有挖掘出来的，因此，我们在 $U_1$ 和 $U_2$ 之外再考虑 $r'$ 个隐式变量 $H_1\in\mathbb{R}_+^{m\times r'}$ 和 $H_2\in\mathbb{R}_+^{n\times r'}$，并最终用分解因子矩

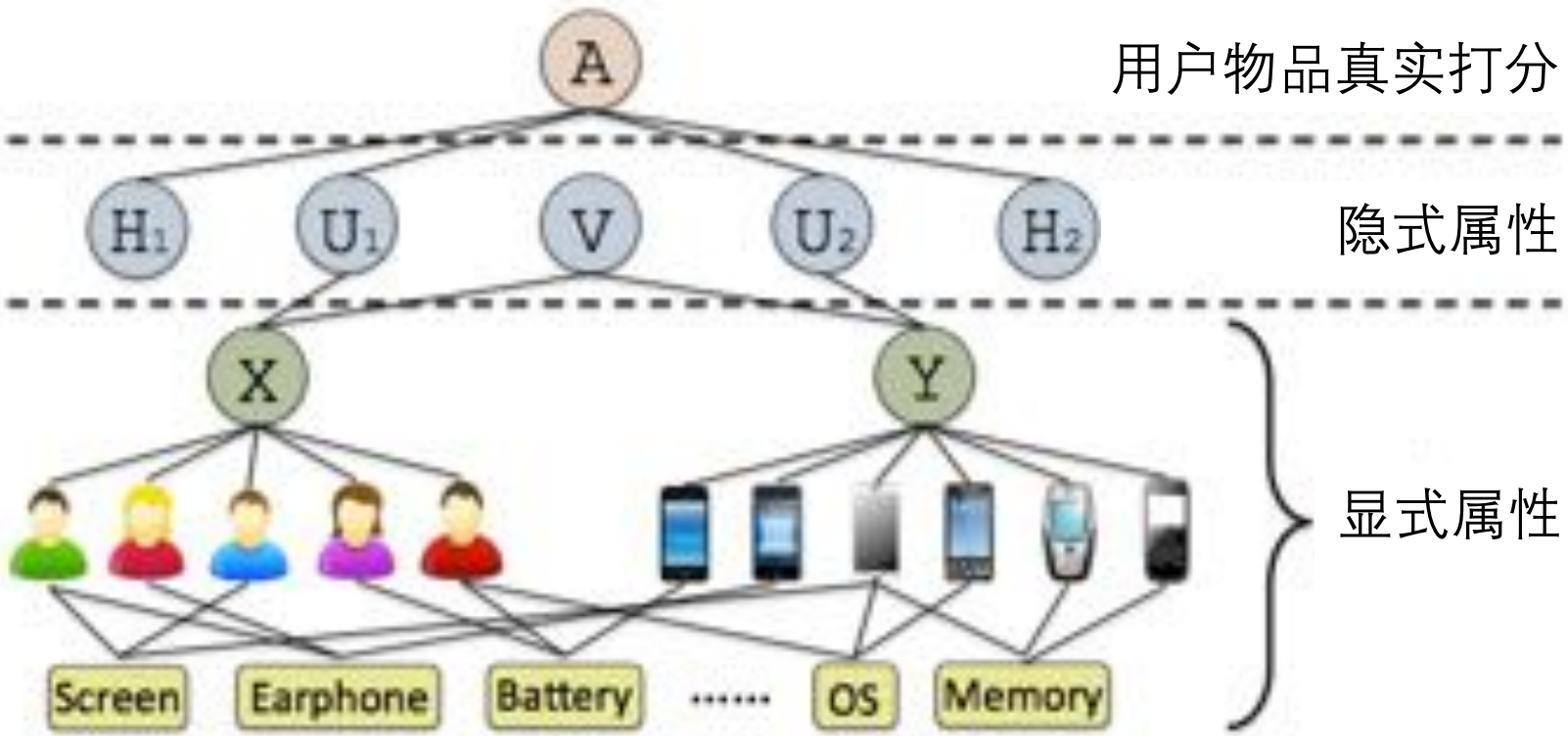

图 4.6　显式变量分解模型中产品属性词、用户属性关注度矩阵 $X$、物品属性好评度矩阵 $Y$、用户物品真实打分矩阵 $A$ 以及分解因子矩阵之间的关系示意图

阵 $P = [U_1\ H_1]$ 和 $Q = [U_2\ H_2]$ 来近似用户物品打分矩阵 $A$：

$$\underset{P,Q}{\text{minimize}}\left\{\|PQ^T - A\|_F^2\right\} \tag{4-5}$$

综合公式 (4-4) 和公式 (4-5)，我们构建最终的显式变量分解模型。图4.6描述了产品属性词、用户属性关注度矩阵 $X$、物品属性好评度矩阵 $Y$，以及用户物品评分矩阵 $A$ 之间的关系，模型参数可以通过如下的优化问题得到：

$$\begin{aligned}
&\underset{U_1,U_2,V,H_1,H_2}{\text{minimize}}\Big\{\|PQ^T - A\|_F^2 + \lambda_x\|U_1V^T - X\|_F^2 + \lambda_y\|U_2V^T - Y\|_F^2 \\
&\quad +\lambda_u(\|U_1\|_F^2 + \|U_2\|_F^2) + \lambda_h(\|H_1\|_F^2 + \|H_2\|_F^2) + \lambda_v\|V\|_F^2\Big\} \\
&\quad \text{s.t. } U_1 \in \mathbb{R}_+^{m\times r}, U_2 \in \mathbb{R}_+^{n\times r}, V \in \mathbb{R}_+^{p\times r}, H_1 \in \mathbb{R}_+^{m\times r'}, H_2 \in \mathbb{R}_+^{n\times r'} \\
&\qquad P = [U_1\ H_1],\ Q = [U_2\ H_2]
\end{aligned} \tag{4-6}$$

当 $r = 0$ 时，模型退化为传统的基于用户物品评分矩阵 $A$ 的隐变量分解模型，此时我们没有使用任何来自显式产品属性词的信息。我们用公式 (4-6) 的优化结果来获得近似和补全之后的矩阵 $A$、$X$ 和 $Y$，并据此为用户提供个性化的推荐结果和属性级推荐理由。

像很多矩阵分解优化问题一样，问题 (4-6) 不存在封闭解，受 Ding[279] 和 Hu[231] 等工作的启发，我们构建交替最小化算法来学习模型参数 $U_1, U_2, V, H_1, H_2$。当给定其它参数时，公式 (4-6) 相对于参数 $U_1$ 的形式为：

$$\min_{U_1\geq 0}\left\{\|U_1U_2^T + H_1H_2^T - A\|_F^2 + \lambda_x\|U_1V^T - X\|_F^2 + \lambda_u\|U_1\|_F^2\right\} \tag{4-7}$$

令 $\Lambda$ 为优化约束 $U_1 \geq 0$ 所对应的拉格朗日系数，则如上优化问题的拉格朗日

**Algorithm 6:** Explicit Factor Model

**Input:** $A, X, Y, m, n, p, r, r', \lambda_x, \lambda_y, \lambda_u, \lambda_h, \lambda_v, T$

**Output:** $U_1, U_2, V, H_1, H_2$

$U_1 \leftarrow \mathbb{R}_+^{m\times r},\ U_2 \leftarrow \mathbb{R}_+^{n\times r},\ V \leftarrow \mathbb{R}_+^{p\times r}$;

$H_1 \leftarrow \mathbb{R}_+^{m\times r'},\ H_2 \leftarrow \mathbb{R}_+^{n\times r'}$; //随机初始化

$t \leftarrow 0$;

**repeat**

　　$t \leftarrow t + 1$;

　　更新矩阵：$V_{ij} \leftarrow V_{ij}\sqrt{\frac{[\lambda_x X^T U_1+\lambda_y Y^T U_2]_{ij}}{[V(\lambda_x U_1^T U_1+\lambda_y U_2^T U_2+\lambda_v I)]_{ij}}}$

　　更新矩阵：$U_{1ij} \leftarrow U_{1ij}\sqrt{\frac{[AU_2+\lambda_x XV]_{ij}}{[(U_1U_2^T+H_1H_2^T)U_2+U_1(\lambda_x V^T V+\lambda_u I)]_{ij}}}$

　　更新矩阵：$U_{2ij} \leftarrow U_{2ij}\sqrt{\frac{[A^T U_1+\lambda_y YV]_{ij}}{[(U_2U_1^T+H_2H_1^T)U_1+U_2(\lambda_y V^T V+\lambda_u I)]_{ij}}}$

　　更新矩阵：$H_{1ij} \leftarrow H_{1ij}\sqrt{\frac{[AH_2]_{ij}}{[(U_1U_2^T+H_1H_2^T)H_2+\lambda_h H_1]_{ij}}}$

　　更新矩阵：$H_{2ij} \leftarrow H_{2ij}\sqrt{\frac{[A^T H_1]_{ij}}{[(U_2U_1^T+H_2H_1^T)H_1+\lambda_h H_2]_{ij}}}$

**until** 算法收敛 **or** $t > T$;

**return** $U_1, U_2, V, H_1, H_2$;

形式为：

$$L(U_1) = \|U_1U_2^T + H_1H_2^T - A\|_F^2 + \lambda_x\|U_1V^T - X\|_F^2 + \lambda_u\|U_1\|_F^2 - \mathrm{tr}(\Lambda U_1) \tag{4-8}$$

对参数 $U_1$ 的梯度为：

$$\nabla_{U_1} = 2(U_1U_2^T + H_1H_2^T - A)U_2 + 2\lambda_x(U_1V^T - X)V + 2\lambda_u U_1 - \Lambda \tag{4-9}$$

令 $\nabla_{U_1} = 0$，我们有：

$$\Lambda = 2(U_1U_2^TU_2 + H_1H_2^TU_2 + \lambda_x U_1V^TV + \lambda_u U_1) - 2(AU_2 + \lambda_x XV) \tag{4-10}$$

根据约束 $U_1 \geq 0$ 所对应的 KKT 条件，我们有 $\Lambda_{ij} \cdot U_{1ij} = 0$，因此：

$$[-(AU_2 + \lambda_x XV) + (U_1U_2^TU_2 + H_1H_2^TU_2 + \lambda_x U_1V^TV + \lambda_u U_1)]_{ij} \cdot U_{1ij} = 0 \tag{4-11}$$

由此，我们可以得到参数 $U_1$ 的迭代优化公式如下：

$$U_{1ij} \leftarrow U_{1ij}\sqrt{\frac{[AU_2 + \lambda_x XV]_{ij}}{[(U_1U_2^T + H_1H_2^T)U_2 + U_1(\lambda_x V^T V + \lambda_u I)]_{ij}}} \tag{4-12}$$

参数 $U_2, V, H_1, H_2$ 的迭代公式可以用类似的方法得到。因此，我们给出如下基于迭代最小化的优化算法，如算法6所示，其核心思想为在固定其它四组参数不变的情况下对一组参数进行优化。算法不断地对每组参数依次进行优化直到收敛或者达到预先指定的最大迭代次数。

### 4.1.5　推荐列表的构建

给定算法6的优化结果，我们可以对用户属性关注度矩阵 $X$、物品属性好评度矩阵 $Y$，以及用户物品评分矩阵 $A$ 中的缺失值进行预测和估计，分别为 $\tilde{X} = U_1V^T$，$\tilde{Y} = U_2V^T$，以及 $\tilde{A} = U_1U_2^T + H_1H_2^T$。我们基于这些信息为用户提供个性化推荐列表并进行推荐理由的构建。

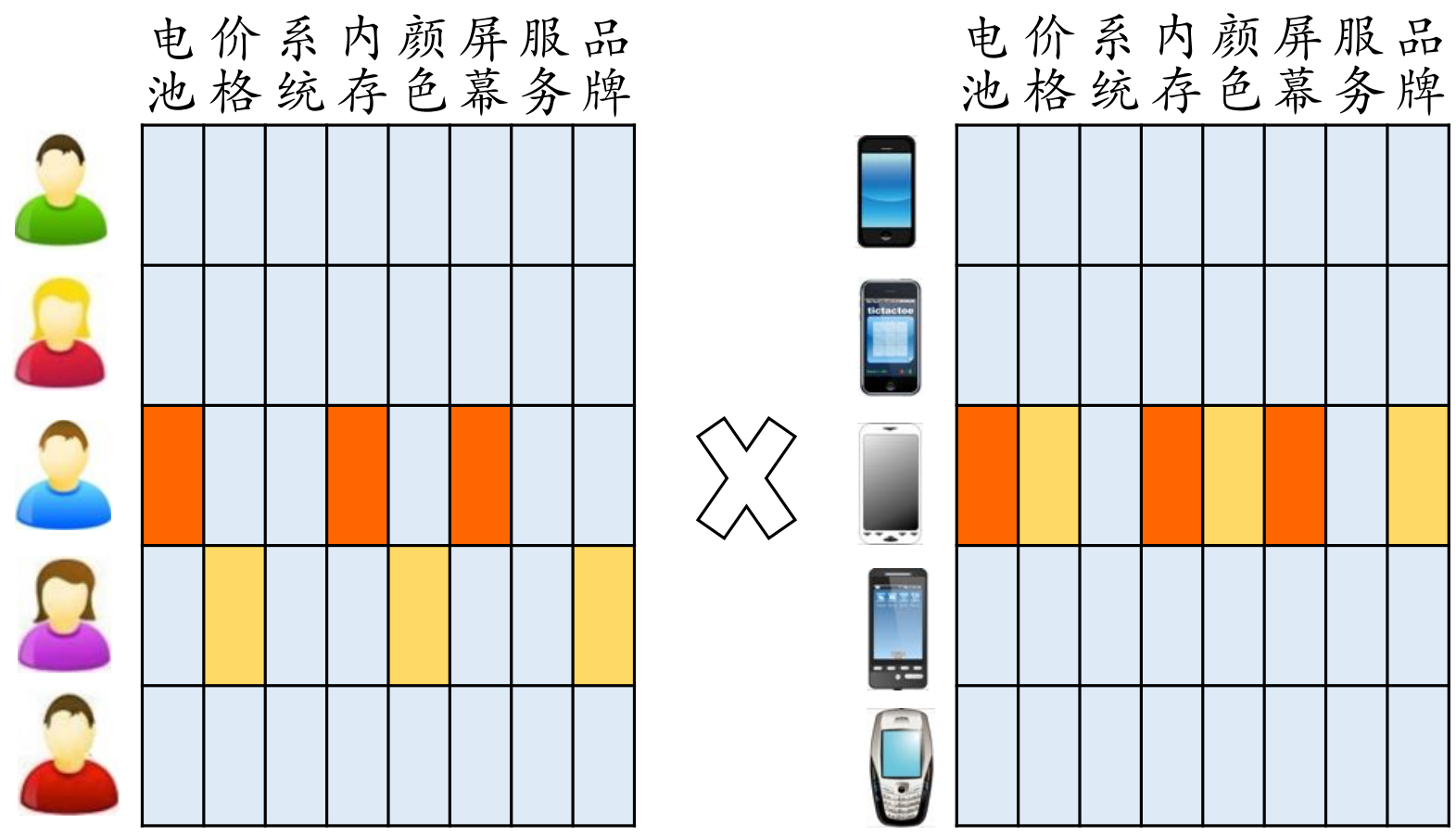

图 4.7　不同用户所最关心的产品属性有所不同，因此即使在计算同一个产品上的预测值时，不同用户也有可能采用不同的产品属性（即该用户最关心的前 $k$ 个属性词显式变量维度）进行向量乘法

对于不同的用户而言，决定用户在特定物品上的喜好的因素往往是不同的，用户往往对其最为关心的几个属性进行主要的考量，并由此决定对物品的整体态度，而非对全部的上百个属性进行一一考虑。因此，当我们为用户提供推荐结果时，考虑其最为关心的前 $k$ 个产品属性而非全部上百个属性，如图4.7所示例。对于用户 $u_i(1 \leq i \leq m)$，$\tilde{X}_{i\cdot}$ 表示补全的关注度矩阵中相应于该用户的行向量，$C_i = \{c_{i1}, c_{i2}, \cdots, c_{ik}\}$ 表示该行向量中前 $k$ 个最大值所对应的列编号，则我们按照如下

的方式计算用户 $u_i$ 与每一个物品 $p_j(1 \le j \le n)$ 之间的相关性：

$$R_{ij} = \alpha \cdot \frac{\sum_{c \in C_i} \tilde{X}_{ic} \cdot \tilde{Y}_{jc}}{kN} + (1 - \alpha)\tilde{A}_{ij} \tag{4-13}$$

其中 $N = \max(A_{ij})$ 用来进行预测值的正规化，在大多数系统中（包括本研究所使用的数据集中）我们有 $N = 5$。在公式 (4-13) 中，第一项表示基于用户最关心的前 $k$ 个属性词计算得到的用户物品相似度，$0 \le \alpha \le 1$ 为该相似度与用户物品预测打分之间的权重。对于用户 $u_i$，我们按照 $R_{ij}$ 由高到低对候选物品进行排序，并构建 Top-K 推荐列表。

### 4.1.6　属性级个性化推荐理由的构建

传统矩阵分解隐变量模型难以为推荐结果给出直观的推荐理由，从而让系统人员和用户理解为什么一个物品得到推荐或者未被推荐。而本工作基于物品属性词的显式变量分解模型其重要优势就在于能够让我们很容易地理解每一个产品属性在生成推荐列表时的作用。除了个性化推荐之外，我们在本工作中还利用显式变量分解模型考察一个全新的应用场景——个性化“不”推荐——即当系统分析发现用户当前浏览的物品不适合购买时，则告诉用户当前物品不推荐并给出不推荐的理由。通过为用户提供不推荐物品，我们期望增强系统的可信度和说服力，并帮助用户实现更好的消费决策。

在显式物品属性的基础上，推荐理由的构建有可能是多种多样的。在本工作中，我们为用户提供最直接的属性级个性化推荐理由，并为用户展示被推荐物品的属性词词云，从而为用户提供产品的直观属性。对于被推荐物品，属性级个性化推荐理由的构建模板如下：

| |
|---|
| 您可能对 [属性] 感兴趣，而该产品在 [属性] 上表现不错。 |

而对于不推荐的物品，属性级个性化不推荐理由的构建模板如下：

| |
|---|
| 您可能对 [属性] 感兴趣，而该产品在 [属性] 上表现欠佳。 |

对于每一个用户 $u_i$ 以及一个被推荐的物品 $p_j$，我们在该用户历史上所提到的物品属性中，选择那个在物品属性好评度矩阵中预测打分最高的属性 $F_c$ 来构造推

荐理由，并在该预测打分高于给定阈值时展示推荐理由：

$$c = \text{argmax}_{c \in C_i} \tilde{Y}_{jc} \tag{4-14}$$

而对于每一个不推荐物品 $p_j$，相应的用来构建推荐理由的属性词 $F_c$ 为用户所提到的属性中预测打分最低的，并且在该预测分低于给定阈值时展示推荐理由：

$$c = \text{argmin}_{c \in C_i} \tilde{Y}_{jc} \tag{4-15}$$

基于属性词云的推荐理由则进一步展示被推荐物品的历史评论中所命中的属性情感词对及其频率，从而为用户提供产品整体属性的直观展示。在性能评测环节，我们将对推荐界面以及推荐理由在用户接受度方面的作用进行系统的研究和报告。

## 4.2　动态化时序推荐模型

在本节，我们进一步分析产品属性词随时间变化的动态特性，从而对用户偏好进行动态建模，并实现动态的个性化推荐。本节主要内容包括基于时间序列的产品属性词时间特性分析、天级别的属性词流行度预测，以及基于流行度预测的天级别个性化推荐[229,280]。

### 4.2.1　本节引言

在实际系统中，用户兴趣并非是一成不变的，而很有可能随着时间的推移发生漂移、改变甚至反转[162,163,281]。因此，如果简单地将用户或物品的所有历史记录中所挖掘出的属性看成是其静态、长期、不变的性质，将对实际系统中的用户物品建模带来偏差。例如在电子商务网站的化妆品购物领域，用户在夏季更注重产品的“防晒”性能，而在秋冬季节则重视“营养保湿”性能，如图4.8所示。如果系统简单地静态地利用用户的历史购买记录对其偏好进行建模，则很有可能仅仅因为用户在夏天购买了防晒霜而在冬季仍然频繁地给出防晒霜推荐。类似的，系统中的物品也在不断地收到新的用户评论，因而其特征描述的分布也有可能随着时间而不断变化。如果系统仍然想用户推荐已经无法满足其个性化偏好的物品，将对系统的有效性带来负面影响。

用户偏好和物品属性的动态性为推荐系统的时间敏感建模提出了要求。近年来，研究人员已经开始对如何将时间因素纳入推荐系统中而展开了相关的研究，并

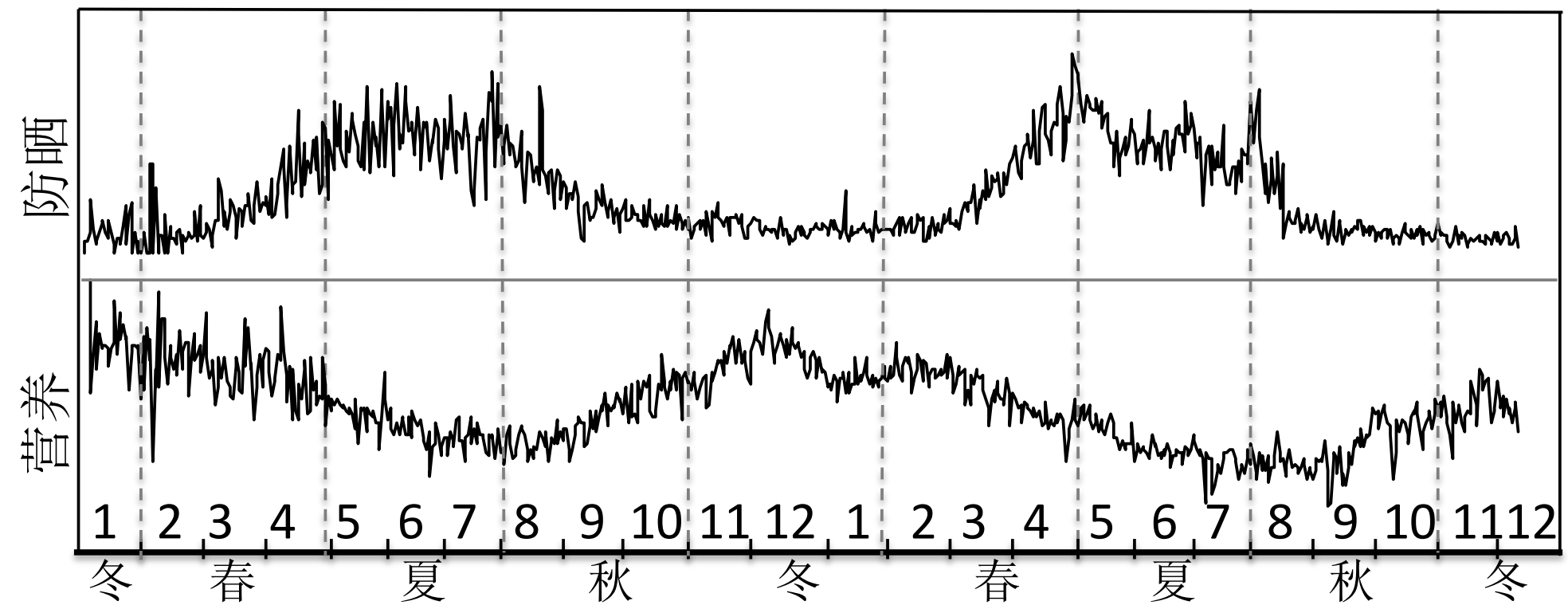

图 4.8　京东商城化妆品领域属性词每天被提及的频次占系统属性词每日总频次的百分比随时间变化的序列，可见属性词“防晒”周期性地在夏季得到关注，而属性词“营养”周期性地在冬季得到关注

提出了时间敏感推荐的研究方向[282]，并作为上下文敏感（Context-Aware）的推荐研究[283] 子方向而受到了越来越多的重视。

时间敏感推荐的研究最初主要集中在用户或物品画像的动态建模方面[284–286]，然而基于内容画像的方法无法充分利用用户的群体智慧来提高推荐效果。为了解决相关问题，学术界发展了两条相对独立的研究路线：Koren[162] 和 Xiang[281] 等采用了显式反馈的动态时间窗口（time bins）建模和隐式反馈会话（session）建模的方法，并进一步被 Koenigstein 等[163] 结合应用在了雅虎音乐推荐数据集中而在 KddCup-2011 取得了很好的效果。与此同时，Lu[287]、Shi[288]、Karatzoglou[170] 以及 Gantner[289] 等采用矩阵或张量分解的方式对时间进行建模。最近，Wang 等[290,291] 利用条件机会模型对用户的后续购买行为进行建模和预测。

然而，不管是基于用户物品画像的方法，还是基于时间敏感动态建模的方法，由于都利用用户购买和评价信息在物品空间上对用户的偏好进行建模，因此都存在数据系数的问题，这主要是由于单个用户的物品购买数量相比于庞大的物品空间而言小之又小。因此，绝大多数目前的方法无法直接建模用户的动态偏好，而是假设用户偏好满足由特定参数决定的分布函数，并通过分布函数的参数学习对用户偏好进行建模。

在本工作中，我们在属性词级别上对用户偏好的动态性进行建模。具体而言，我们对抽取得到的物品属性词进行动态时间序列分析和流行度预测，从而对用户偏好和物品属性的变化进行动态模拟。由于用户在一条评论中可能提到多个属性词，因此相比于物品级别上历史行为的稀疏性而言，每个用户在属性词级别上的历史行为更加充分，从而使得我们可以不必对用户关注度做额外的模型假设，而是直接采用时间序列分析和预测的方法进行建模，并给出时间敏感的个性化推荐。进一步，通过对属性词的季节性、周期性和趋势进行分析，我们可以更好地理解用

户偏好随时间变化的直观原因。

为了尽可能实时地描述动态偏好并提供及时的推荐，我们采用精确到天级别的时间序列分析和属性词流行度预测。而这意味着长达一年的（365 天）的时间序列周期，使得经典的移动平均自回归（Auto-Regressive Integrated Moving Average，ARIMA）模型在计算上难以完成多达 365 个参数的同时估计。为了解决这一问题，我们构造了基于傅里叶辅助项的移动平均自回归模型（Fourier-assisted ARIMA，FARIMA），首先利用傅里叶级数对原始时间序列进行粗略建模，并进一步利用低阶 ARIMA 模型对残差部分进行精细建模，从而较好地还原和预测原始时间序列。在属性词流行度预测的基础上，我们进一步构建条件机会模型（Conditional Opportunity Model）为用户提供时间敏感的个性化推荐。在性能评测部分，我们将对属性词流行度的时间特定进行具体直观的分析，并对流行度预测效果和时序推荐效果进行定量评测。

### 4.2.2　相关工作

实际系统中用户偏好和物品属性往往随时间而动态变化[282]，因此绝对静态将用户历史记录看做用户偏好的长期描述并不能及时地反映用户最新当前的偏好。例如，虽然用户在夏季购买了防晒产品，但是在冬季仍然购买防晒产品的可能性会变小；再比如如果用户已经在最近购买了一款单反相机，那么在接下来的短时间内再次购买单反相机的概率也会比较小[282]。

在时间敏感推荐研究的初期，研究人员尝试构建用户或物品的动态画像。Chen 等[284] 构建了个性化画像的动态发现、存储和更新系统来进行时间敏感的推荐；Baltrunas 等[285] 等构建用户的微画像（Micro-profile），从而将用户描述分解为几个时间敏感的子方面来进行建模；Chu 等[286] 利用动态双线性模型（Bilinear modeling）将用户画像和物品画像进行融合并提供个性化推荐。Gauch 等[292] 对基于用户和物品画像的个性化推荐方法进行了综述。

然而画像的构建往往对领域知识具有较高的要求，而在实际系统中领域知识往往很难获得，另外，相关方法也难以对群体的智慧进行充分的利用。为了改进这些问题，研究人员开始更多地关注基于协同过滤的时间敏感推荐相关研究。

Oku[293] 和 Yuan[294] 等将时间因素加入到基于近邻方法中，提出动态相似度的计算方法，从而进行时间敏感的协同过滤；Koren 等[162] 将时间因素加入到基于矩阵分解的协同过滤中并提出了 SVD++ 算法，从而对 SVD 模型中的用户和物品偏置进行动态建模；Xiang 等[281] 进一步基于随机游走模型对用户隐式反馈中的动态特性进行建模；Lu[287] 和 Shi[288] 等进一步对基于矩阵分解的动态建模进行了系统

性的研究，Karatzoglou[170] 和 Gantner[289] 等则对基于张量分解的动态个性化推荐进行建模；Chen[295] 和 Vaca[296] 等基于话题模型对用户评论进行动态化建模。

目前的方法往往在物品级别对用户的偏好进行建模，然而单个用户所购买或评论过的物品相对于系统中总的物品集合而言少之又少，因而面临着严重的数据稀疏性问题。考虑产品的内容信息可以在一定程度上缓解数据稀疏性的问题[283,286,290]，但在实际系统中恰当的内容信息往往往难以获得，并且往往需要大量的人工处理和结构化。

近年来，短语级情感分级技术的发展[210,211,213,297] 为用户评论自由文本中属性词和情感词的抽取、摘要、结构化提供了可能，并开始在推荐系统中得到应用[194,195,298,299]，这也为属性级用户偏好的动态建模带来了新的思路。与物品级建模相比，属性级建模可以做到更加细粒度，因为即便一个用户在历史上只对一个物品有过评论，该评论中也很有可能包含一个以上的物品属性。另外，对具体的物品属性进行时间动态分析可以帮助我们更为直观地了解用户兴趣到底发生了什么变化以及为什么会发生变化，相比于以往参数意义不甚明确的复杂模型，这是该方法的一大优点。

我们采用时间序列分析[300,301] 进行精确到天级别的属性词流行度预测。时间序列分析技术已经被广泛应用于诸多数据分析任务中，例如计量经济学、生物信息学、物理学和天文学[301] 等等。最近，研究人员也开始将时间序列分析技术应用于数据挖掘任务中，例如流行病预测[302]、谷歌趋势中的经济指标预测[303]，等等。然而由于物品级用户行为数据的缺乏，时间序列分析技术尚未被应用于个性化推荐任务当中。然而正如我们接下来将要介绍的，用户评论文本的逐渐积累和短语级情感分析技术的逐渐成熟使得用户偏好的直接时间序列分析和预测成为可能，并为精确到天级别的动态个性化推荐带来了新的思路。

### 4.2.3　用户偏好的时序性质

我们首先对购物网站中物品属性词的流行度进行实际分析，从而对实际系统中的时间序列给出一个直观的介绍。我们采用京东商城化妆品领域 2011 年 1 月 1 日到 2014 年 3 月 31 日共 3 年 1 个季度的用户评论数据，共包括 1,844,569 个用户对 53,188 物品的 5,524,491 条评论。我们首先仍然利用如上节所述的短语级情感分析技术抽取该领域的物品属性词集合 $\mathcal{F} = \{f_1, f_2, \cdots, f_r\}$，包括“价格”、“防晒”、“营养”、“触感”等等。该属性词集合将作为显式属性空间被用于接下来的时间序列分析、流行度预测，以及个性化推荐等环节。

我们分析了三种可能的时间序列来确定适合用来进行流行度预测和用户兴趣

动态建模的时间序列，以属性词“营养”为例，三种时间序列示例如图4.9所示。

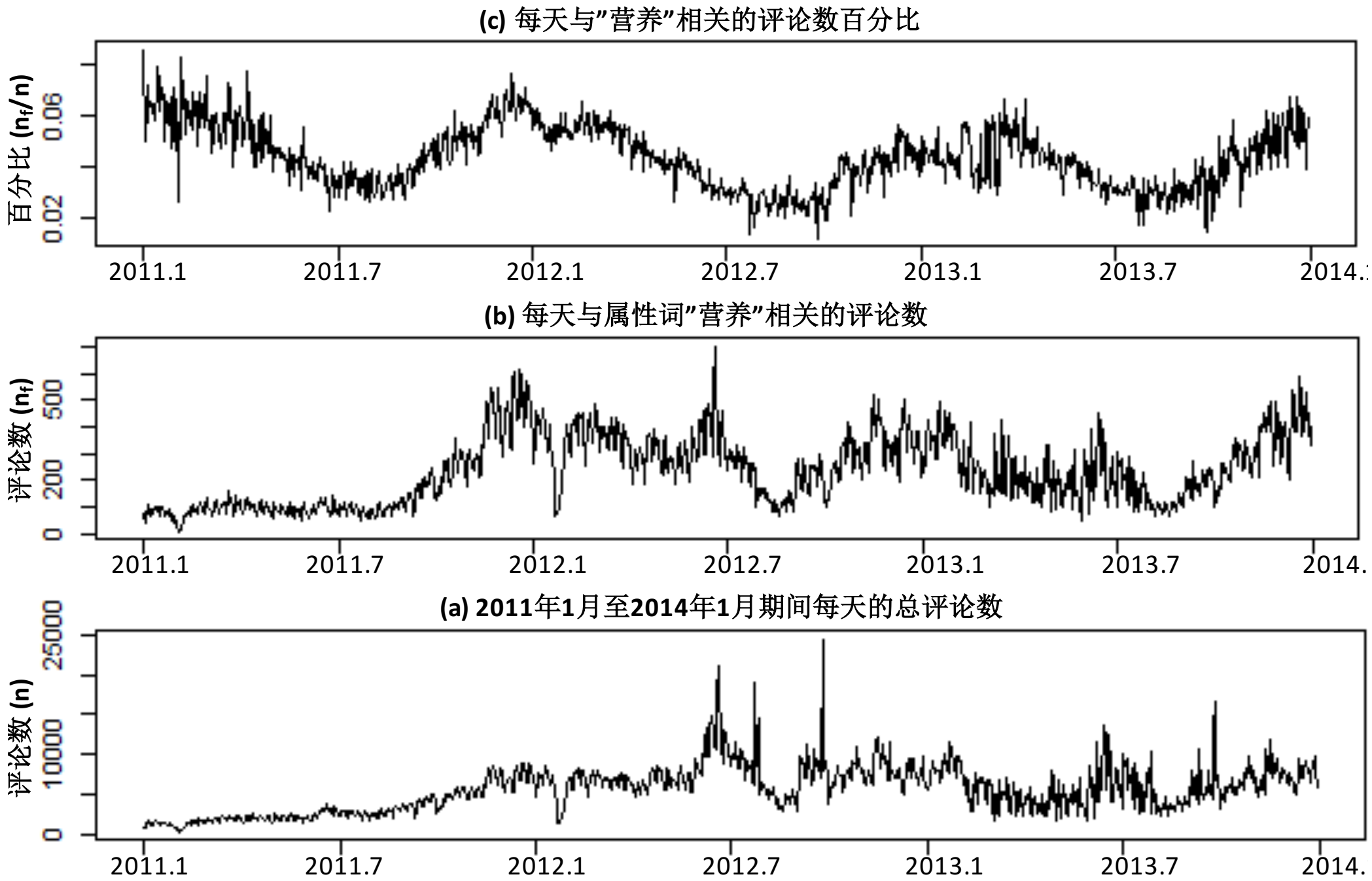

图 4.9　三种被考察的时间序列：(a) 每天的总评论数时间序列 $N(t)$；(b) 每天与特定属性词 $f$（以“营养”为例）相关的评论数时间序列 $N_f(t)$；(c) 每天与属性词 $f$ 相关的评论数百分比时间序列 $N_f(t)/N(t)$

我们首先观察每天生成的评论数时间序列 $N(t)$，如图4.9(a) 所示，该时间序列表述了系统中每天总的购买量信息，与特定的用户和物品无关。经过分析发现，$N(t)$ 时间序列中的尖峰对应于系统中特定的促销活动，例如在 2012 和 2013 年的十月一日，网站进行了针对国庆假期的促销活动。

由于我们需要预测每一个属性词各自的流行度，我们进一步分析每一个属性词 $f$ 每天被提及的评论的数量 $N_f(t)$，如图4.9(b) 所示。我们希望能够从各个属性词的时间序列中挖掘特定的周期性和季节性规律从而完成时间序列的模拟和预测，然而属性词直接的评论数序列 $N_f(t)$ 并没有明显的周期性信息，这主要是由于随着京东商城系统的不断扩张和用户数的逐渐增加，使得每日用户活跃量的增加掩盖了系统中隐含的周期性与规律性，庞大的数据量使得每个属性词各自的时间序列 $N_f(t)$ 展示出与总体时间序列 $N(t)$ 相似的模式。

因此，我们进一步分析了提及某个属性词的用户评论的百分比时间序列 $X_f(t) = N_f(t)/N(t)$，如图4.9(c) 所示。如我们所期望的，百分比时间序列在三个连续年度上展示出明显的季节性和周期性规律。

我们进一步验证百分比时间序列的定量性质，也为了验证除示例属性词“营养”之外所有属性词的总体统计性质，我们对 $r$ 个属性词中的每一个属性词 $f \in$

$\mathcal{F} = \{f_1, f_2, \cdots, f_r\}$ 所对应的百分比时间序列 $X_f(t)$ 进行分解。不失一般性，我们用 $X(t)$ 表示任意一个属性词的百分比时间序列。与通用的时间序列分解方式一样[300,301]，我们将一个时间序列 $X(t)$ 分解为三个部分的累加，分别为趋势项 $T(t)$、周期项 $S(t)$，以及随即项 $R(t)$：

$$X(t) = T(t) + S(t) + R(t) \tag{4-16}$$

趋势项用来描述时间序列中长期的增减趋势以及与周期无关的长期起伏变化，并由如下的移动平均过程获得[300]，其中移动平均窗口的长度为 $2s+1$：

$$T(t) = \frac{1}{2s+1}\sum_{j=-s}^{s} X(t+j) \tag{4-17}$$

周期项 $S(t)$ 用来描述时间序列中蕴含的周期性变化，周期可能是季度、月、周甚至天，等等。为了获得周期项 $S(t)$，我们首先在原始时间序列中将趋势项剔除，得到 $X(t) - T(t) = S(t) + R(t)$；我们进一步尝试以一年为周期构建周期项，即将时间序列 $X(t) - T(t)$ 按照年度分为三份。通过将这三份时间序列进行平均，我们得到季节项 $S(t)$ 一年的值[300]。该构造方式是基于如下的两个性质：首先，按照定义周期项 $S(t)$ 是按照周期不断重复的；其次，白噪声随机项 $R(t)$ 可以通过平均法得到消除。在趋势项和周期项的基础上，我们最终通过在原始时间序列中将两者剔除得到随机项 $R(t) = X(t) - T(t) - S(t)$。

图4.10显示了属性词“营养”的百分比时间序列分解结果，其中在计算趋势项 T(t) 时所使用的滑动窗口长度为一年（即公式 (4-17) 中 $s = 182$）。趋势项的结果显示，用户对该属性词的长期关注程度先下降后上升，而周期项则随时间显示出该属性词关注度明显的周期性和季节性。

在百分比时间序列中，周期项 $S(t)$ 占据了时间序列的绝大部分能量，而这作为时间序列动态预测的基础具有重要意义。为了研究周期项的规则程度，我们用一个简单的周期为 $m = 365$ 的一阶傅里叶级数来模拟每一个属性词 $f$ 的周期项时间序列 $S(t)$：

$$S(t) \approx \hat{S}(t) = a + A\sin\left(\frac{2\pi t}{m} + \varphi\right) \tag{4-18}$$

我们利用无偏赤池信息量准则（Akaike Information Criterion corrected，AICc）最小化来估计拟合的参数，AICc 是时间序列分析中通用的用来描述统计模型拟合

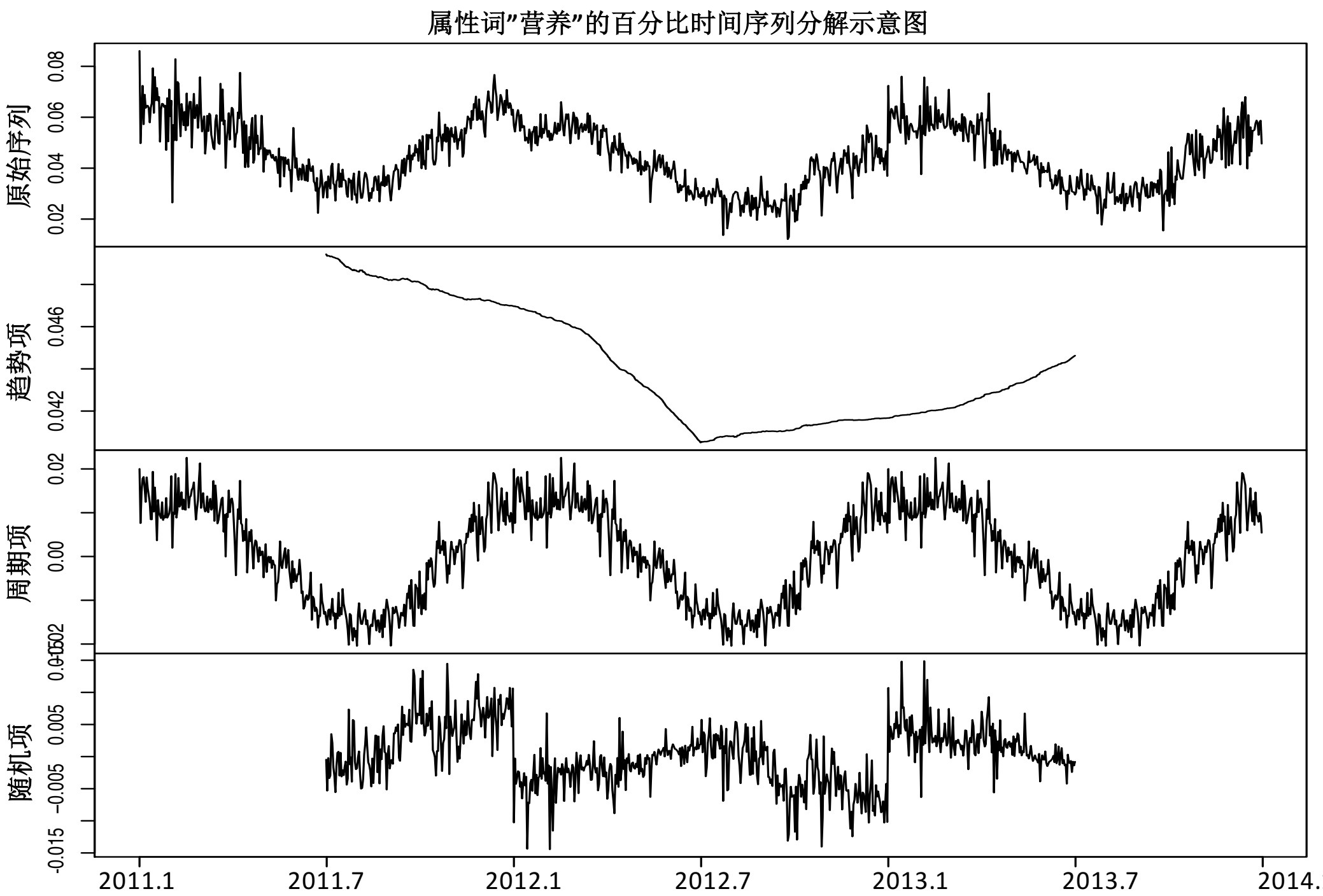

图 4.10　百分比时间序列的分解示例，从上到下分别为原始百分比时间序列、趋势项、周期项和随机项

优度的指标[304]，它是赤池信息量准则（Akaike Information Criterion，AIC）在考虑参数复杂度时的无偏纠正版本：

$$\text{AICc} = \underbrace{2k + n\ln\left(\frac{1}{n}\sum_{t=1}^{n}\left(S(t) - \hat{S}(t)\right)^2\right)}_{\text{AIC}} + \frac{2k(k+1)}{n-k-1} \tag{4-19}$$

其中 $k$ 为参数的个数，$n$ 为时间序列中总的观测点数。由于我们采用了包含三个参数的一阶傅里叶项，因此在本部分我们有 $k = 3$。

对于每一个属性词 $f$ 所对应的周期项时间序列 $S_f(t)$，对其进行一阶傅里叶近似，并计算这样一个简单的一阶傅里叶级数所能够描述的原时间序列全部能量的百分比：

$$p_f = \sum_{t=1}^{n}|\hat{S}_f(t)|^2 \Big/ \sum_{t=1}^{n}|S_f(t)|^2 \tag{4-20}$$

根据 Parseval 定理，一个满阶的傅里叶级数能够完备描述从时域转化到频域的全部能量，即 $p_f = 1$，而一个低秩的傅里叶级数近似将带来信息的损失，即 $p_f < 1$。在本工作中，我们通过 $F_1$ 值过滤共筛选了 $r = 58$ 个高质量属性词，并将在性能评测部分对属性词的选择及其性质进行介绍。在这 58 个属性词上 $p_f$ 的分布情况如

表4.2所示。

表 4.2　各个属性词周期项时间序列的一阶傅里叶近似所能保持的能量百分比分布情况

| 能量百分比 $p_f$ | 70% ~ 80% | 80% ~ 90% | 90% ~ 100% |
|---|---|---|---|
| 该区间属性词个数 | 7 | 35 | 16 |
| 所占百分比 (%) | 12.1 | 60.3 | 27.6 |

我们发现，一个简单的一阶傅里叶近似就能在所有属性词上保持 70% 以上的能量，而有约 88%（60.3%+27.6%=87.9%）的属性词保持了 80% 以上的能量。

通过以上的时间序列分解分析可知，百分比时间序列能够较好地描述属性词流行度随时间变化的趋势性、周期性、季节性和随机性，并且其中所包含的周期项能够被傅里叶近似项较好地近似。这是接下来动态时间序列建模和预测的重要基础。在下面的部分，我们采用该百分比时间序列进行用户偏好的动态建模和个性化推荐。

### 4.2.4　属性词流行度的动态预测

在本节，我们介绍精确到天级别的属性词流行度动态预测算法，我们首先介绍传统的移动平均自回归模型（Auto Regressive Integrated Moving Average，ARIMA）并指出它在本任务中的不足之处，进而在此基础上提出我们的基于傅里叶辅助项的移动平均自回归模型（Fourier-assisted Auto Regressive Integrated Moving Average，FARIMA）。

#### 移动平均自回归模型

一个 $(p,q)$-阶的移动平均回归模型（Auto Regressive Moving Average，ARMA）利用一个 $p$ 阶的回归项（Auto Regressive，AR）和一个 $q$ 阶的移动平均项（Moving Average，MA）对时间序列进行拟合：

$$\begin{aligned} X(t) &= \alpha_1 X(t-1) + \cdots + \alpha_p X(t-p) \\ &\quad + Z(t) + \beta_1 Z(t-1) + \cdots + \beta_q Z(t-q) \end{aligned} \tag{4-21}$$

其中 $Z(t) \overset{iid}{\sim} \Phi(0,\sigma^2)$ 为从一个零期望正态分布中独立采样得到的随机变量。

例如，在最简单的移动平均回归模型 $X(t) \sim \mathrm{ARMA}(1,0)$ 中，我们有 $X(t) = \alpha_1 X(t-1) + Z(t)$，即模型将时间序列中的每一个观测点建模为它前面的观测点加上一个白噪声。

移动回归模型的有效使用要求时间序列 $X(t)$ 为一个平稳过程（stationary process），即对于每一个正整数 $k \geq 1$ 以及每一个整数 $\tau$，如下的两个离散过程：

$$\{X(t_1), X(t_2), \cdots, X(t_k)\}和\{X(t_1+\tau), X(t_2+\tau), \cdots, X(t_k+\tau)\} \tag{4-22}$$

必须具有相同的点分布：

$$F_{t_1,t_2,\cdots,t_k}(X_1, X_2, \cdots, X_k) = P(X(t_1) < X_1, \cdots, X(t_k) < X_k) \tag{4-23}$$

实际上，多数实际中的时间序列并不一定满足如上的平稳过程要求；然而多数时间序列的 $d$-阶差分序列 $W(t) \doteq \nabla^d X(t)$ 满足如上的要求。其中，一阶差分序列定义为 $\nabla X(t) = X(t) - X(t-1)$，更高阶的差分序列迭代地定义为 $\nabla^d X(t) = \nabla(\nabla^{d-1} X(t))$。

当我们用如公式 (4-21) 所示的 ARMA$(p,q)$ 过程来近似原始时间序列的 $d$-阶差分时间序列 $W(t) \doteq \nabla^d X(t)$ 时，我们称原始时间序列 $X(t)$ 被建模为 $(p,d,q)$-阶的移动回归自平均（Auto Regressive Integrate Moving Average，ARIMA）过程 ARIMA$(p,d,q)$。

### 基于傅里叶辅助项的移动平均自回归模型

在 ARIMA$(p,d,q)$ 模型中，对未来时间点的预测需要借助于至少一个周期的历史信息才能正确地对时间序列中的周期性进行恰当的模拟，这意味着模型所包含的参数个数（尤其是回归项的个数 $p$）需要至少达到时间序列的周期长度量级。对于以季度、月度或者周为周期的时间序列而言，它们的周期分别包含约 4 个、12 个或者 52 个数据点，较小的参数规模使得这些时间序列可以较为容易地在 AICc 下进行拟合。而本工作力图在精确到天级别上对属性词流行度和用户偏好进行动态预测，因而需要处理周期长达一年（365 个时间点）的时间序列，大量的参数规模使得模型参数估计在实际算法中几乎不可行。

幸运的是，如上节所述的时间序列能量分析使得我们可以借助简单的低阶傅里叶级数对时间序列进行粗略的预测，并进一步利用一个低阶的 ARIMA 模型对残差项进行估计，从而使精确到天级别的时间序列预测成为可能，如下所示：

$$X(t) \approx \underbrace{a + \sum_{k=1}^{K}\left[\alpha_k \sin\left(\frac{2\pi kt}{m}\right) + \beta_k \cos\left(\frac{2\pi kt}{m}\right)\right]}_{F(t)} + \underbrace{\mathrm{ARIMA}(p,d,q)(t)}_{E(t)} \doteq \hat{X}(t) = F(t) + E(t) \tag{4-24}$$

其中 $m = 365$ 为时间序列的周期，$F(t)$ 为一个 $K$-阶的傅里叶级数，$E(t)$ 为一个低阶的 ARIMA 过程用来描述原始时间序列去除傅里叶级数项之后的残差部分。该模型的直观原始在于，由于时间序列的周期项是年度稳定的，因此我们可以用具有固定周期的傅里叶级数对时间序列的周期框架进行近似；我们进一步从原始时间序列中减去傅里叶项从而使残差时间序列的周期大大降低，进而可以用一个低阶的 ARIMA 模型来进行估计。图4.11通过直观的例子描述了该模型的直观意义，其中的傅里叶项 $F(t)$ 和 ARIMA 项 $E(t)$ 之和为对原始时间序列 $X(t)$ 的估计。

在本工作中，我们采用 AICc 最小化来确定傅里叶项 $F(t)$ 的阶数 $K$ 以及

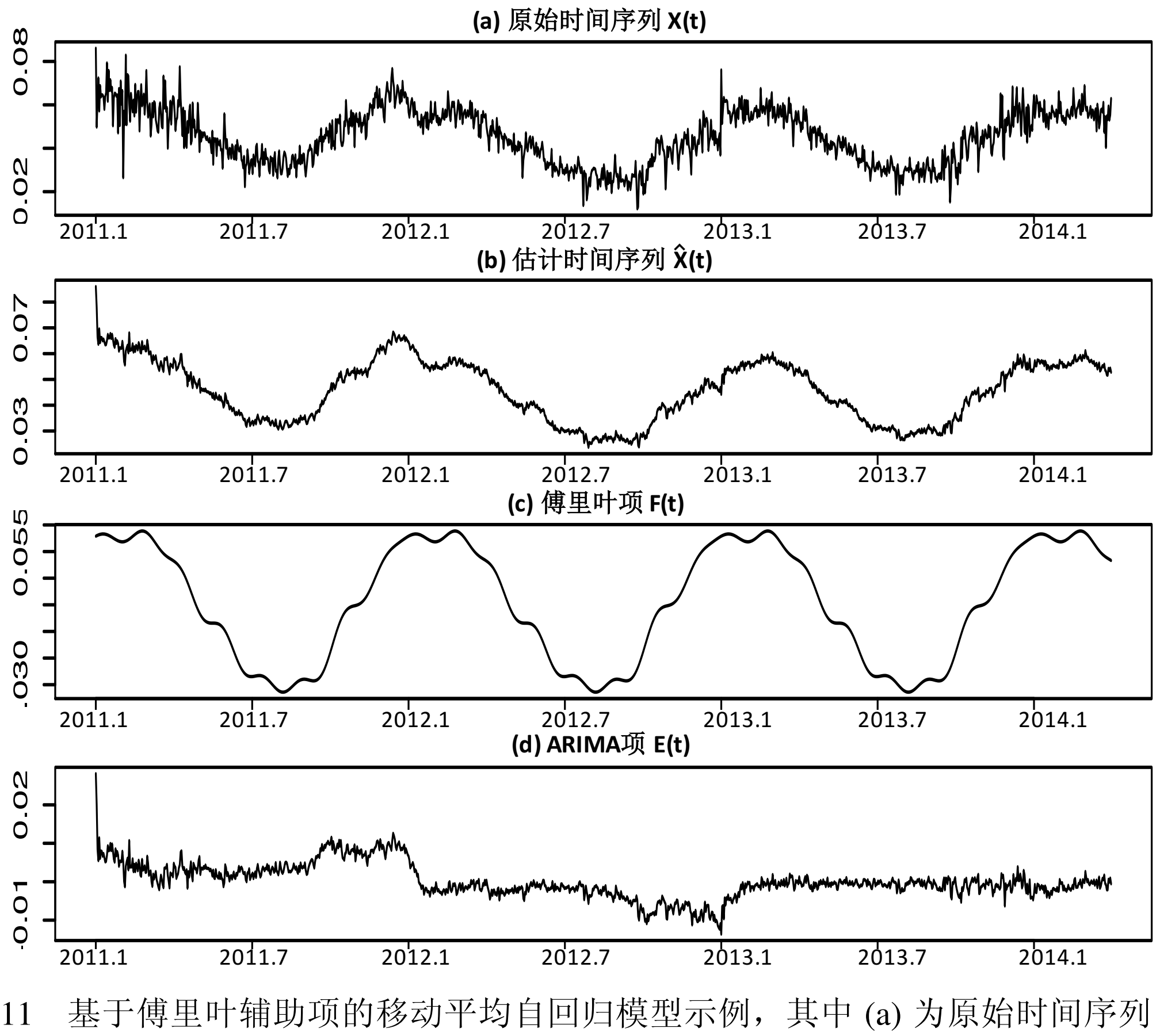

图 4.11　基于傅里叶辅助项的移动平均自回归模型示例，其中 (a) 为原始时间序列 $X(t)$，(b) 为估计的时间序列 $\hat{X}(t) = F(t) + E(t)$，(c) 为傅里叶项 $F(t)$，(d) 为 ARIMA 项 $E(t)$

ARIMA 项的阶数 $(p, d, q)$。算法7给出了时间序列模拟的参数估计算法，其中的子进程 Fourier 和 Arima 分别在指定的阶次 $K$ 和 $(p, d, q)$ 下利用 AICc 最小化对时间序列进行拟合，并返回拟合模型及其对应的 AICc 值。在本工作中，我们采用数据分析工具 R 语言[①] 中的 Fourier 和 Arima 进程作为具体实现。

**Algorithm 7:** Farima$(X(t), m, \bar{K}, \bar{p}, \bar{d}, \bar{q})$

**Input:** 时间序列 $X(t)$, 周期 $m = 365$,
傅里叶项的最大阶数：$\bar{K}$，ARIMA 项的最大阶数：$\bar{p}, \bar{d}, \bar{q}$

**Output:** 拟合时间序列 $\hat{X}(t)$

$minAICc \leftarrow \infty; \hat{F}(t);$
**for** $K \leftarrow 1$ **to** $\bar{K}$ **do**
  $\{F(t), AICc\} \leftarrow$ Fourier$(X(t), m, K)$;
  **if** $AICc < minAICc$ **then**
    $\hat{F}(t) \leftarrow F(t);$
    $minAICc \leftarrow AICc;$
  **end**
**end**
$minAICc \leftarrow \infty; \hat{E}(t);$
**for** $(p, d, q)$ ***in*** $[0, \bar{p}] \times [0, \bar{d}] \times [0, \bar{q}]$ **do**
  $\{E(t), AICc\} \leftarrow$ Arima$(X(t) - \hat{F}(t), p, d, q)$;
  **if** $AICc < minAICc$ **then**
    $\hat{E}(t) \leftarrow E(t);$
    $minAICc \leftarrow AICc;$
  **end**
**end**
**return** $\hat{X}(t) \leftarrow \hat{F}(t) + \hat{E}(t);$

### 4.2.5 基于条件机会估计的时序推荐模型

在本节，我们利用如上所预测的百分比时间序列（表示用户的全局偏好信息）、用户的个性化偏好，以及物品的个性化的属性来实现精确到天级别的个性化推荐。表4.3列出了本节的主要符号。

① http://www.r-project.org

表 4.3　本节符号表

| | |
|---|---|
| $\mathcal{F}$ | $\mathcal{F}=\{f_1,f_2,\cdots,f_r\}$ 表示属性词集合 |
| $\hat{X}_f(t)$ | 对时间 $t$ 上属性词 $f$ 的百分比时间序列预测值 |
| $\mathbf{x}(t)$ | 包含全局动态偏好、用户偏好和物品属性的表示向量 |
| $\lambda,\gamma$ | Weibull 分布的幅度参数和形状参数 |
| $\boldsymbol{\theta}$ | 表示向量的拟合参数 |
| $\boldsymbol{\mu},\boldsymbol{\Sigma}$ | 对拟合参数 $\boldsymbol{\theta}$ 进行采样的高斯分布参数 |
| $a,b$ | 对形状参数 $\gamma$ 进行采样的 $\Gamma$ 分布参数 |
| $(y\|u,i,t)$ | 包含用户 $u$、物品 $i$、打分 $y$ 和时间 $t$ 的一个样本点 |
| $\xi_u$ | $\xi_u=\{\boldsymbol{\theta}_u,\gamma_u\}$ 为用户 $u$ 的隐变量参数 |
| $D$ | $D=\{(y\|u,i,t)\}_{k=1}^{n}$ 为包含 $n$ 个数据点的样本集合 |
| $y_u$ | $y_u=\{y_{u,1},y_{u,2},\cdots,y_{u,N_u}\}$ 为数据集 $D$ 中用户 $u$ 的打分集合 |
| $\hat{y}_{ui}(t)$ | 对用户 $u$ 在物品 $i$ 和时间 $t$ 上的打分预测 |

### 特征空间的表示

对于每一个属性词 $f$，算法所预测的百分比时间序列 $\hat{X}_f(t)$ 是系统中用户群体在时间 $t$ 对属性词 $f$ 可能的关注程度，因此描述了系统用户群的全局偏好。在百分比时间序列预测的基础上，我们构造如下的全局偏好表示向量：

$$\mathbf{x}^g(t)=[x_1^g(t),x_2^g(t),\cdots,x_r^g(t)]^{\mathrm{T}} \tag{4-25}$$

其中第 $k(1\le k\le r)$ 个元素 $x_k^g(t)=\frac{1}{1+\exp(-\hat{X}_{f_k}(t))}$ 是在 Sigmoid 函数下第 $k$ 个属性词 $f_k\in\mathcal{F}$ 的预测流行度 $\hat{X}_{f_k}(t)$ 的归一化值。

另外，由于不同的用户在属性词上的偏好不同，我们仍然采用用户 $u$ 的历史评论数据来对该用户的偏好进行建模。假设在时间 $t$ 上用户 $u$ 的历史评论数据中对第 $k$ 个属性词 $f_k$ 提及了 $w_k^u(t)$ 次，则我们用如下的用户偏好向量对该用户的个性化信息进行建模：

$$\mathbf{x}^u(t)=[x_1^u(t),x_2^u(t),\cdots,x_r^u(t)]^{\mathrm{T}} \tag{4-26}$$

其中 $x_k^u(t)=\frac{1}{1+\exp(-w_k^u(t))}$ 为 $w_k^u(t)$ 的 Sigmoid 归一化。

类似的，我们同样利用一个物品在时间 $t$ 之前的历史评论对其构建物品属性向量：

$$\mathbf{x}^i(t)=[x_1^i(t),x_2^i(t),\cdots,x_r^i(t)]^{\mathrm{T}} \tag{4-27}$$

其中 $x_k^i(t) = \frac{1}{1+\exp(-w_k^i(t))}$，$w_k^i(t)$ 为物品 $i$ 的历史评论中属性词 $f_k$ 被提及的次数。

最终，我们通过融合全局偏好向量 $\mathbf{x}^g(t)$、个性化用户偏好向量 $\mathbf{x}^u(t)$，以及物品属性向量 $\mathbf{x}^i(t)$，在时间 $t$ 上构建用户-物品对 $(u, i)$ 的描述向量 $\mathbf{x}(t)$：

$$\mathbf{x}(t) = \left[\mathbf{x}^g(t)^{\mathrm{T}}, \mathbf{x}^u(t)^{\mathrm{T}}, \mathbf{x}^i(t)^{\mathrm{T}}\right]^{\mathrm{T}} \tag{4-28}$$

在实际应用中，如上描述向量的构建实际上具有很强的灵活性，我们不仅可以利用属性词流行度来构建用户物品对 $(u, i)$ 的时间敏感描述向量，还可以考虑系统中存在的其它可行信息，例如用户相关的人口统计学信息（Demographics），以及物品相关的内容信息，等等。由于描述向量的灵活性和可扩展性，这些信息可以很容易地被加入到描述向量中并参与如下的用户行为建模和预测。

### 条件机会模型

我们在动态表示向量的基础上构建如下的条件机会模型[290] 来实现打分预测，其中数值打分 $y$ 的分布函数以标准 Weibull 分布来描述：

$$p(y) = \gamma\lambda y^{\gamma-1}\exp(-\lambda y^{\gamma}) \tag{4-29}$$

给定一个动态表示向量 $\mathbf{x}(t)$（其中包含全局、用户以及物品信息），我们对 Weibull 分布中的幅度系数 $\lambda$ 利用回归参数 $\boldsymbol{\theta}$ 和表示向量 $\mathbf{x}(t)$ 进行重参数化，即 $\lambda = \exp(\boldsymbol{\theta}^{\mathrm{T}}\mathbf{x}(t))$，进而得到：

$$p(y) = \gamma\exp\left(\boldsymbol{\theta}^{\mathrm{T}}\mathbf{x}(t)\right)y^{\gamma-1}\exp\left(-\exp\left(\boldsymbol{\theta}^{\mathrm{T}}\mathbf{x}(t)\right)y^{\gamma}\right) \tag{4-30}$$

对于每一个用户 $u$，我们从高斯分布 $\boldsymbol{\theta}_u \sim \Phi(\boldsymbol{\mu}, \boldsymbol{\Sigma})$ 中采样回归参数 $\boldsymbol{\theta}_u$，并从伽马分布 $\gamma_u \sim \Gamma(a, b)$ 中采样形状参数 $\gamma_u$。令 $D = \{(y|u, i, t)\}_{k=1}^{n}$ 表示 $n$ 个用于训练的用户物品打分样本点，其中的每个样本点 $(y|u, i, t)$ 表示用户 $u$ 在时间 $t$ 向物品 $i$ 打了分数 $y$。我们利用如下的条件机会模型对打分 $y$ 的条件概率分布进行建模：

$$\begin{aligned} p(y|u,i,t) &= p\left(y|\boldsymbol{\theta}_u, \gamma_u, \mathbf{x}(t)\right) \\ &= \gamma_u\exp\left(\boldsymbol{\theta}_u^{\mathrm{T}}\mathbf{x}(t)\right)y^{\gamma_u-1}\exp\left(-\exp\left(\boldsymbol{\theta}_u^{\mathrm{T}}\mathbf{x}(t)\right)y^{\gamma_u}\right) \end{aligned} \tag{4-31}$$

令 $\phi = (\boldsymbol{\mu}, \boldsymbol{\Sigma}, a, b)$ 和 $\xi_u = \{\boldsymbol{\theta}_u, \gamma_u\}$ 表示用户 $u$ 的隐变量，$\mathbf{y}_u = \{y_{u,1}, y_{u,2}, \cdots, y_{u,N_u}\}$ 表示用户 $u$ 在训练集中的打分，则训练集 $D$ 的似然函数可

以表示为如下的 $\phi$ 的函数：

$$p(D|\phi) = \prod_{u=1}^{|\mathcal{U}|} p(y_u|\phi) = \prod_{u=1}^{|\mathcal{U}|} \int p(\xi_u, y_u|\phi) d\xi_u \tag{4-32}$$

其中 $\mathcal{U}$ 为系统中的用户集合。最大化似然函数 $p(D|\phi)$ 等价于最大化其对数似然函数 $L(\phi) = \ln p(D|\phi)$。在本工作中，我们 Wang 等[290] 的 EM 算法进行最大化和参数估计。

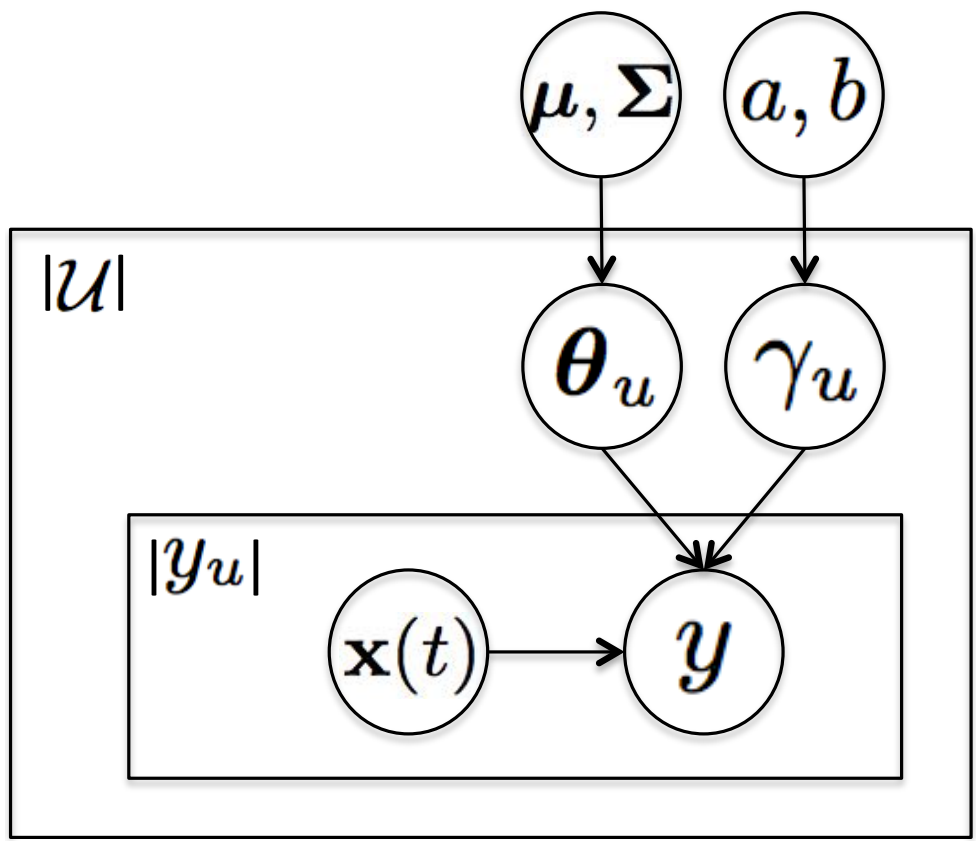

图 4.12　基于条件机会模型的个性化推荐参数依存关系，用户 $u$ 在物品 $i$ 上的打分 $y_{u,i}$ 依存于用户 $u$ 的条件机会模型参数 $\xi_u = \{\theta_u, \gamma_u\}$ 以及整合的表示向量 $\mathbf{x}(t)$；不同用户的条件机会模型通过共同的先验概率参数 $\phi = (\mu, \Sigma, a, b)$ 共享信息。

### 打分预测和个性化推荐

对于每一个用户 $u$ 和物品 $i$，给定由 FARIMA 模型预测得到的表示向量 $\mathbf{x}(t)$，以及由条件机会模型估计得到的回归参数 $\theta_u$ 和形状参数 $\gamma_u$，我们选择使得公式 (4-31) 中的条件概率 $p(y|\theta_u, \gamma_u, \mathbf{x}(t))$ 最大化的打分 $y$ 作为预测打分。打分预测值的封闭解可以由公式 (4-29) 中 Weibull 分布的导数很容易地得到：

$$p'(y) = \gamma\lambda y^{\gamma-2} \exp(-\lambda y^{\gamma})(\gamma - 1 - \lambda\gamma y^{\gamma}) \tag{4-33}$$

通过令 $p'(y) = 0$ 以及 $\lambda = \exp(\theta_u^{\mathrm{T}}\mathbf{x}(t))$，用户 $u$ 在物品 $i$ 上的预测打分将为：

$$\hat{y}_{ui}(t) = \left(\frac{\gamma_u - 1}{\gamma_u \exp(\theta_u^{\mathrm{T}}\mathbf{x}(t))}\right)^{\frac{1}{\gamma_u}} \tag{4-34}$$

在实际中，我们将用户 $u$ 未购买过的物品按照预测打分 $\hat{y}_{ui}(t)$ 降序排列，从而为用户 $u$ 提供天级别的动态个性化推荐列表。

## 4.3　性能评测

在本节，我们对模型的可解释性进行评测，包括打分预测和推荐列表构建等线下任务和评测指标，以及真实用户线上评价指标。

### 4.3.1　基于显式变量模型的可解释性推荐评测

我们对基于显式变量分解模型的可解释性推荐进行评测。我们采用 Yelp① 和 Dianping②（大众点评）数据集，如表4.4所示。Yelp 数据集由美国凤凰城的地点及其用户评论组成，主要为英文评论；大众点评数据集与上一章所介绍的相同，包括中国几大主要城市的餐厅等地理位置及其用户评论，内容主要为中文评论，其中为了避免冷启动的问题，我们选择了那些评论次数大于 20 次的用户进行实验。

表 4.4　Yelp 和大众点评数据集基本统计数据

| 数据集 | 用户数 | 物品数 | 评论数 | $\frac{\text{评论数}}{\text{用户数}}$ |
|---|---|---|---|---|
| Yelp | 45,981 | 11,537 | 229,907 | 5.00 |
| Dianping | 11,857 | 22,365 | 510,551 | 43.06 |
| Yelp10 | 4,393 | 10,801 | 138,301 | 31.48 |

作为自然语言处理的一个分支，短语级情感分析具有一定的语言相关性，因此，我们分别采用了两种语言环境（Yelp 的英文和点评的中文）对模型效果进行研究。通过两种不同语言的实验结果研究，我们试图分析模型在不同环境下的效果和差异。

需要指出的是，Yelp 数据本身具有非常强的稀疏性，实际上统计发现数据中有 49% 的用户只有一条评论记录，这使得该数据集上无法进行 top-K 推荐的评价。为了解决该问题，我们进一步从 Yelp 数据集中选出了那些打分次数超过 10 次的用户并构建了“Yelp10”数据集，如表4.4所示，该数据集用于对 Yelp 进行 top-K 推荐的评测。

在本实验中，我们设置算法6的最大迭代次数为 100，利用网格搜索确定超参数的最优取值，并利用五折交叉验证进行算法运行和评价。

① `http://www.yelp.com/dataset_challenge`
② `http://www.dianping.com`

### 短语级情感词典构建

在短语级情感分析的构建中，我们往往需要在准确率和召回率之间进行权衡：较高的阈值要求可以带来较高的准确率，但是召回率会降低；而较低的阈值要求可以抽取出更多的属性观点词对从而召回率较高，但是准确率会相应降低。作为显式变量分解模型的第一步，我们在本工作中采用较为严格的参数设定获得准确率较高的属性观点词对，从而尽可能降低词典构建精度对预测和推荐效果带来的影响。随着情感分析技术的不断发展和成熟，情感词典构建的性能越来越高，我们的推荐算法也可以从中受益。

与 Lu 等[210] 相同，我们请三位标注人员对情感分析算法所构建的情感词典进行人工标注和评价。对于每一个属性词和情感词，如果有超过两位标注人员认为是正确抽取的词语，则我们认为该词语抽取正确。为了对属性情感词对的极性进行标注，我们将算法所给出的连续数值情感得分 $S \in [-1, 1]$ 按照正负分别映射为“正面情感”和“负面情感”，并请标注人员进行正确性标注，由于算法没有给出恰好为 0 的情感得分，所以结果中没有“中性情感”分类。对情感词典的评测统计结果如表4.5所示。

表 4.5　短语级情感词典构建的主要统计信息及效果评测

| 数据集 | 属性词数 | 情感词数 | 词条数 | 属性词精度 | 情感词精度 | 极性精度 |
|---|---|---|---|---|---|---|
| Yelp | 96 | 155 | 845 | 92.71% | 91.61% | 94.91% |
| Dianping | 113 | 284 | 1,129 | 89.38% | 89.79% | 91.41% |

表 4.6　Yelp 和点评数据集上的属性观点词对样例

| Yelp 数据集 | | 大众点评数据集 | |
|---|---|---|---|
| 属性词相关 | 情感词相关 | 属性词相关 | 情感词相关 |
| (decor, cool, +) | (price, fire, +) | (服务质量, 高, +) | (上菜, 快, +) |
| (service, cool, -) | (price, high, -) | (价格, 高, -) | (上菜, 慢, -) |
| (price, high, -) | (parking space, plenty, +) | (有效期, 长, +) | (环境, 优雅, +) |
| (service quality, high, +) | (parking space, limit, -) | (等位时间, 长, -) | (环境, 吵, -) |

统计结果可见，我们利用短语级情感分析技术在每个数据集上各抽取了约 100 个高质量的属性词，其中在两个数据上的一些词条样例如表4.6所示，该表通过样例展示了属性词和情感词的上下文相关性。

### 数值打分预测

我们首先对用户物品评分矩阵 $A$ 上的数值打分预测任务进行评价。我们采用如下的方法作为基线方法进行对比：

- NMF：非负矩阵分解[62]，我们采用对数预测函数和 Frobenius 范数正则化。
- PMF：概率化矩阵分解，采用基于马尔科夫蒙特卡洛（MCMC）的贝叶斯概率分解模型[68]。
- MMMF：最大间隔矩阵分解，采用了快速最大间隔矩阵分解算法[64]。
- HFT：基于话题模型的隐变量法[191]，将矩阵分解中的隐变量映射为利用评论文本挖掘得到的话题，是目前基于话题模型的评分预测最好方法之一。

对于每种算法，我们采用网格搜索的方法确定超参数的最优取值。对于 HFT 方法，我们设置话题个数为 5，与论文原文报告的最好设定相一致[191]，并且实验发现设定更多的话题个数并不能进一步提高预测精度。我们仍然采用跟均方差（RMSE）作为评测指标。

在显式变量分解模型（公式 (4-6)）中，显式变量 $U$ 和隐式变量 $H$ 分别用来刻画短语级情感分析所挖掘的显式属性词和算法所设定的未知隐式属性词所带来的影响。为了研究两者对预测效果各自的影响，我们首先分析显式变量的个数 $r$（分解因子矩阵 $U$ 的维度）和隐式变量的个数 $r'$（分解因子矩阵 $H$ 的维度）之比对预测效果的影响。我们设定两张变量总个数 $r + r' = 100$，并调节两者的比例。对于基线矩阵分解算法 NMF、PMF 和 MMMF，我们同样设定矩阵分解的维度为 100 以保证比较的公平性，对于 HFT 算法，如上所述，我们设定话题个数为 5。

实验结果如图4.13所示，其中横坐标表示我们的显式变量分解模型（EFM）中所使用的显式分解变量的个数 $r$，纵坐标表示预测精度指标 RMSE，其中 RMSE 越小，表示预测精度越高。在五折交叉验证中，基线算法和我们的显式变量分解模型中 RMSE 的标准差小于 0.002。

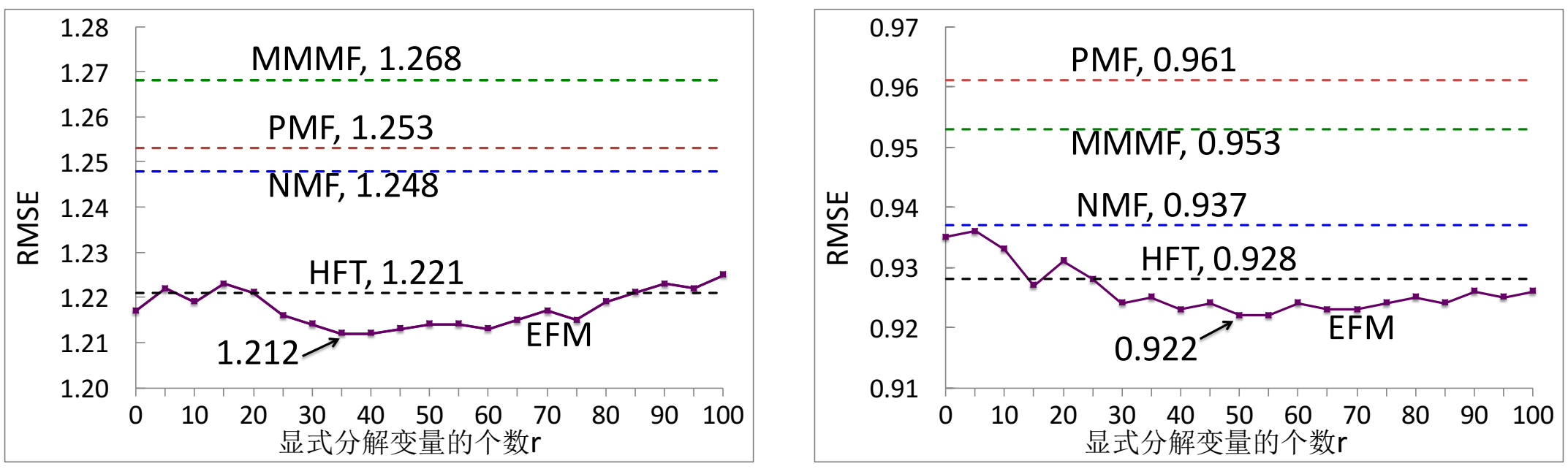

(a) Yelp 数据集　　　　(b) 点评数据集

图 4.13　Yelp 和大众点评数据集上显式分解变量 $U$ 的个数对预测精度的影响。

由实验结果可见，当我们仅使用少量的显式分解变量和多数的隐式分解变量时，我们的方法与基于话题的隐变量法以及非负矩阵分解法具有类似的预测效果。然而当算法使用恰当个数的显式变量时（约占变量总数的 30% ~ 80%），我们的显式变量分解模型能够获得更好的预测效果。在 Yelp 数据集上，当显式分解变量的个数 $r = 35$ 时，我们获得最好的预测效果 RMSE=1.212；在点评数据集上，当 $r = 50$ 时获得最好的预测效果 RMSE=0.922。我们也发现，当算法使用过多的显式分解变量时会对预测精度带来负面影响。由此可见，由于显式变量并不能完全刻画用户所考虑的全部信息，在模型中保留合适个数的隐式变量对于保证模型的灵活性和预测效果具有重要作用。

我们进一步研究算法在不同的总分解因子个数 $r + r'$ 下的性能表现，在本实验中，我们固定显式分解变量的占比为在如上实验中获得最优性能的 40%，并调节总的分解变量个数 $r + r'$。作为对比，我们选择同样适用用户评论文本信息的 HFT 算法，以及如上实验中效果最好的隐变量矩阵分解算法 NMF 作为基线算法，效果如图4.14所示。

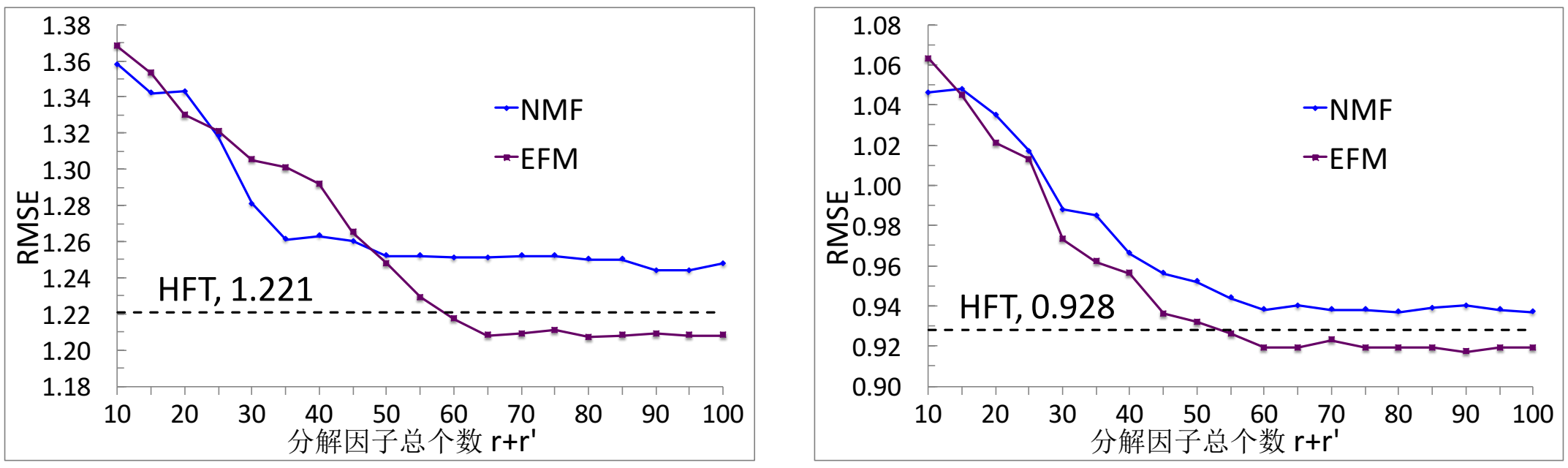

(a) Yelp 数据集　　(b) 点评数据集

图 4.14　Yelp 和大众点评数据集上显式分解变量 $U$ 和隐式分解变量 $H$ 的分解因子总个数 $r + r'$ 对预测精度的影响。

实验结果显示，对于非负矩阵分解和显式变量分解模型，预测精度都随着分解变量总个数 $r + r'$ 的增加而提高。然而，只有当分解因子总个数 $r + r'$ 足够多时，我们的方法才能获得比 NMF 和 HFT 更好的预测效果。这意味着我们的显式变量分解模型需要足够数量的分解因子才能较好地刻画用户偏好，并对用户打分的规律性进行正确的建模。然而只要我们选择恰当的分解因子个数，显式变量分解模型就能够在比较大的分解因子个数范围内提高打分预测精度。

### Top-K 推荐任务评测

在本部分，我们在 Top-K 推荐列表构建任务上对显式变量分解模型进行性能评测。在此基础上，我们进一步对属性词在个性化推荐中的作用和影响进行分析，

并指出什么样的属性词在个性化推荐任务中起重要作用。

我们仍然将显式分解因子的个数设定为 40%；在如上的实验中，我们发现分解因子总个数 $r + r'$ 为 65 时在两个数据集上都足以取得理想的效果，并且进一步增加分解因子个数并不能显著提高预测小姑偶，因此在本实验中我们设置分解因子总个数为 65，包括 25 个显式分解因子和 40 个隐式分解因子。我们采用如下的基线算法对效果进行评价：

- MostPopular：非个性化的推荐列表构建方法之一，对于每一个用户，推荐列表的构建方法均为按照物品的流行度（评论用户总数）由高到低进行排序。
- SlopeOne：实际系统中广泛使用的基于近邻的个性化推荐算法，相比于基于用户的协同过滤和基于物品的系统过滤而言实现更为简单，且具有更好的推荐效果[46]。
- NMF：基于对数预测函数和 Frobenius 范数正则化的非负矩阵分解[62]，在如上的实验中，该方法在所有基于隐变量矩阵分解的算法中取得最好效果。
- BPRMF：基于矩阵分解（Matrix Factorization，MF）的贝叶斯个性化排序（Bayesian Personalized Ranking，BPR）算法[168]，是目前最好的基于数值评分的 Top-K 推荐算法之一。
- HFT：基于话题模型的隐变量法[191]，将矩阵分解中的隐变量映射为利用评论文本挖掘得到的话题，是目前基于话题模型的评分预测最好方法之一。

由于 Yelp10 和点评数据集上用户的最少打分个数分别为 10 个和 20 个，因此我们在两个数据集上分别进行 Top-5（$K = 5$）和 Top-10（$K = 10$）推荐列表的构建；对于每一个用户，我们按照评论时间对其隐藏最后 $K$ 个物品进行评测，并用剩下的打分记录进行训练和 top-K 推荐列表的构建。由于隐藏的物品数和推荐列表所推荐的物品数一致，所以推荐列表在精度指标上的最优值为 100%。为了将被推荐物品在推荐列表中的位置信息纳入考虑，我们采用标准化折扣累计增益（Normalized Discounted Cumulative Gain，NDCG）和 ROC 特征曲线下面积（Area Under the ROC Curve，AUC）对排序性能进行评价：

$$\mathrm{NDCG} = \frac{1}{|U|}\sum_{i=1}^{|U|}\frac{1}{\mathrm{IDCG}}\sum_{j=1}^{K}\frac{2^{\delta_{ij}}-1}{\log_2(j+1)} \tag{4-35}$$

其中当用户 $u_i$ 的推荐列表在位置 $j$ 的物品确实被该用户评论过时，我们有 $\delta_{ij} = 1$，否则 $\delta_{ij} = 0$；IDCG 为推荐全部正确时（对于所有的位置 $\delta_{ij} = 1$）$\sum_{j=1}^{K}\frac{2^{\delta_{ij}}-1}{\log_2(j+1)}$ 的理想值，用于对公式进行归一化以保证 NDCG 的最优值为 1。

我们同时使用考虑物品相对排序的 ROC 曲线下面积（Area Under the ROC

Curve，AUC）指标来评价算法的排序性能：

$$\mathrm{AUC} = \frac{1}{|U|}\sum_{i=1}^{|U|}\frac{1}{|E(u_i)|}\sum_{(j,k)\in E(u_i)}\delta(r_{ij} < r_{ik}) \tag{4-36}$$

其中 $r_{ij}$ 和 $r_{ik}$ 分别为物品 $i_j$ 和物品 $i_k$ 在用户 $u_i$ 的推荐列表中的排序位置，对于用户 $u_i$ 而言，用于进行评价的物品二元组的集合 $E(u_i)$ 为：

$$E(u_i) = \{(j,k)|(i,j) \in \mathcal{S}_{test} \wedge (i,k) \notin (\mathcal{S}_{train} \cup \mathcal{S}_{test})\} \tag{4-37}$$

直观而言，我们希望那些在测试集中的真正被用户打过分的物品相对于用户没有过任何行为的物品排在前面；一个理想的随机推荐算法的 AUC 值为 0.5，而 AUC 的最高值为 1，即推荐列表将测试集中的物品全部推荐出来。

我们研究在推荐列表的构建排序公式 (4-13) 中所使用的用户最关心的属性词个数 $k$ 对排序性能的影响。我们首先固定 $k = 10$ 并调节公式 (4-13) 中的权重参数 $\alpha$，发现当 $\alpha = 0.85$ 时算法在 NDCG 上取得最好效果，且调节 $\alpha$ 的取值对排序性能的影响并不显著。因此我们在接下来的实验中固定 $\alpha = 0.85$ 并集中于对用户关心的属性词个数 $k$ 的研究。

在 Yelp 和点评两个数据集上，我们以 5 为步长，分别从 $k = 5$ 到其可能的最大取值（即对应数据集上属性词的个数，Yelp10 数据集为 96，点评数据集为 113）对 $k$ 进行调节，算法在 Yelp10 数据集上 NDCG 和 AUC 的性能表现如图4.15所示。

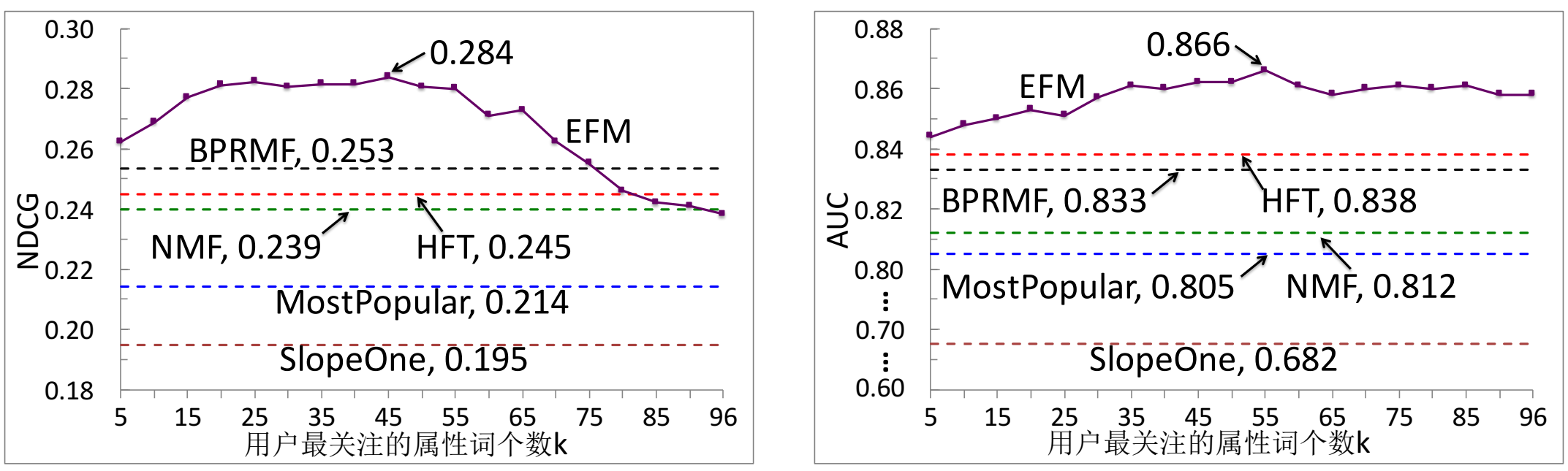

(a) NDCG 随 $k$ 的变化　　　　(b) AUC 随 $k$ 的变化

图 4.15　推荐算法排序性能（NDCG 和 AUC）随所使用的用户最关心的属性词个数 $k$ 的变化情况。

结果显示，在 $k = 15$ 之前，显式变量分解模型的 NDCG 排序性能随着所使用属性词的个数 $k$ 增加而增加，在 $k = 45$ 之前保持稳定，而当 $k > 45$ 则性能开始下降。当 $k < 70$ 时，显式变量分解模型的排序性能在 NDCG 指标上优于其它基线算

法，在取得最优值的 $k = 45$ 处，显式变量分解模型比最好的基线算法 BPRMF 性能提高 12.3%。然而，当 $k$ 的选择过高时，算法在 NDCG 指标上的性能开始下降，在点评数据集上有类似的实验观察。算法在 AUC 和 NDCG 上的实验结果性能具有稳定性，在每个实验点上的五次等价实验标准差小于 0.01。

该实验结果进一步验证了考虑过多的属性词将有可能对推荐算法引入噪声，从而降低推荐性能，这与基于话题模型的隐变量方法[191] 的实验观察和结论一致。然而算法在 AUC 上的结果显示我们的显式变量分解模型相比于基线算法具有更好的排序效果且在 $k$ 的取值上保持稳定。由于 AUC 只考虑被推荐物品与未推荐物品之间的相对排序，而不考虑它们的绝对位置，这表明当算法考虑过多的无关属性词时，对被推荐物品的绝对排序影响较大，而对正负样本之间的相对排序影响不大。

#### 对属性词的进一步分析

虽然实验观察到算法在 NDCG 上的排序性能在 $k \geq 15$ 时开始不再提升，但是相关实验结果仍然超出我们的预期，因为在实际中我们认为用户在形成对物品的观点和看法时，往往最多考虑几个自己关心的重要属性，而很少会考察十几个甚至几十个属性。为了进一步研究用户对属性词的使用行为及其对推荐性能的影响，我们分析在每个用户的历史评论中，该用户最关心的前 $k$ 个属性词相对于全部属性词在词频上的覆盖率，如公式 (4-38) 所示，其中 $t_{ij}$ 为属性词 $F_j$ 在用户 $u_i$ 的历史评论中出现的词频，$C_i$ 为用户 $u_i$ 最关心的前 $k$ 个属性词的集合，$p$ 为全部属性词的个数。覆盖率与我们所选用的用户最关心的属性词个数 $k$ 之间的关系如图4.16所示。

$$\text{Coverage@}k = \frac{1}{|U|} \sum_{i=1}^{|U|} \frac{\sum_{j \in C_i} t_{ij}}{\sum_{j=1}^{p} t_{ij}} \tag{4-38}$$

实验发现，一小部分用户最关心的属性词实际上占据了用户词频的大部分。例如在 Yelp10 数据集上，96 个属性词当中的 24 个就占据了 80% 的词频，这意味着用户在评论中经常将大部分的注意力集中于自己最关心的属性词上，而则在一定程度上佐证了我们的假设，并为算法使用用户最关心的前 $k$ 个属性词而非全部的属性词提供了依据，因为考虑过多的无关属性词有可能为用户兴趣偏好的建模带来负面影响。

然而在点评数据集上则需要考虑用户最关心的 46 个属性词才能达到 80% 的覆盖率，这几乎是 Yelp 数据集上所需属性词的两倍。为了进一步理解其中的内在

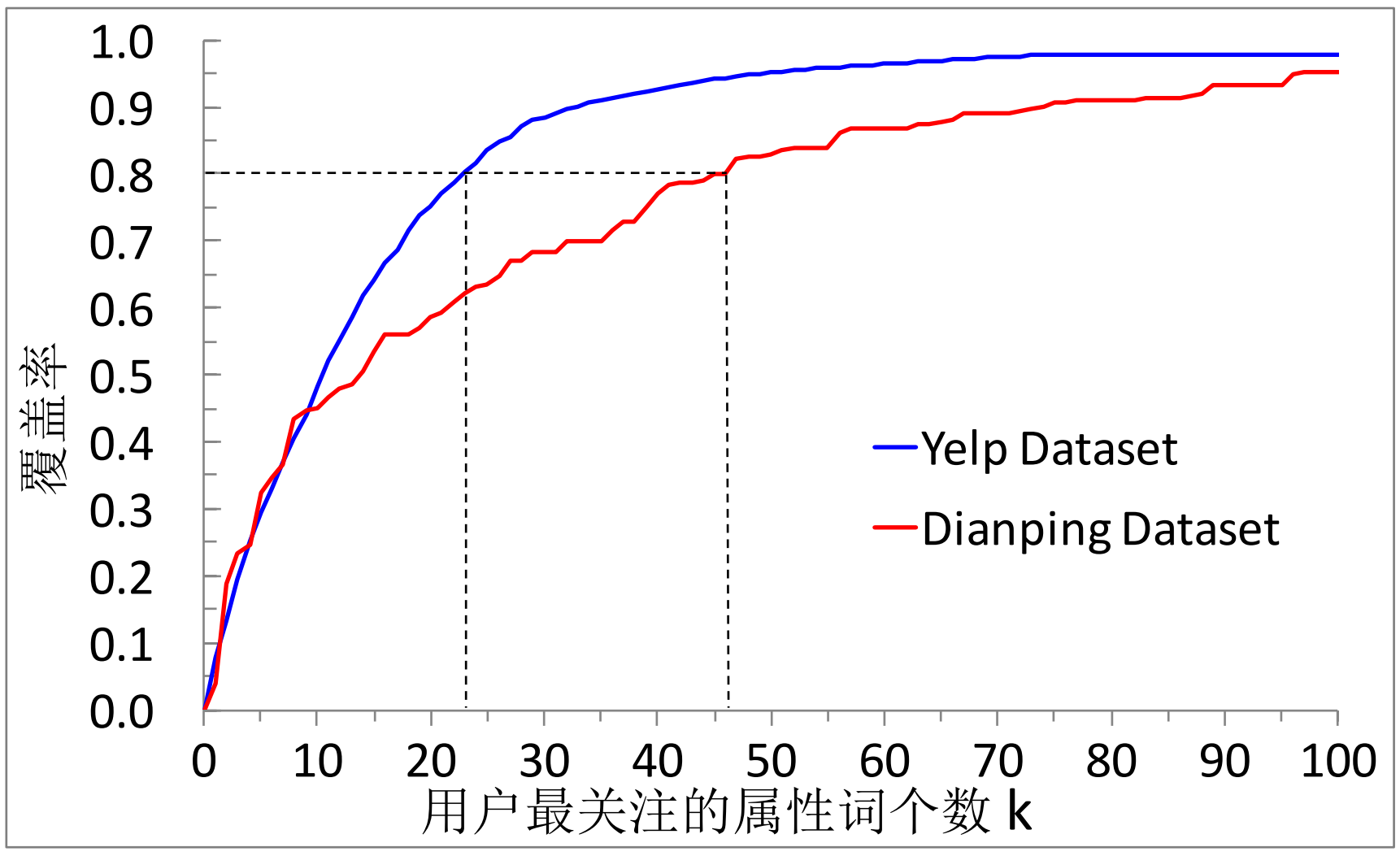

图 4.16　覆盖率与用户最关心的属性词个数 $k$ 的关系

原因，我们分别对 Yelp10 和点评数据集中的属性词利用 WordNet① 和 HowNet② 进行同义词聚类，因为不同的属性词可能实际上表达的是相同的概念。例如属性词“价格”和“费用”将会被聚类为同一个概念，表4.7给出了同义词聚类的一些基本统计信息。

表 4.7　同义词聚类的主要统计信息

| 数据集 | 属性词个数 | 聚类个数 | 平均每类属性词数 |
|---|---|---|---|
| Yelp10 | 96 | 31 | 3.10 |
| 大众点评 | 113 | 26 | 4.35 |

实验结果显示，平均而言在中文语料上同义词的现象更为常见（平均每类的属性词更多），而同义词的存在实际上稀释了属性词空间。例如，在 Yelp 上表示价格的属性词集合包含两个属性词，分别为“Price”和“cost”，而在点评数据集上，表示同样意义的属性词集合包含四个属性词，分别为“价格”、“费用”、“消费”、“花费”。这表明在不同语言环境下我们所需要考虑的用户最关心的属性词个数可能是不同的。

例如在 Yelp10 数据集上，流行度最高的前 15 个属性词被聚到了 7 个同义词集合中，如表4.8所示，其中黑体表示流行度最高的前 15 个属性词，其它属性词为 15 个之外的同义词。

结果显示，虽然在 $k \leq 15$ 时 NDCG 随着 $k$ 的增加而提高，但是实际上这些属性词背后所隐含的实际概念仍然限制在 7 个以内。另外，在模型中考虑 15~45 个

① http://wordnet.princeton.edu
② http://www.keenage.com

表 4.8　前十五个属性词所对应的同义词集合

| | | | |
|---|---|---|---|
| 1 | **place**, **restaurant**, **location**, area, way | 2 | **food**, **menu**, **lunch**, **pizza**, dinner |
| 3 | **service**, **time**, **staff**, order | 4 | **experience**, quality |
| 5 | **room**, **atmosphere**, decor | 6 | **price**, cost |
| 7 | **beer**, wine, drink, water, coffee | | |

属性词实际上只是向前面已经存在的概念集合中添加同义词而已，而并没有考虑更多新的属性词概念，因此模型在 NDCG 指标上的表现并没有明显的提高。

该实验结果进一步验证了我们只使用用户最关心的前 $k$ 个属性词进行个性化推荐的合理性。更重要的是，通过在个性化推荐算法中考虑显式属性词，使得各种潜在的自然语言处理基础在推荐系统中有用武之地，从而在更细的粒度上分析用户的兴趣偏好。

### 4.3.2　基于浏览器的真实用户线上评测

在本部分，我们利用搜狗浏览器真实用户在电子商务环境下进行线上实验，以分析个性化属性级推荐理由的作用。我们主要集中于研究个性化推荐理由如何影响用户的购买决策，即对推荐系统说服力的研究。

#### 实验设置

在业务合作的基础上，我们基于拥有上亿用户群和 26% 月度活跃用户的搜狗浏览器进行 A/B 实验。在该实验中，我们为在京东商城[①] 中浏览电子产品的用户提供相关产品推荐，并记录和分析用户的行为信息。

图4.17展示了我们的推荐界面，在该过程中我们根据用户的历史评论信息和当前的浏览物品给出个性化推荐。需要指出的是，推荐列表并不是由网站给出，而是由我们的浏览器插件弹出。图4.17(a) 中所示的推荐界面主要包含两个部分：一是最上方的“推荐指数”面板，该面板显示当前用户浏览的物品是否被推荐，并给出推荐或不推荐的理由；二是”推荐列表”面板，该面板给出了我们对当前用户提供的个性化推荐及其推荐理由。图4.17(b) 给出了物品属性词云的示例，对于每一个物品，我们从其所有历史评论文本数据中抽取所有命中的属性观点词对进行词云展示，其中绿色表示正面评价、蓝色表示负面评价，词条的大小正比于其在评论中出现的词频。

我们根据浏览器日志选择那些曾经进行过十次以上评论的用户作用目标用户

① http://www.jd.com

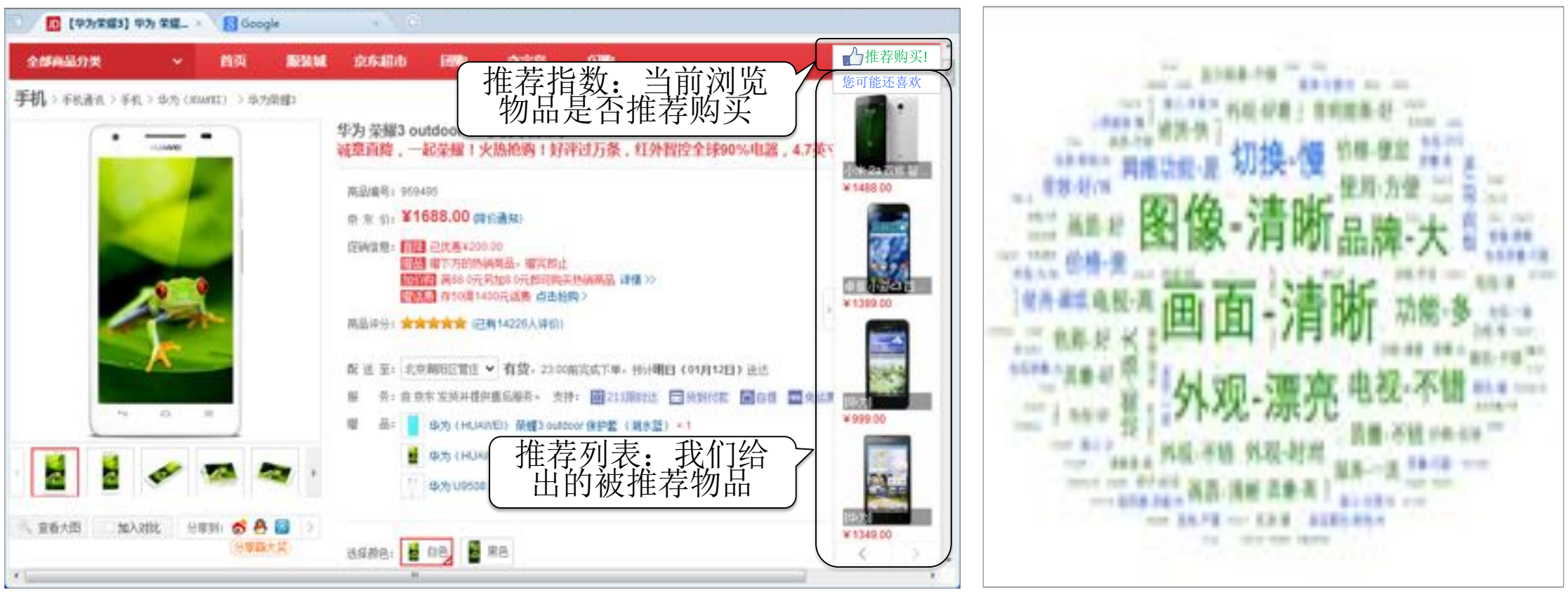

(a) 基于搜狗浏览器插件的推荐　　　　　　　　(b) 产品属性词词云

图 4.17　基于搜狗浏览器插件的推荐系统，当用户在购物网站（本实验以京东为例）中进行购物浏览时插件从右侧弹出并给出 Top-4 推荐；当用户鼠标在被推荐物品上悬停时展示属性级推荐理由和产品属性词云。

进行评测，并用我们的显式变量分解模型为用户提供推荐和（不）推荐理由，其中不推荐理由仅在“推荐指数”面板给出不推荐当前浏览物品时展示。另外，我们仅当用户鼠标在被推荐物品或推荐指数上悬停时才显示推荐理由，通过这样的方法我们跟踪用户是否确实查看了我们的推荐理由，从而减少实验误差。在接下来的部分，我们分别对推荐列表和推荐指数中推荐理由的作用进行评测。

### 推荐列表评测

我们随机地将受试者分为以下三组：A 组（实验组）的用户展示我们的短语级个性化推荐理由及词云；B 组（对照组）的用户收到简单的“其它人也查看了”推荐理由；C 组（控制组）的用户不展示任何推荐理由。在如下的数据分析中，我们只考虑那些有过查看推荐理由行为（鼠标悬停）的用户。另外为了实验的公平性，我们只考察那些在 A、B、C 三个组的推荐列表中共同出现的物品，这包括 944 个物品上的 44,681 条用户查看推荐的记录。表4.9展示了点击率（Click Through Rate，CTR）指标的评测结果。

表 4.9　三组实验用户中用户查看推荐的记录数、点击推荐的次数，以及点击率

| 用户组 | A | | B | | C | |
|---|---|---|---|---|---|---|
| 查看推荐 | 记录数 | 点击数 | 记录数 | 点击数 | 记录数 | 点击数 |
| 记录数 | 15,933 | 691 | 11,483 | 370 | 17,265 | 552 |
| 点击率 | 4.34% | | 3.22% | | 3.20% | |

在双边 $t$ 检验下，实验组 A 的点击率分别在 $p = 0.033$ 和 $p = 0.041$ 水平上显著高于对照组 B 和控制组 C，表明在电子商务应用场景下，我们基于文本和词云

的短语级个性化推荐理由更能有效说服用户对被推荐物品进行点击和查看。

### 推荐指数评测

为了评测推荐指数中（不）推荐理由对用户的影响，我们对推荐指数使用 A 组（实验组）和 B 组（控制组）两个组别，其中两者的区别在于 A 组用户可以看到我们提供的短语级推荐理由，而对于 B 组用户我们不展示任何推荐理由。在本实验中我们没有设置对照组，因为在目前所有的推荐算法中均无法给出不推荐理由。由于网站安全保护方面的限制，我们没有办法通过浏览器捕获用户是否最终为物品成功付款的信息，因此我们在本实验中采用用户加入购物车的行为来代替购买行为，从而对推荐效果进行评价。

该组线上实验共收集了来自 1,328 个 A、B 组共同出现物品上的 53,372 次用户查看推荐的记录，其中 A 组包括 582 个用户的 20,735 条行为记录，B 组包括 733 个用户的 32,637 条行为记录。表4.10展示了是否推荐购买和用户是否加入购物车的混淆矩阵。其中 AddToCart%=$\frac{x_{11}+x_{21}}{x_{11}+x_{12}+x_{21}+x_{22}}$ 表示用户将当前浏览物品加入购物车的比例；Agreement%=$\frac{x_{11}+x_{22}}{x_{11}+x_{12}+x_{21}+x_{22}}$ 表示用户的最终行为与推荐理由一致的比例（推荐购买则加入购物车、不推荐购买则未加入购物车）；RecAgree%=$\frac{x_{11}}{x_{11}+x_{12}}$ 和 DisRecAgree%=$\frac{x_{22}}{x_{21}+x_{22}}$ 分别表示当前物品为被推荐和不被推荐时，用户行为与推荐理由一致的比例。

表 4.10　A 组（实验组）和 B 组（对照组）共现的 1328 个物品上用户行为混淆矩阵

| 混淆矩阵 | A 组 | | | | B 组 | | | |
|---|---|---|---|---|---|---|---|---|
| | 加入购物车 | | 未加入购物车 | | 加入购物车 | | 未加入购物车 | |
| 推荐购买 | $x_{11}$ | 1,261 | $x_{12}$ | 16,572 | $x_{11}$ | 1,129 | $x_{12}$ | 28,218 |
| 不推荐购买 | $x_{21}$ | 72 | $x_{22}$ | 2,830 | $x_{21}$ | 541 | $x_{22}$ | 2,749 |
| AddToCart% | 6.43% | | | | 5.12% | | | |
| Agreement% | 19.73% | | | | 11.89% | | | |
| RecAgree% | 7.07% | | | | 3.85% | | | |
| DisRecAgree% | 97.5% | | | | 83.56% | | | |

实验发现，实验组（A 组）用户加入购物车的比例（AddToCart%）和行为一致性的比例（Agreement%）都显著高于对照组（B 组），其中双边 $t$ 检验的显著性系数分别为 $p = 0.0374$ 和 0.0068。这表明通过向用户提供个性化推荐理由可以显著提高用户对推荐的接受度，同时在实际电子商务应用中提高推荐系统的转化率。更为重要的是，我们发现当推荐系统认为当前浏览物品并不适合购买时，通过向用户展示不推荐理由可以有效防止用户加入购物车的行为。该实验结果表明属性

级的个性化推荐理由具有较强的说服力和可行性。

### 4.3.3　基于属性词流行度的动态推荐评测

在本节，我们对基于属性词流行度的动态个性化推荐进行评测，内容包括京东商城化妆品领域数据描述和属性词抽取评测、精确到天级别的动态属性词流行度预测、动态个性化推荐的打分预测精度与排序性能评测，以及算法在解决冷启动用户问题上的效果。

#### 数据集描述

我们收集了中国主要的电子商务网站之一“京东商城”化妆品领域自 2011 年 1 月 1 日至 2014 年 3 月 31 日共三年零一个季度的用户评论信息，数据的时间跨度足以满足时间序列分析和预测的需求。表4.11描述了该数据集的统计信息，其中每个季度的统计数据为截止到该季度末（包含该季度）系统中总的用户数、物品数和评论数统计信息。

表 4.11　数据分季度的累计统计信息，其中每个季度的数据为系统中截止到该季度末的统计数据

| 年度 | 2011 | | | | 2012 | | | |
|---|---|---|---|---|---|---|---|---|
| 季度 | $Q_1$ | $Q_2$ | $Q_3$ | $Q_4$ | $Q_1$ | $Q_2$ | $Q_3$ | $Q_4$ |
| 用户数 | 72,403 | 164,256 | 280,918 | 500,619 | 701,975 | 925,649 | 1,070,542 | 1,239,967 |
| 物品数 | 3,559 | 5,474 | 7,961 | 11,235 | 14,265 | 20,444 | 28,746 | 37,380 |
| 评论数 | 119,517 | 305,974 | 571,602 | 1,108,673 | 1,667,607 | 2,409,277 | 3,015,849 | 3,656,338 |

| 年度 | 2013 | | | | 2014 |
|---|---|---|---|---|---|
| 季度 | $Q_1$ | $Q_2$ | $Q_3$ | $Q_4$ | $Q_1$ |
| 用户数 | 1,355,395 | 1,417,551 | 1,575,223 | 1,710,040 | 1,844,569 |
| 物品数 | 42,184 | 46,877 | 49,710 | 52,117 | 93,243 |
| 评论数 | 4,100,668 | 4,253,294 | 4,656,628 | 4,963,927 | 5,524,491 |

为了在时间尺度上尽可能地模拟系统在实际中的运行情况，也为了在一年中的不同季节评测算法的性能，我们在原始数据集的基础上根据最后的四个季度分别构建四个时间相关的子数据集 $D_1 \sim D_4$。具体而言，我们采用最后四个季度的数据（2013 年的 $Q_2 \sim Q_4$，以及 2014 年的 $Q_1$）分别作为测试集，并对这四个季度中的每一个采用它前面时间上的所有数据作为训练集，如表4.12所示，这四个子数据集中的测试集分别对应了一年中的四个季度和季节。在接下来的实验中，我们在这四个子数据集中的每一个上面分别评测时间序列预测和动态个性化推荐的性能，

并通过这四个子数据集上的平均性能来刻画算法在一年时间段内的总性能。

表 4.12　基于三年零一季度的京东商城化妆品领域点评数据构建的四个事件相关子数据集及其统计信息

| 数据集编号 | 训练集 | 测试集 |
|---|---|---|
| $D_1$ | $2011 + 2012 + Q_1, 2013$ | $Q_2, 2013$ |
| $D_2$ | $2011 + 2012 + Q_1, Q_2, 2013$ | $Q_3, 2013$ |
| $D_3$ | $2011 + 2012 + Q_1, Q_2, Q_3, 2013$ | $Q_4, 2013$ |
| $D_4$ | $2011 + 2012 + 2013$ | $Q_1, 2014$ |

在接下来的实验中，我们设定算法7中傅里叶项的最大可能阶数为 $\bar{K} = 10$，ARIMA 项的最大可能阶数为 $(\bar{p}, \bar{d}, \bar{q}) = (10, 3, 10)$，实验发现这样的阶数设定就足以获得较好的时间序列预测效果。

### 属性词抽取

我们同样需要首先在京东商城化妆品评论数据集上利用短语级情感分析技术抽取产品属性词。与上一小节在 Yelp 和点评数据集上的情感词典构建过程相同，我们仍然需要在准确率和召回率之间权衡。在本实验中，我们首先随机选取了 1,000 条评论并展示给三位标注人员从中进行人工属性词抽取，并进一步选择三位标注人员共同的属性词作为正确属性词集合，该属性词集合被用户对算法抽取的属性词进行评价。我们利用如上节所述的短语级情感分析工具包① 并调节阈值获得六组不同的属性词集合，这六组属性词的准确率、召回率和 $F_1$ 值如表4.13所示。

表 4.13　在不同的准确率和召回率权衡下所抽取的六组属性词集合及其评价

| 组号 | 1 | 2 | 3 | 4 | 5 | 6 |
|---|---|---|---|---|---|---|
| 属性词个数 | 32 | 44 | 58 | 66 | 79 | 95 |
| 准确率 | **0.9063** | 0.8864 | 0.8621 | 0.7879 | 0.6962 | 0.6316 |
| 召回率 | 0.3867 | 0.5200 | 0.6667 | 0.6933 | 0.7333 | **0.8000** |
| $F_1$ 值 | 0.5421 | 0.6555 | **0.7519** | 0.7376 | 0.7143 | 0.7059 |

在综合考虑属性词个数和效率的前提下，如果没有特别说明，我们选择 $F_1$ 值最高的一组属性词（第三组）进行流行度预测并用来实现动态的个性化推荐，其中共包括 58 个属性词。总体而言，我们倾向于在属性词个数足够多的前提下选择准确率足够高的属性词集合，从而减小在推荐算法中由错误属性词引入噪声的可

① http://yongfeng.me/software

能性。我们不加修改地使用算法直接得到的属性词集合进行后续实验，以保证我们的实验结果没有引入人工的干预。

#### 精确到天级别的属性词流行度评测

为了对基于傅里叶辅助项的移动平均自回归模型（FARIMA）进行评测，我们在四个子数据集 $D_1 \sim D_4$ 上分别对每个属性词所对应的百分比时间序列进行预测和评价。对于每一个数据集，我们利用算法7在该数据集的训练集上进行模型训练，预测下一个季度的时间序列并在相应的测试集上进行测试。图4.18展示了四个子数据集上的时间序列预测结果，其中每个数据集上最终选定的傅里叶阶数 $K$ 和 ARIMA 阶数 $(p, d, q)$ 由算法在最小化 AICc 原则下自动确定。由图可见，FARIMA 模型不仅可以预测时间序列在未来的升降趋势（如图4.18(a) 和 (c) 所示），还可以预测时间序列未来的拐点（如图4.18(b) 和 (d) 所示）。

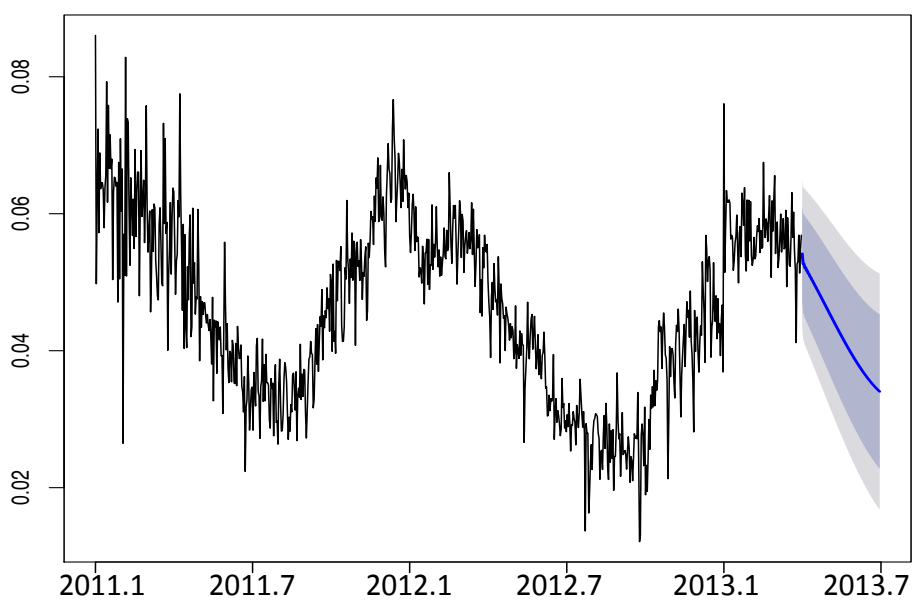

(a) $D_1 : K, p, d, q = 6, 4, 1, 3$

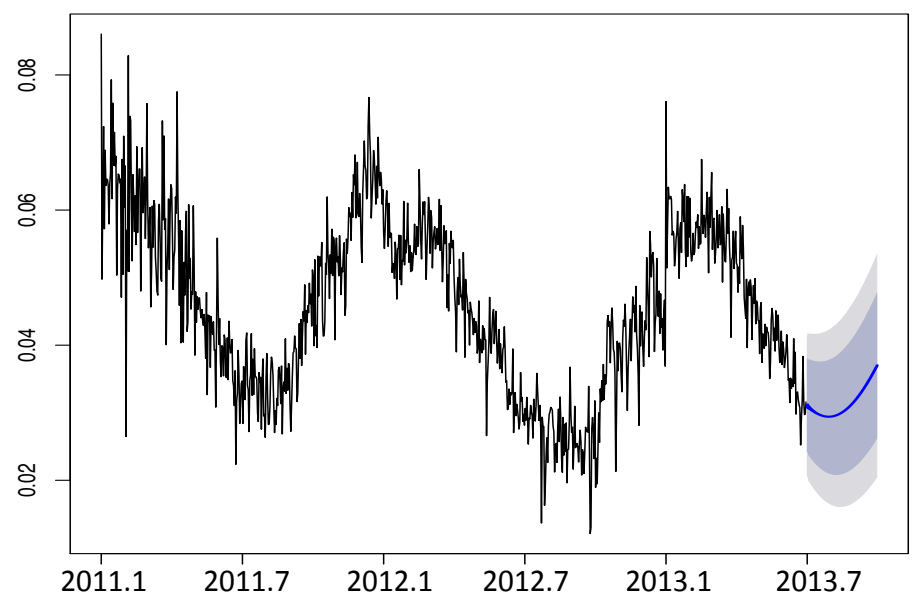

(b) $D_2 : K, p, d, q = 6, 4, 1, 5$

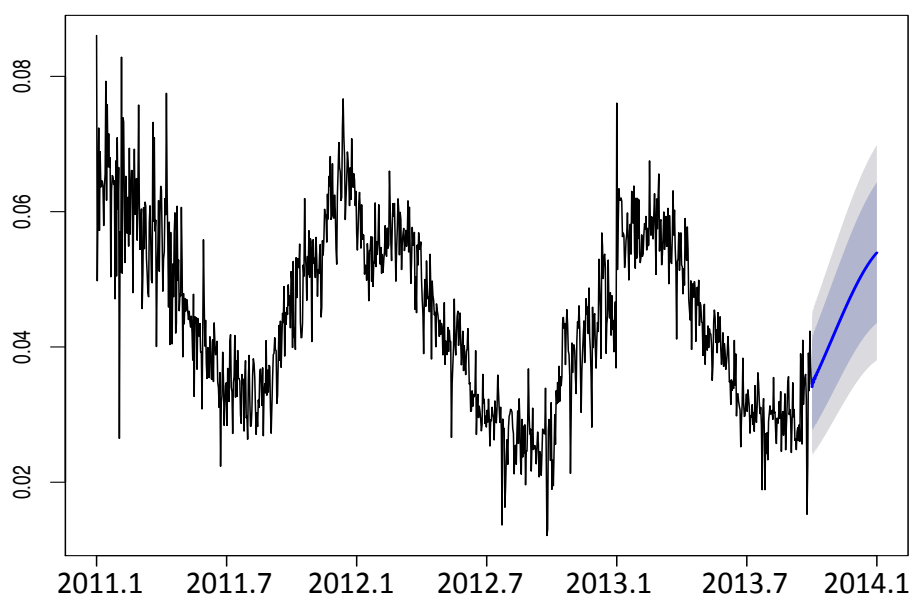

(c) $D_3 : K, p, d, q = 6, 3, 1, 2$

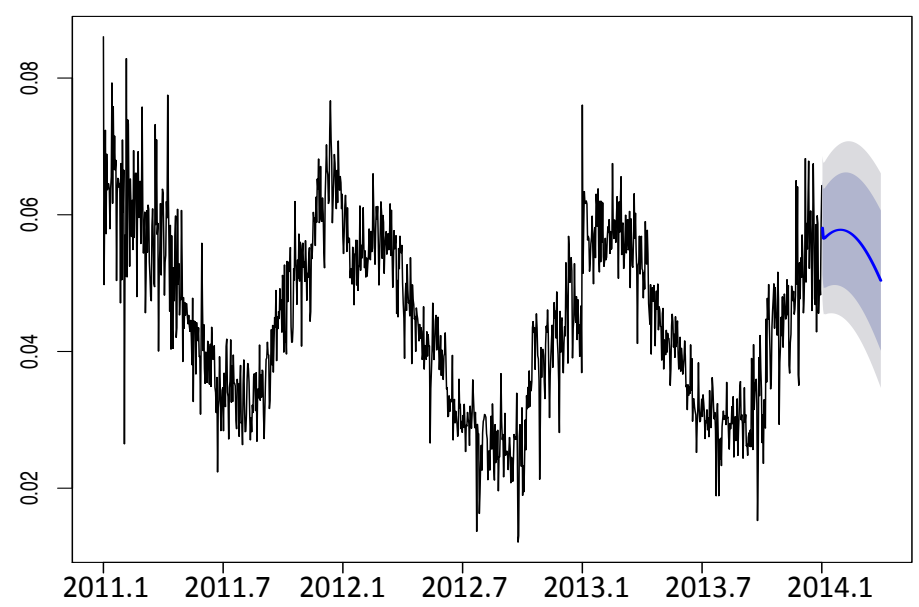

(d) $D_4 : K, p, d, q = 7, 3, 1, 2$

图 4.18　四个数据集 $D_1 \sim D_4$ 上属性词“营养”的百分比时间序列预测示例，其中 $K$ 为算法最终确定的傅里叶项阶数，$(p, d, q)$ 为算法最终选定的 ARIMA 项阶数；蓝线为在每一个季度上的预测时间序列深灰色和浅灰色分别为 80% 和 95% 置信区间。

为了对时间序列预测进行定量的评价，我们采用根均方差（Root Mean Square Error，RMSE）和平均百分比绝对误差（Mean Absolute Percentage Error，MAPE）

作为评价指标：

$$\mathrm{RMSE}=\sqrt{\frac{\sum_{t=1}^{n}(X(t)-\hat{X}(t))^2}{n}},\mathrm{MAPE}=\frac{1}{n}\sum_{t=1}^{n}\left|\frac{X(t)-\hat{X}(t)}{X(t)}\right| \tag{4-39}$$

其中 $n$ 为测试集中的样本数，$X(t)$ 为实际流行度数值，$\hat{X}(t)$ 为预测的流行度数值。

RMSE 描述了预测值与真实值之间的绝对误差，并在很多推荐相关的预测任务中得到广泛的应用。然而在本任务中，由于百分比时间序列中流行度的绝对数值较小且数值范围不固定，因而 RMSE 的参考意义并不权威。因此，我们进一步采用了平均百分比绝对误差 MAPE 进行评价，它描述了预测值相比于真实值绝对误差的百分比。我们在每一个子数据集上将每一个属性词的 RMSE 或 MAPE 作平均以描述该数据集上总体的预测误差，并在所有四个数据集上求平均作为整体的预测误差，结果如表4.14所示。

表 4.14　流行度百分比时间序列预测评价

| 数据集 | $D_1$ | $D_2$ | $D_3$ | $D_4$ | 平均值 |
|---|---|---|---|---|---|
| 数据点数 | 4,253,294 | 4,656,628 | 4,963,927 | 5,524,491 | 4,849,585 |
| RMSE | 0.00406 | 0.00453 | 0.00704 | 0.00603 | 0.00541 |
| MAPE | 0.08012 | 0.11962 | 0.12580 | 0.07770 | 0.10081 |
| ~MAE | 0.3205 | 0.4785 | 0.5032 | 0.3108 | 0.4032 |

RMSE 和 MAPE 上的评价结果显示我们的 FARIMA 模型能够获得较好的时间序列预测效果，整体而言预测值相对于真实值的误差在 10% 左右。为了更直观地理解该评测结果，我们通过将平均百分比绝对误差（MAPE）乘以长度 4 从而将其归一化到常见的 1 ~ 5 星打分区间上并转化为估计的绝对值误差（Mean Absolute Error，MAE），如表4.14中最后一行所示。结果显示，在常见的五星打分区间上我们的时间序列预测算法可以达到约 0.4 的误差。在接下来的实验中，我们就将在 FARIMA 所给出的时间序列预测值的基础上进行天级别的动态个性化推荐并与基线算法进行比较。

#### 动态打分预测精度评价

在属性词流行度预测的基础上，我们采用4.2.5小节所述的条件机会模型进行动态打分预测和推荐。

在实际系统中，用户和物品的数量随着系统的业务增长而不断增长，这在表4.11中用户数、物品数和评论数的增长上可以看出。这为推荐系统带来了冷启

动的问题：新加入的用户由于历史行为相对较少而难以为其进行个性化建模和个性化推荐。而在时间敏感的个性化推荐中，冷启动的问题显得更为突出，因为用户或物品可能仅有几个历史评分记录，而我们却不得不将这些记录进一步划分到不同的时间段内，这进一步加重了数据稀疏性和冷启动问题，甚至使得一些用于比较的基线算法根本无法正确执行。

因此在本实验中，对于每一个子数据集 $D_i$，我们选择那些在训练集中确实拥有训练数据（历史行为）的用户进行测试，因为只有对于这些用户我们才能对其进行个性化偏好建模。对于在训练集中没有任何历史行为记录的冷启动用户，我们在后面的实验中单独进行冷启动推荐的评测。我们采用如下的基线算法进行实验对比：

- NMF：非负矩阵分解算法[166]，该算法在如上实验中取得了矩阵分解类算法的最好效果，在本实验中我们同样采用基于对数预测函数和 Frobenius 正则化的非负矩阵分解算法。该方法属于时间无关的推荐算法。
- timeSVD++：受到广泛应用的时间敏感协同过滤算法[162]，我们采用了开源工具包 MyMedialite[179] 中的实现。
- Tensor：时间敏感的张量分解模型[170]，是上下文相关的推荐中经常使用的算法之一。为了实现精确到天级别的动态预测，我们将一年的中的 365 个日期分别作为张量中时间维度上的一层，如果一个用户物品打分的时间是一年中的第 $i$ 天，则我们将其填入张量中第 $i$ 个矩阵上该用户和物品所对应的位置。
- EFM：我们的显式变量分解模型，如4.1小节所述，该方法为时间无关的个性化推荐，但与其它基线算法不同的是，该方法也利用了文本评论及显式属性词信息。

为了研究属性词个数 $r$ 对预测效果的影响，我们利用表4.13中所示的六组属性词集合 $\mathcal{F} = \{f_1, f_2, \cdots, f_r\}$ 分别进行预测，它们对应的属性词个数分别为 $r = 32, 44, 58, 66, 79, 95$。为了保证模型对比的公平性，我们对于基线算法 NMF、timeSVD++ 和张量分解也采用同样个数的隐变量来进行模型学习，对于显式变量分解模型 EFM 我们则采用同一组属性词集合进行模型学习。

我们仍然采用根均方差 RMSE 进行效果评测，并采用网格搜索的方法确定每个算法的最优参数值。图4.19展示了每个方法在四个子数据集上和不同的属性词个数下各自的平均 RMSE，而每个方法在每个数据集上所能取得的最好效果如表4.15所示，其中 FTSA 表示我们的属性级时间序列分析（Feature-level Time Series Analysis）方法。

由实验结果可见，对于基于隐变量分解的方法（NMF，timeSVD++ 和张量分

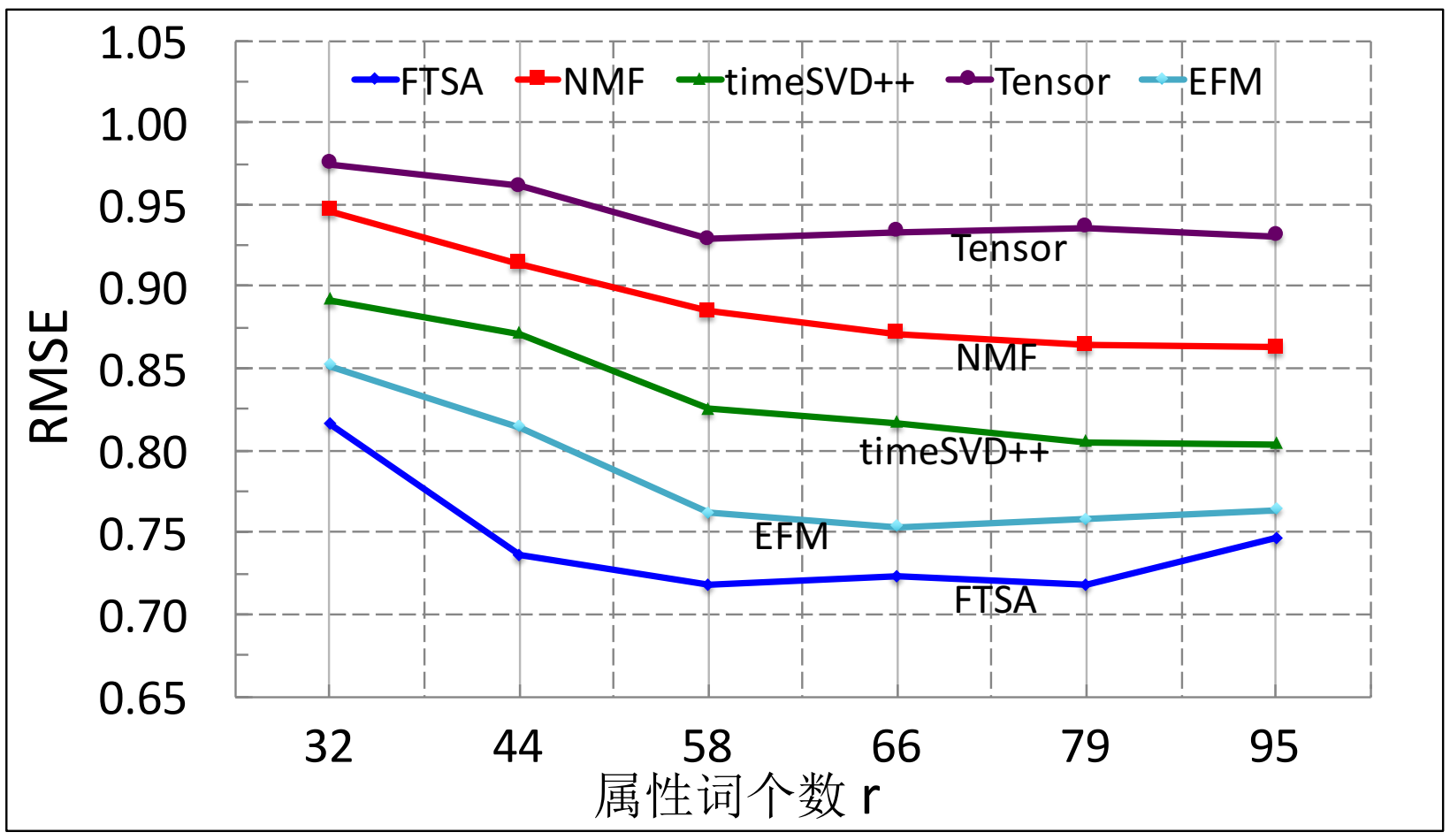

图 4.19　四个数据集 $D_1 \sim D_4$ 上的平均 RMSE 与属性词/隐变量个数之间的关系

解）而言，RMSE 随着算法所使用的隐变量个数 $r$ 的增加而降低。同时，我们发现考虑到时间因素在内的 timeSVD++ 算法比简单的 NMF 算法取得了更好的预测精度，然而同样考虑时间因素在内的张量分解方法却并没有比非负矩阵分解算法取得更好的预测效果，一个可能的原因是将原本就比较稀疏的用户打分数据进一步分割到一年的 365 天中会进一步加重数据稀疏性，从而影响预测精度。

表 4.15　各方法在四个数据集上所能取到的最好效果以及四个数据集上最好效果的平均值，每个方法在每个数据集上运行的五次标准差 $\leq 0.05$

| 数据集 | $D_1$ | $D_2$ | $D_3$ | $D_4$ | Average |
|---|---|---|---|---|---|
| NMF | 0.857 | 0.866 | 0.851 | 0.878 | 0.863 |
| timeSVD++ | 0.782 | 0.813 | 0.795 | 0.826 | 0.804 |
| Tensor | 0.912 | 0.937 | 0.916 | 0.951 | 0.929 |
| EFM | 0.733 | 0.768 | 0.742 | 0.769 | 0.753 |
| FTSA | **0.706** | **0.729** | **0.711** | **0.726** | **0.718** |
| FTSA′ | 0.716 | 0.732 | 0.722 | 0.735 | 0.726 |

通过如上实践结果，我们进一步发现将用户评论数据和情感信息考虑在内的非动态推荐（EFM）要好于只使用数值打分信息的动态推荐算法（timeSVD++），这在一定程度上显示了用户评论和情感信息对提高推荐效果的重要作用。另外，通过进一步在使用用户评论信息的基础上加入时间信息，我们的属性级时间序列分析（FTSA）方法能够在 EFM 的基础上取得更好的预测效果。

为了进一步验证我们的 FARIMA 模型给出的属性词流行度预测值在打分预测中的作用，我们将公式 (4-25) 中的预测流行度 $\hat{X}_{f_k}(t)$ 用当天的真实流行度 $X_{f_k}(t)$ 来代替，并同样用我们的条件机会模型进行打分预测，结果如表4.15中的最后一行（FTSA′）所示。我们发现利用预测流行度得到的打分预测精度甚至要高于使用真

实流行度得到的结果，这表明 FARIMA 模型给出的预测流行度对时间序列的平滑效果对推荐效果的提升有一定的帮助，这有助于对属性词流行度中噪音扰动的剔除。例如，某一属性词在某一天的流行度可能受当天活跃用户数的影响而较低，但是通过利用往年同期该属性词的流行度信息，我们可以对当天该属性词的流行度给出更有代表性的估计，从而帮助我们对用户偏好进行更准确地建模。

### 动态 Top-K 推荐排序性能评价

我们同样关心在实际系统中更为重要的 top-K 推荐列表排序性能评价。在本实验中我们对于每一个子数据集 $D_i$ 选择那些在测试集中拥有 10 个或以上购买记录的用户，并对他们进行 top-10 推荐和评测。我们同样使用广为采用的标准化折扣累计增益（Normalized Discounted Cumulative Gain，NDCG）进行性能评测，并仍然使用上节介绍的 NMF、timeSVD++、Tensor 和 EFM 方法进行性能比较。对于我们在本节的属性级时间序列分析法（Feature-level Time Series Analysis，FTSA）和上节的显式变量分解模型（EFM），我们采用 $F_1$ 值最高的 $r = 58$ 个属性词，对于隐变量分解算法（NMF、timeSVD++ 和 Tensor）我们同样采用 $r = 58$ 个隐变量以保证性能比较的公平性。

表4.16显示了四个数据集上各自的评测效果及它们平均值。我们发现，时间敏感的方法（timeSVD++ 和 Tensor）整理好于静态的不考虑时间的模型（NMF 和 EFM），表明在 top-K 排序推荐任务中时间因素的重要性相比于在数值打分预测任务上要更为重要。通过既考虑时间因素，又考虑显式属性词信息，本节基于属性词时间序列分析的动态个性化推荐取得了最好的排序性能。

表 4.16　每个数据集上测试集评论数 ≥ 10 的用户数及 NDCG 评测结果，每个算法在每个数据集上运行五次的标准差 ≤ 0.02

| 数据集 | $D_1$ | $D_2$ | $D_3$ | $D_4$ | 平均值 |
|---|---|---|---|---|---|
| 用户数 | 122,476 | 146,099 | 159,747 | 167,370 | 148,923 |
| NMF | 0.154 | 0.136 | 0.163 | 0.147 | 0.150 |
| timeSVD++ | 0.228 | 0.241 | 0.236 | 0.233 | 0.235 |
| Tensor | 0.207 | 0.192 | 0.196 | 0.211 | 0.202 |
| EFM | 0.186 | 0.194 | 0.208 | 0.195 | 0.196 |
| FTSA | **0.254** | **0.267** | **0.258** | **0.271** | **0.263** |

考虑时间因素对推荐性能的提升在一定程度上验证了用户倾向于在不同时间购买不同产品的假设。仅仅基于数值打分的推荐算法由于数值打分区分度不高而难以获得较好的效果，而考虑时间因素和显式属性词的时间敏感推荐则因使用了

更好的区分信息而获得较好的推荐效果。为了进一步分析用户在不同产品上的购买行为，我们对具有不同流行度的产品分别进行排序性能的评测。具体而言，我们从数据集 $D_i$ 中选择那些拥有至少 $L$ 条评论信息（包括训练集和测试集）的产品。在这些选出来的产品基础上，我们同样选则那些在测试集上有至少十条评论的用户为其进行 top-10 推荐。在不同的过滤阈值 $L$ 下，四个数据集上的平均 NDCG 如图4.20所示。

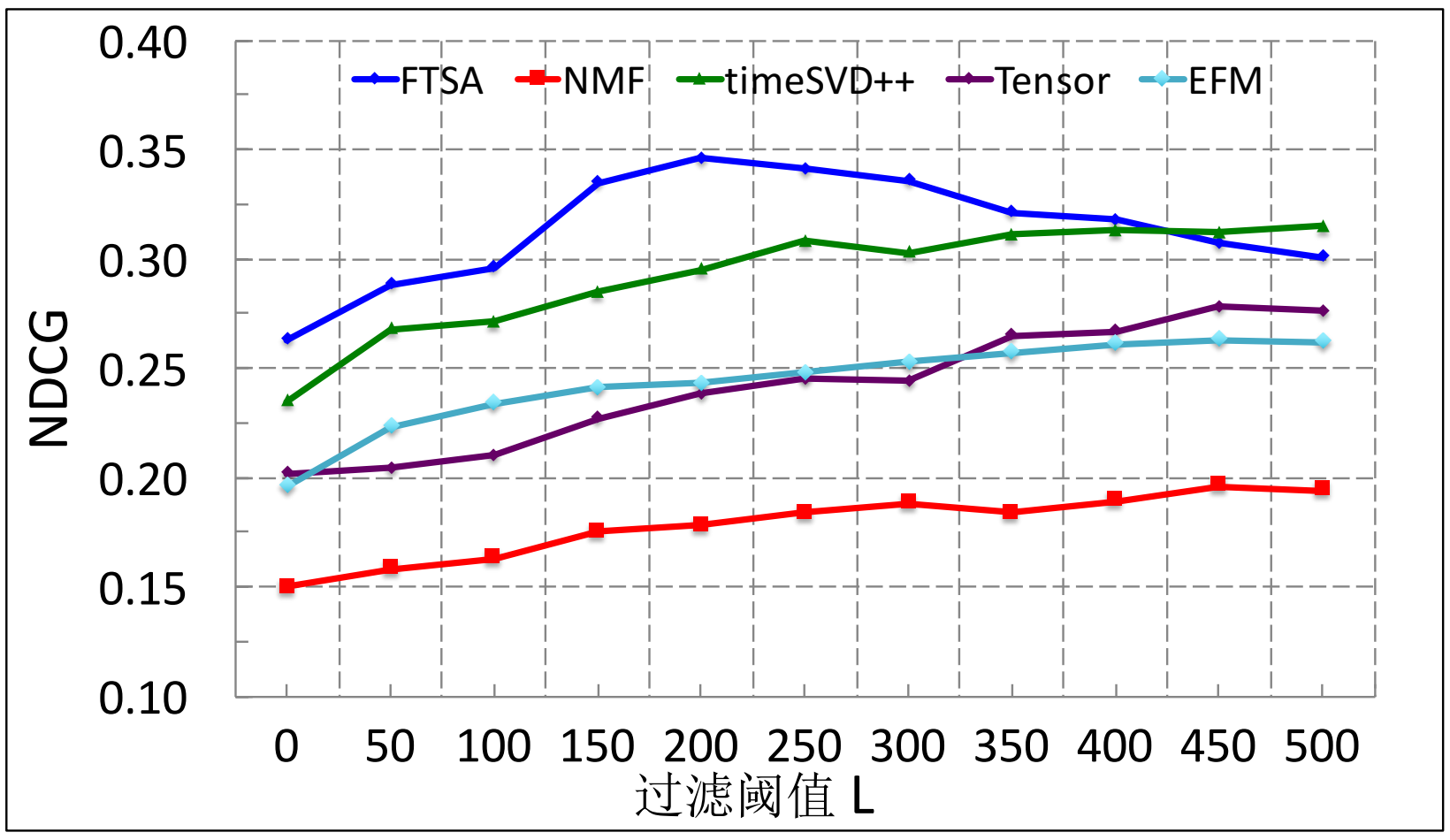

图 4.20　排序性能 NDCG 随用户过滤阈值 $L$ 的变化示意图

由结果可见，随着过滤阈值 $L$ 的增加（即选出更为高频的产品）动态推荐算法（timeSVD++ 和 Tensor）和静态推荐算法（NMF 和 EFM）在 NDCG 指标上的性能都逐渐提高，这主要得益于通过筛选出高频产品而使得训练数据变得更为密集，降低数据稀疏性对推荐性能的影响，另外推荐热门物品本身也比推荐长尾的小众物品更为容易。然而需要注意的是，对于我们的基于属性级时间序列预测的动态推荐（FTSA）而言，当过滤阈值 $L$ 取值过高时，算法的排序性能反而会下降。

这表明我们的算法整体上在低频或中频的物品推荐上比基线算法要好，而对于高频热门的物品而言，我们的方法未必有较大的优势。经过对高频物品进行仔细的分析发现，高频物品往往是那些与时间无关，而在一年中的各个季节上都有较多购买量的产品，例如其中有大量产品为口红类产品，因此在这些物品上强行添加周期性时间限制可能反而会为推荐性能带来负面的影响。然而我们的方法仍然能在较大范围的 $L$ 阈值内取得最好的排序效果，另外为了提高推荐系统的新颖性和良性循环的建立，我们在实际系统中会综合考虑全部的物品而非仅限于一直推荐热门物品。

### 冷启动用户推荐性能评测

在以往的工作中，如果不是特别地研究推荐系统的冷启动问题，算法在相关性能评测中往往滤除冷启动用户（例如评论数少于 10 或 20 的用户），并在系统的成熟用户（非冷启动用户）上进行算法评测。然而这对于实际系统中的应用而言或许并不一定完全合适，因为对于实际系统而言最难处理的用户反而被忽略了。实际上，这样的冷启动用户甚至会占到系统用户总数的大部分，例如在 Yelp 数据集中，有 49% 的用户仅有一个历史评论记录。因此，在本小节中，我们对冷启动用户的推荐性能进行专门的评测，以分析算法在解决冷启动问题上的效果。

对于每一个数据集 $D_i$，我们仍然利用其中的训练集进行模型训练，但是我们仅对测试集中那些在训练集上至多有一个历史记录的用户进行评价。该实验设置与推荐系统的实际运行情况相吻合，在实际应用中我们可以使用系统中当前拥有的所有记录进行模型训练，并对未来的用户行为进行预测，包括冷启动用户在内。在每个数据集上选出的测试用户数及其相应的 RMSE 打分预测评价和 NDCG 排序性能评价结果如表4.17所示。

表 4.17　每个数据集上用来进行测试的冷启动用户数及相应的 RMSE 与 NDCG 值，其中 RMSE 指标的五次标准差 $\leq 0.05$，NDCG 指标的五次标准差 $\leq 0.02$

| 评价指标 | 数据集 | $D_1$ | $D_2$ | $D_3$ | $D_4$ | 平均值 |
|---|---|---|---|---|---|---|
| | 测试用户数 | 527,328 | 637,965 | 613,904 | 712,003 | 622,800 |
| RMSE | NMF | 1.729 | 1.693 | 1.732 | 1.706 | 1.715 |
| | timeSVD++ | 1.441 | 1.438 | 1.506 | 1.475 | 1.465 |
| | Tensor | 1.763 | 1.779 | 1.725 | 1.769 | 1.759 |
| | EFM | 1.383 | 1.362 | 1.360 | 1.389 | 1.374 |
| | FTSA | **1.344** | **1.287** | **1.278** | **1.326** | **1.309** |
| NDCG | NMF | 0.067 | 0.073 | 0.061 | 0.066 | 0.067 |
| | timeSVD++ | 0.084 | 0.075 | 0.088 | 0.092 | 0.085 |
| | Tensor | 0.046 | 0.054 | 0.044 | 0.048 | 0.048 |
| | EFM | 0.093 | **0.107** | 0.095 | 0.087 | 0.096 |
| | FTSA | **0.118** | 0.098 | **0.122** | **0.115** | **0.113** |

总体而言，算法在冷启动用户上的打分预测精度性能（RMSE）与在全体用户上（表4.15）具有相似的表现。将用户评论考虑在内的算法（EFM 和 FTSA）要好于仅考虑数值打分的算法（NMF，timeSVD++ 和 Tensor）。其原因在于仅使用数值打分的算法由于数据稀疏性的问题而难以对用户和物品的打分偏好进行精确的建模，因此相关算法倾向于将冷启动用户的打分预测为全局平均值附近的值。例如在 NMF、timeSVD++ 和 Tensor 等方法中，正则项实际上是在预测打分之上添加

了一个以零值或全局平均值为中心的高斯先验分布，从而使得对冷启动用户的打分预测变得不是那么个性化。然而通过考虑用户文本评论中所蕴含的属性词信息，我们可以进一步降低数据稀疏性并提高打分预测精度。统计发现，对于冷启动用户而言平均每个用户有 0.66 个评分，而平均每个冷启动用户评论过的属性词个数则达到 2.56 个，这是因为用户在一条历史评论中可能提到多个属性词。在使用属性词的基础上，我们的 FTSA 算法由于进一步考虑了时间因素而获得了最好的冷启动预测效果。

在排序性能指标 NDCG 上，我们发现静态的非负矩阵分解算法（NMF）比动态的张量分解方法（Tensor）要好，这与表4.16中的“热启动”情况有所不同。这主要是由于张量分解方法限于更为严重的数据稀疏性（每个测试用户在时间维度上至多只有一个打分记录）而无法对用户和物品偏好的时间漂移进行较好的建模。通过利用产品属性词减少数据稀疏性，我们基于属性级时间序列分析的推荐算法（FTSA）在多数数据集上取得了较好的效果。

## 4.4　本章小结

本章以模型的可解释性为核心，分析了显式属性词在个性化推荐模型的直观意义和推荐效果中的应用。为了解决传统的基于隐变量的个性化推荐模型难以为推荐给出直观解释的问题，我们利用短语级情感分析技术从大规模用户评论语料中抽取产品属性词、用户情感词，并构建由属性词、情感词和情感极性三元组组成的情感词典。在此基础上，我们提出了基于多矩阵分解的显式变量分解模型，使得矩阵分解的维度具有直观的物理意义，从而一方面使得模型的优化结果可以得到直观解释，另一方面使得模型可以给出短语级的个性化推荐理由甚至不推荐理由。

在基于显式属性词的静态推荐的基础上，我们进一步分析了属性词的时间动态性，提出了基于属性词时间序列分析的动态个性化推荐模型。为此我们首先分析了实际系统中精确到天级别的属性词流行度时间序列，提出了百分比时间序列并分析了其较好的时序周期性。为了解决精确到天级别的时间序列预测中参数复杂度的问题，我们设计了基于傅里叶辅助项的移动平均自回归模型。在属性词流行度预测的基础上，我们进一步利用条件机会模型设计了动态时序推荐算法。

大规模真实数据上的实验结果显示，基于属性词的显式变量分解模型和动态化建模能够在打分预测和排序列表的构建等线下任务中显著提高推荐系统的性能；属性级个性化推荐理由的构建则在线上评测中提高推荐列表的点击率和说服力。

本章工作发表于 CCF-A 类国际会议 SIGIR 2014a、SIGIR 2014b、WWW 2015、IJCAI 2015，以及 CCF-B 类会议 WSDM 2015 中。所提出的显式变量分解模型相关

论文被来自中国、美国、加拿大、新加坡、欧洲的高校和研究组引用，其中清华大学、新加坡南洋理工大学、新加坡国立大学、爱尔兰都柏林大学等相关研究机构对模型进行重现，并作为基线算法进行比较、改进和提升。

# 第 5 章　推荐的经济学解释

个性化推荐技术的主要使用场景是各种各样的互联网应用，如网络电子商务、社交网站、在线金融服务、在线工作平台，等等，这些网络应用构成了互联网经济系统的主体。无处不在的个性化推荐系统通过向用户推荐可能感兴趣的物品的形式，实际上是在这一经济系统中起到资源分配的核心作用，从而影响着互联网经济系统的运行。在本章，我们以互联网经济系统为背景，对个性化推荐在互联网中的经济学意义进行解释。我们提出互联网环境下成本、效用和福利的基本概念与定义，并以此为基础提出基于福利最大化的网络服务匹配框架。我们分析福利最大化与服务匹配和推荐的关系，并在不同的应用场景下对框架进行特殊化，包括在电子商务、P2P 网络借贷和在线众包网站中的具体应用与效果评测。

## 5.1　互联网福利的最大化

在本节，我们提出基于互联网福利最大化的网络服务匹配框架，并将其应用于个性化推荐任务中。在此基础上，我们对个性化推荐在互联网环境中的经济学意义进行解释。

### 5.1.1　本节引言

互联网的迅速发展正使得越来越多的线下人类活动线上化，例如电子商务网站中的在线购物、互联网金融服务，甚至在线人力资源市场中的线上工作，等等。在这一背景下，互联网已不再仅仅是信息交流的平台，而是正在变成一个综合的线上经济系统。

互联网经济系统及各种网络应用的一个基本功能是在线服务分配（Online Service Allocation，OSA），即将网络服务生产者（Producer）所提供的线上物品（Goods）分配给合适的网络服务消费者（Consumer），其中的线上物品包括各种可能的在线消费品，例如电子商务网站中实实在在的产品、互联网金融网站中的资金、在线人力资源网站中的工作项目，等等。由于消费者往往具有选择消费或者不消费、以及具体消费哪种物品的自由，因此这样的在线服务分配过程往往并不是强制地发生，而是通过个性化推荐或搜索的方式来实现的。搜索引擎（如百度、谷歌或亚马逊购物搜索等）在用户（大概）清楚自己的需求时通过搜索和排序来发挥作用，而个性化推荐（如产品推荐、好友推荐等）则在用户没有显式地提供自己的查询需求

时发挥作用。因此，个性化推荐系统实际上以隐性的方式调节着网络产品或服务与消费者之间的匹配过程，如图5.1示例。

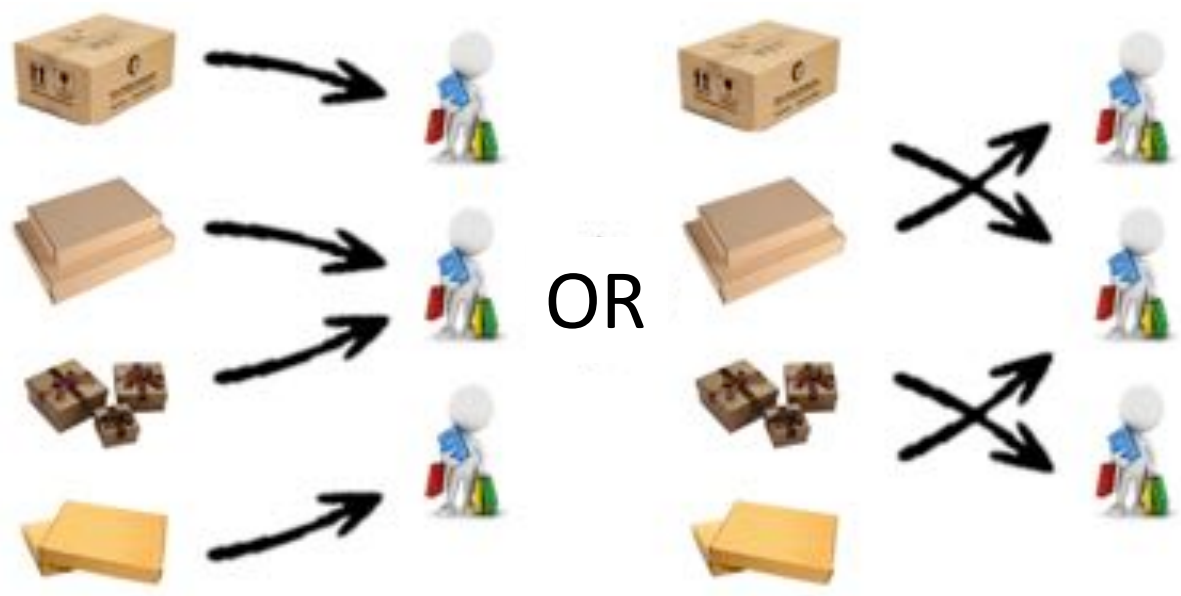

图 5.1　个性化推荐在系统背后隐形地影响和改变在线商品或服务与消费者之间的匹配

由定义可见，在线服务分配是一个生产者和消费者双边匹配的过程，并且资源分配是任何一个经济系统中最为重要的基本问题之一。自从亚当・斯密（Adam Smith）时代以来，经济学家已经开始从一个双边的平衡的角度来看待经济系统的资源分配问题，而其核心则是系统总福利（消费者净收益和生产者净收益之和）。一个最有效的经济系统（市场）应当最大化系统的总福利，使得消费者和生产者中的任何一方都比在单边失衡的市场中要获得更好的市场体验，而这样的经济系统才是社会最希望看到的。

然而由于生产者和消费者收益本质上的互斥性[305]，目前的个性化推荐系统往往在设计之初就只关注于服务生产者或消费者中的一方，而另一方的体验或收益往往被忽略甚至牺牲。例如在最为广泛使用的基于协同过滤的个性化推荐中，推荐算法更多地试图去满足消费者的个性化需求，而几乎没有明确考虑生产者的潜在收益；而在一些在线金融业务（如 P2P 借贷）中，系统则更倾向于为贷款者提供尽可能高的资金收益而弱化了对借款者的关注。然而，这样的系统性设计失衡往往导致实际系统的不可持续性，一个可持续的网络经济系统应当综合考虑系统双方的收益和体验，否则系统可能会因失衡而导致一方的损失过度甚至退出。

本章就以该问题为核心，对个性化推荐在互联网经济系统中深刻的经济学意义进行解释。我们将借助经济学家对经济系统的深刻认识来解读个性化推荐在网络服务分配过程中的作用。具体而言，我们提出互联网总福利最大化（Total Surplus Maximization，TSM）框架，从而将生产者福利和消费者福利的整体最大化理念融入到个性化推荐系统的设计中。通过网络总福利最大化，系统为生产者和消费者提供更大的蛋糕（总福利）进行分享，而不是在既有的总福利中进行争夺。实际上，传统的经济学研究范式往往面向特定商品的对包含众多生产者的有效竞争市场或包含特定生产者的寡头垄断市场进行研究，而互联网应用中众多的商品和非

常个性化的用户需求与传统的经济学范式之间具有很大的不同。为了桥接两者之间的间隔，我们在多个在线市场应用中对如上的总福利最大化框架进行模型具体化，从而展示在不同的应用场景下互联网经济系统平衡的思想是如何为系统使用者（包括生产者和消费者双方）服务的。

在真实数据（包括电子商务、P2P 借贷、在线众包平台）上的实验结果显示，我们的互联网总福利最大框架可以在提高推荐系统用户体验（传统的推荐评价指标）的同时，也提高系统使用双方的总福利，从而做到在不损害消费者体验的情况下提高整个经济系统的运行效率。

### 5.1.2　相关工作

在现代主流经济学中，经济福利（Surplus）[305–307] 往往用来表示经济系统的运行总效率，它主要包括三个紧密相关的概念：消费者福利（Consumer Surplus，CS）、生产者福利（Producer Surplus，PS）和社会总福利（Total/Social Surplus），其中总福利为系统中生产者和消费者所获得的福利之和。社会福利的研究已经拥有很长的历史，最早可以追溯到 19 世纪经济学家对剩余价值的理解[308,309]，在当时，第一次和第二次工业革命的兴起趋势经济学家探索经济增长的本质是什么[310]。

经济福利理论研究的里程碑之一是经济学家 Paul A. Baran 在现代供需关系框架内重新解释了经济福利的概念[307]。他进一步阐述了经济福利的概念对经济学的基础性作用，及其与传统的劳动经济学剩余价值理论的一致性[311]。

在现代经济学中，经济福利的概念在经济系统分析和机制设计中得到经济学家的广泛应用，它经常作为一种对社会福利（Social Good）的直接计量指标而在经济系统优化中得到使用[305,312,313]。然而，虽然线下活动的不断线上化使得互联网像线下经济一样已经越来越成为一个完整的经济系统，但是目前学术界还没有对网络经济协同的福利本质进行深入的研究。

实际上，大部分的互联网应用都可以被形式化为生产者-消费者模型[314–316]，包括我们经常使用的电子商务网站[317]、在线金融服务[318,319]、众包系统[320,321]，甚至社交网络[322,323]，等等。在这些系统中，消费者分别从相应的生产者处消费普通商品、金融产品、在线任务、以及新闻信息，等等。

这些网络应用为在线服务由生产者到消费者的匹配提出了新的问题，而与这样的网络服务匹配过程最为相关的互联网技术就是个性化推荐[1,7,324] 与信息检索[325,326]，它们分别通过推荐列表和搜索列表的方式来满足用户隐式和显式的信息需求。

然而，目前的推荐和检索解决方案往往在设计之初就集中于满足供需双方其

中一方的利益而没有将整个互联网经济系统作为整体进行考虑，从而忽略甚至牺牲了另一方面的潜在收益。例如，最为广泛使用的基于协同过滤的个性化推荐算法[7] 试图基于用户的兴趣偏好而尽可能地最大化用户满意度。尽管消费者满意度的提升可以通过提高点击率而潜在地提高生产者的福利，但是并没有明确的保证使得我们可以通过一个单方面的优化问题而提高整个系统双方面的收益。在本节工作中，我们在成熟的经济学理论和获得广泛认可的经济学基本概念的基础上，将整个互联网系统看成一个完整的经济体系，从而对个性化推荐在其中的作用进行深刻的解释和优化。

### 5.1.3　互联网成本效用与福利

在本节，我们对本工作中所用到的基本经济学概念和定义进行介绍和形式化，它们将成为本工作后续部分的基础。

#### 效用

在经济学中，效用（Utility）是用来描述消费者在消费一个或者一组商品或服务上所获得的满足的度量，它是构成经济学理性选择理论（Rational Choice Theory）[327] 的基础性概念。

效用 $U(q)$ 一般是消费量 $q$ 的函数，$U(q)$ 函数的根本性质由边际效用递减率（Law Of Diminishing Marginal Utility）来决定[328]，即随着消费者消费某一商品的数量的增加，消费者从单位商品中所获得的效用递减，从数学意义上来表示即 $U(q)$ 的一阶导数为正 $U'(q) > 0$，而二阶导数为负 $U''(q) < 0$。为了解释其直观意义，经济学家经常使用的例子是，一位饥饿的消费者可以从消费第一份面包中获得极大的满足感，而当他/她继续消费面包以至于有饱涨感时，再消费一份面包所获得的满足感就会大大减小。

目前，经济学家已经提出了各种各样的效用函数模型，不失一般性，我们在本节介绍两种简单且常用的效用函数以帮助理解效用的直观经济学意义及其性质。两者分别为指数效用函数：

$$U(q) = \frac{1 - \exp(-aq)}{a} \tag{5-1}$$

以及对数效用函数（也称 King-Plosser-Rebelo，KPR 效用函数）：

$$U(q) = a\ln(1 + q) \tag{5-2}$$

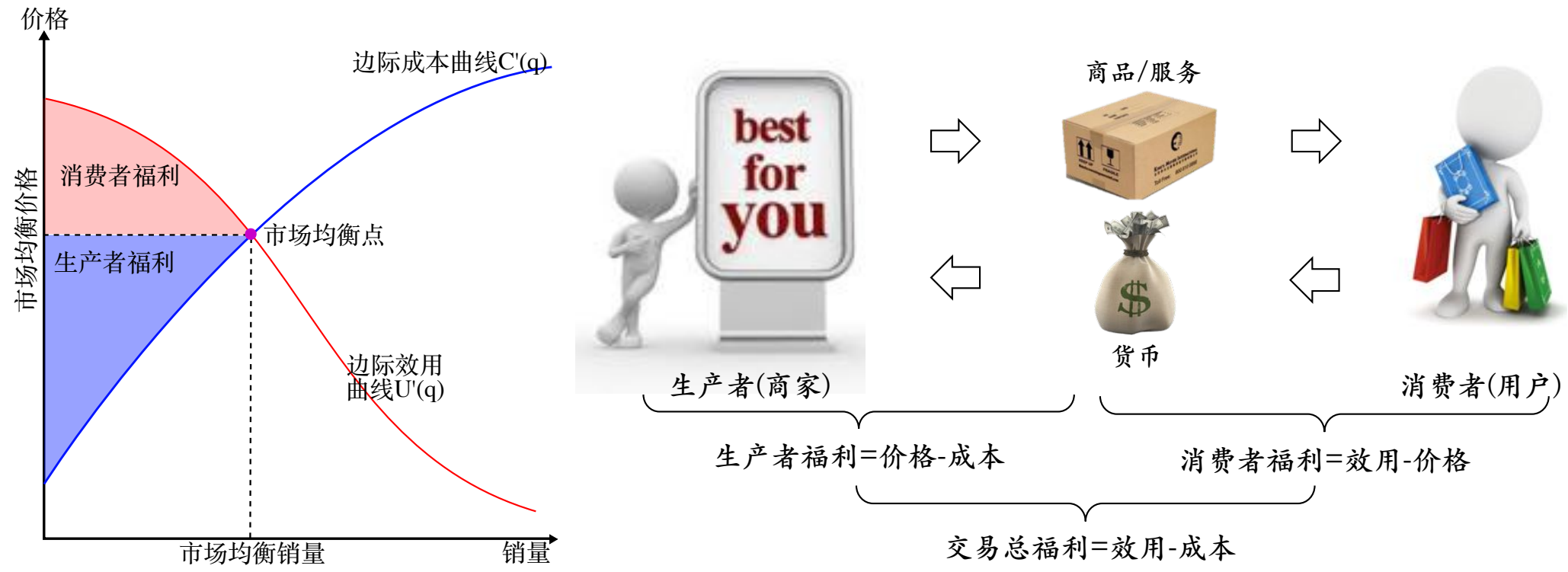

图 5.2　图左为从边际效用曲线和边际成本曲线导出消费者福利和生产者福利示意图；图右为实际网络经济系统的生产者-消费者模型，其中在线商品或服务由生产者（商家）流向消费者（用户），而货币由消费者流向生产者，直观而言，消费者福利为被购买的产品在消费者眼中的实际效用减去消费者购买该产品所需要付出的价格，生产者福利为生产者在消费者那里得到的价格减去生产成本，而一个完整交易所带来的系统总福利为两者之和，即消费者效用减去生产者成本

其中指数效用函数中的参数 $a$ 描述了消费者的风险厌恶程度，对数效用函数中的参数 $a$ 则一方面描述了消费者的风险厌恶程度，另一方面决定了效用曲线的幅度。两者都符合边际效用递减率，且当消费数量 $q = 0$ 时，两者都有 $U(0) = 0$。

进一步，效用可以看做是对价格概念的扩展，在很多实证研究中，经济学家利用消费者对产品的支付意愿（Willingness To Pay，WTP）来刻画该产品对消费者的效用[305,329]，即消费者乐意为多消费一单位的商品所愿意付出的价格 $U'(q)$。这使得我们可以将难以直观度量的效用用与价格一样的货币单位进行度量，从而可以进一步对效用进行定量的计算和比较。

## 福利

福利为货币意义上消费者（或生产者）从消费（或卖出）一个商品或服务中所获得的额外收益。直观而言，消费者福利（Consumer Surplus，CS）为消费者从消费商品中所获得的效用超出他所需要付出的价格之外的部分，而生产者福利（Producer Surplus，PS）为生产者从售出商品中所获得价格超出其成本的部分。图5.2显示了消费者福利和生产者福利与消费者需求曲线和生产者供给曲线之间的关系。其中消费者需求曲线表示消费者在给定数量下，再多消费一单位数量所愿意付出的价格，它由边际效用函数 $U'(q)$ 决定，根据边际效用递减定律，该曲线单调递减。同时，生产者供给曲线表示在给定数量下，生产者再多生产一单位的产品所需要付出的成本，它由边际成本曲线 $C'(q)$ 来决定，根据边际收益递减率[328]，成本曲线是单调递增的。

在自由竞争市场中，市场均衡价格由供需曲线的均衡点（交点）决定，在此基

础上，给定消费者的消费数量 $q = q_c$，消费者福利 $CS$ 由从 0 到 $q_c$ 对需求曲线超出价格的部分进行积分得到：

$$CS = \int_0^{q_c} (U'(q) - P)\,\mathrm{d}q = U(q_c) - Pq_c \tag{5-3}$$

同理，生产者福利 $PS$ 由价格超出成本的部分积分而得到：

$$PS = \int_0^{q_c} (P - C'(q))\,\mathrm{d}q = Pq_c - C(q_c) \tag{5-4}$$

进一步，该生产者和消费者的交易所带来的总福利由生产者和消费者的福利之和决定：

$$TS = CS + PS = U(q_c) - C(q_c) \tag{5-5}$$

最终，整个经济系统的总福利为该经济系统中所有交易中消费者和生产者的总福利之和。

有公式 (5-5) 可知，在总福利中来自生产者福利和消费者福利的价格因素相互抵消，因而不影响整个交易或系统的总福利。这在经济学上是具有重要意义的基本性质，它表明在一个确定的交易上，生产者或消费者一方福利的增加必定是基于另一方福利的牺牲，而价格决定了两者对总福利的分配情况。因此要想同时增加消费者和生产者的福利从而使经济系统的各方都获得更好的体验，唯一的方法是增加经济系统的总福利，从而使得两者有更大的蛋糕可以分配。

### 协同过滤

为了对符号进行清晰的定义和下文使用的方便，我们再次对协同过滤进行简单的介绍和符号化。很多网络应用允许消费者对生产者提供的服务进行满意度打分，例如在亚马逊等电子商务网站中，用户可以对商家进行 1 到 5 星级的评分。我们用数值打分 $r_{ij}$ 表示用户（消费者）$u_i$ 对商家（生产者）$p_k$ 提供的商品或服务 $g_j$ 的满意程度。由于每一个单独的消费者只能消费众多网络商品或服务中的一小部分，因此绝大多数的 $r_{ij}$ 值我们都是未知的，协同过滤的一个重要目的就是去预测这些未知的打分会是多少，从而基于这些预测打分为消费者提供恰当的产品推荐。

最“标准”也最有代表性的协同过滤算法是基于矩阵分解的隐变量模型（La-

tent Factor Models，LFM）[59]，该方法通过用户和物品偏置来预测消费者打分 $\hat{r}_{ij}$：

$$\hat{r}_{ij} = \alpha + \beta_i + \gamma_j + \vec{x}_i^T \vec{y}_j \tag{5-6}$$

其中 $\alpha$ 为全局偏置，$\beta_i$ 和 $\gamma_j$ 分别为消费者和物品偏置，$\vec{x}_i$ 和 $\vec{y}_j$ 为 $K$ 维的用户 $u_i$ 表示向量和商品 $g_j$ 表示向量。在一组已观测打分（训练集）$\mathcal{R}$ 的基础上，模型目标致力于给出尽可能精确的预测打分，即通过最小化如下的损失函数来决定模型参数 $\Theta = \{\alpha, \beta_i, \gamma_j, \vec{x}_i, \vec{y}_j\}$ 的取值：

$$\Theta = \underset{\Theta}{argmin} \sum_{r_{ij} \in \mathcal{R}} (r_{ij} - \hat{r}_{ij})^2 + \lambda \Omega(\Theta) \tag{5-7}$$

其中 $\Omega(\Theta)$ 为 $\ell_2$-范数正则化项。公式 (5-7) 的最小化可以通过随机梯度下降算法（Stochastic Gradient Descent，SGD）或交替最小化算法（Alternating Least Squares，ALS）来获得。

### 5.1.4　基于福利最大化的个性化推荐框架

在本小节，我们提出基于互联网总福利最大化（Total Surplus Maximization，TSM）的在线服务分配（Online Service Allocation，OSA）框架。为了表述的清晰和理解的方便，我们在接下来的部分按照逻辑顺序依次介绍框架的主要组成部分，并在本小节最后给出最终模型。

#### 在线服务分配的问题形式化

我们考虑如下的问题：如何对给定的网络商品在消费者（网络用户）之间进行分配，以最大化网络经济系统的总福利。

假设系统中存在 $m$ 个消费者 $\{u_1, u_2, \cdots, u_m\}$ 和 $n$ 个物品 $\{g_1, g_2, \cdots, g_n\}$，这里的“物品”有可能是电子商务网站中的商品、在线众包网站中的工作，或者 P2P 借贷网站中的贷款理财产品，等等。这些商品由 $r$ 个生产者 $\{p_1, p_2, \cdots, p_r\}$ 提供，其中每一个物品背后有一个确定的生产者，而每个生产者有可能提供多种物品。在接下来，我们用 $1 \le i \le m$，$1 \le j \le n$，和 $1 \le k \le r$ 来表示消费者、物品和生产者。

我们令 $M = [M_1, M_2, \cdots, M_n]$ 表示商品数量向量，其中 $M_j \ge 0$ 为系统中物品 $g_j$ 可以被其生产者所提供的最大数量。例如在亚马逊 MTurk① 或猪八戒网② 等在

① https://www.mturk.com
② http://www.zbj.com

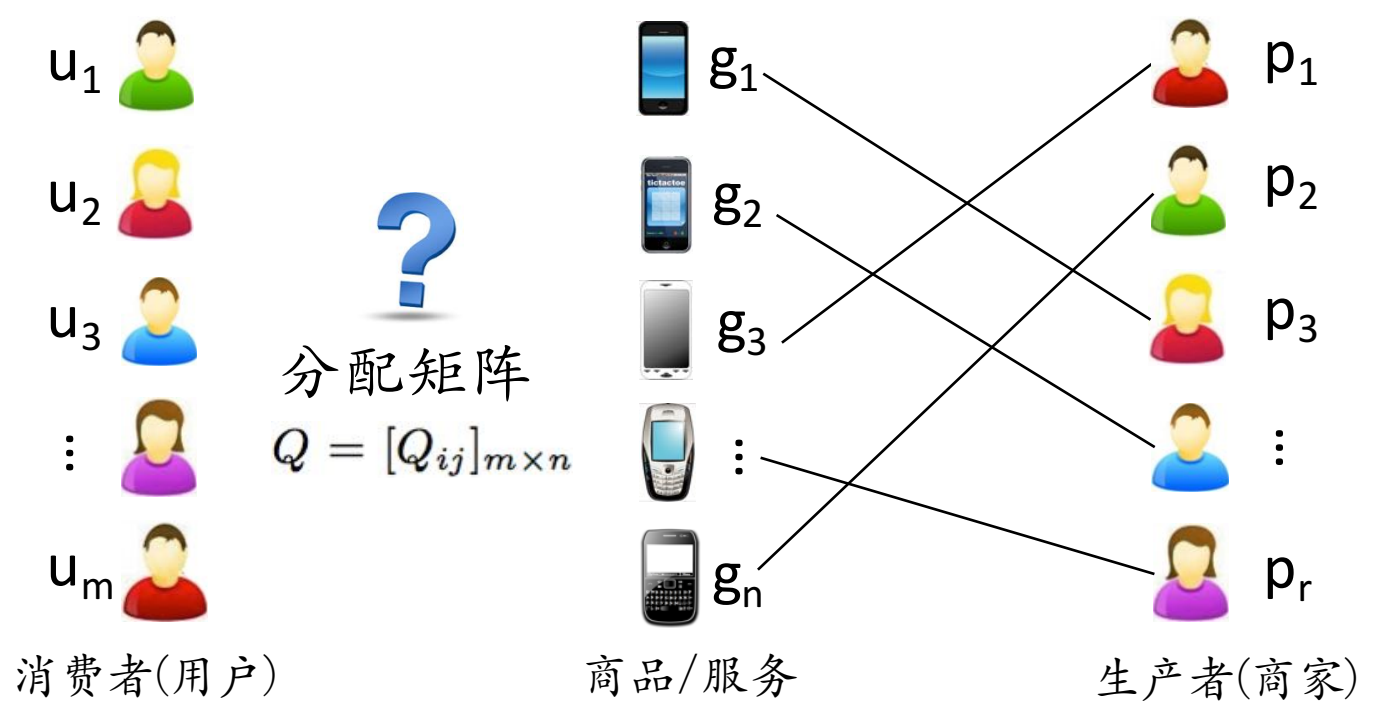

图 5.3　在线服务分配问题示例：在典型的网络应用中，每个产品或服务在背后所对应的生产者或提供者是确定的（即图的右半部分是已知的），而在线服务分配的核心问题为寻找恰当的分配矩阵 $Q = [Q_{ij}]_{m\times n}$，从而确定每个用户 $u_i$ 应当在产品 $g_j$ 上被分配的数量 $Q_{ij}$

线工作任务众包网站中我们有 $M_j = 1$，因为一项任务可以且只能够分配给一个工作者；而在 P2P 借贷网站中我们则有 $0 < M_j < \infty$，即每一项贷款业务所需要募集的贷款总额，这是一个正实数；对于亚马逊等电子商务网站而言，我们则可以认为 $M_j = \infty$，因为卖家（生产者）理论上可以在库存不足的时候随时进货以补充库存。

**定义 5.1：　在线服务分配**（Online Service Allocation，OSA）

在线服务分配任务试图寻找一个分配矩阵（Allocation Matrix）$Q = [Q_{ij}]_{m\times n}$，其中 $Q_{ij} \geq 0$ 为商品 $g_j$ 分配给消费者 $u_i$ 的数量，如图5.3所示。

为了满足商品可提供总量 $M$ 的要求，对于每一个商品 $g_j$ 我们有 $\sum_i Q_{ij} \leq M_j$，即 $\mathbf{1}^T Q \leq M$，其中 $\mathbf{1}$ 为一个元素全部为 1 的列向量。在不同的应用场景下，我们可能对商品的分配数量 $Q$ 指定额外的取值范围限制，以满足特定的实际需求。例如，对于电子商务网站而言我们有 $Q_{ij} \in \mathbb{N}$，因为网络商品的个数只能为整数；而对于在线任务众包应用中的一个工作，我们则有 $Q_{ij} \in \{0, 1\}$，因为一项具体的任务或者分配给一个用户、或者不分配给该用户，而一般不能被拆分并部分地分配给某个用户。

在线服务分配任务在众多网络或移动应用中广泛存在，只要涉及到产品或服务的消费，就存在产品或服务的分配问题。除了电子商务、P2P 借贷、网络众包等应用之外，其它应用包括 Uber 和 Lyft 等在线打车服务、美团网等在线团购服务，甚至 Airbnb 等在线租房业务，等等。

### 个性化效用函数

即便对于同样数量的同一种产品或服务，不同的消费者因其喜好的不同也可能获得不同的效用。例如在电子商务网站中，一个单反镜头可能对于一位已经购

买了单反相机的消费者而言具有较高的效用，而对于一位没有单反相机的消费者而言则可能几乎没有效用。类似的在 P2P 借贷中，同样数量的贷款对于一位具有紧急资金需求的借贷者而言具有很高的效用，因此该借贷者更愿意接受较高的利率；而对于一位对货币需求不是那么紧急的借贷者而言则未必具有很高的效用，因而该借贷者可能更会坚持一个较低的利率。

效用的这一“个性化”本质是服务分配得以有实际作用的根本驱动力，如果所有用户对所有产品的效用是完全一样的，则不同的分配方式对最终系统的福利没有什么影响。正是因为个性化效用的不同，使得我们可以设计合适的分配方案让产品找到效用最高的消费者，从而提高经济系统的福利和效用。而这样的分配过程则是隐形地以推荐系统等商业智能手段来实现的。

在本工作中，我们采用消费者到商品级别的个性化效用函数 $U_{ij}(q)$，即：

$$U_{ij}(q) = \frac{1 - \exp(-a_{ij}q)}{a_{ij}}, \text{ or } U_{ij}(q) = a_{ij}\ln(1+q) \tag{5-8}$$

其中 $U_{ij}(q)$ 表示向消费者 $u_i$ 提供 $q$ 单位的商品 $g_j$ 所带来的效用，该函数的性质由个性化风险厌恶参数 $a_{ij}$ 来决定。

在不同的应用场景中，根据可用数据的不同以及具体适用的经济学理论不同，个性化参数 $a_{ij}$ 可能由不同的方法来估计。例如对于电子商务网站我们利用最后一单位商品的*零收益定理*（Law of Zero Surplus for the Last Unit）[330] 对该参数进行估计，而在网络众包网站中，我们采用*相对福利比*的性质进行参数估计。在接下来的模型具体化部分，我们将对相关内容进行具体的介绍。

#### 网络系统总福利最大化框架

在消费者-商品个性化效用函数 $U_{ij}(q)$ 和每个商品 $g_j$ 成本函数 $C_j(q)$ 的基础上，最直接的总福利最大化建模试图在给定的数量约束条件下找到一个精确的分配矩阵 $Q$，从而使得该分配矩阵下系统的总福利取得最大值：

$$\begin{gathered}\underset{Q}{\text{maximize}} \sum_i \sum_j \left(U_{ij}(Q_{ij}) - C_j(Q_{ij})\right) \\ s.t.\ \mathbf{1}^T Q \le M,\ Q_{ij} \in \mathbb{S}\end{gathered} \tag{5-9}$$

其中 $\mathbb{S}$ 表示在一个特定应用场景下 $Q$ 的合法取值集合，例如对于电子商务应用有 $\mathbb{S} = \mathbb{N}$，而对于在线众包平台中的单个任务分配，我们则有 $\mathbb{S} = \{0, 1\}$。

然而在处理实际数据中，我们需要进一步考虑用户的真实行为模式。实际上，

经济学中的纯理性人假设[327] 在实际中并不一定总是成立，因为消费者并不能总是精确地知道自己对于每个商品最优的消费量是多少。因此，如果我们将分配矩阵 $Q$ 中的值强制限定为具体的实数，则有可能损害模型对真实数据的拟合能力。实际上，用户确实有可能以更大的概率购买最优的商品数量，但是他们也有可能以一定的概率选择最优数量周围但不是最优的购买量，这是由于人类无法在实际生活中对所有的消费进行精确的计算造成的。

为了在模型中对用户的“非理性”进行建模，我们将分配矩阵 $Q$ 中的元素 $Q_{ij}$ 放松为满足一定分布的随机变量。例如当其服从正态分布 $Q_{ij} \sim \mathcal{N}(\mu_{ij}, \sigma_{ij})$ 时，消费者 $u_i$ 对商品 $g_j$ 确实以最大的概率选择消费数量 $\mu_{ij}$，但是他也有可能以较低的概率选择其它可能的消费量。在这个意义上，我们基于总福利最大化的网络服务分配模型最大化如下福利的期望值：

$$\begin{aligned} \underset{\Theta(Q)}{\text{maximize}} \sum_i \sum_j \int & (U_{ij}(Q_{ij}) - C_j(Q_{ij}))p(Q_{ij})\mathrm{d}Q_{ij} \\ s.t.\ \mathbf{1}^T \int & Qp(Q)\mathrm{d}Q \le M,\ Q_{ij} \in \mathbb{S} \end{aligned} \tag{5-10}$$

其中 $p(Q_{ij})$ 为分配矩阵 $Q$ 中每一个元素 $Q_{ij}$ 的密度函数，矩阵 $Q$ 的密度函数为 $p(Q) = \left[p(Q_{ij})\right]_{m\times n}$，且在矩阵 $Q$ 上的积分是元素级别的。另外，$\Theta(Q)$ 为所有 $Q_{ij}$ 分布的参数。如上的概率化建模同时也在一些具体应用场景中起到简化计算的作用，这在接下来的模型具体化中同样会做详细介绍。该模型的最终输出为分配矩阵 $Q$ 的最优密度函数 $p(Q)$，在此基础上，我们将概率密度函数的期望值 $\bar{Q} = \int Qp(Q)\mathrm{d}Q$ 作为最终的分配矩阵，并用来进行个性化推荐。

## 5.2　典型网络平台中的福利最大化

如前所述，基于总福利最大化的网络服务分配框架可以针对具体的网络应用及可用的数据进行不同的模型具体化。为了对框架的实际使用进行具体的解释，我们在本节将该框架具体化到三种不同的网络应用中，分别为电子商务网站、P2P 贷款网站，以及在线众包网站，它们对应于公式 (5-10) 的模型框架中不同的参数取值。为了更好的比较，表5.1汇总了三种应用中模型的具体参数取值。

### 5.2.1　电子商务网站

我们首先基于用户历史购买记录对个性化效用函数 $U_{ij}(q)$ 进行估计。虽然 $U_{ij}(q)$ 无法直接从历史购买数据中得到观察和估计，但是我们知道个性化效用函

表 5.1　基于网络福利最大化的在线服务分配模型在三种不同应用场景下参数的具体选择

| 应用场景 | $CS_{ij}(Q_{ij})$ | $PS_{ij}(Q_{ij})$ | $\mathbb{S}$ |
| --- | --- | --- | --- |
| 电子商务 | $\hat{a}_{ij}\ln(1+Q_{ij})-P_jQ_{ij}$ | $(P_j-c_j)Q_{ij}$ | $\mathbb{N}$ |
| P2P 借贷 | $(r_j-\hat{r})Q_{ij}$ | $(r_j^{max}-r_j)Q_{ij}$ | $\mathbb{R}_+$ |
| 在线众包 | $h(\hat{r}_{ij})s_jQ_{ij}$ | $h(\hat{r}_{kj})s_jQ_{ij}$ | $\{0,1\}$ |
| 应用场景 | $M$ | $p(Q_{ij})$ | $\bar{Q}_{ij}$ |
| 电子商务 | $M_j=\infty$ | $p(Q_{ij}=q)=\lambda_{ij}^{q}e^{-\lambda_{ij}}/q!$ | $\lambda_{ij}$ |
| P2P 借贷 | $0<M_j<\infty$ | $Q_{ij}\sim\mathcal{N}(\mu_{ij},\sigma_{ij})$ | $\mu_{ij}$ |
| 在线众包 | $M_j=1$ | $p(Q_{ij}=1)=\alpha_{ij},P(Q_{ij}=0)=1-\alpha_{ij}$ | $I_{\alpha_{ij}=\max\{\alpha_{i'j}\}_{i'=1}^{m}}$ |

数的经济学本质使用户购买记录满足最后一单位商品的零收益定理[330]。令 $q_{ij}$ 为训练数据集中用户 $u_i$ 对商品 $g_j$ 的真实购买数量，并令 $CS_{ij}(q_{ij})=U_{ij}(q_{ij})-P_jq_{ij}$ 为用户从一次购买行为中所获得的消费者福利，则零收益定理由如下的数学性质所刻画：

$$\begin{aligned}\Delta CS_{ij}(q_{ij})&=CS_{ij}(q_{ij})-CS_{ij}(q_{ij}-1)>0\\ \Delta CS_{ij}(q_{ij}+1)&=CS_{ij}(q_{ij}+1)-CS_{ij}(q_{ij})\leq 0\end{aligned} \tag{5-11}$$

其经济学直观意义在于，随着商品边际效用的逐渐递减，一位消费者之所以对某商品最终购买了数量 $q_{ij}$，是因为当其购买最后的一单位商品时，仍然能够获得一部分正的收益（福利），而如果他再多购买一单位商品，就无法再从中获得正的福利了，甚至有可能减少他的福利（负福利）。

根据协同过滤的思想，我们对公式 (5-8) 中的个性化参数进行如下的建模：

$$a_{ij}=\alpha+\beta_i+\gamma_j+\vec{x}_i^T\vec{y}_j \tag{5-12}$$

其中 $\vec{x}_i$ 为 $K$ 维的用户 $u_i$ 表示向量，$\vec{y}_j$ 为 $K$ 维的商品 $g_j$ 表示向量。因此，消费者风险厌恶参数 $a_{ij}$ 成为中间变量，而模型的直接优化变量为 $\Theta=\{\alpha,\beta_i,\gamma_j,\vec{x}_i,\vec{y}_j\}$。

在此基础上，我们优化如下的训练集对数似然函数：

$$\begin{aligned}
&\underset{\Theta}{\text{maximize}}\ \log p(D)\\
&=\sum_{i=1}^{m}\sum_{j=1}^{n} I_{ij}\log\left(Pr(\Delta CS_{ij}(q_{ij})\geq 0)Pr(\Delta CS_{ij}(q_{ij}+1)<0)\right)\\
&-\lambda(\alpha^2+\sum_{i=1}^{m}\beta_i^2+\sum_{j=1}^{n}\gamma_j^2+\sum_{i=1}^{m}\|\vec{x}_i\|_2^2+\sum_{j=1}^{n}\|\vec{y}_j\|_2^2)\\
&\quad s.t.\ \vec{x}_i,\vec{y}_j\geq 0,\ \forall\ 1\leq i\leq m,\ 1\leq j\leq n
\end{aligned} \tag{5-13}$$

其中 $I_{ij}$ 为示性函数，当消费者 $u_i$ 购买了产品 $g_j$ 时它的值为 1，否则值为 0；以正则化系数 $\lambda>0$ 为权重的正则化项用来防止模型过拟合。我们对隐变量 $\{\vec{x}_i\}_{i=1}^m$ 和 $\{\vec{y}_j\}_{j=1}^n$ 采用经常使用的非负约束，并采用如下的 Sigmoid 函数对条件进行概率化：

$$Pr(\Delta CS_{ij}(q_{ij})\geq 0)=\frac{1}{1+\exp(-\Delta CS_{ij}(q_{ij}))} \tag{5-14}$$

以及，

$$Pr(\Delta CS_{ij}(q_{ij}+1)<0)=1-Pr(\Delta CS_{ij}(q_{ij}+1)\geq 0) \tag{5-15}$$

公式 (5-13) 的最优解可以利用风险厌恶参数 $a_{ij}$ 上的随机梯度下降来求解。为了简化计算并使得模型的参数估计变得可行，我们采用 KPR 对数效用函数 $U_{ij}(q)=a_{ij}\ln(1+q)$。通过随机梯度下降得到参数集 $\Theta$ 的估计值后，我们则可以直接得到风险厌恶参数 $\hat{a}_{ij}$ 的估计值，从而有如下的个性化效用函数估计：

$$U_{ij}(q)=\hat{a}_{ij}\ln(1+q)=(\alpha+\beta_i+\gamma_j+\vec{x}_i^T\vec{y}_j)\ln(1+q) \tag{5-16}$$

为了简单起见，我们令网络商品销售的成本函数为销售量 $q$ 的线性函数 $C(q)=c_jq$，其中 $c_j$ 为商品 $g_j$ 每单位的成本。

最后，由于电子商务应用场景中有 $Q_{ij}\in\mathbb{N}$，我们令公式 (5-10) 中分配矩阵 $Q$ 的元素 $Q_{ij}$ 满足泊松分布 $Q_{ij}\in\mathbb{N}$，即 $p(Q_{ij}=q)=\lambda_{ij}^q e^{-\lambda_{ij}}/q!$，其中 $\lambda_{ij}$ 为分布参数。在此基础上，公式 (5-10) 基于总福利最大化的在线服务分配框架可以具体化为如下的优化问题：

$$\underset{\Lambda}{\text{maximize}}\sum_i\sum_j\sum_{q=0}^{\infty}\frac{\lambda_{ij}^q e^{-\lambda_{ij}}}{q!}(\hat{a}_{ij}\ln(1+q)-c_jq)-\eta\sum_i\sum_j I_{ij}(\lambda_{ij}-q_{ij})^2 \tag{5-17}$$

其中 $\Lambda = [\lambda_{ij}]_{m\times n}$ 为模型参数集；$\eta > 0$ 为正则化系数；$I_{ij}$ 同样为描述训练集中消费者 $u_i$ 是否购买过商品 $g_j$ 的示性函数；$q_{ij}$ 为实际购买量。由于在电子商务应用场景下有 $M_j = \infty$，因此公式 (5-10) 中的数量约束被忽略。

在实际计算泊松分布下的期望福利时，我们不需要对数量 $q$ 从 0 加到正无穷，而只需要加到足够大的 $q$ 值即可。在本工作中，我们选择从 $q = 0$ 一直加到 $q = 10$，因为根据泰勒展开的残差性质，$10! = 3,628,800$ 已经足以让我们忽略残差项。

公式 (5-13) 和公式 (5-17) 的最优解均可以通过随机梯度下降法获得。一旦我们得到公式 (5-17) 中的分布参数 $\Lambda$，即可得到如下的期望分配矩阵 $\bar{Q}$：

$$\bar{Q}_{ij} = \sum_{q=0}^{\infty} q \cdot \frac{\lambda_{ij}^{q} e^{-\lambda_{ij}}}{q!} = \sum_{q=0}^{\infty} \frac{\lambda_{ij}^{q} e^{-\lambda_{ij}}}{(q-1)!} = \lambda_{ij} \tag{5-18}$$

我们最终采用该期望分配矩阵进行商品分配和个性化推荐。需要指出的是，根据泊松分布的性质（公式 (5-18)），在公式 (5-17) 的正则化项中，参数 $\lambda_{ij}$ 实际上就是泊松分布下消费数量 $Q_{ij}$ 的期望值。因此，正则化项实际上是对模型的学习过程施加了一个指导条件，使得模型对那些训练集中已观测购买历史记录的估计购买量不会偏离真实购买量太大。

### 5.2.2　P2P 网络贷款

在 P2P 网络贷款应用中，借方为贷款请求的生产者，而这里的贷款请求可以看成是网络贷款应用中的理财产品。贷方为理财产品的消费者，他们通过对自己贷款资金的分配购买理财产品，并从中获得收益。因此，在 P2P 网络贷款应用中，我们所关心的网络服务分配问题的提法为：贷方（消费者）应当如何在理财产品中对资金进行合理的分配（即确定分配矩阵 $Q$），才能使得整个借贷系统的总福利最大化。

在一个标准的网络借贷流程中，借方（贷款请求的生产者）$p_k$ 通过指定如下的两个基本参数来发起一项请求：一是期望的贷款数额 $M_j$，二是他/她在该贷款上可以接受的最高贷款利率 $r_j^{max}$。一旦一个新的贷款请求被创建，则贷方（贷款请求的消费者）$u_i$ 开始对贷款请求进行竞拍。在竞拍中，每一位贷方给出自己希望贷出的金额以及自己希望获得的利率，而该利率必须小于或等于借方所指定的最大可接受利率 $r_j^{max}$。如果在指定的时间段内竞拍者的总金额达到或超过了借方所需要的金额，则该贷款请求就可以生效，并且系统对参与竞拍的贷方按照他们期望的利率由低到高进行排序，进而选择那些要求的利率最低且总金额达到借方需求金额的贷款者作为胜出者进行借贷。最后，系统在这些胜出的贷方中选择那个

最高的期望利率 $r_j$ 作为该贷款 $g_j$ 的最终利率。通过这样的竞拍机制，系统一方面通过引入自由竞争使得借方的利益受到保护（不超过其可接受的最高利率且通过贷方竞争尽可能降低利率），另一方面使得贷方的利益也受到保护（在成功竞拍人中选择最高利率以防止恶意竞拍）。

因此，作为消费者的贷方从该贷款中所收获的福利为其获得的贷款利息 $r_jQ_{ij}$ 减去其机会成本 $\hat{r}Q_{ij}$，即将该笔资金用于一个无风险投资所可能获得的收益，例如将该笔资金简单地存入银行所必定可以获得的利息。在这里 $\hat{r}$ 为无风险利率。因此，我们有：

$$CS_{ij}(Q_{ij}) = (r_j - \hat{r})Q_{ij} \tag{5-19}$$

类似的，作为生产者的借方所收获的福利为他乐意付出的最高利息 $r_j^{max}Q_{ij}$ 减去他实际上所需要付出的利息 $r_jQ_{ij}$，即：

$$PS_{ij}(Q_{ij}) = (r_j^{max} - r_j)Q_{ij} \tag{5-20}$$

因此，在该贷款所带来的系统总福利为：

$$TS_{ij}(Q_{ij}) = CS_{ij}(Q_{ij}) + PS_{ij}(Q_{ij}) = (r_j^{max} - \hat{r})Q_{ij} \tag{5-21}$$

由于 $Q_{ij}$ 所表示的货币量为一个连续值，因此我们对 $Q_{ij}$ 施加一个高斯分布函数，即 $Q_{ij} \sim \mathcal{N}(\mu_{ij}, \sigma_{ij})$。

最终，P2P 网络贷款环境下的在线服务分配将公式 (5-10) 的框架具体化为如下的优化问题：

$$\begin{aligned} &\underset{U,\Sigma}{\text{maximize}} \sum_i \sum_j \int \frac{(r_j^{max} - \hat{r})Q_{ij}}{\sqrt{2\pi}\sigma_{ij}} \exp\left(-\frac{(Q_{ij} - \mu_{ij})^2}{2\sigma_{ij}^2}\right) \mathrm{d}Q_{ij} \\ &s.t.\ \mathbf{1}^T \int \frac{Q}{\sqrt{2\pi}\Sigma} \exp\left(-\frac{(Q - U)^2}{2\Sigma^2}\right) \mathrm{d}Q \le M, Q_{ij} \in \mathbb{R}_+ \end{aligned} \tag{5-22}$$

其中 $U = [\mu_{ij}]_{m\times n}$ 和 $\Sigma = [\sigma_{ij}]_{m\times n}$ 为模型的优化参数，该式可以进一步简化为如下的优化问题：

$$\begin{aligned} &\underset{U,\Sigma}{\text{maximize}} \sum_i \sum_j \mu_{ij}(r_j^{max} - \hat{r}) \\ &s.t.\ \mathbf{1}^T U \le M,\ \mu_{ij} \in \mathbb{R}_+ \end{aligned} \tag{5-23}$$

而公式 (5-23) 可以通过简单的线性规划求得最优解。同样，我们采用高斯分布下分配矩阵的期望值作为最终的分配矩阵，即：

$$\bar{Q}_{ij} = \mu_{ij} \tag{5-24}$$

如上的结果具有重要的直观意义，它实际上允许我们依据贷款请求的单位资本福利 $(r_j^{max} - \hat{r})$ 以贪心的模式对资金进行分配，而这在实际投资问题中是一种直观简单且常用的策略。

### 5.2.3　在线众包平台

在亚马逊 Mturk 和猪八戒网等在线众包平台中，雇佣方（工作任务的生产者）$p_k$ 将工作任务 $g_j$ 公布在平台中，自由职业者（工作任务的消费者）$u_i$ 则申请完成他们感兴趣的工作。由于单项具体的工作任务只能分配给一位工作者，而每一个工作者只能决定是否接受某项工作任务、而不能接受某工作任务的一部分，因此在该场景中分配矩阵 $Q$ 中的元素 $Q_{ij}$ 只能是二值的，即 $Q_{ij} \in \{0, 1\}$。

雇佣方和工作者通过谈判或由雇佣方直接指定某工作任务 $g_j$ 的工资 $s_j$。当工作任务完成时，工作双方通过为对方打分的方式来表示自己是否对另一方感到满意。我们用 $r_{ij}$ 表示工作者 $u_i$ 在任务 $g_j$ 上的打分，用 $r_{kj}$ 表示雇佣方 $p_k$ 在任务 $g_j$ 上的打分。

为了对雇佣双方在某项工作上所获得的福利，我们采用如下的经济学假设，即雇佣双方在每单位价格上所获得的福利正比于双方在该任务上的归一化打分[329,330]，即越高的打分就意味着从中获得越高的福利。

为此，我们首先采用如公式 (5-7) 所示的协同过滤算法对工作者-任务打分 $\hat{r}_{ij}$ 和雇佣者-任务打分 $\hat{r}_{kj}$ 进行估计和补全。在 Sigmoid 函数 $h(x) = \frac{2}{1+\exp(-x)} - 1$ 的基础上，我们将工作者的单位价格福利建模如下：

$$\frac{U_{ij}(Q_{ij}) - s_j}{s_j} = h(\hat{r}_{ij})Q_{ij} = \left(\frac{2}{1 + e^{-\hat{r}_{ij}}} - 1\right)Q_{ij} \tag{5-25}$$

并将雇佣者的单位价格福利建模如下：

$$\frac{s_j - C_j(Q_{ij})}{s_j} = h(\hat{r}_{kj})Q_{ij} = \left(\frac{2}{1 + e^{-\hat{r}_{kj}}} - 1\right)Q_{ij} \tag{5-26}$$

其中 $Q_{ij} \in \{0, 1\}$ 可以看作表示一项工作是否被分配给某个工作者的二值示性函数。

因此，将任务 $g_j$ 分配给工作者 $u_i$ 所带来的消费者福利、生产者福利、和系统总福利为：

$$\begin{aligned} CS_{ij}(Q_{ij}) &= U_{ij}(Q_{ij}) - s_j = h(\hat{r}_{ij})s_jQ_{ij} \\ PS_{ij}(Q_{ij}) &= s_j - C_j(Q_{ij}) = h(\hat{r}_{kj})s_jQ_{ij} \\ TS_{ij}(Q_{ij}) &= \left(h(\hat{r}_{ij}) + h(\hat{r}_{kj})\right) s_jQ_{ij} \end{aligned} \tag{5-27}$$

考虑到在该场景下 $Q_{ij}$ 为二值的，我们利用伯努利分布来刻画其随机性：

$$p(Q_{ij} = 1) = \alpha_{ij}, P(Q_{ij} = 0) = 1 - \alpha_{ij} \tag{5-28}$$

其中 $0 \le \alpha_{ij} \le 1$。令 $A = [\alpha_{ij}]_{m\times n}$ 为参数集，由于每一项具体的工作任务只被提供一次，因此我们进一步令 $M_j = 1$。在此基础上，在线众包平台中的任务分配可以具体化为如下的优化问题：

$$\begin{aligned} &\underset{A}{\text{maximize}} \sum_i \sum_j \left(h(\hat{r}_{ij}) + h(\hat{r}_{kj})\right) s_j\alpha_{ij} \\ &s.t.\ \mathbf{1}^T A \le \mathbf{1},\ 0 \le \alpha_{ij} \le 1 \end{aligned} \tag{5-29}$$

公式 (5-29) 同样可以利用线性规划求得最优解。一旦我们在训练集上学习得到参数集合 $A = [\alpha_{ij}]_{m\times n}$，则将工作任务 $g_j$ 分配给在所有的工作者所对应的概率 $\alpha_{i'j}$ 中具有最大分配概率 $\alpha_{ij}$ 的工作者 $u_i$：

$$\bar{Q}_{ij} = \begin{cases} 1,\ \text{if } \alpha_{ij} = \max\{\alpha_{i'j}\}_{i'=1}^{m} \\ 0,\ \text{otherwise} \end{cases} \tag{5-30}$$

该结论同样具有重要的直观意义，它可以通过将公式 (5-29) 中的参数 $\alpha_{ij}$ 替换为 $Q_{ij}$ 并同样采用贪婪策略进行任务分配而获得。在该策略下，我们将工作 $g_j$ 分配给在 $\left(h(\hat{r}_{ij}) + h(\hat{r}_{kj})\right) s_j$ 上具有最大值的工作者 $u_i$，而这实际上是公式5-9所描述的非概率化福利最大化模型的具体化。进一步，这实际上是对传统上基于协同过滤的个性化推荐算法基于福利最大化的增强版，在下面的讨论中，我们将对此进行更具体的分析介绍。

### 5.2.4　小结与讨论

经过如上的模型具体化，我们在这里对基于福利最大化的服务分配和推荐与传统的个性化推荐算法进行对比，以指出两者的不同和联系。

在 $M_j = \infty$ 所描述的数量无限制的情况下，公式 (5-10) 中的数量约束 $\mathbf{1}^T \int Qp(Q)\mathrm{d}Q \leq M$ 可以不必考虑，因此我们得到无约束的优化目标，如公式 (5-17) 所示。在这样的情况下，每一个消费者所对应的总福利实际上是相互独立的，因此每一个消费者的最优分配也是相互无关的。此时，公式 (5-17) 可以被分解为对每一个用户 $u_i$ 分别进行最大总福利优化：

$$\underset{\{\lambda_{ij}\}_{j=1}^{n}}{\text{maximize}} \sum_j \left( \sum_{q=0}^{\infty} \frac{(\hat{a}_{ij} \ln(1+q)) \lambda_{ij}^q e^{-\lambda_{ij}}}{q!} - \lambda_{ij} c_j \right) - \eta \sum_j (\lambda_{ij} - q_{ij})^2 \tag{5-31}$$

这与传统的考虑每个用户的个性化偏好来提供最相关物品推荐的推荐算法在本质上是相同的，只不过这里的“个性化偏好”以效用和福利来描述。在这里协同过滤的设计思想体现在公式 (5-16) 的个性化效用函数中，其中 $\hat{a}_{ij} = \alpha + \beta_i + \gamma_j + \vec{x}_i^T \vec{y}_j$ 描述了利用全部用户的历史行为以群体智能的方式对用户物品偏好进行学习。

同样，对于公式 (5-29) 所描述的在线众包应用而言，我们看到对于一个给定的工作任务 $g_j$，雇佣者对任务的打分 $h(\hat{r}_{kj})$（通过协同过滤预测得到）以及该工作的工资 $s_j$ 将会是定值。因此，贪心分配策略下每个工作任务的权重 $\left(h(\hat{r}_{ij}) + h(\hat{r}_{kj})\right) s_j$ 只取决于工作者 $u_i$。在这个意义上，我们最终是将工作 $g_j$ 分配给具有最大 $h(\hat{r}_{ij})s_j$ 值的工作者 $u_i$。这实际上是对基于协同过滤的个性化推荐的算法的扩展，在传统的协同过滤中，算法将任务 $g_j$ 推荐给具有最大预测打分 $\hat{r}_{ij}$ 值的工作者 $u_i$，唯一的区别在于我们在推荐中进一步考虑了工资因素 $s_j$，从而在货币意义上取得最大的总福利。

另外，从公式 (5-31) 非无穷最优解的存在我们可以得到如下直观的结论：尽管我们不对生产者的销售数量进行限制，但是算法自然的最优结果表明并非系统内的总销售量越多则总福利越高。实际上，这是由消费者的边际效用递减来决定的，而在实际人类社会的经济系统中，如上结论也被倾销对经济的危害性所证实。

然而，当模型的数量限制条件存在时，消费者福利是互相联系在一起的，因此取得全局总福利最大化的分配矩阵并非意味着每一个消费者福利都取得可能的最大值。我们将在实验环节研究总福利最大化与推荐系统的消费者体验之间的关系。

## 5.3　性能评测

在本节，我们在对基于总福利最大化的服务分配和个性化推荐框架进行性能评测，这既包括传统的个性化推荐指标，又包括经济系统福利最大化指标。我们在电子商务、P2P 借贷，以及在线众包平台三种不同的网络应用环境下对互联网福利最大化进行研究。

### 5.3.1　电子商务网站

我们采用电子商务网站 Shop.com① 的用户历史购买数据进行实验。在我们的总福利最大化框架中，一个重要的信息为用户在每一次购买行为中对商品的具体购买数量，然而在很多其它电子商务网站数据集中购买量信息是缺失的，而在 Shop.com 数据集中，我们不仅知道用户购买了什么商品，还知道具体的购买数量，因此可以用来进行建模。

为了避免冷启动问题的干扰而专注于福利最大化问题的研究，我们从原始数据集中选择那些至少拥有五次购买或被购买记录的用户和商品，这在个性化推荐研究中是经常使用的数据预处理技术[6,331,332]。表5.2展示了 Shop.com 的基本统计数据信息。

表 5.2　Shop.com 数据集基本统计信息

| 消费者数 | 商品数 | 购买记录数 | 密度 | 训练集/测试集 |
|---|---|---|---|---|
| 34,099 | 42,691 | 400,215 | 0.03% | 75%/25% |

可见，该数据集与其它电子商务网站数据集一样非常稀疏，只有约 0.03% 的密度。我们进一步对每一个用户随机选择 75% 的历史购买记录（Transaction）并构建训练集，并将剩下的 25% 作为测试集。最终测试集中包括约 34$k$ 用户到 30$k$ 商品的 100$k$ 历史购买记录。

#### 参数选择与实验设置

在公式 (5-16) 的个性化 KRP 效用函数 $U_{ij}(q)$ 中，唯一的模型参数为风险厌恶系数 $a_{ij}$，而风险厌恶系数 $a_{ij}$ 的估计则进一步退化为公式 (5-13) 中对消费者和商品偏置以及表示向量的学习。

在该实验中，我们通过网格搜索确定公式 (5-13) 中的超参数 $\lambda$ 以及公式 (5-17) 中的超参数 $\eta$ 的最优值。在接下来的实验中，除非我们对两个参数进行专门的调节以研究它们对模型性能的影响，否则我们令它们等于网格搜索得到的最优值

① http://www.shop.com

$\lambda = 0.05$ 以及 $\eta = 5$。在整个实验中，我们令公式 (5-13) 中的隐变量个数（即表示向量 $\vec{x}_i$ 和 $\vec{y}_j$ 的维度）为 $K = 20$，因为实验发现 20 个维度的分解变量足以获得稳定的最优解，使用更多的分解变量并不能进一步提高效果。

一旦我们利用公式 (5-13) 获得了风险厌恶系数的估计值 $\hat{a}_{ij}$，则可以对每一个消费者在每一个产品上的效用 $U_{ij}(q)$ 进行定量计算，使得我们可以进一步对公式 (5-17) 中的消费者商品期望分配数量 $\lambda_{ij}$ 进行估计。考虑到每一个 $\lambda_{ij}$ 与 $\hat{a}_{ij}$ 一样是和一个消费者 $u_i$ 及一个商品 $g_j$ 相关的，我们同样采用协同过滤的方式对 $\lambda_{ij}$ 进行参数化以方便模型的学习。我们令 $\lambda_{ij} = \alpha' + \beta'_i + \gamma'_j + \vec{x}_i'^T \vec{y}_j'$，因此 $\lambda_{ij}$ 可以作为中间变量而通过对实际变量 $\Theta' = \{\alpha', \beta'_i, \gamma'_j, \vec{x}'_i, \vec{y}'_j\}$ 的随机梯度下降而学习出来。

由于缺乏每一个商品的真实成本数据，为了简单起见，我们令数据集中的每一个商品的成本为价格的一般，即 $c_j = 0.5P_j$，其中 $P_j$ 为商品 $g_j$ 的价格。

需要指出的时，当公式 (5-17) 中的正则化系数 $\eta$ 足够大时，公式中的总福利项的作用将会变得足够小，此时公式退化为一个预测 $q_{ij}$ 的简单协同过滤算法，在接下来的实验中，我们将该协同过滤算法作为基线算法进行比较，以研究总福利项对个性化推荐和福利最大化的作用。

算法最终给出每一个用户 $u_i$ 对每一个物品 $g_j$ 的期望分配数量 $\lambda_{ij}$；在此基础上，对于每一个用户 $u_i$，我们将所有的未购买物品按照 $\lambda_{ij}$ 由大到小进行排序，并从上到下依次取出前 $N$ 个物品从而构建 top-N 推荐列表。表5.3集中展示了模型所有超参数（Hyper-Parameter）的取值。

表 5.3　模型参数的选择，其中隐变量个数 $K$ 和协同过滤项系数 $\lambda$ 在整个实验中为固定值，我们在实验章对 $\eta$ 的取值进行调节以观察它的影响，$c_j$ 为商品 $g_j$ 的成本

| 隐变量个数 $K$ | 公式 (5-13) 中的 $\lambda$ | 公式 (5-17) 中的 $\eta$ | 公式 (5-17) 中的 $c_j$ |
|---|---|---|---|
| 20 | 0.05 | 5 | $0.5P_j$ |

### 购买量预测与个性化推荐的评价

我们首先在传统的购买量预测和个性化推荐任务上对我们的总福利最大化框架进行评测。作为基线算法，我们选择广泛使用的协同过滤算法（如公式 (5-6) 和公式 (5-7) 所示）对用户购买量进行直接预测，而数据中用户的真实购买量为分散在 1 ~ 20 范围内的整数。为了模型比较的公平性，我们在协同过滤算法中对超参数 $K$ 和 $\lambda$ 使用如表5.3所示的同样的设定。

与我们的总福利最大化框架相同，一旦协同过滤算法给出了购买量预测，则我们对测试集中的商品按照预测购买量由高到低进行排序并构建个性化推荐列

表。我们采用在实际电子商务推荐系统中最为关心的转化率（Conversion-Rate@N，CR@N）指标对 top-N 推荐列表的排序性能进行评测[333]。对于给定的测试用户集合以及每个测试用户的 top-N 推荐列表，CR@N 为那些在相应用户的测试集中命中了购买商品的推荐列表占所有推荐列表（即全部测试用户数）的百分比。在实验中，我们令 $N$ 的取值从 1 到 100；对于测试集中的每一个用户，用户推荐的候选商品约为 30$k$ 个，所有这些候选商品都在训练集中出现过，因此没有完全冷启动的商品。

在不同的正则化系数取值 $\eta = 0.1, 1, 5, 10$ 下，传统的协同过滤算法（CF）以及我们的总福利最大化模型（TSM）在推荐列表长度分别为 $N = 5, 10, 20$ 时的效果如表5.4所示，其中黑体数字为双边 t 检验下显著性系数达到 0.05 的结果。在 $N$ 的取值为 1 到 100 范围内更详细的评测结果如图5.4所示。

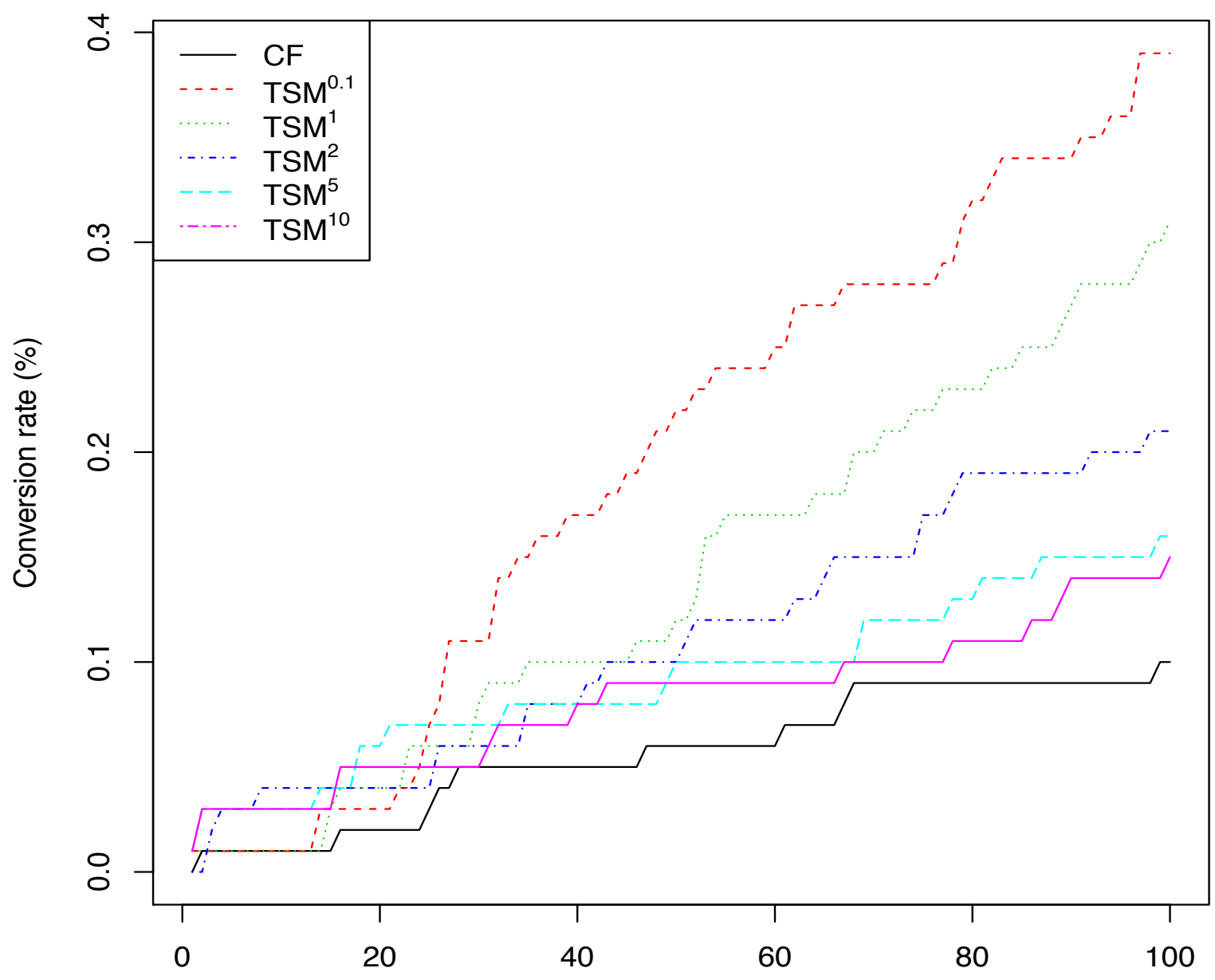

图 5.4　协同过滤 CF 与总福利最大化模型 TSM$^\eta$ 效果对比，其中纵坐标为转化率，横坐标为推荐列表的长度 $N$

结果显示，我们的总福利最大化模型 TSM 在多数正则化系数 $\eta$ 和推荐列表长度 $N$ 的取值上都好于协同过滤 CF 基线算法。另外，一个比较有趣的实验结果发现是，在较长的推荐列表上，TSM$^\eta$ 模型的推荐效果随着 $\eta$ 的增加而不断减小，一直减小到与协同过滤基线算法相同。如前所述，这体现了 TSM 模型内在的合理性，因为当 $\eta \to \infty$ 时，TSM 实际上在模型上退化到协同过滤算法 CF，因此它的实验效果也退化到 CF 上。该实验观察进一步验证了在我们的 TSM 模型中总福利项的重要作用，并且该实验结果表明最大化系统总福利能够提升用户在推荐系统上的体验。

表 5.4　在转化率（Conversion Rate，CR@N）和总福利（Total Surplus，TS@N）指标上对 Top-N 推荐的评测结果，其中 $TSM^*$ 表示我们的总福利最大化模型（Total Surplus Maximization）在正则化系数 $\eta = *$ 时的结果，如公式 (5-17) 所示

| $N$ | 5 | | | | |
|---|---|---|---|---|---|
| 方法 | CF | $TSM^{0.1}$ | $TSM^1$ | $TSM^5$ | $TSM^{10}$ |
| CR (%) | 0.10 | 0.10 | 0.10 | **0.30** | **0.30** |
| TS ($) | 33.05 | **1009.45** | **1009.45** | 422.01 | 24.48 |
| $N$ | 10 | | | | |
| 方法 | CF | $TSM^{0.1}$ | $TSM^1$ | $TSM^5$ | $TSM^{10}$ |
| CR (%) | 0.10 | 0.10 | 0.10 | **0.30** | **0.30** |
| TS ($) | 57.89 | **2278.36** | 2208.50 | 807.56 | 213.45 |
| $N$ | 20 | | | | |
| 方法 | CF | $TSM^{0.1}$ | $TSM^1$ | $TSM^5$ | $TSM^{10}$ |
| CR (%) | 0.20 | 0.30 | 0.40 | **0.60** | 0.50 |
| TS ($) | 98.09 | **2892.03** | 3135.35 | 1137.89 | 676.65 |

另外，实验结果显示 $\eta$ 的值也不能太小，否则优化目标 (5-17) 中的数量约束的作用就会被掩盖，而这尤其会影响排在前面的商品的推荐性能，从而对较短的推荐列表排序性能影响较大。这是由于如果不限制算法在训练样本上的预测购买量与真实购买量接近，那么算法对购买量的预测就只由个性化 KPR 效用函数 $U_{ij}$ 来决定。而由于 KPR 效用函数在拟合灵活度上的局限性（简单的对数函数），使其可能无法对一些商品的用户效用函数进行正确的拟合，从而降低预测效果。概括而言，$\eta$ 通过对总福利项和数量约束项进行权衡而影响推荐效果，并且在实际系统中，$\eta$ 的取值应当通过实验进行合理的设定，过高或过低都有可能降低推荐效果。

### 总福利效果评测

在本小节，我们在总福利指标上对模型的性能进行评测。在如上算法对每个消费者所构建的 top-N 个性化推荐列表的基础上，我们假设每一个用户接受该推荐列表，并以此计算平均每个消费者购买被推荐的 N 个商品所带来的系统总福利：

$$TS@N = \frac{1}{M}\sum_{i=1}^{M}\sum_{j\in\Pi_{i,N}}\left(\hat{a}_{ij}\ln(1+\lambda_{ij}) - c_j\lambda_{ij}\right) \tag{5-32}$$

其中 $i$ 和 $M$ 分别为测试用户的下标和总数，$N$ 为推荐列表的长度，$\Pi_{i,N}$ 为算法为第 $i$ 个用户构建的 top-N 个性化推荐列表。

在推介列表长度为 $N$ 的情况下系统平均总福利 $TS@N$ 如表5.4所示，而图5.5给

出了在 $N$ 的不同取值下更完整的实验结果。

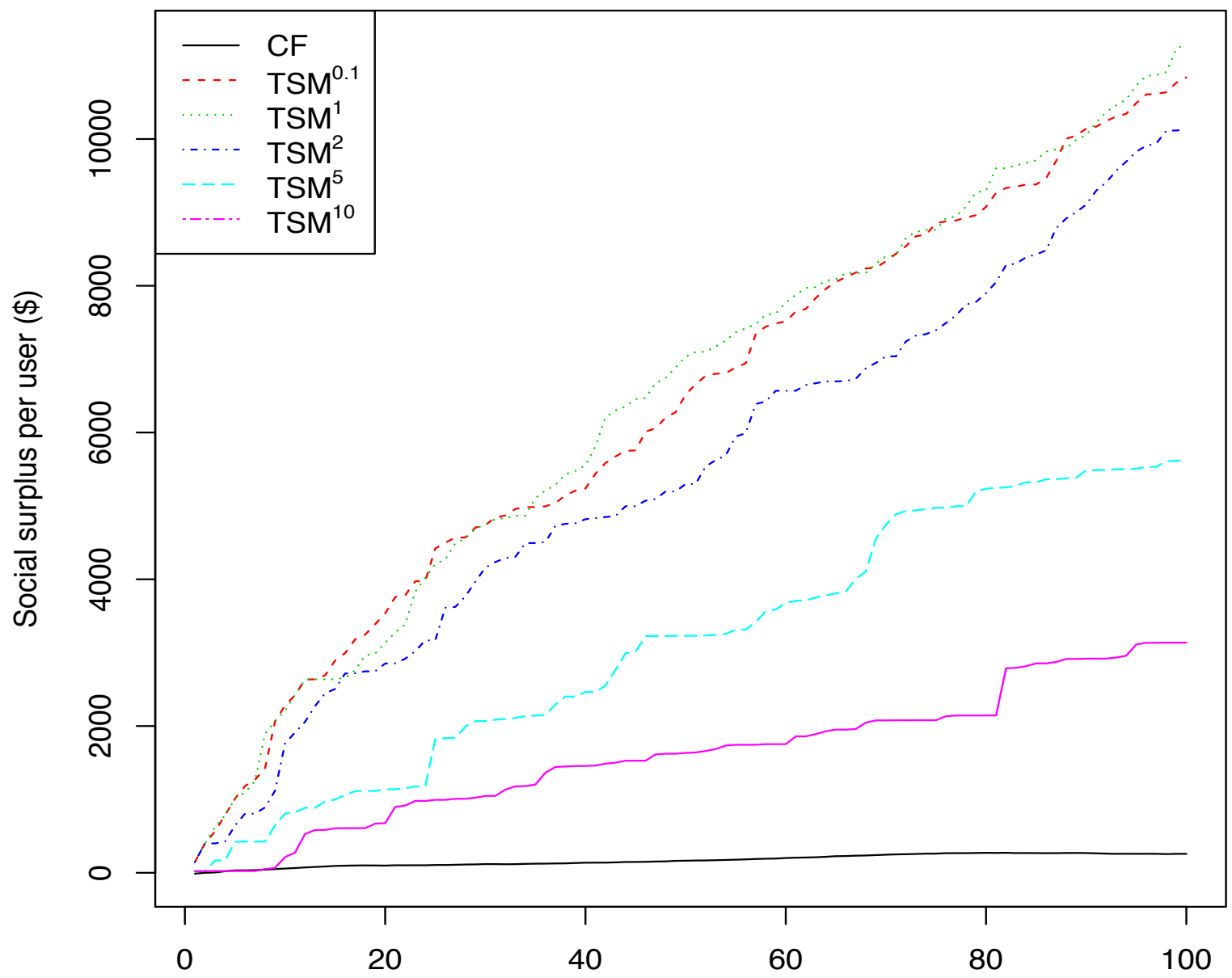

图 5.5　在总福利（TS）指标上我们的总福利最大化框架（TSM$^\eta$）与协同过滤算法（CF）的对比，需要注意的是 $TS@N$ 表示 top-N 推荐列表被用户完全接受时所产生的总福利

由结果可见我们的总福利最大化模型 TSM 要一致地好于协同过滤基线算法 CF。这实际上是在意料之中的结果，因为我们的框架本身就设计为对总福利进行最大化（如公式5-17）。另外，我们发现 $\eta$ 越小，我们的总福利最大化模型越能得到更高的总福利值。参数 $\eta$ 的这一性质进一步验证了公式5-17中总福利项和数量约束项各自的作用。

综合本小节对总福利的观测结果和上一小节对推荐效果的评测结果，可见当模型 $\eta$ 取得合适值时，算法可以同时在推荐效果和总福利指标上取得更好的结果。这表明我们基于总福利最大化的个性化推荐模型可以在提高推荐系统对用户的使用体验的同时，增加整个系统的社会福利，从而提高系统的经济效率。

### 5.3.2　P2P 网络贷款

为了对 P2P 在线网络贷款场景下的模型效果进行评测，我们采用著名的 P2P 借贷网站 Prosper① 公开的数据集[319]。从 2009 年第三季度起，Prosper 推出了自动竞拍系统，即贷方可以设定一定的条件，当新的贷款请求发出时系统会代替贷方自动对贷款请求进行竞拍。然而，我们希望在网络经济系统中研究生产者和消费者自身的直接行为，而不是智能算法代替他们所作出的决定，因此我们选用在这

① http://www.prosper.com

一机制上线之前的数据集进行算法评价，这包括了从 2005 年 11 月 9 日到 2009 年 5 月 8 日约三年半的数据。

另外，由于我们在本工作中并不考虑金融风险因素对用户行为的影响，因此我们选择那些状态不是“违约（Defaulted）”、“取消（Cancelled）”或者“坏账冲销（Charge-off）”的贷款请求进行实验，因为这些贷款请求在实际系统中会被系统的风险控制模块自动滤除。最终，我们的实验数据集包括那些状态为“进行中（Current）”、“偿付中（Payoff in Progress）”或“已付清（Paid）”的贷款请求，这一共包括 46,680 人次贷款、1,814,503 个竞拍记录、以及美元 $157,845,684 的货币资本。表5.5给出了实验数据的基本统计信息，需要指出的是，由于我们只关心一个资本分配方案的安全性和所带来的总福利，因此在本实验中不需要训练集和测试集，而只需要对总资本在贷方和借方之间进行分配即可。

表 5.5　Prosper 实验数据统计信息

| 贷款次数 | 贷方人数 | 竞拍次数 | 总资本 |
|---|---|---|---|
| 46,680 | 49,631 | 1,814,503 | $157,845,684 |
| 最小利率 | 最大利率 | 平均利率 | 平均贷款额 |
| 0.0001 | 0.4975 | 0.1662 | $3,381.44 |

为了计算任一个资金分配方案 $Q = [Q_{ij}]_{m\times n}$ 的总福利，我们采用美国年平均银行存款利率 $\hat{r} = 0.01$ 作为无风险利率，P2P 贷款环境下的总福利（Total Surplus）如下所示：

$$TS_{P2P} = \sum_i \sum_j Q_{ij}(r_j^{max} - \hat{r}) \tag{5-33}$$

在此基础上，系统中实际的资金分配方案所带来的总福利以及我们的 TSM 框架给出的分配方案所带来的总福利如表5.6所示。

表 5.6　系统实际分配方案和总福利最大化框架所给出的分配方案下系统总福利对比

| | 总福利 ($) | 福利/每单位贷款 ($) | 福利/每单位资本 ($) |
|---|---|---|---|
| 实际 | 25,174,131 | 539.29 | 0.1595 |
| TSM | 33,838,364 | 724.90 | 0.2144 |

由实验结果可见，我们的总福利最大化框架给出的资本分配方案能够获得比实际分配方案高出 34.42% 的系统总福利和每单位资本福利，从每单位资本 $0.16

提高到每单位资本 \$0.21。这对于在线经济系统的资本效率和社会福利而言是重要的提升。基于双边 $t$ 测验的显著性检验结果显示效果提升在 $p = 0.01$ 水平上显著。

实际上，模型在总福利上的效果提升是在预期之内的，因为与上一小节的结果一样，我们的模型本身就试图对总福利进行最大化。然而虽然算法给出的分配方案可以获得较高的总福利，我们需要进一步验证算法所给出的分配方案对于实际的贷方而言是可以接受的。因此，我们进一步计算在我们的分配方法下那些得到满足的贷款请求最终的偿还率是多少（Percentage of Paid，PoP），偿还率表示了某种分配方案对投资者的安全性。

结果显示，在真实分配方案中得到满足的贷款请求的最终偿还率为 69.37%，而在我们的总福利最大化框架所给出的分配方案中，得到满足的贷款请求最终偿还率为 73.32%，并不比真实分配方案的偿还率低。这意味着我们的总福利最大化框架可以在保证投资者的资金安全性体验的前提下提高整个经济系统的总福利。

### 5.3.3　在线自由职业与众包平台

我们采用在线众包平台猪八戒网[①]（ZBJ）的数据进行实验验证，每条工作记录包括雇佣方 ID、工作方 ID、任务 ID、工资，以及雇佣方和工作方分别对该任务在 0 ~ 5 打分区间上的评分。表5.7给出了数据集的基本统计信息。

表 5.7　猪八戒数据集的基本统计信息

| 雇佣方人数 | 工作者人数 | 任务数 | 平均工资 |
|---|---|---|---|
| 40,228 | 46,856 | 296,453 | ￥21.68/小时 |
| 雇佣方打分数 | 工作方打分数 | 雇佣方平均打分 | 工作方平均打分 |
| 276,103 | 241,638 | 2.336 | 2.405 |

与电子商务网站的实验相同，我们利用算法最终给出的分配矩阵对工作者进行任务推荐，并在推荐性能和总福利指标上进行评测。本实验同样采用协同过滤（CF）作为基线算法——我们利用所有的工作者对任务的打分构建工作者-任务打分矩阵，并在该矩阵上进行基于协同过滤的打分预测（公式 (5-6) 和 (5-7)），并最终将工作 $g_j$ 分配给预测打分 $\hat{r}_{ij}$ 最高的用户 $u_i$；而在我们的总福利最大化（TSM）框架下，一项工作被分配给 $Q_{ij} = 1$ 的工作者，如公式 (5-30) 所示。

对于每一个方法，我们采用五折交叉验证的方法对训练集和测试集进行分割并进行实验，同样采用转换率（Conversion Rate，CR）进行性能评价，即在所有的任务分配中正确的任务分配（与测试集中的实际任务分配一致）所占的百分比。

① http://www.zbj.com

表5.8展示了 TSM 和 CF 算法在不同分解因子个数 $K$（如公式 (5-6) 所示）下的推荐性能。

表 5.8　工作任务分配的转化率

| $K$ | 5 | 10 | 20 | 30 | 40 | 50 |
|---|---|---|---|---|---|---|
| CF(%) | 0.165 | 0.216 | 0.244 | 0.258 | 0.262 | 0.266 |
| TSM(%) | 0.384 | 0.421 | 0.453 | 0.486 | 0.507 | 0.512 |

结果显示，我们的总福利最大化框架 TSM 在工作任务分配问题上能够比协同过滤算法取得更好的结果，在全部的 $K$ 取值下性能提升均在双边 $t$ 检验中达到 0.01 的显著性水平。根据5.2.4小节的讨论，该性能提升主要是源于总福利最大化模型对每一项任务的工资的考虑，表明工资在工作者对工作任务进行选择的过程中起重要作用。另外，我们发现当 $K \geq 40$ 时算法的性能表现趋于稳定，这表明 40 个左右的分解因子足以用来刻画工作者在选定任务时所可能考虑的因素。

我们进一步在不同的分解因子个数 $K$ 的取值下，对 TSM 和 CF 算法给出的任务分配方案所对应的系统总福利进行评测。给定一个任意的任务分配方案 $Q = [Q_{ij}]_{m\times n}$，该方案所对应的系统总福利为：

$$TS_{Fr} = \sum_i \sum_j \left(h(\hat{r}_{ij}) + h(\hat{r}_{kj})\right) s_j Q_{ij} \tag{5-34}$$

我们在五折交叉验证所对应的五个测试集中分别计算总福利，平均每个测试集中包含 59,291 个任务分配。表5.9展示了五个测试集上的平均总福利，其中福利用人民币（￥）计量、‘$m$’表示“百万”，“真实任务分配”为测试集上真实的任务分配方案的总福利平均值。

表 5.9　在不同的分解因子个数 $K$ 下在线任务分配的总福利

<table>
<tr><td>K</td><td>5</td><td>10</td><td>20</td><td>30</td><td>真实任务分配</td></tr>
<tr><td>CF(￥)</td><td>1.562m</td><td>1.758m</td><td>1.824m</td><td>1.860m</td><td rowspan="2">2,593,618</td></tr>
<tr><td>TSM(￥)</td><td>3.235m</td><td>3.862m</td><td>4.270m</td><td>4.336m</td></tr>
</table>

由实验结果可见，我们的总福利最大化算法在不同的分解因子个数 $K$ 的取值下都取得比协同过滤更高的总福利值，总福利最大化框架给出的分配结果所得到的福利值甚至高于数据集中真实的分配结果。当 $K = 30$ 时，TSM 框架给出的任务分配结果可以得到￥73.13 每任务的福利值，而协同过滤算法和系统中真实的任务分配结果只能达到￥31.37 和￥43.74 每任务的福利值。

实际任务分配结果的福利值小于总福利最大化得到的福利值这一实验结果验证了市场中均衡失效现象的存在，而经济学家在反垄断和管制经济学的相关理论和实证研究中已经对这一现象进行了深入的探讨。在本工作所研究的在线众包应用中，这一现象主要源于雇佣双方对任务的信息不对等，因为工作者几乎不可能将系统中全部的上百万个任务全部浏览一遍，并经过仔细对比后在作出工作决定。而这进一步显示了推荐系统在网络服务分配中的重要作用，推荐系统了解系统中每个项目相关的全部信息（用户评论、工资情况，等等），从而可以在全局最优的视角上对系统进行效率优化，并将最恰当的物品推荐给最恰当的用户，从而避免或缓解信息过载（Information Overload）问题给用户带来的困扰。

综合模型在个性化推荐和总福利最大化两个任务上的效果，我们的总福利最大模型可以在提高市场效率的同时也提高推荐效果，使得模型在实际系统中比现有推荐算法具有更强的实用性和更重要的经济学意义。

## 5.4　本章小结

我们在本章对个性化推荐系统的经济学意义进行解释，为此我们首先介绍了互联网经济系统分析的基本概念和定义，包括效用、消费者福利、生产者福利，以及系统总福利。经济系统的总福利是衡量一个经济系统的市场效率的核心指标，因此在此基础上，我们提出了基于总福利最大化的个性化推荐框架，从而对个性化推荐对经济系统市场效率的影响进行研究。

为了对基于总福利最大化的个性化推荐模型进行实证研究，我们采用了电子商务、P2P 网络贷款和在线众包平台三个非常不同的网络应用场景进行实验，并在个性化推荐系统的排序性能这一传统指标，以及互联网经济系统的总福利这一新指标下分别对算法进行性能评测。结果显示，基于总福利最大化的个性化推荐框架能够在提高推荐系统排序性能的同时也提高经济系统的总福利，从而在提升用户体验的同时提高系统的市场效率。本章工作发表于 CCF-A 类会议 WWW 2016。

# 第 6 章　总结与展望

本文以个性化推荐的可解释性为核心，对数据的可解释性、模型的可解释性，以及推荐系统的经济学解释分别进行了研究。本章对全文研究内容进行总结，并对未来的研究工作进行展望。

## 6.1　研究工作总结

随着互联网应用的不断丰富和网络智能的不断发展，个性化推荐系统已经成为各种互联网应用和平台中必不可少的组成部分，例如电子商务网站中的商品推荐、社交网络中的好友推荐、在线门户网站中的新闻推荐，等等。个性化推荐技术是网络系统用于理解和分析用户的个性化需求并做出响应的基础性技术，在互联网的智能化过程中具有关键作用。

本文围绕个性化推荐技术的可解释性展开研究，分别从数据和模型两个基本层面上对可解释性推荐进行了分析，并进一步上升到整个网络应用经济系统的层面对个性化推荐的经济学意义进行阐述。

*在数据的可解释性方面*，本文对用户物品打分稀疏矩阵内在的社区结构进行了分析。用户物品打分的稀疏矩阵一直以来是个性化推荐问题最基本的输入数据形式之一，而长期以来以协同过滤为代表的矩阵预测算法对矩阵内在的群组结构少有关注。本工作以用户物品打分稀疏矩阵和用户物品二部图之间的等价关系为基础，对稀疏矩阵的内在群组结构进行分析。本文提出了矩阵的（近似）双边块对角结构，并证明了该结构与基于二部图的社区发现之间的内在联系；在此基础上，本文进一步提出了基于块对角子矩阵的协同过滤算法；本文给出和证明了双边块对角型矩阵在精确矩阵分解问题上可拆分性，分析了常见的基于机器学习的实用矩阵分解算法的可拆分性，并据此提出了局部化矩阵分解算法。局部化矩阵分解给出了一种通用的兼容常见矩阵分解算法的并行化框架，使得协同过滤算法一方面可以充分利用矩阵内在的群组结构和同现性质提高预测精度，另一方面可以通过将大规模矩阵转化为数个小规模高密度矩阵来提高矩阵分解算法对数据规模的可扩展性，最后还可以通过简单的多线程并行化提高运行效率。

*在模型的可解释性方面*，本文对隐变量模型的未知维度和学习过程进行解释。基于矩阵分解的隐变量模型在打分预测等推荐系统基本任务上取得了理想效果，因而在实际系统中受到广泛应用。然而正是由于变量本质上的未知性，使得模型的学习过程和最终的推荐结果难以给出直观有效的解释，降低了推荐系统对用户

的可信度。本文从文本评论出发对推荐模型给出直观的解释，利用短语级情感分析技术从大规模评论文本中抽取（属性词，情感词，情感极性）三元组，并利用抽取的具体产品属性构建显式变量分解模型，从而使模型的学习具有直观意义，并通过模型的优化结果直接给出个性化的属性级推荐理由。在静态模型之外，本文进一步对属性词的动态时间周期性进行分析，通过预测属性词流行度构建动态时序推荐模型。本文提出了基于傅里叶辅助项的移动平均自回归模型，从而解决长周期下时间序列分析参数估计可行性的问题，并进一步提出了基于条件机会模型的动态个性化推荐算法。实验结果显示，本文模型在线下真实数据集评测和线上真实用户评测两个方面均取得较好的效果。

*在推荐的经济学解释方面*，本文对个性化推荐系统在互联网应用环境下的经济学意义进行解释。随着越来越多的人类活动不断地从线下走到线上，互联网已经不再仅仅是一个信息传输的平台，而更是一个完整的线上经济活动系统，电子商务、社交网站、在线众包、网络打车、在线传媒等网络应用囊括了人们日常生活的方方面面。典型的网络应用包括生产者和消费者两个方面的信息交互或在线交易，而传统的个性化推荐系统往往将两者割裂开来，只考虑其中一方的体验或收益。例如典型的推荐系统致力于对用户（消费者）的个性化偏好进行尽可能精确的建模并为其推荐最有可能感兴趣的物品，而少有对生产者的关注。本工作将互联网应用看做一个完整的包含供需双方的网络经济系统，提出了基于总福利最大化的个性化推荐框架，通过个性化效用函数和成本函数对消费者和生产者的收益同时纳入考虑。在该框架的基础上，本工作分别针对电子商务、P2P 贷款、在线众包三种不同的网络应用进行模型实例化和真实数据实验验证。实验结果表明基于总福利最大化的个性化推荐模型可以在提高用户（消费者）对推荐系统的体验的同时也提高经济系统的总福利，从而提高系统的经济效益。

对于实际的个性化推荐系统而言，本文在数据、模型和经济意义三个方面上的工作具有依次递进的内在联系。

在系统最底层，用户物品评分数据是推荐算法最主要的数据输入形式之一。对稀疏矩阵数据的处理使得我们从原本杂乱无章的用户物品评分中抽取可解释的用户物品社区结构，从而减小数据稀疏性对预测效果的影响；同时，通过将具有相似兴趣偏好的用户及其打分聚在一起，使得后续算法能够更好地对用户偏好进行建模，从而提高个性化推荐效果。

在中间的个性化推荐模型层面，通过引入从文本中抽取的显式属性词信息构建用户和物品在属性词上的偏好矩阵，从而使得原本的隐变量分解模型变为显式变量分解模型。与原始的用户物品评分矩阵一起，该方法使得分解模型在具备变

量可解释性的同时，也能给出直观可理解的推荐理由，从而提高推荐系统对用户的友好度。

在实际系统的最上层，推荐算法本质上是以推荐的方式影响着电子商务网站等经济系统中资源（物品）与消费者（用户）之间的分配方式，从而也影响着经济系统最终所实现的经济效益，这也是个性化推荐在实际系统中的经济意义所在。通过对生产者、消费者和系统总的福利进行建模，基于总福利最大化的个性化推荐在提高用户体验（推荐效果）的同时，也提高整个经济系统最终所实现的总福利，从而在最终提高整个系统的经济效益。

## 6.2　未来工作展望

本工作从数据、模型和经济学意义三个方面对个性化推荐的可解释性进行研究，作者认为，未来需要进一步研究的内容包括：

1. *考虑用户物品属性和打分数值的二部图社区发现及其预测性能*。本工作利用稀疏矩阵的双边块对角结构抽取高密度子矩阵，减小了数据稀疏性对打分预测效果的影响，并通过将具有相似购买行为的相似用户和物品聚在一起而提升矩阵分解的预测精度。为了构建矩阵的双边块对角结构，本工作构建等边权的用户物品打分二部图并基于图分割算法进行社区挖掘。然而基于二部图的社区发现算法多种多样且受到了学术界的广泛深入研究，未来工作将着力于构建包含用户物品打分具体数值甚至用户物品节点异质属性的二部图，并研究不同的二部图社区发现算法所给出的矩阵双边块对角结构的异同，及其在矩阵分解打分预测上的效果。
2. *其它个性化推荐模型的可解释性分析*。本工作以用户文本评论的充分利用为基础，对基于矩阵分解的隐变量模型这一最为广泛使用的个性化推荐模型进行了可解释性研究。而推荐系统的研究经历多年的发展，已经衍生出各种不同的个性化推荐模型以适应不同的具体应用场景，包括基于近邻的推荐、基于话题模型的推荐、基于频繁项数据挖掘的推荐，等等。在实际系统应用中，不同的推荐模型相辅相成、互为补充，共同为用户提供高质量的推荐，如何对其它常用的推荐模型进行可解释性研究，并对不同模型给出的可解释性推荐结果进行整合，将是可解释性推荐在实际系统中提高可用性的重要问题。
3. *消费者个性化效用函数和生产者成本函数的细化*。本工作从消费者和生产者共同的视角将网络应用看做一个完整的互联网经济系统，给出消费者效用和生产者成本的计算方法，并在此基础上提出基于总福利最大化的网络服务分配和个性化推荐框架，在提高推荐系统体验的同时提升系统总福利。为了便

于理解和简化计算，本工作使用了简单的对数效用函数来刻画物品对消费者的效用，并采用正比关系来刻画生产者的成本。然而在实际中消费者可能具有更复杂的效用函数形式，并且不同的消费者其具体的效用函数甚至是不同的。除了对数效用函数之外，经济学家提出了多项式展开效用函数、常数替代弹性效用函数（Constant Elasticity of Substitution，CES）等不同的效用函数形式以在不同情况下对消费者偏好进行恰当的建模。未来工作将采用不同的效用和成本函数对消费者偏好和生产者收益进行更精确的建模，甚至在无数量约束指导项的条件下对消费者购买量进行预测，从而进一步提高预测精度、推荐效果，以及系统总福利。

# 参考文献